\documentclass[paperpaper,superscriptaddress,english,floatfix, showpacs, amsfonts, amssymb]{revtex4}
\usepackage{epsfig,graphicx,psfrag,amsmath,amssymb,float}
\usepackage[T1]{fontenc}
\usepackage[latin9]{inputenc}
\usepackage{amsmath}
\usepackage{amssymb}
\usepackage{bbold}
\usepackage{amscd}
\usepackage{bm}
\usepackage{psfrag}
\usepackage{epsfig}
\usepackage{graphicx}
\usepackage{dcolumn}
\usepackage{bm}
\usepackage{subfigure}
\usepackage{babel}

\begin{document}
\title{Absence of ballistic charge transport in the half-filled 1D Hubbard model}

\author{J. M. P. Carmelo}
\affiliation{Department of Physics, University of Minho, Campus Gualtar, P-4710-057 Braga, Portugal}
\affiliation{Center of Physics of University of Minho and University of Porto, P-4169-007 Oporto, Portugal}
\affiliation{Beijing Computational Science Research Center, Beijing 100193, China}
\author{S. Nemati}
\affiliation{Beijing Computational Science Research Center, Beijing 100193, China}
\affiliation{Center of Physics of University of Minho and University of Porto, P-4169-007 Oporto, Portugal}
\author{T. Prosen}
\affiliation{Department of Physics, FMF, University of Ljubljana, Jadranska 19, 1000 Ljubljana, Slovenia}

\date{14 November 2017; Published online in Nuclear Physics B: 28 March 2018}

\begin{abstract}
Whether in the thermodynamic limit of lattice length $L\rightarrow\infty$, hole concentration 
$m_{\eta}^z = -2S_{\eta}^z/L = 1- n_e\rightarrow 0$, nonzero temperature $T>0$,
and $U/t > 0$ the charge stiffness of the 1D Hubbard model with first neighbor transfer integral $t$ 
and on-site repulsion $U$ is finite or vanishes and thus whether there is or there is no ballistic charge
transport, respectively, remains an unsolved and controversial issue, as different approaches 
yield contradictory results. (Here $S_{\eta}^z = -(L- N_e)/2$ is the $\eta$-spin projection and $n_e = N_e/L$ 
the electronic density.) In this paper we provide an upper bound on the charge stiffness 
and show that (similarly as at zero temperature), for $T >0$ and $U/t>0$ it vanishes for $m^z_{\eta}\rightarrow 0$
within the canonical ensemble in the  thermodynamic limit $L\to\infty$. Moreover, we show that 
at high temperature $T\rightarrow\infty$ the charge stiffness vanishes as well within the grand-canonical ensemble 
for $L\to\infty$ and chemical potential $\mu\rightarrow \mu_{u}$ where $(\mu -\mu_{u})\geq 0$
and $2\mu_{u}$ is the Mott-Hubbard gap. The lack of charge ballistic transport indicates that charge 
transport at finite temperatures is dominated by a diffusive contribution. Our scheme uses a suitable {\it exact} representation 
of the electrons in terms of rotated electrons for which the numbers of singly occupied and doubly
occupied lattice sites are good quantum numbers for $U/t>0$. In contrast to often less controllable 
numerical studies, the use of such a representation reveals the carriers that couple to the charge probes 
and provides useful physical information on the microscopic processes behind the exotic charge 
transport properties of the 1D electronic correlated system under study. 
\end{abstract}


\maketitle

\section{Introduction}
\label{Introduction}

Likely, the most widely studied correlated electronic model on a lattice in one (spatial) dimension (1D)
is the Hubbard model with first neighbor transfer integral $t$ and on-site repulsion $U$. In spite 
of being solvable by the Bethe ansatz (BA) \cite{Lieb,Lieb-03,Ovchi,Takahashi,Call-74,Woy,Woy-82,Martins,Deguchi-00,1D-05}, 
in the case of electronic density $n_e=N_e/L=1$ its unusual charge transport properties remain poorly 
understood at finite temperatures $T>0$ \cite{ZP-96,Kawa-98,PDSC-00,PSZL-04,ANI-05,HPZ-11,Ilievski-17A,Ilievski-17}. 
This includes, specifically, some of the behaviors of the real part of charge conductivity at finite
temperature $T$ whose general form reads,
\begin{equation}
\sigma (\omega,T) = 2\pi\,D (T)\,\delta (\omega) + \sigma_{reg} (\omega,T) \, .
\label{sigma}
\end{equation}
Even for the $T=0$ Mott-Hubbard insulating quantum phase, the related charge dynamic structure factor is
a complex problem that is only partially understood \cite{PPWSC}.

The charge stiffness or Drude weight $D(T)$ in Eq. (\ref{sigma}) characterizes the response to a static field
and $\sigma_{reg} (\omega,T)$ describes the absorption of light of frequency $\omega$. 
For $T>0$ these quantities can be written as,
\begin{equation}
D (T) = {1\over 2T L}\sum_{\nu} p_{\nu}
\sum_{\substack{\nu'\\ (\epsilon_{\nu}=\epsilon_{\nu'})}} 
\vert\langle\nu,u\vert\hat{J}\vert\nu',u\rangle\vert^2 \, ,
\label{DT-gen}
\end{equation}
and
\begin{eqnarray}
\sigma_{reg} (\omega,T) & = & {\pi\over L}{1-e^{-{\omega\over T}}\over\omega}\sum_{\nu} p_{\nu} 
\sum_{\substack{\nu'\\ (\epsilon_{\nu}\neq\epsilon_{\nu'})}} 
\vert\langle \nu,u\vert\hat{J}\vert \nu',u\rangle\vert^2
\delta (\omega - \epsilon_{\nu'} + \epsilon_{\nu}) \, , 
\label{sigma-reg}
\end{eqnarray}
respectively. In these equations and elsewhere in this paper units of Boltzmann constant $k_B$, Planck constant
$\hbar$, and lattice spacing $a$ one are generally used. Moreover, $L\to\infty$ denotes the system length in the 
thermodynamic limit (TL), which within the units of lattice constant one equals the (even) number of  lattice sites $N_a$, 
$\vert\nu,u\rangle$ are energy and momentum eigenstates, $\nu$ stands for all quantum numbers other than the parameter, 
\begin{equation}
u=\frac{U}{4t} \, ,
\end{equation}
needed to uniquely specify each such a state, the sum runs over states with the same energy eigenvalue, 
$\epsilon_{\nu}=\epsilon_{\nu'}$, $p_{\nu}=e^{-\epsilon_{\nu}/T}/Z$ is the usual Boltzmann weight, 
$Z = \sum_{\nu} e^{-\epsilon_{\nu}/T}$, and $\hat{J}$ is the charge current operator. (Its specific expression for 
the present model is given below in Section \ref{modelDT}.)

The studies of this paper rely in part on the BA solution of the 1D Hubbard model. It was solved first
by the so-called coordinate BA \cite{Lieb,Lieb-03}, which provided 
the ground state energy and revealed that the model undergoes a Mott metal-insulator transition at electronic density $n_e =1$ 
whose corresponding critical onsite interaction is $U = 0$. Which are the effects of a finite
temperature on such a transition is one of the issues studied in this paper.

Following the coordinate BA solution, the ground state properties \cite{Takahashi-69,Shiba-72,Penc-91}
and the excitation spectrum \cite{Ovchi-70,Coll-74,Woy-82,Woy,Woynarovich-83,Woynarovich-83-B,Klumper-90}
were studied by several authors. The 1D Hubbard model thermodynamic Bethe ansatz (TBA) 
and corresponding ideal strings have been 
proposed in Ref. \cite{Takahashi}. This has allowed the study of the thermodynamic 
properties of the model \cite{Kawakami-89,Usuki-90}. The energy spectra of its elementary excitations 
can be obtained from the TBA equations in the zero temperature limit \cite{Deguchi-00}. 

An important property of the 1D Hubbard model is that its spectrum becomes conformal
invariant in the low-energy limit. The corresponding finite-size corrections 
were obtained in Refs. \cite{Woynarovich-87,Woynarovich-89}. The relation between the finite-size spectrum and the 
asymptotic behavior of correlation functions was used to calculate the critical 
exponents of the general two-point correlation functions \cite{Frahm-90,Frahm-91}.
The corresponding conformal dimensions have been expressed in terms of dressed
phase shifts associated with a preliminary pseudoparticle representation 
\cite{Carmelo-92,Carmelo-92-C,Carmelo-93,Carmelo-93-B,Carmelo-94,Carmelo-94-B,Carmelo-97}.

The conformal approach is not applicable to the zero-temperature model Mott insulating phase at half filling.
In the small-$U$ and scaling limits, dynamical correlation functions at low energies \cite{Controzzi-02,Essler-02,Essler-03,Jeckelmann-00}
can though be computed relying on the methods of integrable quantum field theory \cite{Melzer-95,Woynarovich-97,Woynarovich-99}. 
The wave functions of the energy eigenstates can be extracted from the coordinate BA solution. 
An explicit representation for the wave functions was given in Ref. \cite{Woy-82}. 

In the $u=U/4t\rightarrow\infty$ limit the dynamical correlation functions can be computed at zero temperature for 
all energy scales relying on the simplified form that the BA equations acquire. This was achieved by a combination 
of analytical and numerical techniques for the whole range of electronic densities
\cite{Ogata,Ogata-91,Parola-90,Parola-92,Weng,Weng-94,Karlo-95,Karlo-96,Karlo-97,Wang-95,Gallagher-97,Gebhard-97}.
In the case of the one-electron spectral function studies of Refs. \cite{Karlo-95,Karlo-96,Karlo-97},
the method relies on the spinless-fermion phase shifts imposed by $XXX$ chain physical spins $1/2$. Such 
fractionalized particles naturally arise from the zero spin density and $u\rightarrow\infty$ electron wave-function factorization 
\cite{Woy,Woy-82,Ogata}. A related pseudofermion dynamical theory relying on a representation of the model BA solution in terms of the 
pseudofermions generated by a unitary transformation from the corresponding pseudoparticles considered
in Ref. \cite{Carmelo-04} was introduced in Ref. \cite{Carmelo-05}. It is an extension of the $u\rightarrow\infty$ method
of Refs. \cite{Karlo-95,Karlo-96,Karlo-97} to the whole $u>0$ range of the 1D Hubbard model. 
The use of the mobile quantum impurity model \cite{Glazman-09,Glazman-12}, which has been developed to also
tackle the high-energy physics of both integrable and non-integrable 1D correlated quantum problems, 
leads in the case of the 1D Hubbard model to the same results as the pseudofermion dynamical theory \cite{Essler-10,Seabra-14}.
Further general information on the 1D Hubbard model is given in Ref. \cite{1D-05}.

Provided that the energy eigenstates $\vert\nu,u\rangle$ are as well momentum eigenstates,
it is well known \cite{ZNP-97,PDSC-00} that for $u>0$, $T>0$, and in the TL 
the charge stiffness expression, Eq. (\ref{DT-gen}), further simplifies to,
\begin{equation}
D (T) = {1\over 2T L}\sum_{\nu} p_{\nu}
\vert\langle\nu,u\vert\hat{J}\vert\nu,u\rangle\vert^2
\hspace{0.20cm}{\rm for}\hspace{0.20cm}u>0 \, .
\label{DT-thermo}
\end{equation}
Within that limit, this expression is not valid in the $T=0$ regime though. The $T=0$ charge stiffness
is actually known \cite{Shastry-90,Carmelo-92-C}, reading $D(0) = (2t/\pi)\,\delta_{U,0}$ at 
hole concentration $m_{\eta}^z = -2S_{\eta}^z/L = 1- n_e=0$ (half filling) where the $\eta$-spin
$z$ component $S_{\eta}^z = -(L- N_e)/2$ is the eigenvalue of the diagonal generator of the global 
$\eta$-spin $SU(2)$ symmetry. Hence at $T>0$ it is 
finite at $U=0$ and vanishes for the whole $u>0$ range.	

A finite $D(T)$ for $T>0$ value would imply the occurrence of ballistic charge transport.
At $T>0$ the model can behave as an ideal conductor with ballistic charge transport
and thus $D(T)>0$ or a system without such a ballistic transport, so that $D(T)=0$.
In the latter case there are two scenarios, the system behaving as a normal resistor 
if $D(T)=0$ and the diffusive conductivity contribution $\sigma_0 = \lim_{\omega\rightarrow 0}\sigma_{reg} (\omega,T)$ 
is finite or as an ideal insulator with $D(T)=\sigma_0 = 0$ \cite{ZP-96,ZNP-97,CZP-95}.

On the one hand, a $D(T)$ inequality, which is derived from 
the more general Mazur's inequality \cite{Mazur,Suzuki}, provides for hole concentrations $m_{\eta}^z\neq 0$
a finite lower bound for its value \cite{ZNP-97,SPA-11}. This reveals that $D(T)>0$ for 
$m_{\eta}^z\neq 0$ and finite temperature \cite{PDSC-00,PSZL-04}. On the other hand, at $m_{\eta}^z=0$
that lower bound vanishes, so that the inequality is inconclusive. Whether in the TL and for $u>0$ the charge stiffness $D(T)$ 
vanishes or is finite for $T>0$ and $m_{\eta}^z = 0$ remains actually an open and controversial issue, as different approaches 
yield contradictory results \cite{ZP-96,Kawa-98,PDSC-00,PSZL-04,ANI-05,HPZ-11,Ilievski-17A,Ilievski-17}. 

The results of this paper provide strong evidence that the predictions of Ref. \cite{Kawa-98} for the 
1D Hubbard model charge stiffness for $T>0$ and $m_{\eta}^z=0$ are not correct. This is consistent with the 
numerical results of Ref. \cite{PSZL-04} and the large-$u$ studies of Ref. \cite{PDSC-00}. 
The latter results are reached by two completely different methods: an {\it exact} method that 
does not rely on the BA and a TBA calculation \cite{Takahashi}, respectively. These studies
reveal that the finite charge stiffness expression found in Ref. \cite{Kawa-98} for $m_{\eta}^z = 0$ and $T>0$
{\it cannot} be correct for large $u>0$. The results of Refs. \cite{PSZL-04,PDSC-00}
agree with some preliminary conjectures by Zotos and Prelov\v{s}ek according to which 
$\lim_{u\rightarrow\infty}D(T)$ should be zero for the 1D Hubbard model at $m_{\eta}^z = 0$ and $T>0$.

Recently, a general formalism of hydrodynamics for the 1D Hubbard model and other integrable
models was introduced in Refs.~\cite{Ilievski-17A,Ilievski-17}. By linearizing hydrodynamic equations, the
closed-form expressions for the stiffnesses that were conjectured to be valid on the hydrodynamic scale have been accessed. 
The stiffness is then calculated from the stationary currents generated in an inhomogeneous quench 
from bipartitioned initial states \cite{Ilievski-17A}. Within such an hydrodynamic ansatz for the stiffnesses, the studies of
Refs.~\cite{Ilievski-17A,Ilievski-17} clearly established vanishing at finite temperature of charge or spin Drude
weights when the corresponding chemical potentials vanish, irrespective of the interaction strength.
In our work we, however, take a different perspective. We start from the standard linear-response expressions 
for the charge and spin Drude weights and reach conclusions that are consistent with the results of Ref.~\cite{Ilievski-17}.
Although there is no reasonable doubt that the hydrodynamics ansatz used in Refs. \cite{Ilievski-17A,Ilievski-17} is correct, 
it has, nevertheless, not yet been rigorously justified. Hence we believe that adding our independent 
and complementary result is a valuable contribution to the solution of this important problem.
Actually, both methods rely on the standard assumptions behind the TBA.

Our previous results reported in Refs. \cite{CPC} and \cite{CP}, which have been obtained by 
the method used in this paper, provide strong evidence that in the case of the spin stiffness of the 
spin-$1/2$ $XXX$ chain the approach of Ref. \cite{Kawa-98} used in the investigations of
Ref. \cite{Zotos-99} leads to correct results. Specifically, that such a stiffness vanishes in the 
limit of zero spin density. (The apparent inconsistency that the use of the approach of Ref. \cite{Kawa-98}
leads to misleading results for the 1D Hubbard model and to correct results for that spin chain
is an issue discussed below in Section \ref{concluding}.)

Our method to compute suitable upper bounds for the charge stiffness
relies in part on the properties of the charge current operator matrix elements between energy
and momentum eigenstates that follow from the $\eta$-spin $SU(2)$ symmetry operator algebra. This
is similar to the method used in Refs. \cite{CPC} and \cite{CP} for the spin stiffness of the
spin-$1/2$ $XXX$ chain in what its relation to its spin $SU(2)$ symmetry operator algebra is
concerned. The method combines the TBA \cite{Takahashi} with stiffness expressions in terms of current operator
expectation values. It accounts though for the effects of complex-rapidity string deviations \cite{Deguchi-00}
and {\it does not} access the charge stiffness through the second derivative of the energy eigenvalues of the TBA  
relative to a uniform vector potential \cite{Kawa-98}. 

In the case of energy eigenstates of spin $S$ of the spin-$1/2$ $XXX$ chain, there
are $L-2S$ spins $1/2$ that are paired within singlet configurations and $2S$
spins $1/2$ that remain unpaired and contribute to the multiplet configurations.
The spin degrees of freedom couple to a vector potential through such $2S$ 
unpaired spins $1/2$, which are those that contribute to the spin currents.

Within the rotated-electron related representation of the 1D Hubbard model used in 
our studies, there emerge from the rotated-electrons $\eta$-spin degrees of freedom basic fractionalized particles
of $\eta$-spin $1/2$ that are associated with $\eta$-spin $SU(2)$ symmetry of the model. Again,
in the case of energy eigenstates of $\eta$-spin $S_{\eta}$ there is a number $2S_{\eta}$
of $\eta$-spin $1/2$ fractionalized particles that couple to a vector potential, which are those 
that participate in $\eta$-spin multiplet configurations and contribute to the charge currents. 

A trivial result is that at $U=0$ the global symmetry of the Hubbard model on a bipartite lattice is 
$O(4)=SO(4)\otimes Z_2$. This thus applies to the 1D Hubbard model. Here the factor $Z_2$ refers 
to the Shiba particle-hole transformation on a single spin under which the Hamiltonian is not invariant for 
$U\neq 0$ and $SO(4)=[SU(2)\otimes SU(2)]/Z_2$ contains the two $SU(2)$ symmetries. 
An exact result of Heilmann and Lieb is that in addition to the spin $SU(2)$ symmetry, also for $u>0$ the model 
has a second global $SU(2)$  symmetry \cite{HL}. It is generally called $\eta$-spin symmetry \cite{Yang,Yang-90}. 
Yang and Zhang considered the most natural possibility that the $SO(4)$ symmetry inherited from 
the $U=0$ Hamiltonian $O(4)=SO(4)\otimes Z_2$ symmetry is the model's global symmetry for $U>0$ \cite{Yang-90}.
The energy and momentum eigenstates are either lowest weight states (LWSs) or highest weight states
with respect to the two $SU(2)$ symmetry algebras \cite{HL,Yang,Yang-90,Lieb-89}. The non-LWSs  
can be generated from the LWSs explicitly accounted for by the BA solution, which confirmed the completeness 
of the quantum problem \cite{Completeness,Complete2,Complete3}. 

At half-filling and zero spin density the 1D Hubbard model TBA dressed phase shifts and the corresponding
$S$-matrices have been associated with fractionalized particles called holon, antiholon, and spinon. 
The holon and antiholon have been inherently constructed to have zero spin and charge $+e$ and $-e$, respectively. 
The spinon has been inherently constructed to have no charge and to have spin $1/2$ \cite{Essler-94-A,Essler-94}. The model 
$SO(4)$ symmetry group state representations were identified with occupancy configurations
of such fractionalized particles. 

The solution of the model by the quantum inverse scattering method has provided further information
on its symmetries. The first steps to obtain such a solution 
were made in Refs. \cite{Shastry-86,Shastry-86A,CM}. The model Hamiltonian was mapped under a Jordan-Wigner 
transformation into a spin Hamiltoninan. It commutes with the transfer matrix of a related covering vertex model 
\cite{Shastry-86}. The $R$-matrix of the spin model was also derived \cite{Shastry-86A,CM}. Alternative derivations 
were carried out by several authors \cite{Olmedilla-87,Olmedilla-88,Wadati-87}. 
The $R$-matrix was later shown to satisfy the Yang-Baxter equation \cite{Shiroishi-95}. 
An algebraic BA having as starting point the results of Refs. \cite{Shastry-86,Shastry-86A,CM} was afterwards constructed 
in Refs. \cite{Martins-97,Martins-98}. The expressions for the eigenvalues of the transfer 
matrix of the two-dimensional statistical covering model were obtained. That problem was also addressed in Ref. \cite{Yue-97}. 

The algebraic BA introduced in Refs. \cite{Martins-97,Martins-98} allowed the quantum transfer matrix approach to 
the thermodynamics of the 1D Hubbard model \cite{Juttner-98}. Within it, the thermodynamic quantities and
correlation lengths can be calculated numerically for finite temperatures \cite{Tsunetsugu-91,Umeno-03}.
The 1D Hubbard model Hamiltonian was found in the TL to be invariant under the direct sum of two $Y(sl(2))$ 
Yangians \cite{Uglov-94}. The relation of these Yangians to the above $R$-matrix and the implications of one 
of these Yangians for the structure of the bare excitations was later clarified \cite{Mura-97,Mura-98}.
More recently, it was demonstrated that the Yangian symmetries of the $R$-matrix specialize to the Yangian symmetry 
of the model and that its Hamiltonian has an algebraic interpretation as the so-called secret symmetry \cite{Leeuw-16}. 

It was found in Ref. \cite{bipartite} that for $u> 0$ the 1D Hubbard model global symmetry is 
actually larger than $SO(4)$ and given by $[SO(4)\otimes U(1)]/Z_2$, and thus equivalently to 
$[SU(2)\otimes SU(2)\otimes U(1)]/Z_2^2$. (This applies as well to the model
on any bipartite lattice.) Consistently with the model's extended global symmetry, the quantum 
inverse scattering method spin and charge monodromy matrices were found to have different ABCD and ABCDF 
forms, respectively. Those are actually associated with the spin $SU(2)$ and charge $U(2)=SU(2)\otimes U(1)$ symmetries, respectively \cite{Martins-98}. 
The latter matrix is larger than the former and involves more fields \cite{Martins-98}. If the global symmetry was 
only $SO(4)=[SU(2)\otimes SU(2)]/Z_2$, the charge and spin monodromy matrices would have the same traditional 
ABCD form, which is that of the spin-$1/2$ $XXX$ chain \cite{Faddeev-81}.

The {\it exact} rotated-electron representation used in our studies is that suitable for the
further understanding of this basic similarity between the spin $SU (2)$ symmetry degrees of freedom
of the spin-$1/2$ $XXX$ chain type of configurations that contribute to spin transport and the 1D Hubbard model
$\eta$-spin $SU(2)$ symmetry degrees of freedom type of configurations that contribute to charge transport. 
The rotated electrons are inherently constructed to their numbers of singly occupied 
and doubly occupied lattice sites being good quantum numbers for $u>0$.
As further discussed below in Section \ref{Introduction2},
the form of the 1D Hubbard model energy and momentum eigenstates wave function for $u\rightarrow\infty$ derived
in Ref. \cite{Woy-82} reveals that in that limit such a model corresponds 
to a spin-$1/2$ $XXX$ chain, an $\eta$-spin-$1/2$ $XXX$ chain, and a quantum problem with simple lattice $U(1)$ 
symmetry, respectively. In terms of the rotated electrons, whose relation to the electrons has been uniquely defined
in Ref. \cite{Carmelo-17}, the energy and momentum eigenstates wave function has that form for the whole $u>0$ range.

The degrees of freedom of the rotated electrons naturally separate into two fractionalized particles with
spin $1/2$ and $\eta$-spin $1/2$, respectively, plus one basic fractionalized particle without internal 
degrees of freedom \cite{Carmelo-17,Carmelo-16}. (The $\eta$-spin projections $+1/2$ and $-1/2$ refer to the 
$\eta$-spin degrees of freedom of the rotated-electron unoccupied and doubly-occupied sites, respectively.) 
The occupancy configurations of these three basic fractionalized particles 
generate exact state representations of the group associated with the spin $SU(2)$ symmetry,
$\eta$-spin $SU(2)$ symmetry, and $c$ lattice $U(1)$ symmetry, respectively, in the global 
$[SU(2)\otimes SU(2)\otimes U(1)]/Z_2^2$ symmetry of the model \cite{bipartite}. 

In the case of the spin-$1/2$ $XXX$ chain, the translational degrees of freedom of the $2S$
unpaired spins $1/2$ that contribute to the spin currents are described by an average number $2S$
of holes in each TBA $n$-band with finite occupancy. Here $n=1,...,\infty$ is the number
of singlet pairs bound within each of the $n$-band pseudoparticles considered in
Ref. \cite{CP} that populate such a band. The $n$-band pseudoparticles occupancies generate the singlet configurations of
the spin $SU(2)$ symmetry group state representations.

Also in the case of the 1D Hubbard model charge transport, the translational degrees of freedom of the $2S_{\eta}$
unpaired $\eta$-spin $1/2$ fractionalized particles that contribute to the charge currents are found in this paper to
be described by an average number $2S_{\eta}$ of holes in each TBA $\eta n$-band with finite occupancy.
For that model, the corresponding $\eta n$ pseudoparticles occupancies generate the $\eta$-spin singlet 
configurations of the $\eta$-spin $SU(2)$ symmetry group state representations. 
The difference relative to the spin-$1/2$ $XXX$ chain refers to contributions from the holes in the charge band of the above mentioned basic fractionalized particles without internal degrees of freedom whose occupancy configurations generate state representations of the
group associated with the $c$ lattice $U(1)$ symmetry. Indeed, an average number $2S_{\eta }$ of such holes holes
also contributes to the translational degrees of freedom of the $2S_{\eta }$
unpaired $\eta $-spin $1/2$ fractionalized particles that couple to the charge probes.
This is related to the above mentioned $U(2)=SU(2)\otimes U(1)$ symmetry in the model's
$[SU(2)\otimes SU(2)\otimes U(1)]/Z_2^2$ global symmetry referring to the charge degrees of freedom.
(The remaining $SU(2)$ symmetry refers to the spin degrees of freedom.)
Indeed, that charge $U(2)=SU(2)\otimes U(1)$ 
symmetry includes the $\eta$-spin $SU(2)$ symmetry and the $c$ lattice $U(1)$ symmetry 
beyond $SO(4)=[SU(2)\otimes SU(2)]/Z_2$.

The use of the above mentioned holon and spinon representations  \cite{ANI-05,Essler-94-A,Essler-94} 
provides a suitable description of the model both at low excitation
energy relative to a ground state and more generally in subspaces spanned by energy
and momentum eigenstates described by a vanishing density of both
TBA complex rapidities and $\eta$-spin strings of length one \cite{Takahashi}. In the case of the 1D Hubbard model, such holons 
and spinons are different from the three fractionalized particles that naturally emerge from the exact 
rotated-electron degrees of freedom separation. The latter have operators that have simple expressions
in terms of rotated-electron operators and are defined for the 1D Hubbard model in its full Hilbert space \cite{Carmelo-17}. 

The charge stiffness problem under study in this paper involves summations that run over all energy and momentum eigenstates. This is why
the holon and/or spinon (and anti-spinon) representations are not suitable to study it.
For instance, the phenomenological method in terms of a spinon and anti-spinon particle basis 
used in Ref. \cite{ANI-05} leads to a misleading large spin stiffness for the spin-$1/2$ $XXX$ chain in
the limit of zero spin density. The validity of that result is excluded by the careful investigations of Ref. \cite{SPA-11}, 
which indicate that transport at finite temperatures is dominated by a diffusive contribution, the spin stiffness 
being very small or zero. They are also excluded by the studies of Refs. \cite{CPC,CP} and the TBA results of Ref. \cite{Zotos-99},
which find a vanishing spin stiffness within the zero spin density limit in the TL.

We emphasize that the electrons and the rotated electrons are for $u>0$ related by a mere
unitary transformation under which the electronic charge and spin degrees of freedom
remain invariant. Hence a rotated electron carries the same charge and has the
same spin $1/2$ as an electron. Indeed, such a unitary transformation only changes 
the lattice occupancies and corresponding spatial distributions of the 
charges and spins $1/2$. The relation of the rotated electrons to the rotated spins $1/2$, 
rotated $\eta$-spins $1/2$, and $c$ pseudoparticles is direct. It is uniquely defined for the 
full Hilbert space spanned by a complete set of $4^L$ energy and momentum eigenstates \cite{Carmelo-17,Carmelo-16}. 
The corresponding representation of the 1D Hubbard model in terms of such fractionalized particles is thus faithful 
in that space. 

The holons and spinons are related to such fractionalized particles for the
1D Hubbard model in some reduced subspaces mentioned above for which they correspond
as well to a faithful representation. However, the representation in terms of holons and spinons
is only defined for the model in such subspaces. This is why in our studies 
we rather use the representation in terms of the fractionalized particles that
naturally emerge from the separation of the rotated-electrons degrees of freedom.

In the $u\rightarrow\infty$ limit the rotated electrons become electrons and
the $c$ pseudoparticles and rotated spins $1/2$ of the representation used
in the studies of this paper become 
the spinless fermions and $XXX$ chain spins $1/2$, respectively,
of Refs. \cite{Ogata,Ogata-91,Karlo-95,Karlo-96,Karlo-97}. As mentioned above,
such fractionalized particles naturally emerge from the $u\rightarrow\infty$ 
electron wave-function factorization \cite{Woy,Woy-82}. That
factorization includes a third factor \cite{Woy-82} associated with the $\eta$-spin $SU(2)$
symmetry. It corresponds to the $u\rightarrow\infty$ limit of the
rotated $\eta$-spins $1/2$ of the $u>0$ representation used in this paper.

In summary, there are two main reasons why we use in our study the representation
of the rotated-electron related three fractionalized particles. Given their
simple and direct relation to the rotated-electrons charge and spin degrees
of freedom, it allows a more clear physical description of the microscopic processes that
control the charge properties under study. This is consistent with each of the
set of $4^L$ energy and momentum eigenstates that span the model
Hilbert space being generated from the electron and rotated-electron vacuum
by occupancy configurations of the three types of fractionalized particles
under consideration that are much simpler than those in terms of electrons. A second
reason is that, in contrast to the usual holon and spinon representation,
that representation is defined for the model in its full Hilbert space.
The holon and spinon representation applies for instance to low-energy problems
whereas here we consider all ranges of temperatures.

Our study refers to zero spin projection, $S_s^z=0$. It addresses the problem of the charge stiffness 
of the 1D Hubbard model in the TL within the canonical ensemble 
at hole concentration $m_{\eta}^z = 0$ and for $m_{\eta}^z\rightarrow 0$ at temperatures $T>0$. 
Within that ensemble for $T>0$ we find that the charge stiffness vanishes as $m_{\eta}^z\to 0$  
for fixed total $\eta$-spin projection $S_{\eta}^z$, including $S_{\eta}^z=0$, at least as fast as, 
\begin{equation}
D (T) \le  \frac{c_{\rm c}\,t^2 L}{2T} (m_{\eta}^z)^2 \, , 
\label{Dcm0}
\end{equation}
where $c_{\rm c}$ is a $L$--independent constant that smoothly varies as a function of $u$
for the whole $u>0$ range. A similar result is also reached for a canonical ensemble near 
the $\eta$-spin fully polarized sector of maximal hole concentration $m_{\eta}^z=1$,
\begin{equation}
D (T) \le \frac{c_{\rm c}'\,t^2 L}{2T} (1-m_{\eta}^z)^2 \, , 
\label{Dclinem1}
\end{equation}
where $c_{\rm c}'$ is found to be independent of $u$ for $u>0$.

That for finite temperatures our results {\it partially} resolve the charge stiffness behavior of the 1D Hubbard model as 
$m_{\eta}^z\rightarrow 0$ stems in part from the fact that they leave out, marginally, the grand canonical ensemble
in which $\langle (m_{\eta}^z)^2\rangle = {\cal O}(1/L)$. 
(While for a canonical ensemble one considers that the $\eta$-spin density $m_{\eta}^z$ is kept constant, 
in the case of a grand-canonical ensemble it is the chemical potential $\mu$ that is fixed.)

However, for the canonical ensemble our study relies on a charge stiffness upper
bound whose derivation involves a large overestimation of the elementary charge currents of the
energy and momentum eigenstates. Hence accounting for the usual expectation of the equivalence of 
the canonical and grand canonical ensembles in the TL, one would expect that our results remain valid in the 
latter grand canonical case for any finite temperature $T>0$. 
The canonical-ensemble and grand-canonical ensembles lead indeed in general to the same results 
in the TL {\it except} near a phase transition or a critical point. Since a quantum phase
transition from a metallic state to a Mott-Hubbard insulator occurs in the $u>0$ 1D Hubbard model as 
$m_{\eta}^z\rightarrow 0$ and $\mu\rightarrow \mu_{u}$ for $(\mu -\mu_{u})\geq 0$
where $2\mu_{u}$ is the Mott-Hubbard gap \cite{Lieb,Lieb-03,Ovchi}, Eq. (\ref{2mu0}) of Appendix \ref{Appendix1}, 
this issue deserves the careful analysis in these limits carried out in this paper. 

We have addressed such an issue in the limit of high temperatures $T\rightarrow\infty$ for which
strong evidence is provided that the charge stiffness indeed also vanishes 
within the grand-canonical ensemble for chemical potential $\mu$ such that
$(\mu -\mu_{u})\geq 0$ in the $\mu\rightarrow \mu_{u}$ limit.
Specifically, within that ensemble for $T\rightarrow\infty$ we find that the charge stiffness vanishes 
as $m_{\eta}^z\to 0$, at least as fast as, 
\begin{equation}
D (T) \le  \frac{c_{\rm gc}\,t^2}{2T} (m_{\eta}^z)^2 \, , 
\label{Dcm0GC}
\end{equation}
where $c_{\rm gc}$ is again a $L$--independent constant that smoothly varies as a function of $u$. 
A similar result is also reached for a grand-canonical ensemble near 
the $\eta$-spin fully polarized sector of maximal hole concentration $m_{\eta}^z=1$,
\begin{equation}
D (T) \le \frac{c_{\rm gc}'\,t^2}{2T} (1-m_{\eta}^z) \, , 
\label{Dclinem1GC}
\end{equation}
where $c_{\rm gc}'$ is found to be independent of $u$ up to ${\cal{O}}(u^{-2})$ order. That the
upper bounds on the right-hand side of Eqs. (\ref{Dcm0}) and (\ref{Dclinem1}) have an
extra factor $L$ as compared to those on the right-hand side of Eqs. (\ref{Dcm0GC}) and (\ref{Dclinem1GC})
confirms the large overestimation of the elementary charge currents used in the
case of the canonical ensemble. The found lack of ballistic transport in the half-filled 1D Hubbard model 
indicates that charge transport at finite temperatures is dominated by a diffusive contribution \cite{Medenjak}.

The paper is organized as follows. The 1D Hubbard model, its energy and momentum eigenstates, 
symmetry, and the rotated-electron representation are the topics addressed in Section \ref{modelDT}.
In Section \ref{curr-m-el-exp-val} useful subspaces for our charge current absolute values upper bounds
and charge stiffness upper bounds studies are considered and
expressions for the charge current operator expectation values are obtained. 
Useful upper bounds for absolute values of the charge current are then introduced in Section \ref{UecUB}. 
In Section \ref{two} a related charge stiffness upper bound is constructed within the canonical ensemble.
Moreover, a charge stiffness upper bound is introduced in Section \ref{UPTinf}
within the grand-canonical ensemble for $T\rightarrow\infty$. Finally, the concluding remarks
are presented in Section \ref{concluding}.

\section{The model, energy eigenstates, the rotated-electron representation, and symmetry}
\label{modelDT}

The goal of this section is the introduction of the rotated-electron related representation used in our study of the expectation values of the
charge current operator and charge stiffness in the 1D Hubbard model. 
Its relatively large length is justified by the complexity of the problem. However, the use of the representation
introduced in this section simplifies the description in later sections of the model charge transport properties.
Importantly, it has been inherently constructed to be that suitable to clarify the issue of the 
microscopic mechanisms behind such exotic properties.

\subsection{The 1D Hubbard model, its energy eigenstates, and the rotated-electron representation}
\label{modelDT1}

We consider the 1D Hubbard model Hamiltonian under periodic boundary conditions in the TL and in
a chemical potential $\mu$, 
\begin{equation}
\hat{H} = -t\sum_{\sigma}\sum_{j=1}^{L}\left[c_{j,\sigma}^{\dag}\,c_{j+1,\sigma} + 
{\rm h.c.}\right] + U\sum_{j=1}^{L}\hat{\rho}_{j,\uparrow}\hat{\rho}_{j,\downarrow} 
+ 2\mu\,{\hat{S}}_{\eta}^{z} \, .
\label{H}
\end{equation}
It describes $N_e$ electrons in a lattice with $N_a = L$ sites. Here
$c_{j,\sigma}^{\dag}$ creates one electron of spin projection $\sigma = \uparrow,\downarrow$ at site $j = 1,...,L$,
$\hat{\rho}_{j,\sigma}= (\hat{n}_{j,\sigma}-1/2)$, $\hat{n}_{j,\sigma}=c_{j,\sigma}^{\dag}\,c_{j,\sigma}$,
and ${\hat{S}}_{\eta}^{z}=-{1\over 2}\sum_{j=1}^{L}(1-\hat{n}_{j})$ with $\hat{n}_{j}=\sum_{\sigma}\hat{n}_{j,\sigma}$
is the diagonal generator of the global $\eta$-spin $SU(2)$ symmetry.

The $z$-component $\eta$-spin current operator $\hat{J}_{\eta}^z$ and charge current operator 
$\hat{J}_{\rho}$ are closely related as follows,
\begin{eqnarray}
\hat{J}_{\eta}^z & = & (1/2)\,\hat{J}\hspace{0.20cm}{\rm and}\hspace{0.20cm}\hat{J}_{\rho} = (e)\,\hat{J}  \, ,
\nonumber \\
{\rm where} & & \hat{J} = -i\,t\sum_{\sigma}\sum_{j=1}^{L}\left[c_{j,\sigma}^{\dag}\,c_{j+1,\sigma} - 
c_{j+1,\sigma}^{\dag}\,c_{j,\sigma}\right] \, ,
\label{c-s-currents}
\end{eqnarray}
and $e$ denotes the electronic charge. Hence, except for a constant pre-factor, the charge 
current operator $\hat{J}_{\rho}$ equals the $\eta$-spin current operator $\hat{J}_{\eta}^z$.
For simplicity, in several general expressions we use units such that
$\hat{J}_{\eta}^z=\hat{J}_{\rho}=\hat{J}$. We thus call $\hat{J}$, Eq. (\ref{c-s-currents}), the
charge current operator.

Within the exact representation of the $u>0$ 1D Hubbard model in terms of rotated electrons used in our study, 
the operators that create and annihilate such rotated electrons are related to the 
corresponding electron operators as follows,
\begin{equation}
{\tilde{c}}_{j,\sigma}^{\dag} =
{\hat{V}}^{\dag}\,c_{j,\sigma}^{\dag}\,{\hat{V}}
\, ; \hspace{0.50cm}
{\tilde{c}}_{j,\sigma} =
{\hat{V}}^{\dag}\,c_{j,\sigma}\,{\hat{V}} 
\, ; \hspace{0.50cm}
{\tilde{n}}_{j,\sigma} = {\tilde{c}}_{j,\sigma}^{\dag}\,{\tilde{c}}_{j,\sigma}
\hspace{0.20cm}{\rm where}\hspace{0.20cm}j = 1,...,L
\hspace{0.20cm}{\rm and}\hspace{0.20cm}\sigma = \uparrow, \downarrow \, .
\label{rotated-operators}
\end{equation} 
Here ${\hat{V}}$ is the electron - rotated-electron unitary operator uniquely defined 
in Eq. (11) of Ref. \cite{Carmelo-17} in terms of the  $4^L\times 4^L$ matrix elements between 
a complete set of $4^L$ $u>0$ energy and momentum eigenstates of the 1D Hubbard model. 
For all these $4^L$ states the number $N^R_{s,\pm 1/2}$ of spin projection $\pm 1/2$ rotated-electron singly occupied 
sites, $N^R_{\eta,-1/2}$ of $\eta$-spin projection $-1/2$ rotated-electron doubly occupied sites, and $N^R_{\eta,+1/2}$
of $\eta$-spin projection $+1/2$ rotated-electron unoccupied sites are good quantum numbers for $u>0$ \cite{Carmelo-17,Carmelo-16}. 
Hence the number $N^R_{s} = N^R_{s,+1/2}+N^R_{s,-1/2}$ of rotated-electron singly occupied sites and 
$N^R_{\eta} = N^R_{\eta,+1/2}+N^R_{\eta,-1/2}$ of rotated-electron unoccupied plus doubly occupied sites
are conserved for $u>0$ as well. 

Our choice of energy and momentum eigenstates $\vert\nu,u\rangle$
in Eqs. (\ref{DT-gen})-(\ref{DT-thermo}) is different at $u=0$ and for $u>0$. 
For $u>0$, the energy and momentum eigenstates associated with
the exact BA solution are chosen along with those generated from application onto them of the off-diagonal
generators of the global $\eta$-spin $SU(2)$ and spin $SU(2)$ operator algebras symmetries. As 
reported in Section \ref{Introduction}, for $u>0$ the 1D Hubbard model global symmetry is
$[SU(2)\otimes SU(2)\otimes U(1)]/Z_2^2$. Here
$U(1)$ refers to the global $c$ lattice $U(1)$ symmetry, which 
is associated with the lattice degrees of freedom and is independent from the
two $SU(2)$ symmetries. Its generator is the operator
${\tilde{N}}^R_{\eta} = \sum_{j=1}^{L}(1- \sum_{\sigma =\uparrow
,\downarrow}\,\tilde{n}_{j,\sigma}\,(1- \tilde{n}_{j,-\sigma}))$
that counts the number $N^R_{\eta} = 0,1,...,L$ of rotated-electron unoccupied plus doubly occupied sites. 
(Alternatively, it could be chosen to be the operator ${\tilde{N}}^R_{s} = \sum_{j=1}^{L}\sum_{\sigma =\uparrow
,\downarrow}\,\tilde{n}_{j,\sigma}\,(1- \tilde{n}_{j,-\sigma})$
that counts the number $N^R_{s} =L -N^R_{\eta} = 0,1,...,L$ of rotated-electron singly occupied sites.) 
The generator ${\tilde{N}}^R_{\eta}$ eigenvalues are thus the numbers of rotated-electron unoccupied 
plus doubly occupied sites. As justified in later sections, the role of such an eigenvalue in several physical quantities
that emerge from the interplay of the model's symmetry with its exact BA solution justifies
that it is called in this paper $L_{\eta}$, {\it i.~e.} $L_{\eta}\equiv N^R_{\eta}=0,1,...,L$.

We denote each of the $u>0$ energy and momentum eigenstates that belong to the subset of such states that span the 
$S_s^z =0$ subspace considered here by $\vert l_{\rm r},L_{\eta},S_{\eta},S_{\eta}^z,u\rangle$. Here $l_{\rm r}$ 
stands for all quantum numbers other than $L_{\eta}$, $S_{\eta}$, $S_{\eta}^z$, and $u>0$ 
needed to uniquely specify each such a state. This includes spin $S_s$, spin projection
$S_s^z$, and a well-defined set of $u$ independent TBA quantum numbers. Such states can be written as,
\begin{equation}
\vert l_{\rm r},L_{\eta},S_{\eta},S_{\eta}^z,u\rangle = \left[\frac{1}{
\sqrt{{\cal{C}}_{\eta}}}({\hat{S}}^{+}_{\eta})^{\gamma_{\eta}}\right]\vert l_{\rm r},L_{\eta},S_{\eta},-S_{\eta},u\rangle \, ,
\label{Gstate-BAstate}
\end{equation}
where,
\begin{equation}
\gamma_{\eta} = S_{\eta} + S_{\eta}^z = 0,1,..., 2S_{\eta}\hspace{0.20cm}{\rm and}\hspace{0.20cm}
S_{\eta}^z=-(L-N_e)/2 \, .
\label{n-etan-s}
\end{equation} 
Furthermore, ${\cal{C}}_{\eta} = [\gamma_{\eta}!]\prod_{j=1}^{\gamma_{\eta}}[\,2S_{\eta}+1-j\,]$
is a normalization constant and ${\hat{S}}_{\eta}^{+}$ is the $\eta$-spin $SU(2)$ off-diagonal generator, 
\begin{equation}
{\hat{S}}_{\eta}^{+} = \sum_{j=1}^{L}(-1)^j\,c_{j,\downarrow}^{\dag}\,c_{j,\uparrow}^{\dag} 
\hspace{0.20cm}{\rm and}\hspace{0.20cm}{\rm thus}\hspace{0.20cm}{\hat{S}}_{\eta}^{-} = \left({\hat{S}}_{\eta}^{+}\right)^{\dag} \, .
\label{S+S+}
\end{equation}

Except in the $u\rightarrow\infty$ limit, electron single occupancy, electron double occupancy, 
and electron non-occupancy are not good quantum numbers for the energy and momentum eigenstates 
$\vert l_{\rm r},L_{\eta},S_{\eta},S_{\eta}^z,u\rangle$. For instance, upon decreasing $u$ there emerges for ground states 
for which $m_{\eta}^z\geq 0$ a finite electron double occupancy
expectation value, which vanishes for $u\rightarrow\infty$ \cite{Carmelo-03}. 

We call {\it $\eta$-Bethe states} the $u>0$ energy and momentum eigenstates $\vert l_{\rm r},L_{\eta},S_{\eta},-S_{\eta},u\rangle$
that are LWSs of the $\eta$-spin $SU(2)$ algebra, so that $S_{\eta}^z =-S_{\eta}$ and thus $\gamma_{\eta}=0$
in their expression, Eq. (\ref{Gstate-BAstate}). We call  {\it Bethe states} the $u>0$ energy and momentum eigenstates that
are {\it both} LWSs of the $\eta$-spin and spin $SU(2)$ operator algebras for which $S_{\alpha}^z =-S_{\alpha}$
for $\alpha =\eta,s$. However, the $\eta$-Bethe states considered in this paper can either be spin LWSs or 
spin non-LWSs. The designation LWS and non-LWS refers in general in this paper to the $\eta$-spin $SU(2)$ operator algebra alone. In the case
of the spin $SU(2)$ operator algebra, one always specifies {\it spin LWS} and {\it spin non-LWS}, respectively.

For $\eta$-Bethe states $\vert l_{\rm r},L_{\eta},S_{\eta},-S_{\eta},u\rangle$ and general 
energy and momentum eigenstates 
$\vert l_{\rm r},L_{\eta},S_{\eta},S_{\eta}^z,u\rangle$ the electron numbers are given by,
\begin{equation}
N_e^0 = L - 2S_{\eta}\hspace{0.20cm}{\rm and}\hspace{0.20cm}
N_e = N_e^0 + 2\gamma_{\eta} \, ,
\label{NN-LWS+TOWER}
\end{equation}
respectively, where $\gamma_{\eta} = S_{\eta}+S_{\eta}^{z} = 0,1,..., 2S_{\eta}$, Eq. (\ref{n-etan-s}).

In the case of $\eta$-Bethe states, the $u>0$ charge current operator expectation values 
$\langle l_{\rm r},L_{\eta},S_{\eta},-S_{\eta},u\vert\hat{J}\vert l_{\rm r},L_{\eta},S_{\eta},-S_{\eta},u\rangle$,
which are such states charge currents, can be expressed 
in terms of the BA solution momentum rapidity and rapidity functionals, Eqs. (\ref{J-partDEF}) and
(\ref{jn-fnPSEU}) of Appendix \ref{Appendix1}. 
For each $u>0$ $\eta$-Bethe state, such functions are uniquely defined by the TBA equations, 
Eqs. (\ref{Tapco1}) and (\ref{Tapco2}) of that Appendix. 
Furthermore, we rely on exact symmetry relations to express the charge currents of general
energy and momentum eigenstates $\vert l_{\rm r},L_{\eta},S_{\eta},S_{\eta}^z,u\rangle$, Eq. (\ref{Gstate-BAstate}), in terms of that of the
corresponding $\eta$-Bethe state $\vert l_{\rm r},L_{\eta},S_{\eta},-S_{\eta},u\rangle$ on
the right-hand side of that equation.

A {\it $V$ tower} is within the rotated-electron representation the set of energy eigenstates 
$\vert l_{\rm r},L_{\eta},S_{\eta},S_{\eta}^z,u\rangle$ with exactly the same $u$-independent quantum numbers 
$l_{\rm r}$, $L_{\eta}$, $S_{\eta}$, and $S_{\eta}^z$ and different $u$ values in the range $u>0$ \cite{Carmelo-17}. 
The set of energy and momentum eigenstates $\vert l_{\rm r},L_{\eta},S_{\eta},S_{\eta}^z,u\rangle$ that belong 
to the same $V$ tower are for any $u>0$ value generated by exactly the same occupancy configurations 
of the $u$-independent quantum numbers as the corresponding $u=\infty$ energy and momentum eigenstate
$\vert l_{\rm r},L_{\eta},S_{\eta},S_{\eta}^z,\infty\rangle = \lim_{u\rightarrow\infty}\vert l_{\rm r},L_{\eta},S_{\eta},S_{\eta}^z,u\rangle$. 
Out of the many choices of $u = \infty$ energy and momentum eigenstates, the states
$\vert l_{\rm r},L_{\eta},S_{\eta},S_{\eta}^z,\infty\rangle$ are those obtained from the finite-$u$ energy and momentum eigenstates,
Eq. (\ref{Gstate-BAstate}), whose LWSs are the $\eta$-Bethe states, 
as $\lim_{u\rightarrow\infty}\vert l_{\rm r},L_{\eta},S_{\eta},S_{\eta}^z,u\rangle$. 

The Hilbert space remains the same for the whole $u>0$ range. 
For any fixed $u>0$, there is thus a uniquely defined unitary operator ${\hat{V}}={\hat{V}} (u)$ such that 
$\vert l_{\rm r},L_{\eta},S_{\eta},S_{\eta}^z,u\rangle={\hat{V}}^{\dag}\vert l_{\rm r},L_{\eta},S_{\eta},S_{\eta}^z,\infty\rangle$. 
This operator ${\hat{V}}$ is the electron - rotated-electron unitary operator appearing in Eq. (\ref{rotated-operators}).
It is uniquely defined in Eq. (11) of Ref. \cite{Carmelo-17}.
The $\sigma=\uparrow,\downarrow$ electron single occupancy, electron double occupancy, and electron non-occupancy 
are good quantum numbers for a $u\rightarrow\infty$ energy and momentum eigenstate $\vert l_{\rm r},L_{\eta},S_{\eta},S_{\eta}^z,\infty\rangle$.
This is why for all the finite-$u$ energy and momentum eigenstates $\vert l_{\rm r},L_{\eta},S_{\eta},S_{\eta}^z,u\rangle$ belonging to the same
$V$ tower the rotated-electron numbers $N^R_{s,\pm 1/2}$, $N^R_{\eta,\pm 1/2}$, $N^R_{s} = L - L_{\eta}$,
and $N^R_{\eta} = L_{\eta}$ are conserved as well. 

\subsection{Effects of the symmetry on the charge degrees of freedom}
\label{Introduction1}

One of the few rigorous results for the Hubbard model on any bipartite lattice refers to its global symmetry. 
As was mentioned in Section \ref{Introduction}, it is well known that on such a lattice the Hamiltonian has two global $SU(2)$ symmetries 
\cite{HL,Yang,Yang-90,Lieb-89}. Consistently, in the early nineties of the past 
century it was found that for $u\neq 0$ the Hubbard model on a 
bipartite lattice has at least a $SO(4) = [SU(2)\otimes SU(2)]/Z_2$ symmetry, which contains the 
$\eta$-spin and spin $SU(2)$ symmetries \cite{Yang,Yang-90}. More recently it was found in Ref. \cite{bipartite} 
that for $u\neq 0$ and on any bipartite lattice its global symmetry is actually larger and given by 
$[SO(4)\otimes U(1)]/Z_2=[SU(2)\otimes SU(2)\otimes U(1)]/Z_2^2$. 
(The $1/Z_2^2$ factor in the $u>0$ model global symmetry refers to the number $4^{L}$ of its independent representations 
being four times smaller than the dimension $4^{L+1}$ of the group $SU(2)\otimes SU(2)\otimes U(1)$.) 

The origin of the $u>0$ global $[SU(2)\otimes SU(2)\otimes U(1)]/Z_2^2$ symmetry is a local gauge 
$SU(2)\otimes SU(2)\otimes U(1)$ symmetry of the $U>0$ Hamiltonian $t=0$ term first identified in Ref. \cite{Stellan-91}. 
This $u^{-1}=0$ local gauge symmetry becomes for finite $u=U/4t$ a group of permissible unitary transformations. 
(The corresponding local $U(1)$ canonical transformation is not the ordinary gauge $U(1)$ subgroup of electromagnetism. 
It is rather a ``nonlinear" transformation \cite{Stellan-91}.)  
The related global $c$ lattice $U(1)$ symmetry beyond $SO(4)$ found in Ref. \cite{bipartite}, which 
is associated with the lattice degrees of freedom and does not exist at $U=0$, 
emerges at any arbitrarily small $u$ value. 

Importantly, the rotated-electrons charge and spin $1/2$ are the same as those of the corresponding electrons
and thus remain invariant under the electron - rotated-electron unitary transformation. That 
transformation only changes the lattice occupancies and corresponding spatial
distributions of the charges and spins $1/2$. Furthermore, in the $u\rightarrow\infty$ limit the electron - rotated-electron 
unitary operator ${\hat{V}}$ becomes the unit operator. This is why in such a limit the
rotated electrons become electrons. 

That in the $u^{-1}\rightarrow 0$ limit the rotated electrons become electrons and for $u>0$ 
they have the same charge and spin $1/2$ as the electrons reveals that for finite $u$ they are quasiparticles whose 
``noninteracting'' limit is $u^{-1}=0$. In terms of the onsite repulsion, this is thus a type of
turned upside-down ``Fermi liquid''. Its exotic properties follow in part from at such $u^{-1}=0$
``noninteracting'' point the degrees of freedom of the electron occupancy configurations that generate from the electron vacuum the $u^{-1}=0$ 
energy and momentum eigenstate $\vert l_{\rm r},L_{\eta},S_{\eta},S_{\eta}^z,\infty\rangle$ 
separating into three types of configurations. Those refer to state representations of the two $SU(2)$ 
symmetries and $U(1)$ symmetry in the $u^{-1}=0$ model local gauge $SU(2)\otimes SU(2)\otimes U(1)$ symmetry.
As reported below in Section \ref{Introduction2}, this three degrees of freedom separation persists at finite $u$ in terms of
the rotated-electron occupancy configurations that generate from the electron vacuum the
energy and momentum eigenstate 
$\vert l_{\rm r},L_{\eta},S_{\eta},S_{\eta}^z,u\rangle={\hat{V}}^{\dag}\vert l_{\rm r},L_{\eta},S_{\eta},S_{\eta}^z,\infty\rangle$.
At finite $u$ values this is related though to the three symmetries in the
$u>0$ model global $[SU(2)\otimes SU(2)\otimes U(1)]/Z_2^2$ symmetry that stems from its $u^{-1}=0$
local gauge $SU(2)\otimes SU(2)\otimes U(1)$ symmetry.

Furthermore and as reported above, for $u>0$ and at $U=0$ the global symmetry is different and given by $[SO(4)\otimes U(1)]/Z_2$
and $SO(4)\otimes Z_2$, respectively. The factor $Z_2$ in the $U=0$ global symmetry
corresponds to a discretely generated symmetry associated with a well-known transformation that exchanges 
spin and $\eta$-spin. It is an exact symmetry of the $U=0$ and $t\neq 0$ Hamiltonian. However, it changes the sign of 
$U$ when $U\neq 0$. That the global symmetry is different at $U=0$ and for $u=U/4t>0$
plays an important role in the quantum transition that occurs for $m_{\eta}^z=0$ at $U=U_c=0$.
It separates two qualitatively different types of transport of charge. It may as well play an
important role in the charge transport properties for $T>0$.

Another important symmetry property that has effects on the transport of charge 
is that the $U(2)=SU(2)\otimes U(1)$ and $SU(2)$ symmetries in the  
$[SU(2)\otimes SU(2)\otimes U(1)]/Z_2^2$ global symmetry refer to the charge and spin degrees of freedom, respectively. 
We recall that the charge $U(2)=SU(2)\otimes U(1)$ symmetry includes the $\eta$-spin $SU(2)$ symmetry and the 
$c$ lattice $U(1)$ symmetry beyond $SO(4)$. The state representations of the groups associated with
these two symmetries are found in this paper to contribute to the charge current of the $u>0$ energy and momentum eigenstates.

That the charge and spin global symmetries are $U(2)=SU(2)\otimes U(1)$ and $SU(2)$, respectively, has 
in the present 1D case direct effects on the model's exact BA solution. For instance and as reported in Section
\ref{Introduction}, it is behind the charge and 
spin monodromy matrices of the BA inverse-scattering method \cite{Martins,CM} having different ABCD and 
ABCDF forms \cite{Martins}. 

\subsection{The rotated-electron degrees of freedom separation}
\label{Introduction2}

The rotated-electron degrees of freedom naturally separate for $u>0$ into occupancy configurations of 
three basic fractionalized particles that generate exact state representations of the groups associated
with the independent spin and $\eta$-spin $SU(2)$ symmetries and the $c$ lattice $U(1)$ symmetry, respectively, in the
model global $[SU(2)\otimes SU(2)\otimes U(1)]/Z_2^2$ symmetry. This refers
namely to a number $L_s =N^R_{s}$ of rotated spins $1/2$, $L_{\eta}=N^R_{\eta}$ of rotated $\eta$-spins $1/2$, and
$N_c =L- L_{\eta} =N^R_{s}$ of $c$ pseudoparticles without internal degrees of freedom, respectively.

The corresponding numbers of rotated spins of spin projection $\pm 1/2$ and rotated $\eta$-spins of $\eta$-spin 
projection $\pm 1/2$ are denoted by $L_{s,\pm 1/2}$ and $L_{\eta,\pm 1/2}$, respectively. 
They are determined by corresponding numbers of rotated-electrons as they 
read  $L_{s,\pm 1/2}=N^R_{s,\pm 1/2}$ and $L_{\eta,\pm 1/2}=N^R_{\eta,\pm 1/2}$.
There are in addition $N_c^h=L_{\eta}=N^R_{\eta}$ $c$ pseudoparticle holes.
$L_{\eta}=N^R_{\eta}$ and $L_s =N^R_{s}=L-L_{\eta}$ are as well the number of 
sites of the $\eta$-spin and spin effective lattices, respectively, introduced in the following. 
The state representations of the groups associated with the $\eta$-spin and spin $SU(2)$ symmetries
that are generated by rotated $\eta$-spins $1/2$ and rotated spins $1/2$ occupancy configurations,
respectively, of such effective lattices are similar to those of an $\eta$-spin-$1/2$ and a spin-$1/2$ $XXX$ 
chain on a lattice with $L_{\eta}$ and $L_s $ sites, respectively. This justifies the notations $L_{\eta}$ 
and $L_s $ for $N^R_{\eta}$ and $N^R_{s}$, respectively.

The concept of a squeezed effective lattice is well known in 1D correlated systems \cite{Ogata,Karlo-97,Zaanen}.
In the present case, the rotated $\eta$-spins $1/2$ only ``see'' the set of $L_{\eta}=N^R_{\eta}$ sites unoccupied 
and doubly occupied by rotated electrons. The rotated $\eta$-spins $1/2$ 
thus live in an $\eta$-spin squeezed effective lattice with $L_{\eta}$ sites that corresponds to an $\eta$-spin-$1/2$ $XXX$ chain.
The rotated spins $1/2$ only ``see'' the set of $L_s=L-L_{\eta}$ sites singly occupied by rotated electrons. 
They live in a spin squeezed effective lattice with $L_s=L-L_{\eta}=N^R_{s}$ sites that corresponds to a spin-$1/2$
$XXX$ chain. The $c$ pseudoparticles live on an effective lattice identical to the original model lattice. 
In the case of the electron representation of the 1D Hubbard model, these lattices are known in the $u\rightarrow\infty$ limit in which the 
rotated electrons become electrons \cite{Ogata,Karlo-97,Zaanen}.

The spatial coordinates of the $N_c=L-L_{\eta}=L_s$ sites occupied by $c$ pseudoparticles and those 
of the corresponding $N_c^h = L_{\eta}$ unoccupied sites ($c$ pseudoparticle holes) fully define the relative 
positions in the model's original lattice of the spin squeezed effective lattice sites and $\eta$-spin squeezed 
effective lattice sites, respectively. The role of the $c$ lattice $U(1)$ symmetry is actually to preserve the independence of the 
spin and $\eta$-spin $SU(2)$ symmetries and corresponding squeezed effective lattices 
occupancy configurations, which do not  ``see'' each other. This is fulfilled by the state representations
of the $c$ lattice $U(1)$ symmetry group by storing full information on the relative positions in
the model's original lattice of the spin and $\eta$-spin squeezed effective lattices sites, respectively.

The following relations between the numbers of the three types of fractionalized particles hold,
\begin{eqnarray}
&& L_s = L_{s,+1/2} + L_{s,-1/2} = N_c \, ,
\nonumber \\
&& L_{\eta} = L_{\eta,+1/2} + L_{\eta,-1/2} = L - N_c = N_c^h \, ,
\nonumber \\
&& L_{s,+1/2} - L_{s,-1/2} = -2S_s^z = N_{e\,\uparrow} - N_{e\,\downarrow} \, ,
\nonumber \\
&& L_{\eta,+1/2} - L_{\eta,-1/2} = -2S_{\eta}^z = L - N_e \, ,
\label{severalM}
\end{eqnarray}
where $N_{e\,\sigma}$ is the number of $\sigma = \uparrow,\downarrow$ electrons,
which equals that of $\sigma = \uparrow ,\downarrow $ rotated electrons.
The $u>0$ good quantum numbers $L_{\eta}=N^R_{\eta}$ and $L_{s}=N^R_{s}$ naturally
emerge within the BA solution as $L_{\eta}=L-N_c$ and $L_{s}=N_c$, respectively \cite{Carmelo-17}. 
Here $N_c$ is our notation for the number called $N-M'$ in Ref. \cite{Takahashi}, which 
is the number of real charge rapidities $k_j$ of a Bethe state.

On the one hand, the electron -- rotated-electron unitary transformation changes the lattice occupancies and corresponding spatial
distributions of the rotated-electrons charges and spins $1/2$. On the other hand, the rotated-electrons charge and spin $1/2$ are
the same as those of the corresponding electrons and thus remain invariant under that transformation. This ensures that the
rotated spins $1/2$, which are the spins of the rotated electrons that singly occupy sites, are {\it physical spins $1/2$}.
The same applies to the $c$ pseudoparticles that carry the charges of these rotated electrons
and to the rotated $\eta$-spins $1/2$ of $\eta$-spin projection $+1/2$ and $-1/2$ that describe the $\eta$-spin degrees of freedom of the
rotated-electron unoccupied and doubly occupied sites, respectively. Consistently, the operators associated with 
the rotated spins $1/2$, $c$ pseudoparticles, and rotated $\eta$-spins $1/2$ have explicit expressions
in terms of the rotated-electron  creation and annihilation operators, Eq. (\ref{rotated-operators}).
Moreover, the corresponding electron - rotated-electron unitary operator is uniquely defined in Eq. (11) of 
Ref. \cite{Carmelo-17}. Specifically, the local $SU(2)$ operators associated the rotated $\eta$-spins $1/2$ and  
rotated spins $1/2$ are expressed in terms of rotated-electron creation and annihilation operators
in Eqs. (29)-(31) of that reference. The $c$ pseudoparticle creation and annihilation operators
are expressed in terms of those of the rotated electrons in Eq. (33) and in Eq. (38) for $\beta = c$
of Ref. \cite{Carmelo-17}. 

For $u>0$ energy and momentum eigenstates of $\eta$-spin $S_{\eta}$ and
spin $S_s$, a number $L_{\eta} - 2S_{\eta}$ of rotated $\eta$-spins $1/2$ out of
$L_{\eta}$ such $\eta$-spins and a number $L_{s} - 2S_{s}$ of rotated spins $1/2$ out of
$L_{s}$ such spins are part of $\Pi_{\eta} = (L_{\eta} - 2S_{\eta})/2$
$\eta$-spin singlet pairs and $\Pi_{s} = (L_{s} - 2S_{s})/2$ of spin-singlet pairs, respectively.
Subsets of $n=1,...,\infty$ such pairs refer to the internal structure
of neutral composite $\eta n$ and $sn$ pseudoparticles, respectively \cite{Carmelo-17}. 
The occupancy configurations of the fractionalized particles
and related composite particles that generate the exact energy and momentum eigenstates from
the electron vacuum are found to be labelled by the quantum numbers emerging from the 
model TBA solution \cite{Carmelo-17,Carmelo-16}. This is a generalization of the representation in terms
of spins $1/2$ and $n$-band pseudoparticles used for the spin-$1/2$ $XXX$ chain in Refs. \cite{CPC,CP} 
to address the related problem of that model spin stiffness \cite{Zotos-99}.

The rotated-electron creation and annihilation operators, Eq. (\ref{rotated-operators}), have been inherently 
constructed from those of the electrons to the form of the 1D Hubbard model energy and momentum
eigenstates wave function in terms of rotated electrons being for $u>0$ similar to that of the wave 
function in terms of electrons for $u\rightarrow\infty$. The latter is given in Eq. (2.23) 
of Ref. \cite{Woy-82}. It is a product of an $\eta$-spin $1/2$ $XXX$ chain wave function $\varphi_1$, a 
spin $1/2$ $XXX$ chain wave function $\varphi_2$, and a Slater determinant of fermions without
internal degrees of freedom. Hence this confirms that in the $u\rightarrow\infty$ limit the 1D Hubbard model corresponds 
to an $\eta$-spin-$1/2$ $XXX$ chain, a spin-$1/2$ $XXX$ chain, and a quantum problem with simple $U(1)$ 
symmetry, respectively. The same applies to the whole $u>0$ range within the rotated-electron representation.

Note though that for finite $u$ values this applies only to the $u$-independent 
$l_{\rm r}$, $L_{\eta}$, $S_{\eta}$, and $S_{\eta}^z$ quantum number values
that label the exact energy and momentum eigenstates $\vert l_{\rm r},L_{\eta},S_{\eta},S_{\eta}^z,u\rangle$,
which includes the momentum operator eigenvalues, as well as to
the occupancy configurations that generate such states, which in terms of rotated electrons remain the same
for the whole $u>0$ range, and to the corresponding rotated-electron wave functions. 
The energy eigenvalues, Eqs. (\ref{E}) and (\ref{spectra-E-an-c-0}) of Appendix \ref{Appendix1},
and for instance the charge current operator expectation values, 
Eqs. (\ref{J-partDEF}) and (\ref{jn-fnPSEU}) of that Appendix,
of the energy and momentum eigenstates are though dependent on $u$. They have a different form from the 
corresponding $u\rightarrow\infty$ energy eigenvalues and charge current operator expectation values. 
In the case of the latter this stems from the exotic overlap
that occurs within matrix elements between energy and momentum eigenstates
of the charge current operator expressed in terms of electron creation and
annihilation operators, Eq. (\ref{c-s-currents}), with the rotated-electron occupancy configurations that generate
such states \cite{Carmelo-17}.

\subsection{Relation to the Bethe-anstaz solution quantum numbers}
\label{Introduction3}

The studies of Ref. \cite{Carmelo-17} have considered the relation between the TBA quantum numbers and the three degrees of freedom 
separation of the rotated-electron occupancy configurations. This confirms that such quantum numbers are
directly associated with the occupancy configurations of the above considered three types of fractionalized particles that generate
all Bethe states. Upon application onto those of the off-diagonal generators of the model's two $SU(2)$ symmetries, one
then generates {\it all} $4^L$ $u>0$ energy and momentum eigenstates, as given in Eq. (\ref{Gstate-BAstate}).

The exact Bethe states are populated by $L_s=L-L_{\eta}$
rotated spins $1/2$ and $L_{\eta}$ rotated $\eta$-spins $1/2$. As mentioned above, out of those,
a number $L_{s}-2S_{s}$ of rotated spins $1/2$ are part of a number $\Pi_s=(L_{s}-2S_{s})/2$ of spin-singlet pairs $(\alpha =s)$ 
and a number $L_{\eta}-2S_{\eta}$ of rotated $\eta$-spins $1/2$ are part of a number $\Pi_{\eta}=(L_{\eta}-2S_{\eta})/2$ 
of $\eta$-spin singlet pairs $(\alpha =\eta)$. Such $\Pi_{\alpha}$ spin-singlet $(\alpha =s)$ 
and $\eta$-spin singlet $(\alpha =\eta)$ pairs are bound within a set of $\alpha$ $n$-pairs configurations 
each of which refers to the internal degrees of freedom of one neutral composite $\alpha n$ pseudoparticle. 
Here $n=1,...,\infty$ gives the number of pairs bound within each such pseudoparticles.

Consistently with TBA corresponding results, the following exact sum rules then
hold for all $u>0$ energy and momentum eigenstates, Eq. (\ref{Gstate-BAstate}),
\begin{eqnarray}
\Pi_{\alpha} & = & \sum_{n=1}^{\infty}n\,N_{\alpha n} = {1\over 2}(L_{\alpha} - 2S_{\alpha})
\hspace{0.20cm}{\rm where}\hspace{0.20cm}\alpha = s, \eta \, ,
\nonumber \\
\Pi & \equiv & \sum_{\alpha =\eta,s}\Pi_{\alpha} =
 \sum_{\alpha =\eta,s}\sum_{n=1}^{\infty}n\,N_{\alpha n} = {1\over 2}(L - 2S_s - 2S_{\eta}) \, .
\label{sum-Nseta}
\end{eqnarray}
Here $N_{\alpha n}$ is the number of $\alpha n$ pseudoparticles 
and $\Pi$ denotes the total number of both rotated-spin and rotated-$\eta$-spin pairs.

For a Bethe state, the remaining $M_{\alpha}=2S_{\alpha}$ unpaired rotated spins $(\alpha =s)$
and rotated $\eta$-spins $(\alpha =\eta)$ have spin and $\eta$-spin projection $+1/2$, respectively. 
For general $u>0$ energy and momentum eigenstates, the multiplet configurations 
of these $M_s =2S_s$ unpaired rotated spins and $M_{\eta} =2S_{\eta}$ unpaired rotated $\eta$-spins 
generate the spin and $\eta$-spin, respectively, $SU (2)$ symmetry towers of non-LWSs. The $SU(2)$ 
symmetry algebras off-diagonal generators that flip such unpaired rotated spins and unpaired rotated $\eta$-spins,
which for $\eta$-spin are given in Eq. (\ref{S+S+}),
do not affect though the spin $(\alpha =s)$ and $\eta$-spin $(\alpha =\eta)$ singlet configurations 
of the $\Pi_{\alpha} = \sum_{n=1}^{\infty}n\,N_{\alpha n}$ pairs contained in neutral composite 
$\alpha n$ pseudoparticles. Those remain unchanged. 

For general $u>0$ energy and momentum eigenstates, the number $M_{s,\pm 1/2}$ of unpaired 
rotated spins of projection $\pm 1/2$ and $M_{\eta,\pm 1/2}$ of unpaired rotated $\eta$-spins of 
projection $\pm 1/2$ are good quantum numbers given by,
\begin{equation}
M_{\alpha,\pm 1/2} = S_{\alpha}\mp S_{\alpha}^{z}\hspace{0.20cm}{\rm and}\hspace{0.20cm}
M_{\alpha} = M_{\alpha,-1/2}+M_{\alpha,+1/2} = 2S_{\alpha}\hspace{0.20cm}{\rm where}\hspace{0.20cm}\alpha = \eta,s \, .
\label{L-L}
\end{equation}
For the Bethe states, one has that $M_{\alpha,+1/2} = M_{\alpha} = 2S_{\alpha}$ and 
$M_{\alpha,-1/2} = 0$ for both $\alpha=\eta, s$. 
The rotated $\eta$-spins ($\alpha = \eta$) and rotated spins ($\alpha = \eta$) numbers
$L_{\alpha}$ and $L_{\alpha,\pm 1/2}$ in Eq. (\ref{severalM}) can be written as,
\begin{eqnarray}
L_{\alpha} & = & 2\Pi_{\alpha} + M_{\alpha} = 2\Pi_{\alpha} + 2S_{\alpha} \, ,
\nonumber \\ 
L_{\alpha,\pm 1/2} & = & \Pi_{\alpha} + M_{\alpha,\pm 1/2} = \Pi_{\alpha} + S_{\alpha}\mp S_{\alpha}^{z}
\hspace{0.20cm}{\rm where}\hspace{0.20cm}\alpha = \eta,s \, ,
\label{Mtotal}
\end{eqnarray}
respectively.

Another important symmetry property is that the spatial lattice occupancies of the $M_{\alpha}=2S_{\alpha}$ 
unpaired rotated spins $(\alpha =s)$ and unpaired rotated $\eta$-spins $(\alpha =\eta)$ remain invariant 
under the electron - rotated-electron unitary transformation. This means that their lattice occupancy 
configurations are for the whole $u>0$ range exactly the same as those of the corresponding electrons occupancy configurations.
That invariance plays an important role in the transport of charge
and spin. Indeed and as reported below for the present case of charge transport, 
the electronic degrees of freedom couple to charge and spin probes through {\it only} such
$M_{\eta}=2S_{\eta}$ unpaired physical $\eta$-spins and $M_{s}=2S_{s}$ unpaired physical 
spins, respectively. 

Note though that the paired rotated spins $1/2$ and paired rotated $\eta$-spins $1/2$
are also physical spins $1/2$ and physical $\eta$-spins $1/2$ in what their spin and
$\eta$-spin degrees of freedom, respectively, are concerned. Only their lattice spatial occupancies
are changed under the electron - rotated-electron unitary transformation. Nevertheless, to stress that
the lattice spatial occupancies of the unpaired spins $1/2$ and unpaired $\eta$-spins $1/2$ 
remain invariant under that transformation we omit the term rotated from their designation.
We use more often in the following the designation {\it physical} for them.

The TBA solution contains different types of quantum numbers. Their occupancy configurations are
within the pseudoparticle representation described by corresponding occupancy configurations
of $c$ pseudoparticles with no internal degrees of freedom and composite $\alpha n$ pseudoparticles
plus a number $M_{\eta} = 2S_{\eta}$ of unpaired physical $\eta$-spins and $M_{s} = 2S_{s}$ of 
unpaired physical spins. The $c$ branch momentum rapidity functional $k^c (q_j)$ and set of $\alpha n$ 
branches rapidity functionals $\Lambda^{\alpha n}(q_{j})$ where $\alpha = \eta, s$ and $n =1,...,\infty$ are 
solutions of the coupled TBA integral equations introduced in Ref. \cite{Takahashi}, which are given in functional form in
Eqs. (\ref{Tapco1}) and (\ref{Tapco2}) of Appendix \ref{Appendix1}. For $n>1$, the rapidity functionals 
$\Lambda^{\alpha n}(q_{j})$ are the real part of corresponding $l = 1,...,n$ complex rapidities given below. 

The $c$ branch TBA quantum numbers 
$\{q_j\}$ in the argument of the momentum rapidity functional $k^c (q_j)$ and
corresponding $c$ rapidity functional $\Lambda^{c}(q_{j})\equiv \sin (k^c (q_j))$ and $\alpha n$ branch BA quantum 
numbers $\{q_j\}$ in the argument of the rapidity functionals 
$\Lambda^{\alpha n}(q_{j})$ are given by,
\begin{equation}
q_j = {2\pi\over L}\,I^{\beta}_j\hspace{0.20cm}{\rm for}\hspace{0.20cm}j=1,...,L_{\beta} 
\hspace{0.25cm}{\rm where}\hspace{0.20cm}\beta = c,\eta n,sn\hspace{0.20cm}{\rm and}\hspace{0.20cm}n =1,...,\infty \, .
\label{q-j}
\end{equation} 
Here $\{I_j^{\beta}\}$ are the quantum numbers $\{q_j\}$ in units of $2\pi/L$ that are successive integers or
half-odd integers according to the following boundary conditions,
\begin{eqnarray}
I_j^{\beta} & = & 0,\pm 1,\pm 2,... \hspace{0.50cm}{\rm for}\hspace{0.15cm}I_{\beta}\hspace{0.15cm}{\rm even} \, ,
\nonumber \\
& = & \pm 1/2,\pm 3/2,\pm 5/2,... \hspace{0.50cm}{\rm for}\hspace{0.15cm}I_{\beta}\hspace{0.15cm}{\rm odd} \, .
\label{Ic-an}
\end{eqnarray}
The $\beta = c,\eta n,sn$ numbers $I_{\beta}$ in this equation read,
\begin{eqnarray}
I_c & = & N_{\rm ps} \equiv \sum_{\alpha =\eta,s}\sum_{n=1}^{\infty}N_{\alpha n} \, ,
\nonumber \\
I_{\alpha n} & = & L_{\alpha n} -1\hspace{0.20cm}{\rm where}\hspace{0.20cm}\alpha = \eta, s
\hspace{0.20cm}{\rm and}\hspace{0.20cm}n=1,...,\infty \, .
\label{F-beta}
\end{eqnarray}	
Moreover, $L_{\beta}=N_{\beta} + N^h_{\beta}$ is the number
of $\beta =c,\alpha n$-band discrete momentum values $q_j$
of which for a given state $N_{\beta}$ are occupied and $N^h_{\beta}$ are unoccupied. They read,
\begin{eqnarray}
L_{c} & = & N_{c} + N^h_{c} = L \, ,
\nonumber \\
N^h_{c} & = & L - N_c = L_{\eta} = 2S_{\eta} + \sum_{n=1}^{\infty}2n\,N_{\eta n} \, ,
\nonumber \\
L_{\alpha n} & = & N_{\alpha n} + N^h_{\alpha n}
\hspace{0.20cm}{\rm where}\hspace{0.20cm}\alpha=\eta, s\hspace{0.20cm}{\rm and}\hspace{0.20cm}n =1,...,\infty \, ,
\nonumber \\
N^h_{\alpha n} & = & 2S_{\alpha}+\sum_{n'=n+1}^{\infty}2(n'-n)N_{\alpha n'} 
= L_{\alpha} - \sum_{n' =1}^{\infty}(n +n' - \vert n-n'\vert)N_{\alpha n'} \, .
\label{N-h-an}
\end{eqnarray}

The momentum eigenvalues can be written as,
\begin{equation}
P =\sum_{j=1}^{L} q_j\, N_c (q_j)
+ \sum_{n =1}^{\infty}\sum_{j=1}^{L_{s n}}
q_{j}\, N_{sn} (q_{j}) 
+ \sum_{n =1}^{\infty}\sum_{j=1}^{L_{\eta n}}
(\pi -q_{j})\, N_{\eta n} (q_{j}) + \pi L_{\eta,-1/2} \, ,
\label{P}
\end{equation}
where $N_{\beta} (q_j)$ are for $\beta =c,\eta n, sn$ pseudoparticle branches the $\beta$-band momentum distribution functions.  
They are such that $N_{\beta} (q_j)=1$ and $N_{\beta} (q_j)=0$ for occupied and unoccupied $q_j$ values, respectively.
For the $c$ and $\alpha n$ branches, such values have intervals $q_j \in [q_c^-,q_c^+]$ and $q_j \in [-q_{\alpha n},q_{\alpha n}]$
where ignoring $1/L$ corrections within the TL the $c$-band limiting momentum values are such that $q_c^{\pm}=\pm q_{c}$.
Here the limiting momentum values $q_{c}$ and $q_{\alpha n}$ are given by,
\begin{equation}
q_{c} = \pi\hspace{0.20cm}{\rm and}\hspace{0.20cm}q_{\alpha n} = \pi\,(L_{\alpha n}-1)/L \, ,
\label{qlimits}
\end{equation}
respectively.

That the momentum eigenvalues, Eq. (\ref{P}), are additive in the quantum numbers $q_j$ in Eq. (\ref{q-j})
is consistent with they playing the role of $\beta = c,\alpha n$ band momentum values. The momentum contribution 
$\pi L_{\eta,-1/2}=\pi (\Pi_{\eta} + M_{\eta,-1/2})$ in Eq. (\ref{P}) follows from both the paired and unpaired rotated 
$\eta$-spins of projection $-1/2$ having an {\it intrinsic momentum} given by,
\begin{equation}
q_{\eta,-1/2} = \pi \, .
\label{q-eta-s}
\end{equation}
For a $\eta$-Bethe state one has that $\pi L_{\eta,-1/2}=\pi\,\Pi_{\eta}$.

\subsection{Internal degrees of freedom of the composite $\alpha n$ pseudoparticles and $u\rightarrow 0$ $n>1$ pairs unbinding}
\label{BindingUn}

As for the spin-neutral composite $n$-band pseudoparticles of the spin-$1/2$ $XXX$ chain \cite{CP},
the problem concerning an $\alpha n$ pseudoparticle internal degrees of freedom and 
that associated with its translational degrees of freedom center of mass motion separate
within the TL. 

On the one hand, the $\alpha n$-band momentum $q_j$, Eq. (\ref{q-j}) for $\beta =\alpha n$, is associated with the latter.
On the other hand, for $n>1$ the internal degrees of freedom are related to the imaginary part of the 
$\alpha n$ rapidities,
\begin{equation}
\Lambda^{\alpha n,l}(q_{j}) = \Lambda^{\alpha n} (q_{j}) + i\,(n + 1 - 2l)\,u\hspace{0.20cm}{\rm where}\hspace{0.20cm} l = 1,...,n \, ,
\label{complex-rap}
\end{equation}
$j = 1,...,L_{\alpha n}$, $\alpha = \eta, s$, and $n=1,...,\infty$. 

Each set of $l = 1,...,n$ complex rapidities $\Lambda^{\alpha n,l}(q_{j})$ with the same real part $ \Lambda^{\alpha n} (q_{j})$
is associated with the $l = 1,...,n$ $\eta$-spin-singlet pairs $(\alpha = \eta)$ or spin-singlet pairs $(\alpha = s)$ bound for $n>1$
within a neutral composite $\alpha n$ pseudoparticle. Each of such $l=1,...,n$ rapidities actually describes one of the 
$\Pi_{\alpha} = \sum_{n=1}^{\infty}n\,N_{\alpha n}$, Eq. (\ref{sum-Nseta}), spin-singlet pairs $(\alpha =s)$ or 
$\eta$-spin-singlet pairs $(\alpha =\eta)$ of the Bethe state under consideration. The real part $\Lambda^{\alpha n}(q_{j})$ 
is the rapidity functional that as reported above is for each Bethe state the solution of the coupled Eqs. (\ref{Tapco1}) and (\ref{Tapco2}) 
of Appendix \ref{Appendix1}. 

For $n=1$, the rapidity $\Lambda^{\alpha 1,1}(q_{j})$, Eq. (\ref{complex-rap}) for $l=n=1$, refers to a single pair 
and is real. We call {\it unbound spin-singlet pairs} $(\alpha = s)$ and
{\it unbound $\eta$-spin singlet pairs} $(\alpha = \eta)$ of a Bethe state the corresponding $N_{\alpha 1}$ pairs,
each referring to a single $n=1$ pair configuration. Otherwise, the $n>1$ rapidities $\Lambda^{\alpha n,l}(q_{j})$
imaginary part $i\,(n + 1 - 2l)\,u$ of a $u>0$ Bethe state is finite. The corresponding set of $l = 1,...,n$ complex rapidities with the same real part
then describes the binding of the $n>1$ pairs within an {\it $\alpha n$-pair configuration}. Such a configuration describes the internal 
structure of a neutral composite $\alpha n$ pseudoparticle. We call {\it bound spin-singlet pairs} $(\alpha =s)$ and {\it bound $\eta$-spin singlet pairs}
$(\alpha =\eta)$ the $\Pi_{\alpha}-N_{\alpha 1}$ pairs that are bound within $n>1$ $\alpha n$-pair
configurations. All this is again exactly as for the spin $n$-pairs configurations of the spin $1/2$-$XXX$ chain \cite{CP}.

In contrast to that chain, for $n>1$ the imaginary part $i\,(n + 1 - 2l)\,u$
of each set of the $l = 1,...,n$ rapidities with the same real part depends on the interaction $u=U/4t$
and thus vanishes as $u\rightarrow 0$. As discussed in more detail in Appendices \ref{Appendix2} and \ref{Appendix3},
such an unbinding in that limit of the $l = 1,...,n$ pairs within each
$u>0$ $\alpha n$-pair configuration marks the qualitatively different physics of the $U=0$ and $u>0$ quantum
problems, respectively. It is associated with the rearrangement of the $\eta$-spin and spin
degrees of freedom in terms of the noninteracting electrons occupancy configurations that
generate the $U=0$ common eigenstates of the Hamiltonian, momentum operator, and current operator.
Indeed, as the imaginary part $i\,(n + 1 - 2l)\,u$ of each set of $l = 1,...,n$ rapidities, the commutator
of the charge current operator and the 1D Hubbard model Hamiltonian, 
\begin{equation}
\left[\hat{J},\hat{H}\right] = i\,u\,4t^2\sum_{\sigma}\sum_{j=1}^{N_a}[c_{j,\sigma}^{\dag} (c_{j+1,c_{j,\sigma}} - c_{j-1,\sigma})
+ (c_{j+1,\sigma}^{\dag} - c_{j-1,\sigma}^{\dag})c_{j,\sigma}]\,\hat{n}_{j,-\sigma} \, ,
\label{comm-JH}
\end{equation}
also vanishes as $u\rightarrow 0$. 

The form $i\,(n + 1 - 2l)\,u$ of that imaginary part and of that
commutator, Eq. (\ref{comm-JH}), confirms that the $u>0$ physics survives for any arbitrarily small value of $u$.
Indeed, the $l = 1,...,n$ pairs unbinding and commutator $[\hat{J},\hat{H}]$ vanishing occur only in the $u\rightarrow 0$ limit.
The rearrangement of the $\eta$-spin and spin degrees of freedom in terms of the noninteracting electrons 
occupancy configurations that occurs within the unbinding of the $l = 1,...,n$ pairs within each
$u>0$ $\eta n$-pair configuration has most severe consequences on the transport of charge 
at hole concentration $m_{\eta}^z=0$. The effects of the $u\rightarrow 0$ transition 
on the charge dynamic structure factor at $m_{\eta}^z=0$ is a problem addressed in Ref. \cite{PPWSC}.
The mechanisms behind the corresponding qualitatively different types of transport 
associated with the occurrence at hole concentration $m_{\eta}^z=0$ of charge ballistic transport at $U=0$
and its absence found in this paper for $u>0$ is an interesting issue
discussed in Appendix \ref{Appendix2} for $m_{\eta}^z\rightarrow 0$ and $m_{\eta}^z = 0$
and in Appendix \ref{Appendix3} for $m_{\eta}^z\in [0,1]$.

Within the usual TBA notation, the set of $l = 1,...,n$ complex rapidities $\Lambda^{\alpha n,l}(q_{j})$ with the same real part is called
an $\alpha n$ string. Specifically, a charge $\eta n$ string and a spin $s n$ string.
It thus refers to an $\alpha n$-pair configuration involving $l = 1,...,n$ pairs. 
Hence the number $N_{\alpha} = \sum_{n=1}^{\infty}N_{\alpha n}$ of composite $\alpha n$ pseudoparticles of all
$n=1,...,\infty$ branches of a $u>0$ energy and momentum eigenstate equals that of corresponding TBA $\alpha n$-strings of
all lengths $n=1,...,\infty$. Such a number obeys an exact sum rule given by \cite{Carmelo-17},
\begin{eqnarray}
N_{\alpha} & = & \sum_{n=1}^{\infty}N_{\alpha n} = {1\over 2}(L_{\alpha} - N_{\alpha 1}^h)\hspace{0.20cm}{\rm where}\hspace{0.20cm}\alpha = \eta,s \, , 
\nonumber \\
N_{\rm ps} & = & \sum_{\alpha =\eta,s}\sum_{n=1}^{\infty}N_{\alpha n} = {1\over 2}(L - N_{s 1}^h - N_{\eta 1}^h) \, . 
\label{NpsNapsSR}
\end{eqnarray}
Here $N_{\rm ps}$ is the number of both $\alpha =\eta$ and $\alpha =s$ composite $\alpha n$ pseudoparticles 
of all $n=1,...,\infty$ branches also appearing in Eq. (\ref{F-beta}) and $N_{\alpha 1}^h$ is that of $\alpha 1$-band holes, 
Eq. (\ref{N-h-an}) for $\alpha = \eta,s$ and $n =1$. Hence $N_{\rm ps}$ is as well the number of
both $\eta n$-strings and $s n$-strings of all lengths $n=1,...,\infty$. 

The TBA solution performs the electron - rotated-electron unitary transformation. Consistently, 
it accounts for the 1D Hubbard model related symmetries and conserved rotated-electron numbers
through the set of $\alpha =\eta,s$ TBA $\alpha n$ strings of length $n$ numbers $\{N_{\alpha n}\}$ 
and $c$ branch number $N_c$. This follows from the generators that produce all $4^L$ energy
and momentum eigenstates, Eq. (\ref{Gstate-BAstate}), from the electron and thus rotated-electron
vacuum being naturally expressed in terms of the three fractionalized particles operators that
emerge from the rotated-electron three degrees of freedom separation. The latter is associated with 
the two independent $SU(2)$ symmetries and the independent $U(1)$ symmetry in the 
model's global symmetry.

Specifically, the conserved rotated-electron numbers of unoccupied sites 
$N^R_{\eta,+1/2}$ and of doubly occupied sites $N^R_{\eta,-1/2}$ can be expressed in terms of 
energy and momentum eigenstates $\eta$-spin $S_{\eta}$, $\eta$-spin projection $S_{\eta}^z$, 
and set of TBA charge $\eta n$ strings of length $n$ numbers $\{N_{\eta n}\}$ as
$N^R_{\eta,\pm 1/2} = \sum_{n=1}^{\infty}n\,N_{\eta n} + S_{\eta}\mp S_{\eta}^{z}$.
Similarly, the conserved spin projection $\pm 1/2$ rotated-electron number of singly occupied sites 
$N^R_{s,\pm 1/2}$ can be expressed in terms of the energy and momentum eigenstates spin $S_{s}$,
spin projection $S_{s}^z$, and set of TBA spin $s n$ strings of length $n$ numbers $\{N_{a n}\}$ as
$N^R_{s,\pm 1/2} = \sum_{n=1}^{\infty}n\,N_{s n} + S_{s}\mp S_{s}^{z}$.
Furthermore, the eigenvalue $L_{\eta}=L-L_s$ of the 
generator of the $c$ lattice $U(1)$ symmetry group such that
$L_{\eta}= N^R_{\eta,+1/2}+N^R_{\eta,-1/2}$ and $L_{\eta}= L-N^R_{s,+1/2}-N^R_{s,-1/2}$
appears in Eq. (\ref{N-h-an}) within the TBA solution through the numbers $N_c = L - L_{\eta}$
and $N_c^h = L_{\eta}$. 

As for the spin-$1/2$ $XXX$ chain \cite{CPC,CP}, for a large finite system some of the 1D Hubbard model 
$\alpha n$ strings of length $n>1$ deviate from their TBA ideal form, Eq. (\ref{complex-rap}).
As discussed in Appendix \ref{Appendix4}, the effects of such string deviations \cite{Deguchi-00} are in the TL though not important
for the problem studied in this paper.

On the one hand, for $u>0$ the imaginary part of the $n>1$ rapidities with the same real part, Eq. (\ref{complex-rap}), 
describes the binding of the $l = 1,...,n$ pairs within the corresponding $\alpha n$-pair configuration.
On the other hand, in Ref. \cite{Carmelo-17} it is shown that the configuration of
the two rotated spins within each such unbound spin-singlet pair and that of the
two rotated $\eta$-spins within each unbound $\eta$-spin singlet pair has for $u>0$ a binding and
anti-binding character, respectively.

\section{Charge current operator expectation values and useful subspaces}
\label{curr-m-el-exp-val}

Our study of the charge stiffness refers to the hole concentration interval $m_{\eta}^z \in [0,1]$, 
yet the limit of particular interest for the clarification of the main issue under consideration is that of $m_{\eta}^z\rightarrow 0$.
This applies to that stiffness. In most cases the charge properties of physical systems are studied at zero spin density, $m_s^z = -S_s^z/L =0$.
This is why for simplicity in the remaining of this paper we consider the 1D Hubbard model in the $S_s^z =0$ subspace. For
such quantum problem only $\eta$-spin $SU(2)$ symmetry state representations for which $S_{\eta}=0,1,2,...$ 
is an integer are allowed, so that the results presented in this and following sections refer to integer $\eta$-spin values. However, concerning the charge 
quantities studied in the following similar results are obtained within the TL for $\eta$-spin half-integer 
values and $\vert S_s^z\vert=1/2$.

For the 1D Hubbard model in the $S_s^z=0$ subspace one 
has that $M_{s,+1/2} = M_{s,-1/2}$ in Eq. (\ref{Mtotal}), so that $L_{s,+1/2} = L_{s,-1/2} = \Pi_{s}+S_s$ 
where $\Pi_{s} = \sum_{n=1}^{\infty}n\,N_{s n} = (L-L_{\eta}-2S_s)/2$. 
The dimension of $S_s^z =0$ subspaces spanned by states populated by fixed numbers $L_s = L - L_{\eta}$ of 
rotated spins and $L_{\eta} = N_c^h = L-N_c$ of rotated $\eta$-spins 
is given in Eq. (\ref{NNcharge}) of Appendix \ref{Appendix5}.

\subsection{Three exact properties of the charge current operator expectation values}
\label{curr-val}

The following commutators play a major role in our evaluation of the 
charge current operator off-diagonal matrix elements and expectation values
that contribute to the real part of the charge conductivity, Eq. (\ref{sigma}),
\begin{eqnarray}
\left[\hat{J},\hat{S}_{\eta}^{z}\right] & = & 0 \, ;
\hspace{0.5cm}
\left[\hat{J},(\hat{\vec{S}}_{\eta})^2\right] = \hat{J}^{+}\hat{S}_{\eta}^{-} - \hat{S}_{\eta}^{+}\hat{J}^{-} \, ,
\nonumber \\
\left[\hat{J},\hat{S}_{\eta}^{\pm}\right] & = &
\left[\hat{S}_{\eta}^{z},\hat{J}^{\pm}\right] = \pm \hat{J}^{\pm} \, ;
\hspace{0.5cm}
\left[\hat{J}^{\pm},\hat{S}_{\eta}^{\mp}\right] = \pm 2\hat{J} \, .
\label{comm-currents}
\end{eqnarray}
Here as usual, $(\hat{\vec{S}}_{\eta})^2 = (\hat{S}_{\eta}^{z})^2 + {1\over 2}(\hat{S}_{\eta}^{+}\hat{S}_{\eta}^{-}
+ \hat{S}_{\eta}^{-}\hat{S}_{\eta}^{+})$, and the current operators $\hat{J}^{\pm}$ read, 
\begin{equation}
\hat{J}^{+} = i\,2t\sum_{j=1}^{L}(-1)^j\left(c_{j,\downarrow}^{\dag}\,c_{j+1,\uparrow}^{\dag} +
c_{j+1,\downarrow}^{\dag}\,c_{j,\uparrow}^{\dag}\right)\hspace{0.20cm}{\rm and}\hspace{0.20cm}
\hat{J}^{-} = (\hat{J}^{+})^{\dag} \, .
\label{+-currents}
\end{equation}
They are related to the transverse $\eta$-spin current operators 
as $\hat{J}^{\pm,\eta} = (1/2)\, \hat{J}^{\pm}$. 	
The commutators given in Eq. (\ref{comm-currents}) have exactly the same form as those 
associated with the spin current operator and corresponding spin $SU(2)$ symmetry algebra 
operators considered in related studies of the spin-$1/2$ $XXX$ chain spin stiffness \cite{CPC,CP}.

For simplicity, we denote the $\eta$-Bethe states charge currents by
$\langle\hat{J}_{LWS} (l_{\rm r},L_{\eta},S_{\eta},u)\rangle \equiv
\langle  l_{\rm r},L_{\eta},S_{\eta},-S_{\eta},u\vert\hat{J}\vert  l_{\rm r},L_{\eta},S_{\eta},-S_{\eta},u\rangle$
and the charge currents of general $u>0$ energy and momentum eigenstates by
$\langle\hat{J} (l_{\rm r},L_{\eta},S_{\eta},S_{\eta}^z,u)\rangle \equiv
\langle  l_{\rm r},L_{\eta},S_{\eta},S_{\eta}^z,u\vert\hat{J}\vert  l_{\rm r},L_{\eta},S_{\eta},S_{\eta}^z,u\rangle$.
By combining the systematic use of the commutators given in Eq. (\ref{comm-currents}) with the
transformation laws, 
\begin{equation}
\hat{S}_{\eta}^{-}\vert l_{\rm r},L_{\eta},S_{\eta},-S_{\eta},u\rangle = 0\hspace{0.20cm}{\rm and}\hspace{0.20cm}
\hspace{0.5cm} \hat{S}_{\eta}^{+}\vert l_{\rm r},L_{\eta},0,0,u\rangle = \hat{S}_{\eta}^{-}\vert l_{\rm r},L_{\eta},0,0,u\rangle = 0 \, ,
\label{SS0}
\end{equation}
we reach the following general useful result for the
current operator matrix elements between $S_{\eta}^z=0$ energy and momentum eigenstates,
\begin{equation}
\langle  l_{\rm r},L_{\eta},S_{\eta},0,u\vert\hat{J}\vert  l_{\rm r},L_{\eta},S_{\eta}+\delta S_{\eta},0,u\rangle = 0
\hspace{0.20cm}{\rm for}\hspace{0.20cm}\delta S_{\eta} \neq \pm 1 \, .
\label{matrix-0}
\end{equation}
This selection rule is useful for the discussion in Appendix \ref{Appendix2}
of the $S_{\eta}^z=0$ and $T>0$ transition that is found to occur at $U=U_c=0$, similarly to 
the $T=0$ quantum Mott-Hubbard insulator - metal transition. It separates 
two qualitatively different types of finite-temperature charge transport. For $S_{\eta}^z=0$ energy and momentum states whose generation from 
metallic LWSs involves small $\gamma_{\eta}=S_{\eta}+S_{\eta}^z$ 
values, the calculations to reach the result, Eq. (\ref{matrix-0}), are straightforward. They become lengthly as the $\gamma_{\eta}$ 
value increases, yet remain straightforward.

In the following we report three {\it exact} properties that play a major role in our study. The first property refers to the identification 
of the carriers that within the exact rotated-electron representation couple to charge probes. 
The 1D Hubbard model in a uniform vector potential $\Phi/L$ whose Hamiltonian is given
in Eq. (4) of Ref. \cite{Gu-07A} remains solvable by the BA. 
The TBA equations for the model in a uniform vector potential are given in Eq. (9) of that reference.
The only difference relative to the $\Phi=0$ case is the $c$ band and $\eta n$ band momentum values $q_j$ being
shifted to $q_j+\Phi/L$ and $q_j-2n\Phi/L$, respectively, whereas the $sn$ band momentum values
remain unchanged.

Concerning the coupling of the charge degrees of freedom to the vector potential, one finds that the  
$\eta$-Bethe states momentum eigenvalues, $P (\Phi)$, have the general form,
\begin{eqnarray}
P (\Phi/L) = P (0) - (L_{\eta}-\sum_{n}2n\,N_{\eta n}){\Phi\over L} = P (0) - 2S_{\eta}\,{\Phi\over L} 
= P (0) - M_{\eta}\,{\Phi\over L} \, .
\label{PeffU}
\end{eqnarray}
Here the $\Phi=0$ momentum eigenvalue $P (0)$ is given in Eq. (\ref{P}) with $L_{\eta,-1/2}=\Pi_{\eta}$ for the present
$\eta$-Bethe states. The sum rule $\sum_{n=1}^{\infty}2n\,N_{\eta n} = L_{\eta} -2S_{\eta}$ involving the 
number $L_{\eta} -2S_{\eta}$ of paired rotated $\eta$-spins $1/2$ has been used in Eq. (\ref{PeffU}). 
(Such a sum rule follows from that of the corresponding $\eta$-spin singlet pairs, Eq. (\ref{sum-Nseta}) for $\alpha =\eta$.) 

On the one hand, the charge currents of the $\Phi\rightarrow 0$ $\eta$-Bethe states can be derived from the 
$\Phi/L$ dependence of the energy eigenvalues $E(\Phi/L)$ as $\langle \hat{J}\rangle = - d E(\Phi/L)/d(\Phi/L)\vert_{\Phi = 0}$,
as given in Eqs. (\ref{J-partDEF}) and (\ref{jn-fnPSEU}) of Appendix \ref{Appendix1}.
On the other hand, $d P(\Phi/L)/d(\Phi/L)\vert_{\Phi =0}$ gives the number of charge carriers that couple to the vector potential. 
The natural candidates are the numbers $L_{\eta}=N_c^h$ of rotated $\eta$-spins $1/2$. Within the TBA, their translational
degrees of freedom are described by $c$ band and $\eta n$ bands particle-hole processes. The form of the exact momentum eigenvalues, 
Eq. (\ref{PeffU}), reveals that only the $M_{\eta} =2S_{\eta}$ unpaired physical $\eta$-spins $1/2$ contributing to the $\eta$-spin
multiplet configurations couple to the vector potential $\Phi/L$. 
Since the $L_{\eta} -2S_{\eta}$ rotated $\eta$-spins $1/2$ left over are those within the 
$\Pi_{\eta} = (L_{\eta}-2S_{\eta})/2$ {\it neutral} $\eta$-spin singlet pairs, this exact result is physically appealing.
Consistently with results reported in the following, one finds that in the case of general energy and momentum eigenstates, 
$P (\Phi/L)$ rather reads
\begin{eqnarray}
P (\Phi/L) = P (0) - (M_{\eta,+1/2}-M_{\eta,-1/2})\,{\Phi\over L} \, ,
\label{PeffUGEME}
\end{eqnarray}
with $P (0)$ given now by the general expression provided in Eq. (\ref{P}). This reveals as expected that the coupling
to the vector potential of unpaired physical $\eta$-spins $1/2$ with opposite $\eta$-spin projections $\pm 1/2$ 
has opposite sign.

The total flux $-2S_{\eta}\,\Phi=-M_{\eta}\,\Phi$ in Eq. (\ref{PeffU}), has been found 
within the $u\rightarrow\infty$ limit in Ref. \cite{PDSC-00} directly from the 
solution of the TBA equations of 1D Hubbard model in a uniform vector potential, Eq. (9) of Ref. \cite{Gu-07A}.
Since the lattice occupancy spatial distributions of the $M_{\eta} =2S_{\eta}$ unpaired physical $\eta$-spins $1/2$ 
that couple to the vector potential remain invariant under the electron - rotated-electron unitary transformation,
these results hold as well for the whole $u>0$ range, as found here from the use of the momentum
eigenvalues, Eqs. (\ref{PeffU}) and (\ref{PeffUGEME}).	

A second exact property is related to only the $M_{\eta} = 2S_{\eta}$ unpaired physical $\eta$-spins 
$1/2$ coupling to the charge vector potential also holding for non-LWSs, as given in Eq. (\ref{PeffUGEME}).
For a $\eta$-Bethe state carrying an $\eta$-spin current $\langle\hat{J}_{LWS} (l_{\rm r},L_{\eta},S_{\eta},u)\rangle$ all
$M_{\eta} =2S_{\eta}$ unpaired physical $\eta$-spins have projection $+1/2$. The following exact relation 
that refers to the charge current of general energy and momentum eigenstates, Eq. (\ref{Gstate-BAstate}), holds,
\begin{equation}
\langle\hat{J}  (l_{\rm r},L_{\eta},S_{\eta},S_{\eta}^z,u)\rangle = 
\sum_{\sigma =\pm 1/2} j_{\eta,\sigma}\,M_{\eta,\sigma} \, ,
\label{exp-values}
\end{equation}
where the elementary currents $j_{\eta,\pm 1/2}$ are given by,
\begin{equation}
j_{\eta,\pm 1/2} = \pm {\langle\hat{J}_{LWS} (l_{\rm r},L_{\eta},S_{\eta},u)\rangle\over 2S_{\eta}} = 
\pm {\langle\hat{J}_{LWS} (l_{\rm r},L_{\eta},S_{\eta},u)\rangle\over M_{\eta}} \, .
\label{J-eta-spin-spin}
\end{equation}

The exact expression, Eqs. (\ref{exp-values}) and (\ref{J-eta-spin-spin}), is derived by combining the systematic use of the commutators 
given in Eq. (\ref{comm-currents}) with the energy and momentum eigenstates transformation laws under 
the $\eta$-spin $SU(2)$ symmetry operator algebra, Eq. (\ref{SS0}). After a suitable handling of such
an operator algebra and transformation laws involving commutator manipulations, one finds,
\begin{equation}
\langle\hat{J}  (l_{\rm r},L_{\eta},S_{\eta},S_{\eta}^z,u)\rangle = -{S_{\eta}^z\over S_{\eta}}\,
\langle\hat{J}_{LWS} (l_{\rm r},L_{\eta},S_{\eta},u)\rangle \, ,
\label{currents-gen}
\end{equation}
where as in Eq. (\ref{n-etan-s}), $S_{\eta}^z = -S_{\eta} + \gamma_{\eta}$ and $\gamma_{\eta} = 1,...,2S_{\eta}$.
The relation, Eq. (\ref{currents-gen}), can then be exactly rewritten as given in Eq. (\ref{exp-values}).
For non-LWSs whose generation from $\eta$-Bethe states in Eq. (\ref{Gstate-BAstate}) involves small 
$\gamma_{\eta}=S_{\eta}+S_{\eta}^z$ values the calculations to reach the relation, Eq. (\ref{currents-gen}), 
is straightforward, and remains so as the $\gamma_{\eta}$ value increases yet becomes lengthly. 

The exact relation, Eqs. (\ref{exp-values}), (\ref{J-eta-spin-spin}), and (\ref{currents-gen}),
confirms that also for non-LWSs the $M_{\eta} = M_{\eta,+1/2} + M_{\eta,-1/2}$ unpaired physical $\eta$-spins $1/2$
control the $\eta$-spin current values. For each elementary $\eta$-spin flip process generated by application of the off-diagonal 
$\eta$-spin generator ${\hat{S}}_{\eta}^{+}$, Eq. (\ref{S+S+}), (and ${\hat{S}}_{\eta}^{-}=({\hat{S}}_{\eta}^{+})^{\dag}$) 
onto an energy and momentum eigenstate with finite numbers $M_{\eta,+1/2}$ 
and $M_{\eta,-1/2}$, the $\eta$-spin current exactly changes by a {\it current quantum} $2j_{\eta,-1/2}$ (and $2j_{\eta,+1/2}$.) 
Hence each unpaired physical $\eta$-spin with $\eta$-spin projection $\pm 1/2$ carries an elementary current $j_{\eta,\pm 1/2}$,
Eq. (\ref{J-eta-spin-spin}). For a $\eta$-Bethe state one has that $M_{\eta,+1/2}=2S_{\eta}$ and $M_{\eta,-1/2}=0$, so that
$\langle\hat{J}_{LWS} (l_{\rm r},L_{\eta},S_{\eta},u)\rangle = j_{\eta,+1/2} \times M_{\eta}= j_{\eta,+1/2} \times 2S_{\eta}$.

That in the present case of charge only the $M_{\eta} = m_{\eta}\,L$ unpaired physical $\eta$-spins $1/2$ couple to 
the vector potential implies that all $\eta$-spin currents exactly vanish as $m_{\eta}\rightarrow 0$.
This exact property by itself can be used to confirm that within the canonical ensemble 
at fixed value of $S_{\eta}^z$, in the TL, and for nonzero temperatures the charge stiffness 
vanishes as $m_{\eta}^z\rightarrow 0$. In addition, in the case of high temperature $T\rightarrow\infty$
that result is extended in this paper to the grand-canonical ensemble.

The third exact property concerns the processes that contribute to the charge currents $\langle\hat{J}_{LWS} (l_{\rm r},L_{\eta},S_{\eta},u)\rangle$
on the right-hand side of Eq. (\ref{currents-gen}) of general $\eta$-Bethe states described by groups of charge $c$ band real momentum rapidities,
charge $\eta 1$ real rapidities, and $n>1$ charge $\eta n$ complex rapidities. The third property reported in the following is a direct 
consequence on the $\beta = c,\eta n$ band occupancy configurations that within the TBA
describe such $\eta$-spins $1/2$ translational degrees of freedom of only the $M_{\eta}=2S_{\eta}$ unpaired physical 
$\eta$-spins $1/2$ coupling to charge probes.

It is shown in Appendix \ref{Appendix1} that within the TBA the $\eta$-Bethe states charge currents can be written in the TL in terms of 
$c$-band holes and $\eta n$-band holes occupancies as follows,
\begin{equation}
\langle\hat{J}_{LWS} (l_{\rm r},L_{\eta},S_{\eta},u)\rangle = \sum_{j=1}^{L}\,N_c^h (q_j)\,\,J_c^h (q_j)
+ \sum_{n=1}^{\infty}\sum_{j=1}^{L_{\eta n}}\,N_{\eta n}^h (q_j)\,\,J_{\eta n}^h (q_j) \, ,
\label{J-part}
\end{equation}
where the hole current spectra $J_c^h (q_j)$ and $J_{\eta n}^h (q_j)$ read,
\begin{eqnarray}
J_c^h (q_j) & = & - J_c (q_j) = {2t\sin k^c (q_j)\over 2\pi\rho_c (k^c (q_j))} 
\hspace{0.20cm}{\rm for}\hspace{0.20cm}q_j \in [-\pi,\pi] \hspace{0.20cm}{\rm and}
\nonumber \\
J_{\eta n}^h (q_j) & = & - J_{\eta n} (q_j) = 4nt\sum_{\iota=\pm 1}
{\Lambda^{\eta n} (q_j) -i\,\iota n u \over 2\pi\sigma_{\eta n} (\Lambda^{\eta n} (q_j))
\sqrt{1-(\Lambda^{\eta n} (q_j) -i\,\iota n u)^2}}
\hspace{0.20cm}{\rm for}\hspace{0.20cm}q_j \in [-q_{\eta n},q_{\eta n}] \, ,
\label{jn-fn}
\end{eqnarray}
respectively. Here $N_{c}^h (q_j) = 1 - N_{c} (q_j)$, $N_{\eta n}^h (q_j) = 1 - N_{\eta n} (q_j)$,
and the related $c$- and $\eta n$-bands current spectra $J_c (q_j)$ and $J_{\eta n} (q_j)$,
respectively, are given in Eq. (\ref{jn-fnPSEU}) of Appendix \ref{Appendix1}.
Moreover, $q_{\eta n} = \pi\,(L_{\eta n}-1)/L$, Eq. (\ref{qlimits}), and the rapidity momentum functional $k^c (q_j)$ 
and rapidity functionals $\Lambda^{\eta n} (q_j)$ are obtainable for each $\eta$-Bethe state from solution of the TBA equations, 
Eqs. (\ref{Tapco1}) and (\ref{Tapco2}) of Appendix \ref{Appendix1}.
Such equations also involve spin rapidity functionals $\Lambda^{sn} (q_j)$ associated with
distributions $2\pi\sigma_{sn} (\Lambda_j)$, besides
the distributions $2\pi\rho_c (k_j)$ and  $2\pi\sigma_{\eta n} (\Lambda_j)$ explicitly 
appearing in Eq. (\ref{jn-fn}). The general distributions $2\pi\rho_c (k_j)$ and  $2\pi\sigma_{\alpha n} (\Lambda_j)$
are defined in Eq. (\ref{sigmsrhoderiv}) of Appendix \ref{Appendix1}. (The functionals $q^c (k)$ and $q^{\alpha n} (\Lambda)$ 
in that equation stand for the inverse functions of the rapidity momentum functional 
$k^c (q)$ and rapidity functionals $\Lambda^{\alpha n} (q)$, respectively.)

That the lattice occupancy spatial distributions of the $M_{\eta}=2S_{\eta}$ unpaired physical $\eta$-spins $1/2$
that couple to the charge probes remain invariant under the electron - rotated-electron unitary transformation
implies that such $\eta$-spins with $\eta$-spin projection $+1/2$ and $-1/2$ refer for the
whole $u>0$ range to the $\eta$-spin degrees of freedom of original lattice sites unoccupied by bare electrons and
onsite spin-singlet pairs of bare electrons, respectively. Their translational degrees of freedom 
are within the TBA solution described by an average number $2S_{\eta}$ of $c$ band holes out of
that band $N^h_{c} = 2S_{\eta} + \sum_{n=1}^{\infty}2n\,N_{\eta n}$ holes, Eq. (\ref{N-h-an}),
and by an average number $2S_{\eta}$ of holes out of
the $N^h_{\eta n} = 2S_{\eta}+\sum_{n'=n+1}^{\infty}2(n'-n)N_{\eta n'}$ holes, Eq. (\ref{N-h-an})
for $\alpha =\eta$, of each of the $n=1,...,\infty$ $\eta n$ bands for which $N_{\eta n}>0$
in the energy and momentum eigenstates under consideration.

Hence in terms of the exact solution quantum numbers, the local processes that generate the charge currents 
of the energy and momentum eigenstates refer to the relative occupancy configurations of the  $N_c^h = L_{\eta}$ holes and
corresponding $N_c = L-L_{\eta}$ $c$ pseudoparticles and $N_{\eta n}^h$ holes and corresponding 
$N_{\eta n}$ $\eta n$ pseudoparticles in each $\eta n$ band for which $N_{\eta n}>0$. 
Consistently, the charge currents $\langle\hat{J}_{LWS} (l_{\rm r},L_{\eta},S_{\eta},u)\rangle$ on the
right-hand side of the charge current expression, Eq. (\ref{currents-gen}), of general energy and momentum eigenstates 
can alternatively be expressed in terms of $c$-band and $\eta n$-band holes, as given in Eq. (\ref{J-part}), or of
$c$ and $\eta n$ pseudoparticles, Eq. (\ref{J-partDEF}) of Appendix \ref{Appendix1}.

The third exact property refers to a {\it total} and a {\it partial} virtual elementary current cancelling occurring
in the $\beta = c,\eta n$ bands of $S_{\eta}=0$ and $S_{\eta}>0$, respectively, $\eta$-Bethe states for which
$N_{\beta}^h>2S_{\eta}$ and $N_{\beta} >0$. (Unoccupied $\beta$-bands for which $N_{\beta}=0$ do not
contribute to the charge current.) Such a cancelling is encoded within the interplay
of the current expressions, Eq. (\ref{jn-fn}) and Eq. (\ref{J-partDEF}) of Appendix \ref{Appendix1}, 
with the TBA equations, Eqs. (\ref{Tapco1}) and (\ref{Tapco2}) of that
Appendix. It also affects the charge current expression, Eq. (\ref{currents-gen}), of general energy and 
momentum eigenstates, which in the case of non-LWSs involves the charge currents $\langle\hat{J}_{LWS} (l_{\rm r},L_{\eta},S_{\eta},u)\rangle$
of the $\eta$-Bethe states from which such states are generated in Eq. (\ref{Gstate-BAstate}).

On the one hand, $S_{\eta}=0$ $\eta$-Bethe states whose charge current is zero lack unpaired physical 
$\eta$-spins $1/2$ to couple to charge probes. Their numbers of band holes $N_c^h$ and
$N_{\eta n}^h$ are given by $N_{c}^h = L_{\eta} = \sum_{n=1}^{\infty}2n\,N_{\eta n}$ and
$N_{\eta n}^h = \sum_{n'=n+1}^{\infty}2(n'-n)\,N_{\eta n'}$, respectively, as reported in Eq. (\ref{N-h-an}) for
$S_{\eta} = 0$. Consistently with the lack of unpaired physical $\eta$-spins $1/2$, the virtual elementary currents 
carried by a number $\sum_{n=1}^{\infty}n\,N_{\eta n}$ of $c$ band
holes and $\sum_{n'=n+1}^{\infty}(n'-n)\,N_{\eta n'}$ of $\eta n$-bands holes
exactly cancel those carried by an equal number $\sum_{n=1}^{\infty}n\,N_{\eta n}$ of remaining $c$ band
holes and $\sum_{n'=n+1}^{\infty}(n'-n)\,N_{\eta n'}$ of remaining $\eta n$-band holes, respectively. 
Such two sets of $\beta =c,\eta n$ bands holes describe the translational degrees of freedom of two 
corresponding sets of paired rotated $\eta$-spins $1/2$ of opposite $\eta$-spin projection. Indeed, this exact
total elementary currents cancelling involves the opposite $\eta$-spin projections
within each $\eta$-spin singlet pair. As in the case of the unpaired physical $\eta$-spins $1/2$ 
in the $\eta$-spin multiplet configurations that contribute to the charge currents, Eq. (\ref{exp-values}), 
paired rotated $\eta$-spins with opposite $\eta$-spin projection carry virtual elementary currents of opposite sign.

On the other hand, within the $\beta = c,\eta n$ bands of $S_{\eta}>0$ $\eta$-Bethe states for which
$N_{\beta}^h>2S_{\eta}$ and $N_{\beta} >0$ there is a corresponding partial virtual elementary 
current cancellation. For such $c$ and $\eta n$ bands the number of holes, Eq. (\ref{N-h-an}),
are given by $N_{c}^h = L_{\eta} = 2S_{\eta} + \sum_{n=1}^{\infty}2n\,N_{\eta n}$ and
$N_{\eta n}^h = 2S_{\eta}  + \sum_{n'=n+1}^{\infty}2(n'-n)\,N_{\eta n'}$, respectively.
There is in average in these bands a number $2S_{\eta}$ of $\beta$-band holes that describe the translational degrees of freedom of the
$M_{\eta}=2S_{\eta}$ unpaired physical $\eta$-spins $1/2$. Hence their elementary currents contribute to
the $\eta$-Bethe states charge currents, Eq. (\ref{J-part}). The virtual elementary currents carried by average numbers 
$\sum_{n=1}^{\infty}n\,N_{\eta n}$ and $\sum_{n=1}^{\infty}n\,N_{\eta n}$ of two sets of $c$ band holes 
and $\sum_{n'=n+1}^{\infty}(n'-n)\,N_{\eta n'}$ and $\sum_{n'=n+1}^{\infty}(n'-n)\,N_{\eta n'}$ 
of two sets of $\eta n$-bands holes that describe the translational degrees of freedom of two sets of
paired rotated $\eta$-spins $1/2$ of opposite $\eta$-spin projection remain though cancelling each other. 
 
As mentioned above, the $\eta$-Bethe states virtual elementary charge currents cancellation is encoded in the interplay of the 
current expressions, Eqs. (\ref{J-part}) and (\ref{jn-fn}), with the TBA equations, Eqs. (\ref{Tapco1}) and (\ref{Tapco2}) of 
Appendix \ref{Appendix1}. Only within the present exact rotated-electron related representation is that virtual elementary currents 
cancellation described in terms of explicit physical processes. The main role of such virtual elementary current 
cancelling processes is to control the dependence on the density $m_{\eta}$ of unpaired physical $\eta$-spins $1/2$
of the charge currents of the $\eta$-Bethe states. The $m_{\eta}$ dependence of such charge currents is smooth and continuous. 

The virtual elementary charge currents partial cancelling does not occur within $\beta =c,\eta n$ bands 
occupancies for which $N_{\beta}^h = 2S_{\eta} = M_{\eta}$. For such $\beta = c,\eta n$
bands of a $S_{\eta}>0$ $\eta$-Bethe state all their $N_{\beta}^h = 2S_{\eta} = M_{\eta}$ holes fully contribute to charge currents.
Indeed, all such $\beta$-band holes describe the translational degrees of freedom 
of the corresponding $\eta$-Bethe state $M_{\eta} = 2S_{\eta}$ unpaired physical $\eta$-spins.

\subsection{Simplified stiffness expression and subspaces of the fixed-$S_{\eta}^z$ and $S_s^z=0$ subspaces}
\label{SzSS}

The use of the exact relation, Eq. (\ref{currents-gen}), in the charge stiffness expression, Eq. (\ref{DT-thermo}),
leads to the following simplified stiffness expression in terms of only $\eta$-Bethe states current operator 
expectation values that is exact in the TL and valid for $T>0$ and $u>0$,
\begin{equation}
D (T) = {(2S_{\eta}^z)^2\over 2 L T}\,\,\sum_{L_{\eta}=2\vert S_{\eta}^z\vert}^{L}\sum_{S_{\eta}=\vert S_{\eta}^z\vert}^{L_{\eta}/2}\sum_{l_{\rm r}} 
p_{l_{\rm r},L_{\eta},S_{\eta},S_{\eta}^z}{\vert\langle\hat{J}_{LWS} (l_{\rm r},L_{\eta},S_{\eta},u)\rangle\vert^2\over (2S_{\eta})^2} \, .
\label{D-all-T-simp}
\end{equation}
In the present case of the $S_s^z=0$ subspace, the available $\eta$-spin projection absolute values 
are integers, $\vert S_{\eta}^z\vert=0,1,2,...,L/2$. Hence the summations on the right-hand side
of Eq. (\ref{D-all-T-simp}) run over even integers $L_{\eta} = 2\vert S_{\eta}^z\vert, 2\vert S_{\eta}^z\vert +2,
2\vert S_{\eta}^z\vert + 4,...,L$ and integers 
$S_{\eta}=\vert S_{\eta}^z\vert, \vert S_{\eta}^z\vert + 1,\vert S_{\eta}^z\vert + 2,...,L_{\eta}/2$, respectively.

The charge stiffness upper bounds constructed in this paper rely on the use in the general expression,
Eq. (\ref{D-all-T-simp}), of corresponding upper bounds for the absolute values of $\eta$-Bethe states charge currents 
$\langle\hat{J}_{LWS} (l_{\rm r},L_{\eta},S_{\eta},u)\rangle$. 
Such charge currents result from microscopic processes that are actually easiest to be described 
in terms of original lattice occupancy configurations.
Each $c$ pseudoparticle and $\eta n$ pseudoparticle is associated with the charge degrees of freedom of one and $2n$ sites of that
lattice, respectively. We call {\it charge pseudoparticles} to both the $c$ and $\eta n$ pseudoparticles,
their number reading, 
\begin{equation}
N_{\rho} = N_c + N_{\eta} = L - 2S_{\eta} - \sum_{n=1}^{\infty}(2n-1)N_{\eta n} \, .
\label{NchargeP}
\end{equation}

The charge only flows along the original lattice provided that the unpaired physical $\eta$-spins $1/2$ that couple to
charge probes interchange site positions in it with the charge $c$ and $\eta n$ pseudoparticles. 
This occurs upon the latter moving along the original lattice. Hence one can consider that
such charge pseudoparticles, whose current spectra $J_c (q_j)=-J_c^h (q_j)$ and 
$J_{\eta n} (q_j)=-J_{\eta n}^h (q_j)$ are given in Eq. (\ref{jn-fnPSEU}) of Appendix \ref{Appendix1}, play the role of charge carriers.
This is consistent with, for a $\eta$-Bethe state, the charge pseudoparticles 
carrying {\it all} $L-2S_{\eta}$ electronic charges, with each $c$ pseudoparticle and $\eta n$ pseudoparticle carrying 
one and $2n$ such elementary charges, respectively.

Within the canonical ensemble, the general charge stiffness expression, Eq. (\ref{D-all-T-simp}),
refers to one of the fixed-$S_{\eta}^z$ and $S_s^z=0$ subspaces contained in the larger $S_s^z=0$ subspace. 
The hole concentrations of such subspaces belong to the interval, 
\begin{equation}
m_{\eta}^z = - {2S_{\eta}^z\over L} = {M_{\eta,+1/2}-M_{\eta,-1/2}\over L} \in [0,1] \, .
\label{densitiesSz}
\end{equation}	

It is useful for the study of the charge currents and the introduction of suitable upper bounds for their absolute 
values to consider the subspaces contained in each fixed-$S_{\eta}^z$ and $S_s^z=0$ subspace. The definition
of such subspaces requires a careful account for the summations on the right-hand side of Eq. (\ref{D-all-T-simp}).
The summations $\sum_{L_{\eta}=2\vert S_{\eta}^z\vert}^{L}\sum_{S_{\eta}=\vert S_{\eta}^z\vert}^{L_{\eta}/2}$ 
run in that equation over different $\eta$-spin $SU(2)$ multiplet towers that refer to energy and momentum eigenstates with the same
$S_{\eta}^z$ value and different $S_{\eta} = \vert S_{\eta}^z\vert, \vert S_{\eta}^z\vert+1,\vert S_{\eta}^z\vert+2,...,L_{\eta}/2$ values. Their currents, 
$\langle\hat{J} (l_{\rm r},L_{\eta},S_{\eta},S_{\eta}^z,u)\rangle =(-S_{\eta}^z/S_{\eta})\,
\langle\hat{J}_{LWS} (l_{\rm r},L_{\eta},S_{\eta},u)\rangle$, Eq. (\ref{currents-gen}), are for $S_{\eta}>\vert S_{\eta}^z\vert$ expressed
in terms of $\eta$-Bethe states currents, $\langle\hat{J}_{LWS} (l_{\rm r},L_{\eta},S_{\eta},u)\rangle$, whose
$\eta$-spin projection ${S'}_{\eta}^z=-S_{\eta}$ such that $-{S'}_{\eta}^z>-S_{\eta}^z$ is
different from their $\eta$-spin projection $S_{\eta}^z$. 

The $\eta$-spin flip processes that upon successive applications of the
$\eta$-spin $SU(2)$ symmetry off-diagonal generator ${\hat{S}}^{+}_{\eta}$ onto each 
$S_{\eta}>0$ $\eta$-Bethe state generate the non-LWSs in Eq. (\ref{Gstate-BAstate}) only change
the $\eta$-spin projections of the $\eta$-Bethe state $M_{\eta}=2S_{\eta}$ unpaired physical $\eta$-spins $1/2$.
Such processes do not change the $c$ pseudoparticle occupancies, $\eta$-spin-singlet configurations,
and $S_s^z=0$ spin-singlet and spin-multiplet configurations, which remain those of the $\eta$-Bethe state. 
On the one hand, the summations $\sum _{L_{\eta }=2\vert S_{\eta }^z\vert }^{L}\sum _{S_{\eta }=\vert S_{\eta }^z\vert }^{L_{\eta }/2}$
on the right-hand side of Eq. \ref{D-all-T-simp} run over
energy and momentum eigenstates with the same $S_{\eta }^z$ value. On the other hand and
in spite of that, this symmetry invariance allows that the summation $\sum_{l_{\rm r}} 1=d_{\rm subspace}^{LWS} (L_{\eta},S_{\eta})$
where $d_{\rm subspace}^{LWS} (L_{\eta},S_{\eta})$ is the dimension, Eq. (\ref{dsinglet}) of Appendix \ref{Appendix5},
can run over $c$ pseudoparticle occupancy configurations, $\eta$-spin-singlet configurations,
and spin-singlet and spin-multiplet configurations of $\eta$-Bethe states with the same $L_{\eta}$ and $S_{\eta}$ values that have a 
$\eta$-spin projection ${S'}_{\eta}^z=-S_{\eta}$ different from the $\eta$-spin projection $S_{\eta}^z$
of such energy and momentum eigenstates. Indeed the latter configurations are exactly the same as those of the 
corresponding non-LWSs with fixed $-S_{\eta}^z<-{S'}_{\eta}^z$ that contribute to the charge stiffness, Eq. (\ref{D-all-T-simp}).

An exact property reported above of major importance for our study is that 
only the charge degrees of freedom of the $M_{\eta} = 2S_{\eta}$ unpaired 
physical $\eta$-spins whose lattice spatial occupancy distributions remain invariant under the electron - rotated-electron 
unitary transformation couple to a uniform vector potential. It is thus convenient to divide each fixed-$S_{\eta}^z$ 
and $S_s^z=0$ subspace into a set of fixed-$S_{\eta}$ and $S_s^z=0$ subspaces
that we call S$^z$S subspaces, such that $S_{\eta}\geq - S_{\eta}^z$.
Each $\eta$-spin value $S_{\eta}=\vert S_{\eta}^z\vert, \vert S_{\eta}^z\vert+1, \vert S_{\eta}^z\vert+2,...,L_{\eta}/2$
in the summation $\sum_{S_{\eta}=\vert S_{\eta}^z\vert}^{L_{\eta}/2}$ on the right-hand side of Eq. (\ref{D-all-T-simp})
corresponds to one such a S$^z$S subspace. Its dimension corresponds to
the summation $\sum_{L_{\eta}=2\vert S_{\eta}\vert}^{L}\sum_{l_{\rm r}}$ where $\sum_{L_{\eta}=2\vert S_{\eta}\vert}^{L}$
is for $S_{\eta}>\vert S_{\eta}^z\vert$ only a part of the overall summation $\sum_{L_{\eta}=2\vert S_{\eta}^z\vert}^{L}$ in Eq. (\ref{D-all-T-simp})
and $\sum_{l_{\rm r}}$ is a summation that runs over $c$ pseudoparticle occupancy configurations, 
$\eta$-spin-singlet configurations, and $S_s^z=0$ spin-singlet and spin-multiplet configurations. Those are 
associated with spin values $S_s = 0,1,...,(L-L_{\eta})/2$ of $\eta$-Bethe states with the same $L_{\eta}$ and $S_{\eta}$ values. We emphasize that
although such $\eta$-Bethe states have an $\eta$-spin projection ${S'}_{\eta}^z=-S_{\eta}$ different from the 
$\eta$-spin projection $S_{\eta}^z$ of the corresponding non-LWSs, due to the above reported symmetry
invariance their configurations associated with the summation $\sum_{L_{\eta}=2\vert S_{\eta}\vert}^{L}\sum_{l_{\rm r}}$ are identical
to those of the latter states.

Accounting for that symmetry invariance, the S$^z$S subspaces are defined
here as being spanned by $\eta$-Bethe states with fixed density $m_{\eta}=M_{\eta}/L$ of unpaired
physical $\eta$-spins $1/2$ that belongs to the interval,
\begin{equation}
m_{\eta} = {M_{\eta}\over L} = {2S_{\eta}\over L} \in [m_{\eta}^z,1] \, ,
\label{densitieSzS}
\end{equation}	
where the maximum density is reached for states for which $l_{\eta}=L_{\eta}/L=1$. Each density
$m_{\eta}$ in that interval corresponds to one S$^z$S subspace.

Each S$^z$S subspace can be further divided into smaller subspaces spanned by 
$\eta$-Bethe states with fixed total number $L_{\eta}$ of rotated $\eta$-spins. Such
subspaces have $L_{\eta}$ values in the interval $L_{\eta} \in [2S_{\eta},L]$. 
Their dimension corresponds to the above mentioned summation $\sum_{l_{\rm r}}$ that runs over $c$ pseudoparticle 
occupancy configurations, $\eta$-spin-singlet configurations, and spin-singlet and spin-multiplet configurations 
of $\eta$-Bethe states with the same $L_{\eta}$ and $S_{\eta}$ values.
We call them S$^z$SL subspaces. The S$^z$SL subspaces contained in a given S$^z$S subspace 
are thus spanned by $\eta$-Bethe states with fixed densities $l_{\eta}=L_{\eta}/L$ that vary in the interval,
\begin{equation}
l_{\eta} = {L_{\eta}\over L} \in [m_{\eta},1] \, .
\label{densitiesSzSL}
\end{equation}	
Each pair of densities $m_{\eta},l_{\eta}$ in the ranges, Eqs (\ref{densitieSzS}) and (\ref{densitiesSzSL}), respectively, 
refers to one S$^z$SL subspace.

The related dependent densities associated with the numbers $N_c = L - L_{\eta}$ of $c$ pseudoparticles,
$\Pi_{\eta} = (L_{\eta} - 2S_{\eta})/2$ of $\eta$-spin singlet pairs, and $L_s = L - L_{\eta}$ of rotated spins $1/2$,
are also fixed for a S$^z$SL subspace. For different S$^z$SL subspaces, such densities hence vary in the ranges,
\begin{eqnarray}
n_c & = & {N_c\over L} = 1- l_{\eta} \in [0,(1-m_{\eta})] \, ,
\nonumber \\
\pi_{\eta} & = & {\Pi_{\eta}\over L}  = {1\over 2}(l_{\eta} - m_{\eta}) \in [0,(1 - m_{\eta})/2] \, ,
\nonumber \\
l_{s} & = & {L_s\over L} = 1-l_{\eta} \in [0,(1- m_{\eta})] \, .
\label{Reldensities}
\end{eqnarray}	

Each S$^z$SL subspace can be further divided into smaller subspaces we call S$^z$SLN subspaces. 
They are spanned by $\eta$-Bethe states whose total number of $\eta n$ pseudoparticles 
$N_{\eta}=\sum_{n=1}^{\infty}N_{\eta n}\in [0,\Pi_{\eta}]$ is fixed. (That $N_{\eta}=0$ implies that $\Pi_{\eta}=0$.)
The S$^z$SLN subspaces contained in a given S$^z$SL subspace can have densities $n_{\eta}=N_{\eta}/L$ and
$n_{\rho}=N_{\rho}/L$ in the intervals,
\begin{eqnarray}
n_{\eta} & = & {N_{\eta}\over L} \in [0,\pi_{\eta}] = [0,(l_{\eta} - m_{\eta})/2] \, ,
\nonumber \\
n_{\rho} & = & n_c + n_{\eta} \in [(1 - l_{\eta}),(2 - l_{\eta} - m_{\eta})/2] \, .
\label{densitiesSzSLNN}
\end{eqnarray}	

Each S$^z$SLN subspace can be further divided into smaller subspaces we call S$^z$SLN$_{S}$ subspaces. 
They are spanned by $\eta$-Bethe states with a fixed total number $M_s = 2S_s = L - L_{\eta} - 2\Pi_s$ of unpaired physical spins $1/2$.
The related dependent density $\pi_s = \Pi_{s}/L = (1-l_{\eta}-m_s)/2$ of spin-singlet pairs is also fixed for a S$^z$SLN$_{S}$ subspace. 
The S$^z$SLN$_{S}$ subspaces contained in a S$^z$SLN subspace can have densities $m_{s}=M_{s}/L=2S/L$
and $\pi_{s}$ in the ranges,
\begin{eqnarray}
m_s & = & {M_s\over L} = 1 - l_{\eta} - 2\pi_s \in [0,(1 - l_{\eta})] \, ,
\nonumber \\
\pi_{s} & = & {\Pi_{s}\over L}  = (1-l_{\eta} - m_{s})/2 \in  [0,(1 - l_{\eta})/2] \, .
\label{mspsdensities}
\end{eqnarray}	

Each S$^z$SLN$_{S}$ subspace can be further divided into smaller subspaces that we call S$^z$SLN$_{SN}$ subspaces.
They are spanned by $\eta$-Bethe states with a fixed overall number of
$sn$ pseudoparticles $N_{s}=\sum_{n=1}^{\infty}N_{s n}\in [0,\Pi_s]$.
The S$^z$SLN$_{SN}$ subspaces contained in a S$^z$SLN$_{S}$ subspace can have densities $n_s$ in the interval,
\begin{equation}
n_s = {N_s\over L}  \in [0,(1 - l_{\eta}-m_s)/2] \, .
\label{spindensities}
\end{equation}	

Each S$^z$SLN$_{SN}$ subspace can be further divided into smaller subspaces that are spanned by $\eta$-Bethe states with 
fixed numbers of $\eta n$ and $sn$ pseudoparticles for all $n=1,...,\infty$ branches. (For the subspaces of more interest for 
our study these numbers are finite only for a finite number of $n=1,...,\infty$ branches.)

Within the above notations used in this paper to designate the subspaces contained in each fixed-$S_{\eta}^z$ and 
$S_s^z=0$ subspace of more interest for its studies, S$^z$, S, and L refer to the charge conserved numbers 
$S_{\eta}^z$, $S_{\eta}$, and $L_{\eta}$, respectively, whereas $_S$ and $_N$ refer to the spin conserved numbers 
$S_s$ and $N_s$, respectively. The designations S$^z$S,\, S$^z$SL,\, S$^z$SLN,\, S$^z$SLN$_{S}$, and S$^z$SLN$_{SN}$ 
only include the corresponding subset of these numbers that are fixed for the subspaces under consideration.

\section{Useful current absolute values upper bounds}
\label{UecUB}

The upper bound procedures of our study are initiated in this section. Specifically,
the charge currents $\langle\hat{J}_{LWS} (l_{\rm r},L_{\eta},S_{\eta},u)\rangle$, Eq. (\ref{J-part}), of $\eta$-Bethe states 
in the stiffness expression, Eq. (\ref{D-all-T-simp}), with largest absolute values are identified.
First, the type of $c$ and $\eta n$ bands occupancy configurations that maximize such
absolute values is considered. The second issue addressed in the following is that of
the largest charge current absolute value of the reference S$^z$SLN$_{SN}$ subspaces contained  
in each S$^z$S subspace spanned by $\eta$-Bethe states as defined in Section \ref{SzSS}.
Finally, useful further information about the charge currents of the 
selected reference S$^z$SLN$_{SN}$ subspace is provided.

\subsection{Compact $c$ and $\alpha n$ bands occupancy configurations}
\label{Compactocc}

It is straightforward to confirm from manipulations of the TBA equations, Eqs. (\ref{Tapco1}) and (\ref{Tapco2}) of 
Appendix \ref{Appendix1}, general $\eta$-Bethe-states charge current expression, Eq. (\ref{J-part}), and corresponding $c$ and $\eta n$-band
holes current functional spectra, Eq. (\ref{jn-fn}), that for S$^z$SL subspaces
the class of $\eta$-Bethe states that reach the largest current absolute values $\vert\langle\hat{J}_{LWS} (l_{\rm r},L_{\eta},S_{\eta},u)\rangle\vert$ have 
asymmetrical compact hole $\beta =c,\eta n$ band distributions for $N_{\beta}^h<N_{\beta}$ and
asymmetrical compact pseudoparticle $\beta =c,\eta n$ band distributions for $N_{\beta}^h>N_{\beta}$.
Such a general charge current expression, Eq. (\ref{J-part}), does not directly depend on the type of 
$s n$ band distributions. In general it rather depends on the corresponding densities $m_s$ and $n_s$ through 
the dependence on it of the $c$ and $\eta n$-band holes current functional spectra, Eq. (\ref{jn-fn}).
Hence for simplicity we consider in general symmetrical compact $sn$ bands distributions. 

The general form of the general compact $\beta =c,\eta n,sn$ bands distributions of 
the class of $\eta$-Bethe states with largest current absolute values is thus,
\begin{eqnarray}
& & {\rm For}\hspace{0.20cm}N_{\beta}\leq N_{\beta}^h\hspace{0.20cm}{\rm where}\hspace{0.20cm}\beta = c,\eta n:
\nonumber \\
N_{\beta,A} (q_j) & = & 0 \hspace{0.20cm}{\rm and}\hspace{0.20cm} N_{\beta,A}^h (q_j) = 1
\hspace{0.20cm}{\rm for}\hspace{0.20cm}q_j \in [q_{\beta}^-,q_{F\beta,+}^{-}]
\hspace{0.20cm}{\rm and}\hspace{0.20cm}q_j \in [q_{F\beta,+}^{+},q_{\beta}^+]
\nonumber \\
N_{\beta,A} (q_j) & = & 1 \hspace{0.20cm}{\rm and}\hspace{0.20cm} N_{\beta,A}^h (q_j) = 0
\hspace{0.20cm}{\rm for}\hspace{0.20cm}q_j \in [q_{F\beta,+}^{-},q_{F\beta,+}^{+}]\hspace{0.20cm} 
\nonumber \\
& & {\rm For}\hspace{0.20cm}N_{\beta}^h\leq N_{\beta}\hspace{0.20cm}{\rm where}\hspace{0.20cm}\beta = c,\eta n:
\nonumber \\
N_{\beta,A} (q_j) & = & 1 \hspace{0.20cm}{\rm and}\hspace{0.20cm} N_{\beta,A}^h (q_j) = 0
\hspace{0.20cm}{\rm for}\hspace{0.20cm}q_j \in [q_{\beta}^-,q_{F\beta,-}^{-}]
\hspace{0.20cm}{\rm and}\hspace{0.20cm}q_j \in [q_{F\beta,-}^{+},q_{\beta}^+]
\nonumber \\
N_{\beta,A} (q_j) & = & 0 \hspace{0.20cm}{\rm and}\hspace{0.20cm} N_{\beta,A}^h (q_j) = 1
\hspace{0.20cm}{\rm for}\hspace{0.20cm}q_j \in [q_{F\beta,-}^{-},q_{F\beta,-}^{+}] 
\nonumber \\
& & {\rm For}\hspace{0.20cm} sn \hspace{0.20cm}{\rm bands}:
\nonumber \\
N_{sn,S} (q_j) & = & 1 \hspace{0.20cm}{\rm for}\hspace{0.20cm}q_j \in [q_{Fsn}^{-},q_{Fsn}^{+}]
\hspace{0.20cm}{\rm otherwise}\hspace{0.20cm}N_{sn,S} (q_j)=0 \, .
\label{NN}
\end{eqnarray}
The two limiting occupancy momenta of each band are related to each other and
are such that,
\begin{eqnarray}
q_{F\beta,+}^{-} & \in & [q_{\beta}^-,q_{\beta}^+ - 2\pi n_{\beta}] \, ,
\nonumber \\
q_{F\beta,+}^{+} & = & q_{F\beta,+}^{-} + 2\pi n_{\beta}\hspace{0.20cm}{\rm for}
\hspace{0.20cm}N_{\beta}\leq N_{\beta}^h
\nonumber \\
q_{F\beta,-}^{-} & \in & [q_{\beta}^-,q_{\beta}^+ - 2\pi n_{\beta}^h] \, ,
\nonumber \\
q_{F\beta,-}^{+} & = & q_{F\beta,+}^{-} + 2\pi n_{\beta}^h\hspace{0.20cm}{\rm for}
\hspace{0.20cm}N_{\beta}^h\leq N_{\beta}\hspace{0.20cm}{\rm where}\hspace{0.20cm}\beta = c, \eta n 
\nonumber \\
q_{Fsn}^{\pm} & = & \pm \pi (N_{sn}-1)/L \approx \pm \pi n_{sn} \, .
\label{qFcetanpms}
\end{eqnarray}

The use of both the compact momentum distributions of general form, Eq. (\ref{NN}),
and of the distributions $2\pi\rho_c (k) = \partial q^c (k)/\partial k$
and $2\pi\sigma_{\alpha n} (\Lambda) = \partial q^{\alpha n} (\Lambda)/\partial \Lambda$,
Eq. (\ref{sigmsrhoderiv}) of Appendix \ref{Appendix1}, in the general current expression, Eqs. (\ref{J-part}) and (\ref{jn-fn}),
straightforwardly leads to the following simplified form of the charge currents of the $\eta$-Bethe states 
associated with such compact momentum distributions,
\begin{equation}
\langle\hat{J}_{LWS} (l_{\rm r},L_{\eta},S_{\eta},u)\rangle = {L\,t\over\pi}\sum_{\iota=\pm}
(\iota) \left(\tau\cos k^c (q_{Fc,\tau}^{\iota}) + \sum_{n=1}^{\infty} \tau_n\,2n\sum_{\iota'=\pm 1}
\sqrt{1-(\Lambda^{\eta n} (q_{F\eta n,\tau}^{\iota}) -i\,\iota' n u)^2}\right) \, .
\label{J-comp}
\end{equation}
The indices, 
\begin{eqnarray}
\tau & = & + \hspace{0.20cm}{\rm for}\hspace{0.20cm}N_{c}\leq N_{c}^h
\nonumber \\
& = & - \hspace{0.20cm}{\rm for}\hspace{0.20cm}N_{c}^h\leq N_{c} \, ,
\nonumber \\
\tau_n & = & + \hspace{0.20cm}{\rm for}\hspace{0.20cm}N_{\eta n}\leq N_{\eta n}^h
\nonumber \\
& = & - \hspace{0.20cm}{\rm for}\hspace{0.20cm}N_{\eta n}^h\leq N_{\eta n} \, ,
\label{iotabeta}
\end{eqnarray}
refer here to the $\beta =c,\eta n$ bands particle-like and hole-like asymmetric compact distributions
of such $\eta$-Bethe states. The simplified current expression, Eq. (\ref{J-comp}), involves
the $\beta =c,\eta n$ rapidity functional at merely the two occupancy limiting momenta $q_{F\beta,\tau}^{\pm}$, Eq. (\ref{qFcetanpms}).

Another type of compact distributions considered in our study refers to $\eta$-Bethe states for which they are symmetrical
for all $\beta =c,\eta n,sn$ branches,
\begin{equation}
N_{\beta,S} (q_j) = 1 \hspace{0.20cm}{\rm for}\hspace{0.20cm}q_j \in [q_{F\beta}^{-},q_{F\beta}^{+}]
\hspace{0.20cm}{\rm otherwise}\hspace{0.20cm}N_{\beta,S} (q_j) = 0 
\hspace{0.20cm}{\rm where}\hspace{0.20cm}q_{F\beta}^{\pm} = \pm \pi n_{\beta} \, ,
\label{CompSymm}
\end{equation}
where that $q_{F\beta}^{\pm} = \pm \pi n_{\beta}$ holds in the TL upon ignoring $\pi/L$ corrections.

Useful quantities are the $\beta =c, \eta n$ bands holes elementary currents $j_{\beta }^h (q_j)$ and 
$\beta =c, \eta n$ pseudoparticle elementary currents $j_{\beta } (q_j)=-j_{\beta }^h (q_j)$ of a $\eta$-Bethe state generated from 
a reference $\eta$-Bethe state with compact distributions of form Eq. (\ref{NN}) or (\ref{CompSymm})
by small $\beta =c, \eta n$ band distribution deviations. They are defined as the deviations in the charge current 
$\langle\hat{J}_{LWS} (l_{\rm r},L_{\eta},S_{\eta},u)\rangle$, Eq. (\ref{J-part}), upon addition onto a reference $\eta$-Bethe state 
with compact distributions of form, Eq. (\ref{NN}), of one $\beta$-band hole of momentum $q_j$ and 
one $\beta$ pseudoparticle of momentum $q_j$, respectively.
Relying on techniques similar to those used in Ref. \cite{Carmelo-92-C} for the excited $\eta$-Bethe states
of a ground state, one finds that such $\beta =c, \eta n$ elementary currents read,
\begin{eqnarray}
j_c^h (q_j) & = & - j_c (q_j) = v_{c} (q_j) + {1\over 2\pi}\sum_{\iota =\pm} (\iota) \left(\tau\,f_{c\,c} (q_j,q_{Fc,\tau}^{\iota})
+ \sum_{n=1}^{\infty} \tau_{n}\,2n\,f_{c\,\eta n} (q_j,q_{F\eta n,\tau_{n}}^{\iota})\right) \, ,
\nonumber \\
j_{\eta n}^h (q_j) & = & -j_{\eta n} (q_j) = - 2n\,v_{\eta n} (q_j) -  
{1\over 2\pi}\sum_{\iota =\pm} (\iota) \left(\tau\,f_{\eta n\,c} (q_j,q_{Fc,\tau}^{\iota})
+ \sum_{n'=1}^{\infty}\tau_{n'}\,2n'\,f_{\eta n\,\eta n'} (q_j,q_{F\eta n',\tau_{n'}}^{\iota})\right) \, .
\label{jnvf}
\end{eqnarray}
The expressions of the $\beta =c, \eta n$  $f_{\beta\,\beta'}$ functions and group velocities $v_{\beta} (q_j)$ appearing here are 
defined in Eqs. (\ref{ff}) and (\ref{vel-beta}) of Appendix \ref{Appendix6}, respectively. In that Appendix all quantities
involved in such expressions are also defined.

For a given $\eta$-Bethe state, the $\beta =c, \eta n$ bands holes current spectra $J_{\beta}^h (q_j)=-J_{\beta} (q_j)$, Eq. (\ref{jn-fn}),
and the $\beta =c, \eta n$ bands holes elementary currents $j_{\beta}^h (q_j)=-j_{\beta} (q_j)$, Eq. (\ref{jnvf}),
are related yet in general different quantities. Indeed, they are generated from the
different energy spectra $E_{\beta} (q_j)$, Eq. (\ref{spectra-E-an-c-0}) of Appendix \ref{Appendix1}, and 
$\varepsilon_{\beta} (q_j) = E_{\beta} (q_j) + \varepsilon_{\beta}^c (q_j)$, Eq. (\ref{epsilon-q}) of Appendix \ref{Appendix6}, 
respectively. Therefore, $j_{\beta }^h (q_j)$ can be written as $j_{\beta }^h (q_j) = J_{\beta}^h (q_j) + \delta J_{\beta}^h (q_j)$
where $\delta J_{\beta}^h (q_j)$ is a well-defined quantity that vanishes in some finite-$u$ subspaces and
more generally for $u\rightarrow\infty$ . 

While the $\beta =c, \eta n$ band current spectra $J_{\beta}^h (q_j)=-J_{\beta} (q_j)$, Eq. (\ref{jn-fn}), 
refer to the charge current of a $\eta$-Bethe state, Eq. (\ref{J-part}), the $\beta =c, \eta n$ band elementary currents
$j_{\beta}^h (q_j)=-j_{\beta} (q_j)$, Eq. (\ref{jnvf}), are associated with the difference of the charge currents of
two $\eta$-Bethe states whose $\beta =c, \eta n$ bands occupancies differ in the TL only in those of a
finite number of $\beta =c, \eta n$ pseudoparticles.
In terms of $\beta =c, \eta n$ pseudoparticle elementary currents such current deviations read,
\begin{equation}
\delta\langle\hat{J}_{LWS} (l_{\rm r},L_{\eta},S_{\eta},u)\rangle = 
\sum_{j=1}^{L}\,\delta N_c (q_j)\,\,j_c (q_j)
+ \sum_{n=1}^{\infty}\sum_{j=1}^{L_{\eta n}}\,\delta N_{\eta n} (q_j)\,\,j_{\eta n} (q_j) \, .
\label{deltaJ-part}
\end{equation}
In the present case this refers within the TL to the charge current deviation of
a given $\eta$-Bethe state relative to that of the $\eta$-Bethe state with 
compact $\beta =c, \eta n$ bands distributions from which it is generated 
by a finite number of $\beta =c, \eta n$ pseudoparticle processes.

It follows from the exact properties considered in Section \ref{curr-val} that the charge currents 
vanish both in the $M_{\eta}=2S_{\eta}\rightarrow 0$ and $N_{\rho} = (N_c + N_{\eta})\rightarrow 0$ 
limits, respectively. That $N_{\rho} = (N_c + N_{\eta})\rightarrow 0$ and thus
$N_c + \sum_{n=1}^{\infty}N_{\eta n}\rightarrow 0$ implies that
$N_c + \sum_{n=1}^{\infty}2n\,N_{\eta n}\rightarrow 0$ and thus that
$N_c + 2\Pi_{\eta}=(L-2S_{\eta})\rightarrow 0$. The $\eta$-Bethe states corresponding
to these limits have compact distributions. We consider two types of states. Namely, $\eta $-Bethe 
states that are generated from $m_{\eta }\rightarrow 0$
states by creation of a finite number of unpaired physical $\eta $-spins $1/2$. Moreover,
$\eta $-Bethe states that are generated from $l_{\eta }\rightarrow 1$ states and thus
$m_{\eta }\rightarrow 1$ states by creation of a finite number of charge pseudoparticles. 
One finds within the TL from the use of Eq. (\ref{deltaJ-part}) that the charge currents
of both such two types of states can be written as,
\begin{equation}
\langle\hat{J}_{LWS} (l_{\rm r},L_{\eta},S_{\eta},u)\rangle = \sum_{j=1}^{L}\,N_c (q_j)\,\,j_c (q_j)
+ \sum_{n=1}^{\infty}\sum_{j=1}^{L_{\eta n}}\,N_{\eta n} (q_j)\,\,j_{\eta n} (q_j) \, .
\label{deltaJ-partlim}
\end{equation}
Hence within the TL this charge current expression is valid for both $\eta$-Bethe states
for which (i) $m_{\eta}\ll 1$ and (ii) $m_{\eta}\rightarrow 1$ provided that $l_{\eta}\rightarrow 1$, respectively. Its
validity implies that the $q_j$ sums in Eq. (\ref{J-partDEF}) of Appendix \ref{Appendix1} and Eq. (\ref{deltaJ-partlim}), 
respectively, lead to exactly the same charge current. It does not imply though that the $\beta =c, \eta n$ pseudoparticle current 
spectra $J_{\beta} (q_j)$, Eq. (\ref{jn-fnPSEU}) of Appendix \ref{Appendix1}, and $\beta =c, \eta n$ pseudoparticle elementary currents 
$j_{\beta } (q_j)=-j_{\beta }^h (q_j)$, Eq. (\ref{jnvf}), in these sums, respectively, are equal. 

The ground state associated with each canonical ensemble is not populated by $\eta n$ pseudoparticles and $s n$ pseudoparticles with
$n>1$ spin-singlet pairs. The corresponding distributions refer to a particular case of those
given in Eq. (\ref{CompSymm}). Within the TL, it has $c$ and $s1$ bands compact and symmetrical 
distributions,
\begin{eqnarray}
N_{c}^{GS} (q_j) & = & 1 \hspace{0.20cm}{\rm for}\hspace{0.20cm}q_j \in [q_{Fc}^-,q_{Fc}^+]
\hspace{0.20cm}{\rm otherwise}\hspace{0.20cm}N_{\beta}^A (q_j)=0 
\nonumber \\
N_{s1}^{GS} (q_j) & = & 1 \hspace{0.20cm}{\rm for}\hspace{0.20cm}q_j \in [q_{Fs1}^-,q_{Fs1}^+]
\hspace{0.20cm}{\rm otherwise}\hspace{0.20cm}N_{\beta}^A (q_j)=0 \, .
\label{NGS}
\end{eqnarray}
where, except for $\pi/L$ corrections, $q_{Fc}^{\pm} = \pm 2k_F$ and 
$q_{Fs1}^{\pm }=\pm k_F$ for $m_{\eta}^z\geq 0$ and $m_s^z=0$.

\subsection{The reference subspaces largest charge current absolute value}
\label{SzSLNJmax}

The $T>0$ charge stiffness expression, Eq. (\ref{D-all-T-simp}), depends on the
charge currents of $\eta$-Bethe states belonging to S$^z$S subspaces. For each fixed density 
$m_{\eta}$ in the range $m_{\eta} \in [\vert m_{\eta}^z\vert,1]$ there is a large number of S$^z$SLN$_{S}$ subspaces as
defined in Section \ref{SzSS}. They are spanned by a set of $\eta$-Bethe states with fixed values of $l_{\eta}$,
$n_{\eta}$, $n_s$, and $m_s$ in the intervals $l_{\eta} \in [m_{\eta},1]$,
$n_{\eta} \in [0,(l_{\eta} - m_{\eta})/2]$, $n_s \in [0,(1-l_{\eta})/2]$, and $m_s \in [0,(1-l_{\eta}-n_s)]$, respectively.

The use of the simplified current expression, Eq. (\ref{J-comp}), of the $\eta$-Bethe states with 
compact distributions, Eq. (\ref{NN}), plays a key role in the present analysis. From it
one finds that each S$^z$SLN$_{S}$ subspace largest charge 
current absolute value $\vert\langle\hat{J}_{LWS}^{\rm max}(l_{\rm r},L_{\eta},S_{\eta},u)\rangle\vert$ 
has the general form,
\begin{equation}
\vert\langle\hat{J}_{LWS}^{\rm max}(l_{\rm r},L_{\eta},S_{\eta},u)\rangle\vert_{l_{\eta},n_{\eta},m_{s},n_{s}} 
= C_{l_{\eta},n_{\eta},m_{s},n_{s}}\,t\,L\,m_{\eta}\,(1-m_{\eta}) \, ,
\label{JImax}
\end{equation}
where the coefficient $C_{l_{\eta},n_{\eta},m_{s},n_{s}}$ depends on $u$ and on the densities $l_{\eta}$,
$n_{\eta}$, $m_s$, and $n_s$.

A S$^z$SL subspace contains a set of S$^z$SLN subspaces, one for each density $l_{\eta}$ in the interval $l_{\eta} \in [m_{\eta},1]$.
Within the procedures used in this paper to derive suitable upper bounds, it is convenient to consider three
limiting reference S$^z$SLN subspaces, which we call reference S$^z$SLN subspace 1, 2, and 3, respectively. 
On the one hand, each S$^z$SL subspace only contains one reference S$^z$SLN subspace 1 for which $l_{\eta}\rightarrow m_{\eta}$ and 
thus $n_{\eta}\rightarrow 0$. On the other hand, it contains a set of S$^z$SLN subspaces for which $l_{\eta}\rightarrow 1$,
each corresponding to a fixed $n_{\eta}$ density in the interval $n_{\eta}\in [0,(1 - m_{\eta})/2]$. Out of those, it only contains one 
reference S$^z$SLN subspace 2 and one reference S$^z$SLN subspace 3 for which $n_{\eta}\rightarrow (1 - m_{\eta})/2$
and $n_{\eta}\rightarrow 0$, respectively.

Since $l_{\eta}\rightarrow 1$ implies that $l_s \rightarrow 0$ and thus that $m_s \rightarrow 0$ and $n_s \rightarrow 0$,
the reference S$^z$SLN subspaces 2 and 3 only contain one S$^z$SLN$_{SN}$ subspace each, which we call reference
S$^z$SLN$_{SN}$ subspaces 2 and 3, respectively. Indeed, they are at the same time S$^z$SLN subspaces, 
S$^z$SLN$_{S}$ subspaces, and S$^z$SLN$_{SN}$ subspaces. In contrast, a reference S$^z$SLN subspace 1 contains 
a set of S$^z$SLN$_{S}$ subspaces, one for each density $m_s$ in the interval $m_s \in [0,(1 - m_{\eta})]$. Furthermore, 
inside each of the latter subspaces there is in general a set of S$^z$SLN$_{SN}$ subspaces for each density $n_s$ in the range 
$n_s \in [0,(1 - m_{\eta}-m_s)/2]$. The reference S$^z$SLN$_{SN}$ subspaces 1A and 1B considered here 
have a fixed density $m_s$ in the interval $m_s \in [0,(1 - m_{\eta})]$ and  a maximum and a minimum density 
$n_s\rightarrow (1 - m_{\eta}-m_s)/2$ and $n_s\rightarrow 0$, respectively.

The $\beta =c,\eta n$ bands for which $N_{\beta}>0$ of the $\eta$-Bethe states that 
span the four reference S$^z$SLN$_{SN}$ subspaces 1A, 1B, 2, and 3 under consideration have occupancies such 
that $N_{\beta}^h = 2S_{\eta} = m_{\eta} L$. Hence the 
contributions to the charge current of such $\beta =c,\eta n$ bands are free of virtual elementary charge currents cancelling, which 
much simplifies the calculation of that current. 
The main effect of the virtual elementary current cancelling processes 
occurring for $u>0$ in the remaining set of S$^z$SLN$_{SN}$ subspaces of a S$^z$S subspace
corresponding to intermediate values of the densities $l_{\eta} \in [m_{\eta},1]$, 
$n_{\eta} \in [0,(l_{\eta} - m_{\eta})/2]$, $m_s \in [0,(1-l_{\eta})]$, and $n_s \in [0,(1-l_{\eta}-m_s)/2]$ 
that label these subspaces is that the corresponding set of largest charge current absolute values, Eq. (\ref{JImax}), 
are in the TL continuous functions of such densities. They smoothly vary between the largest charge current absolute values
of the reference S$^z$SLN$_{SN}$ subspaces 1A, 1B, 2, and 3 considered here.

For such limiting reference subspaces the use of the TBA equations, Eqs. (\ref{Tapco1}) and (\ref{Tapco2}) of Appendix \ref{Appendix1}, allows the
derivation of the general simplified current expression, Eq. (\ref{J-comp}), for $\eta$-Bethe states with $\beta = c,\eta n,sn$ bands 
compact distributions of the general form, Eq. (\ref{NN}). Often such current expressions have 
though only simple analytical form in the $u\rightarrow 0$ and $u\gg 1$ limits. 
Within the set of limiting reference S$^z$SLN$_{SN}$ subspaces 1A, 1B, 2, and 3, the problem is most
complex for the reference S$^z$SLN$_{SN}$ subspace 1A, its analysis being addressed in more
detail below in Section \ref{SzSLN1}. In addition to the direct use of the equivalent charge current 
$\langle\hat{J}_{LWS} (l_{\rm r},L_{\eta},S_{\eta},u)\rangle$ expressions,
Eq. (\ref{J-comp}) and Eq. (\ref{J-partDEF}) of Appendix \ref{Appendix1}, to derive the coefficient $C_{l_{\eta},n_{\eta},m_{s},n_{s}}$ 
in Eq. (\ref{JImax}), one can use its limiting expressions, Eq. (\ref{deltaJ-partlim}), which are valid
in the $m_{\eta}\ll 1$ limit and in the $(1-m_{\eta})\ll 1$ limit provided that $l_{\eta}\rightarrow 1$. 

For a reference S$^z$SLN$_{SN}$ subspace 1A of a S$^z$S subspace as defined above one has that
$n_c \rightarrow (1-m_{\eta})$, $n_{\eta n}\rightarrow 0$ for $n=1,...,\infty$, 
$n_{s1} \rightarrow (1- m_{\eta}-m_s)/2$, and $n_{sn}\rightarrow 0$ for $n>1$. Hence the coefficient 
$C_{l_{\eta},n_{\eta},m_{s},n_{s}}$, Eq. (\ref{JImax}), only depends on $u$ and on the subspace fixed densities 
$m_{\eta}$ and $m_{s}$ and is thus called here $C_{m_{\eta},m_{s}}$. As further discussed in 
Section \ref{SzSLN1}, one finds that in this subspace that coefficient has the limiting behaviors,
\begin{eqnarray}
C_{m_{\eta},m_{s}} & = &
4\hspace{0.20cm}{\rm for}\hspace{0.20cm}m_{\eta}\rightarrow 0\hspace{0.20cm}{\rm and}\hspace{0.20cm}m_{s}\rightarrow 0
\nonumber \\
& = & 2\hspace{0.20cm}{\rm for}\hspace{0.20cm}m_{\eta}\in [0,1]\hspace{0.20cm}{\rm and}\hspace{0.20cm}m_{s}\rightarrow 1 - m_{\eta}
\nonumber \\
& = & 2\hspace{0.20cm}{\rm for}\hspace{0.20cm}m_{\eta}\rightarrow 1\hspace{0.20cm}{\rm and}\hspace{0.20cm}m_{s}\rightarrow 0 \, ,
\label{C10lim}
\end{eqnarray}
for $u\rightarrow 0$ and,
\begin{eqnarray}
C_{m_{\eta},m_{s}} & = & 2\left(1+ {1\over 2}\left({\ln 2\over u}\right)^2 + {\cal{O}} (u^{-4})\right) 
\hspace{0.20cm}{\rm for}\hspace{0.20cm}m_{\eta}\rightarrow 0\hspace{0.20cm}{\rm and}\hspace{0.20cm}m_{s}\rightarrow 0
\nonumber \\
& = & 2\left(1+ {1\over 2}\left({1-m_s\over u}\right)^2 + {\cal{O}} (u^{-4})\right) 
\hspace{0.20cm}{\rm for}\hspace{0.20cm}m_{\eta}\rightarrow 0\hspace{0.20cm}{\rm and}\hspace{0.20cm}m_{s}\rightarrow 1 
\nonumber \\
& = & 2\hspace{0.20cm}{\rm for}\hspace{0.20cm}m_{\eta}\rightarrow 1\hspace{0.20cm}{\rm and}\hspace{0.20cm}m_{s}\rightarrow 0 \, ,
\label{C1cuLlimm}
\end{eqnarray}
up to $u^{-3}$ order, which is a good approximation for approximately $u> 3/2$. 
(Corresponding expansions of the coefficient $C_{m_{\eta},m_{s}}$
up to $u^{-3}$ order and valid for the whole $m_{\eta}\in [0,1]$ interval are given in Section \ref{SzSLN1}.)
For $u>0$ and $m_s \in [0,(1 - m_{\eta})]$, the coefficient $C_{m_{\eta},m_{s}}$ smoothly increases upon 
increasing $m_{\eta}$ from $m_{\eta}=0$ until reaching a maximum value at an $u$-dependent 
intermediate density $m_{\eta}$. Upon further increasing $m_{\eta}$, it is a continuous decreasing function of $m_{\eta}$. 
The largest $C_{m_{\eta},m_{s}}$ value refers for any density $m_{\eta}\in [0,1]$ to the reference S$^z$SLN$_{SN}$ subspace 1
for which $m_s\rightarrow 0$.

The reference S$^z$SLN$_{SN}$ subspace 1B of a S$^z$S subspace is such that
$n_c \rightarrow (1-m_{\eta})$, $n_{\eta n}\rightarrow 0$ for $n=1,...,\infty$, 
$N_{sn} = 1$ for $n=(L-2S_{\eta} - 2S_s)/2$,
and $n_{sn'}\rightarrow 0$ for $n'=1,...,\infty$ (including for $n'=n$ within the TL.)
In the case of this subspace, the coefficient $C_{l_{\eta},n_{\eta},m_{s},n_{s}}$, Eq. (\ref{JImax}), 
is for $u>0$, $m_{\eta} \in [0,1]$, and $m_s \in [0,(1-m_{\eta})]$ found to be independent of 
$u$ and $m_s$, so that  it is here denoted by $C_{m_{\eta}}$. It is found to read,
\begin{equation}
C_{m_{\eta}} = {2\over\pi}{\sin (\pi m_{\eta})\over m_{\eta} (1-m_{\eta})}
\hspace{0.20cm}{\rm for}\hspace{0.20cm} m_{\eta} \in [0,1]\hspace{0.20cm}{\rm and}\hspace{0.20cm}m_{s}\in [0,(1 - m_{\eta})] \, .
\label{C2geg}
\end{equation}
It increases and decreases upon increasing $m_{\eta}$ within
the ranges $m_{\eta}\in [0,1/2]$ and $m_{\eta}\in [1/2,1]$, respectively,
reaching a maximum value $8/\pi$ at
$m_{\eta}=1/2$. Its limiting behaviors are,
\begin{eqnarray}
C_{m_{\eta}} & = & 2\hspace{0.20cm}{\rm for}\hspace{0.20cm}m_{\eta}\rightarrow 0
\hspace{0.20cm}{\rm and}\hspace{0.20cm}m_{s}\in [0,1]
\nonumber \\
& = & 8/\pi \hspace{0.20cm}{\rm for}\hspace{0.20cm}m_{\eta} = 1/2
\hspace{0.20cm}{\rm and}\hspace{0.20cm}m_{s}\in [0,1/2]
\nonumber \\
& = & 2\hspace{0.20cm}{\rm for}\hspace{0.20cm}m_{\eta}\rightarrow 1
\hspace{0.20cm}{\rm and}\hspace{0.20cm}m_{s}\rightarrow 0  \, .
\label{C2Llim}
\end{eqnarray}

For the reference S$^z$SLN$_{SN}$ subspace 2  of a S$^z$S subspace 
one has that $n_c \rightarrow 0$, $n_{\eta 1}\rightarrow (1 - m_{\eta})/2$,
$n_{\eta n}\rightarrow 0$ for $n>1$, and $n_{sn}\rightarrow 0$
for $n=1,...,\infty$. The corresponding coefficient $C_{l_{\eta},n_{\eta},m_{s},n_{s}}$, Eq. (\ref{JImax}),
only depends on $u$ and $m_{\eta}$, so that we call it $C_{m_{\eta}}$. Its limiting behaviors
are found to be given by,
\begin{eqnarray}
C_{m_{\eta}} & = & 2\hspace{0.20cm}{\rm for}\hspace{0.20cm}
u\rightarrow 0 \, , \hspace{0.20cm}m_{\eta}\rightarrow 0 \, ,
\hspace{0.20cm}{\rm and}\hspace{0.20cm}m_{s}\rightarrow 0
\nonumber \\
& = & 2\hspace{0.20cm}{\rm for}\hspace{0.20cm}
u\rightarrow 0 \, , \hspace{0.20cm}m_{\eta}\rightarrow 1 \, ,
\hspace{0.20cm}{\rm and}\hspace{0.20cm}m_{s}\rightarrow 0
\nonumber \\
& = & {\pi\over 2u} \hspace{0.20cm}{\rm for}\hspace{0.20cm}
u\gg 1 \, , \hspace{0.20cm}m_{\eta}\rightarrow 0 \, ,
\hspace{0.20cm}{\rm and}\hspace{0.20cm}m_{s}\rightarrow 0
\nonumber \\
& = &  {1\over u} \hspace{0.20cm}{\rm for}\hspace{0.20cm}
u\gg 1 \, , \hspace{0.20cm}m_{\eta}\rightarrow 1 \, ,
\hspace{0.20cm}{\rm and}\hspace{0.20cm}m_{s}\rightarrow 0 \, .
\label{Ceta1}
\end{eqnarray}
For $u>0$ the coefficient $C_{m_{\eta}}$ is a continuous function of 
$m_{\eta}$ with a maximum value at a $u$-dependent 
intermediate density $m_{\eta}$. 

The reference S$^z$SLN$_{SN}$ subspace 3 of a S$^z$S subspace is such 
that $n_c \rightarrow 0$, $N_{\eta n} = 1$ for $n=(L-2S_{\eta})/2$,
$n_{\eta n'}\rightarrow 0$ for $n'=1,...,\infty$ (including for $n'=n$ within the TL),
and $n_{sn}\rightarrow 0$ for $n=1,...,\infty$. The coefficient $C_{l_{\eta},n_{\eta},m_{s},n_{s}}$, Eq. (\ref{JImax}),
again only depends on $u$ and $m_{\eta}$. It is here denoted by $C_{m_{\eta}}$. Its
values in the $u\rightarrow 0$ and $u\gg 1$ limits are for the whole $m_{\eta} \in [0,1]$ range given by,
\begin{eqnarray}
C_{m_{\eta}} & = & 
{2\sin \left({\pi\over 2} m_{\eta}\right)\over m_{\eta}} 
\hspace{0.20cm} {\rm for}\hspace{0.20cm} 
u\rightarrow 0 \, , \hspace{0.20cm} m_{\eta} \in [0,1] \, ,
\hspace{0.20cm}{\rm and}\hspace{0.20cm}m_{s}\rightarrow 0
\nonumber \\
& = & {\cal{O}}(1/L) = 0\hspace{0.20cm} {\rm in}\hspace{0.20cm}\hspace{0.20cm} {\rm the}\hspace{0.20cm}{\rm TL}
\hspace{0.20cm} {\rm for}\hspace{0.20cm} 
u\gg 1 \, , \hspace{0.20cm}m_{\eta} \in [0,1[   \, ,
\hspace{0.20cm}{\rm and}\hspace{0.20cm}m_{s}\rightarrow 0 \, ,
\label{CcfulletanU0}
\end{eqnarray}
respectively. In the $u\rightarrow 0$ limit it is thus a decreasing function of $m_{\eta}$ with limiting values,
\begin{eqnarray}
C_{m_{\eta}} & = & \pi\hspace{0.20cm}{\rm for}\hspace{0.20cm}
u\rightarrow 0 \, , \hspace{0.20cm}m_{\eta}\rightarrow 0 \, ,
\hspace{0.20cm}{\rm and}\hspace{0.20cm}m_{s}\rightarrow 0
\nonumber \\
& = & 2\hspace{0.20cm}{\rm for}\hspace{0.20cm}
u\rightarrow 0 \, , \hspace{0.20cm}m_{\eta}\rightarrow 1 \, , 
\hspace{0.20cm}{\rm and}\hspace{0.20cm}m_{s}\rightarrow 0 \, .
\label{Cetan}
\end{eqnarray}

As mentioned above, the largest charge current absolute value, Eq. (\ref{JImax}), and thus
the corresponding coefficient $C_{l_{\eta},n_{\eta},m_{s},n_{s}}$ of all remaining reference S$^z$SLN$_{SN}$ subspaces	
contained in a S$^z$S subspace is a continuous and smooth function of the densities
$l_{\eta} \in [m_{\eta},1]$, $n_{\eta} \in [0,(l_{\eta} - m_{\eta})/2]$, $m_s \in [0,(1-l_{\eta})]$, and $n_s \in [0,(1-l_{\eta}-m_s)/2]$.
Such a coefficient $C_{l_{\eta},n_{\eta},m_{s},n_{s}}$ varies between the limiting
values of those of the limiting reference S$^z$SLN$_{SN}$ subspaces 1A, 1B, 2, and 3.
From analysis and comparison of the whole corresponding set of largest charge current absolute value 
$\vert\langle\hat{J}_{LWS}^{\rm max}(l_{\rm r},L_{\eta},S_{\eta},u)\rangle\vert$, Eq. (\ref{JImax}),
one finds that the larger coefficients $C_{l_{\eta},n_{\eta},m_{s},n_{s}}$ are, 
on the one hand concerning the density $l_{\eta}\in [m_{\eta},1]$, reached 
for some of the reference S$^z$SLN$_{SN}$ subspaces contained in the reference S$^z$SLN 1
for which $l_{\eta}\rightarrow m_{\eta}$. Concerning the density $n_s \in [0,(1- m_{\eta}-m_s)/2]$ of 
the subset of reference S$^z$SLN$_{SN}$ subspaces contained in the reference S$^z$SLN 1
for which $l_{\eta}\rightarrow m_{\eta}$ and thus $n_{\eta}\rightarrow 0$ and $m_s \in [0,(1-m_{\eta})]$,
one finds in turn that the larger coefficients $C_{l_{\eta},n_{\eta},m_{s},n_{s}}$ in Eq. (\ref{JImax})
are reached for the S$^z$SLN$_{SN}$ subspaces 1A for which $n_s \rightarrow (1- m_{\eta}-m_s)/2$
where $m_s \in [0,(1-m_{\eta})]$.  This largest charge current absolute value of each S$^z$S subspace is thus that used in the upper
bound procedures considered in the following and within the canonical ensemble in Section \ref{two}.
Actually, the absolute largest coefficient $C_{l_{\eta},n_{\eta},m_{s},n_{s}}$ is found to be that of
the S$^z$SLN$_{SN}$ subspace 1A of all S$^z$S subspaces for which $m_s \rightarrow 0$.

In Appendix \ref{Appendix3} the effect of varying $u$ on the physical microscopic processes 
behind the largest charge current absolute value of a S$^z$S subspace reported in this section 
is addressed. (That such an effect is discussed in an Appendix follows from the remaining studies
of this paper accounting for it but not needing its detailed analysis, which requires a relative long account
that would affect the information flow on the main issues addressed in the following. 
The information provided in Appendix \ref{Appendix3} is though physically important, its
presentation in that Appendix contributing to the further understanding of the microscopic
mechanisms behind the 1D Hubbard model charge transport properties.) 

As discussed in Appendix \ref{Appendix2} for $m_{\eta}^z = 0$
and in Appendix \ref{Appendix3} for $m_{\eta}^z\in [0,1]$, the physics is very different (i) at $u=0$ 
and in the $u\rightarrow 0$ limit and (ii) for finite $u$. In the latter Appendix it is shown that due to the 
$u\rightarrow 0$ unbinding of the $\eta$-spin singlet pairs that at finite $u$ are bound within the
composite $\eta n$ pseudoparticles, the charge carriers that interchange position with the $M_{\eta}$
unpaired physical $\eta$-spins $1/2$ that couple to charge probes are different in the $u\rightarrow 0$ limit and for 
finite $u$. Their numbers read $N_c + 2\Pi_{\eta}=L-2S_{\eta}$ and $N_{\rho}=N_c + N_{\eta}$, respectively. 

Another issue discussed in Appendix \ref{Appendix3} refers to the similarities and differences relative to the case 
of the spin stiffness and currents of the spin-$1/2$ $XXX$ chain, which are studied in Refs. \cite{CPC,CP} by the 
upper-bound method used in this paper. There is an advantage of that upper-bound method relative
to more common numerical approximations used to address
the stiffness problem and the charge currents contributing to it. It is that such a method directly refers to a representation 
in terms of which the microscopic mechanisms under consideration in the
discussions of Appendix \ref{Appendix3} refer to elementary processes of the fractionalized 
particles whose configurations generate the {\it exact} energy and momentum eigenstates. 
Indeed, in terms of processes of the physical particles, the electrons, the present quantum problem is non-perturbative, 
so that the microscopic mechanisms that control the charge currents corresponds to a much more complex 
many-particle problem. This is one of the reasons why such mechanisms remain hidden under the use of standard numerical 
techniques, which usually rely on the electron representation of the problem.

Other techniques that rely on the direct use of the BA quantum numbers without accounting for their relation
to the integrable models physical particles \cite{Kawa-98,ANI-05} also pause technical problems. 
This occurs for instance within the use of phenomelogical spinon and antispinon representations of the 
BA quantum numbers whose relation to the physical particles remains undefined \cite{ANI-05}. 
Moreover, in the case of some integrable models such as the present 1D Hubbard model, divergences emerge
at the densities at which the Mazur's inequality is inconclusive in the stiffness expressions obtained from 
the second derivative of the energy eigenvalues relative to the uniform vector potential obtained from
the BA. These divergences can though be avoided. This is accomplished if one rather expresses the stiffness 
in terms of charge current operator expectation values, as within the method used in this paper.

\subsection{The S$^z$S subspaces largest charge current absolute value}
\label{SzSLN1}

In this section we provide further useful information on the largest charge current absolute value
$\vert\langle\hat{J}_{LWS}^{\rm max}(l_{\rm r},S_{\eta},u)\rangle\vert$ of a reference S$^z$SLN$_{SN}$ subspace 1A, 
which for any fixed density $m_s$ is the largest charge current absolute value of the corresponding S$^z$S subspace.
(Here $L_{\eta}$ was removed from $\vert\langle\hat{J}_{LWS} (l_{\rm r},L_{\eta},S_{\eta},u)\rangle\vert$
because $L_{\eta}=2S_{\eta}$ for a reference S$^z$SLN subspace 1 where the S$^z$SLN$_{SN}$ subspaces 1A
are contained.)

For the 1D Hubbard model in a reference S$^z$SLN subspace 1 the general $\eta$-Bethe-state charge 
current expression, Eq. (\ref{J-part}), simplifies to,
\begin{equation}
\langle\hat{J}_{LWS} (l_{\rm r},S_{\eta},u)\rangle = \sum_{j=1}^{L}\,N_c^h (q_j)\,\,J_c^h (q_j)
\hspace{0.20cm}{\rm where}\hspace{0.20cm}
J_c^h (q_j) = {2t\sin k^c (q_j)\over 2\pi\rho_c (k^c (q_j))} 
\hspace{0.20cm}{\rm for}\hspace{0.20cm}q_j \in [-\pi,\pi] \, .
\label{J-partS1}
\end{equation}
Moreover, for $\eta$-Bethe states with $c$ and $sn$ bands compact distributions of general form, Eq. (\ref{NN}), belonging 
to the reference S$^z$SLN subspace 1, the charge current expression, Eq. (\ref{J-comp}),
further simplifies for densities $\vert m_{\eta}^z\vert\in [0,1]$ and $m_{\eta}\in [\vert m_{\eta}^z\vert,1]$, 
\begin{equation}
\langle\hat{J}_{LWS} (l_{\rm r},S_{\eta},u)\rangle = {\tau\,L\,t\over\pi}\sum_{\iota=\pm}
(\iota) \cos k^c (q_{Fc,\tau}^{\iota}) \, .
\label{J-partS1-comp}
\end{equation}
Thus this applies to any S$^z$SLN$_{SN}$ subspace contained in a reference S$^z$SLN subspace 1
whose densities $m_s \in [0,(1-m_{\eta})]$ and $n_s\in [0, (1 - m_{\eta}-m_s)]/2$ have fixed values.

In the following we consider the reference S$^z$SLN$_{SN}$ subspaces 1A of interest for our upper-bound procedures
for which $n_s\rightarrow (1 - m_{\eta}-m_s)/2$ where the density $m_s$ has a fixed value
in the interval $m_s \in [0,(1-m_{\eta})]$. The derivation of the $c$-band current spectrum $J_c^h (q_j)$ in Eq. (\ref{J-partS1}) involves
the solution of the TBA equations, Eqs. (\ref{Tapco1}) and (\ref{Tapco2}), to derive 
the momentum rapidity function $k^c (q_j)$ and related distribution $2\pi\rho_c (k^c (q_j))$
of the $\eta$-Bethe states that span the reference S$^z$SLN subspaces 1A of 
a S$^z$S subspace. This is simplest to be accomplished in terms of $u^{-1}$ expansions
of such a function and distribution, which for densities $m_{\eta}\in [0,1]$ and $m_s \in [0,(1-m_{\eta})]$
leads to the following universal expansion for $J_c^h (q_j)$ up to $u^{-2}$ order,
\begin{equation}
J_c^h (q_j) = 2t \sin q_j - 2t\,{n_{\eta s}\over u}\sin 2q_j + 
6t\,\left({n_{\eta s}\over u}\right)^2\left(1-{3\over 2}\sin^2 q_j\right)\sin q_j \, .
\label{Jchalletas}
\end{equation}
Here,
\begin{eqnarray}
n_{\eta s} & = & (1-m_{\eta}-m_s) \, g_s \hspace{0.2cm}{\rm where}
\nonumber \\
g_s & = & \ln 2 \hspace{0.2cm}{\rm for}\hspace{0.2cm}m_s \rightarrow 0
\hspace{0.2cm}{\rm and}\hspace{0.2cm}
g_s = 1 \hspace{0.2cm}{\rm for}\hspace{0.2cm}m_s \rightarrow 1 - m_{\eta} \, .
\label{Lms}
\end{eqnarray}
$g_s = g_s (m_s) \in [\ln 2,1]$ is in these equations a continuous increasing function of the spin density $m_s$.

For $j>2$ orders $u^{-j}$ the calculations become a more complex technical problem. 
Analysis of the interplay of the TBA equations with those that define the charge operator expectation values 
reveals that the $c$-band current spectrum $J_c^h (q_j)$ expansion terms of such $j>2$ orders
are state dependent. As an example, two current spectra expansions up to $u^{-3}$ order are 
derived in Appendix \ref{Appendix7} for $\eta$-Bethe states that span two reference S$^z$SLN$_{SN}$ subspaces 1A
with limiting spin densities $m_s=0$ and $m_s\rightarrow 1-m_{\eta}$, respectively.
For the former spin density, the obtained expansion refers to $\eta$-Bethe states with compact distributions of general form, 
Eq. (\ref{NN}), belonging to the reference S$^z$SLN$_{SN}$ subspaces 1A under consideration
and $\eta$-Bethe states generated from those by a finite number of $c$-band particle-hole processes. It reads,
\begin{eqnarray}
J_c^h (q_j)  & = & 2t \sin q_j - 2t\,{(1-m_{\eta})\ln 2\over u}\sin 2q_j
+ 6t\,{((1-m_{\eta})\ln 2)^2\over u^2}\left(1-{3\over 2}\sin^2 q_j\right)\sin q_j
\nonumber \\
& - & 4t\,{((1-m_{\eta})\ln 2)^3\over u^3}\left(1-{8\over 3}\sin^2 q_j\right)\sin 2 q_j
\nonumber \\
& + & {3t\zeta (3)\over 16\,u^3}\left((1-m_{\eta})\left(1+{4\over 3}\sin^2 q_j\right)
+ {3\tau\over 2\pi}\sum_{\iota=\pm} (\iota)\cos (q_{Fc,\tau}^{\iota})\left(\sin  q_j
- {1\over 3}\sin (q_{Fc,\tau}^{\iota})\right)\right)\sin 2q_j \, .
\label{Jqugg}
\end{eqnarray}
That its terms up to $u^{-2}$ order and of $u^{-3}$ order do not depend and depend on the limiting momenta
$q_{Fc,\tau}^{\iota}$ associated with the states with compact and asymmetrical $c$-band 
distributions considered here is consistent with their state independence and state dependence, respectively.

The expansion up to $u^{-3}$ order of $J_c^h (q_j)$ obtained in Appendix \ref{Appendix7} for  
the reference S$^z$SLN$_{SN}$ subspaces 1A with spin $m_s\rightarrow 1-m_{\eta}$ is of second 
order in $(1-m_{\eta}-m_s)\ll 1$ and is given by,
\begin{eqnarray}
J_c^h (q_j)  & = & 2t \sin q_j - 2t\,{(1-m_{\eta}-m_s)\over u}\sin 2q_j
+ 6t\,{(1-m_{\eta}-m_s)^2\over u^2}\left(1-{3\over 2}\sin^2 q_j\right)\sin q_j
\nonumber \\
& + & {4t\over 3}\,{(1-m_{\eta}-m_s)\over u^3}\sin^2 q_j\,\sin 2 q_j 
+ {\cal{O}} ((1-m_{\eta}-m_s)^3) \, .
\label{Jquggmsmax}
\end{eqnarray}
The reference S$^z$SLN$_{SN}$ subspaces 1A associated with the $(1-m_{\eta}-m_s)\ll 1$ limit is the only one 
for which the terms up to second order in $(1-m_{\eta}-m_s)$ of $J_c^h (q_j)$ are state independent for any
$u>0$ value. Note that the terms of $u^{-3}$ order in Eqs. (\ref{Jqugg}) and (\ref{Jquggmsmax}),
respectively, have a completely different form for the two reference S$^z$SLN$_{SN}$ subspaces 1A
for which $m_s=0$ and $m_s\rightarrow 1-m_{\eta}$, respectively.

The use in the general expansion, Eq. (\ref{Jchalletas}), of $J_c^h (q_j)$ up to $u^{-2}$ order
valid for the densities intervals $m_{\eta}\in [0,1]$ and $m_s \in [0,(1-m_{\eta})]$
of the reference S$^z$SLN subspaces 1A of the $n_{\eta s} = (1-m_{\eta})\,\ln 2$
and $n_{\eta s} = (1-m_{\eta} -m_s)$ values of the function $g_s$ given in Eq. (\ref{Lms}) specific to its $m_s=0$ and
$m_s\rightarrow 1 - m_{\eta}$ reference S$^z$SLN$_{SN}$ subspaces 1A
recovers the terms up to $u^{-2}$ order in Eqs. (\ref{Jqugg}) and
Eq. (\ref{Jquggmsmax}), respectively. The calculations reported in Appendix \ref{Appendix7} to derive
the expansions given in these equations are more complex for the
$m_s=0$ reference S$^z$SLN$_{SN}$ subspaces 1A than for that for which 
$m_s\rightarrow 1-m_{\eta}$. For the former $m_s=0$ subspace, the distribution $2\pi\rho_c (k)$ in 
the $J_c^h (q_j)$ expression in Eq. (\ref{J-partS1}) and the function $q_c (k)$ that
is the inverse of the momentum rapidity functional $k^c (q)$ also appearing in that
expression are in Appendix \ref{Appendix7} expanded in powers of $u^{-j}$ for {\it all} $j=1,...,\infty$ orders.

From the use of Eq. (\ref{Jchalletas}) in the general expression
for $\langle\hat{J}_{LWS} (l_{\rm r},S_{\eta},u)\rangle$ in Eq. (\ref{J-partS1}) with compact distributions of 
general form, Eq. (\ref{NN}), one finds the following expansion up to $u^{-2}$ order of the 
charge current expression, Eq. (\ref{J-partS1-comp}), valid for the reference S$^z$SLN subspace 1A,
\begin{eqnarray}
\langle\hat{J}_{LWS} (l_{\rm r},S_{\eta},u)\rangle & = & {\tau\,L\,t\over\pi}\sum_{\iota=\pm} (\iota) \cos (q_{Fc,\tau}^{\iota}) 
+ {\tau\,L\,t\over\pi}{n_{\eta s}\over u}\sum_{\iota=\pm} (\iota) \sin^2 (q_{Fc,\tau}^{\iota}) 
\nonumber \\
& - & {\tau\,3\,L\,t\over 2\pi}\left({n_{\eta s}\over u}\right)^2\sum_{\iota=\pm} (\iota) \sin^2 (q_{Fc,\tau}^{\iota}) \cos (q_{Fc,\tau}^{\iota}) \, .
\label{J-partS1-comp3}
\end{eqnarray}
Here $q_{Fc,\tau}^{\iota}$ with $\tau =\pm$ and $\iota=\pm$ are the $c$ band
limiting occupancy momenta in Eq. (\ref{qFcetanpms}) for $\beta =c$. Terms of $u^{-3}$ order 
of the charge current expression, Eq. (\ref{J-partS1-comp}), are
derived in Appendix \ref{Appendix7} for the $m_s=0$
and $m_s\rightarrow 1-m_{\eta}$ reference S$^z$SLN$_{SN}$ subspaces 1A 
and the whole $m_{\eta}\in [0,1]$ range, with the results,
\begin{eqnarray}
\langle\hat{J}_{LWS}^{(3)} (l_{\rm r},S_{\eta},u)\rangle & = & 
{\tau\,2\,L\,t\over\pi}\left({(1-m_{\eta})\ln 2\over u}\right)^3\sum_{\iota=\pm} (\iota)
\left(1-{4\over 3} \sin^2 (q_{Fc,\tau}^{\iota})\right)\sin^2 (q_{Fc,\tau}^{\iota}) 
\nonumber \\
& - & {\tau\,3\zeta (3)\,L\,t\over 32\pi\,u^3}\sum_{\iota=\pm} (\iota)
\{(1-m_{\eta})\left(1+ {2\over 3}\sin^2 (q_{Fc,\tau}^{\iota})\right)
\nonumber \\
& - & {\tau\over 2\pi}\sum_{\iota'=\pm} (\iota')\cos (q_{Fc,\tau}^{\iota'})
\left(\sin (q_{Fc,\tau}^{\iota'}) - 2\sin (q_{Fc,\tau}^{\iota})\right)\}\sin^2 (q_{Fc,\tau}^{\iota}) 
\hspace{0.2cm}{\rm for}\hspace{0.2cm}
m_s = 0 \, ,
\label{J-only3ms0}
\end{eqnarray}
and
\begin{eqnarray}
\langle\hat{J}_{LWS}^{(3)} (l_{\rm r},S_{\eta},u)\rangle & = &
- {\tau\, (1-m_{\eta}-m_s)\,L\,t\over 3\pi\,u^3}\sum_{\iota=\pm} (\iota)
\sin^4 (q_{Fc,\tau}^{\iota})
\nonumber \\
& + & {\cal{O}} ((1-m_{\eta}-m_s)^3)\hspace{0.2cm}{\rm for}\hspace{0.2cm}
(1-m_{\eta}-m_s) \ll 1 \, ,
\label{J-only3msmax}
\end{eqnarray}
respectively. (The expansion term of $u^{-3}$ order, Eq. (\ref{J-only3msmax}), only includes contributions up to
second order in $(1-m_{\eta}-m_s)\ll 1$.)

The following expansion of the limiting occupancy momenta $q_{Fc,\tau}^{\iota}$ maximizes the $u^{-3}$ order expansion
of the charge current absolute value, Eq. (\ref{J-partS1-comp}), for all densities ranges
$\vert m_{\eta}^z\vert\in [0,1]$, $m_{\eta}\in [\vert m_{\eta}^z\vert,1]$, and $m_s \in [0,(1-m_{\eta})]$ of the 
reference S$^z$SLN subspaces 1A, 
\begin{eqnarray}
q_{Fc,-}^{\iota} & = & {\pi\over 2} + \iota\,\pi m_{\eta} + {2n_{\eta s}\over u} + {\cal{O}} (u^{-3})
\hspace{0.20cm}{\rm for}\hspace{0.20cm}m_{\eta}\in \left[0,{1\over 2}-\delta_{\eta s}^u \right]
\nonumber \\
& = & \pi - (1-\iota)\pi m_{\eta} 
\hspace{0.20cm}{\rm for}\hspace{0.20cm}m_{\eta}\in \left[{1\over 2}-\delta_{\eta s}^u,{1\over 2}\right] \, ,
\nonumber \\ 
q_{Fc,+}^{\iota} & = & \pi - (1-\iota)\pi (1-m_{\eta}) 
\hspace{0.20cm}{\rm for}\hspace{0.20cm}m_{\eta}\in \left[{1\over 2},{1\over 2}+\delta_{\eta s}^u\right]
\nonumber \\
& = &  {\pi\over 2} + \iota\,\pi (1-m_{\eta}) + {2n_{\eta s}\over u} + {\cal{O}} (u^{-3})
\hspace{0.20cm}{\rm for}\hspace{0.20cm}m_{\eta}\in \left[{1\over 2}+\delta_{\eta s}^u,1\right] \, ,
\label{qFcpm1iota}
\end{eqnarray}
where $\iota =\pm$ and,
\begin{equation}
\delta_{\eta s}^u = {(1-m_s)\,g_s\over\pi u} + {\cal{O}} (u^{-3}) \, .
\label{deltau}
\end{equation}
The corresponding largest charge current absolute value of general form, Eq. (\ref{JImax}), of such a reference S$^z$SLN subspace
can be written as,
\begin{equation}
\vert\langle\hat{J}_{LWS}^{\rm max}(l_{\rm r},S_{\eta},u)\rangle\vert 
= C_{m_{\eta},m_s}\,t\,L\,m_{\eta}\,(1-m_{\eta}) \, ,
\label{JImaxSeta}
\end{equation}
where $C_{m_{\eta},m_s}$ stands for the coefficient whose
limiting values are provided in Eqs. (\ref{C10lim}) and (\ref{C1cuLlimm}).
It reaches as a function of $m_{\eta}$ and for $u>0$ and $m_s \in [0,(1-m_{\eta})]$ a maximum 
value at an $u$-dependent intermediate density $m_{\eta}$, being a continuous increasing and decreasing
function of $m_{\eta}$ below and above that density, respectively. At fixed $m_{\eta}$, its largest value is reached
for the $m_s= 0$ reference S$^z$SLN$_{SN}$ subspace 1A.

The coefficient $C_{m_{\eta},m_s}$ limiting values valid for $u\rightarrow 0$, which from Eq. (\ref{C10lim}) read,
\begin{eqnarray}
C_{m_{\eta},m_s} & = &
4\hspace{0.20cm}{\rm for}\hspace{0.20cm}m_{\eta}\rightarrow 0\hspace{0.20cm}{\rm and}\hspace{0.20cm}m_{s}\rightarrow 0
\nonumber \\
& = & 2\hspace{0.20cm}{\rm for}\hspace{0.20cm}m_{\eta}\in [0,1]\hspace{0.20cm}{\rm and}\hspace{0.20cm}m_{s}\rightarrow 1 - m_{\eta}
\nonumber \\
& = & 2\hspace{0.20cm}{\rm for}\hspace{0.20cm}m_{\eta}\rightarrow 1\hspace{0.20cm}{\rm and}\hspace{0.20cm}m_{s}\rightarrow 0 \, ,
\label{Cmulim}
\end{eqnarray}
follow from those in Eq. (\ref{JImaxSetaLimu0}) of Appendix \ref{Appendix7} for
the largest charge current absolute value derived in that Appendix 
for $m_{\eta}\ll 1$ and $(1-m_{\eta})\ll 1$. 

Moreover, from the use of the expansion up to $u^{-2}$ order of $\vert\langle\hat{J}_{LWS}^{\rm max}(l_{\rm r},S_{\eta},u)\rangle\vert$,
Eq.  (\ref{JImaxSetaLim}) of Appendix \ref{Appendix7}, one finds that up to that order 
the coefficient $C_{m_{\eta},m_s}$ in Eq. (\ref{JImaxSeta}) is for densities $m_{\eta} \in [0,1]$ 
and $m_s \in [0,(1-m_{\eta})]$ of reference S$^z$SLN subspaces 1A given by,
\begin{eqnarray}
C_{m_{\eta},m_s} & = & 
{2\over\pi}{\sin (\pi m_{\eta})\over
m_{\eta}(1-m_{\eta})}
\left(1 - {7\over 2}\left({n_{\eta s}\over u}\right)^2
\left(1 - {8\over 7}\cos (\pi m_{\eta}) - {3\over 7}\sin^2 (\pi m_{\eta})\right)\right) 
\nonumber \\
& & \hspace{0.20cm}{\rm for}\hspace{0.20cm}m_{\eta} \in \left[0,{1\over 2}-\delta_{\eta s}^u\right] 
\hspace{0.20cm}{\rm and}\hspace{0.20cm} \hspace{0.20cm}m_s \in [0,(1 - m_{\eta})] 
\nonumber \\
& = & {2\over\pi}{\sin^2 (\pi m_{\eta})\over m_{\eta}(1-m_{\eta})}
\left(1 + {2n_{\eta s}\over u}\left(1
+ {3\over 2}{n_{\eta s}\over u}\cos (2\pi m_{\eta})\right)\cos^2 (\pi m_{\eta})\right)
\nonumber \\
& & \hspace{0.20cm}{\rm for}\hspace{0.20cm}m_{\eta} \in \left[{1\over 2}-\delta_{\eta s}^u,{1\over 2}+\delta_{\eta s}^u\right]
\hspace{0.20cm}{\rm and}\hspace{0.20cm} \hspace{0.20cm}m_s \in [0,(1 - m_{\eta})]
\nonumber \\
& = & {2\over\pi}{\sin (\pi m_{\eta})\over m_{\eta}(1-m_{\eta})}
\left(1 - {7\over 2}\left({n_{\eta s}\over u}\right)^2
\left(1 + {8\over 7}\cos (\pi m_{\eta}) - {3\over 7}\sin^2 (\pi m_{\eta})\right)\right) 
\nonumber \\
& & \hspace{0.20cm}{\rm for}\hspace{0.20cm}m_{\eta} \in \left[{1\over 2}+\delta_{\eta s}^u,1\right] 
\hspace{0.20cm}{\rm and}\hspace{0.20cm} \hspace{0.20cm}m_s \in [0,(1 - m_{\eta})] \, .
\label{C1cuLlim}
\end{eqnarray}
Both for $m_{\eta} \in [0,1/2-\delta_{\eta s}^u]$ and $m_{\eta} \in [1/2+\delta_{\eta s}^u,1]$
the terms of orders $u^{-1}$, $u^{-3}$, and remaining odd orders $u^{-j}$ where $j=5,7,...$ 
of this coefficient expansion exactly vanish. 

Finally, from the use of Eqs. (\ref{JImaxSetaLim3}) and (\ref{JImaxSetaLimmsmax3}) of Appendix \ref{Appendix7} 
one finds that the terms of $u^{-3}$ order of the coefficient $C_{m_{\eta},m_s}$ in Eq. (\ref{JImaxSeta}) 
are for the $m_s=0$ and $m_s\rightarrow 1-m_{\eta}$ reference S$^z$SLN$_{SN}$ subspaces 1A 
and the whole density interval $m_{\eta} \in [0,1]$ given by,
\begin{eqnarray}
C_{m_{\eta},m_s}^{(3)} & = & 0 + {\cal{O}} (u^{-4})\hspace{0.20cm}{\rm for}\hspace{0.20cm}m_{\eta} \in \left[0,{1\over 2}-\delta_{\eta s}^u\right] 
\hspace{0.20cm}{\rm and}\hspace{0.20cm}m_s = 0
\nonumber \\
& = & {\sin^2 (2\pi m_{\eta})\over \pi m_{\eta}\,(1-m_{\eta})}\{2\left({(1-m_{\eta})\ln 2\over u}\right)^3
\left(1 - {4\over 3}\sin^2 (2\pi m_{\eta})\right)
\nonumber \\
& + & {3\zeta (3)\over 16\,u^3}\left((1-m_{\eta})\left(1+{2\over 3}\sin^2 (2\pi m_{\eta})\right)
+ {1\over \pi}\left(1 - {1\over 2}\cos (2\pi m_{\eta})\right)\sin (2\pi m_{\eta})\right)\}
\nonumber \\
& + & {\cal{O}} (u^{-4})\hspace{0.20cm}{\rm for}\hspace{0.20cm}m_{\eta} \in \left[{1\over 2}-\delta_{\eta s}^u,{1\over 2}+\delta_{\eta s}^u\right]
\hspace{0.20cm}{\rm and}\hspace{0.20cm}m_s = 0
\nonumber \\
& = & 0 + {\cal{O}} (u^{-4})\hspace{0.20cm}{\rm for}\hspace{0.20cm}m_{\eta} \in \left[{1\over 2}+\delta_{\eta s}^u,1\right] 
\hspace{0.20cm}{\rm and}\hspace{0.20cm}m_s = 0 \, ,
\label{C1cuLlim3A}
\end{eqnarray}
and 
\begin{eqnarray}
C_{m_{\eta},m_s}^{(3)} & = & 0 + {\cal{O}} (u^{-4})\hspace{0.20cm}{\rm for}\hspace{0.20cm}m_{\eta} \in \left[0,{1\over 2}-\delta_{\eta s}^u\right] 
\hspace{0.20cm}{\rm and}\hspace{0.20cm}m_s\rightarrow 1-m_{\eta}
\nonumber \\
& = & {\sin^4 (2\pi m_{\eta}) \over 3\pi u^{-3}}{(1-m_{\eta}-m_s)\over m_{\eta}\,(1-m_{\eta})}+ {\cal{O}} ((1-m_{\eta}-m_s)^3) 
\nonumber \\
& & {\rm for}\hspace{0.20cm}m_{\eta} \in \left[{1\over 2}-\delta_{\eta s}^u,{1\over 2}+\delta_{\eta s}^u\right]
\hspace{0.20cm}{\rm and}\hspace{0.20cm}m_s\rightarrow 1-m_{\eta}
\nonumber \\
& = & 0 + {\cal{O}} (u^{-4})\hspace{0.20cm}{\rm for}\hspace{0.20cm}m_{\eta} \in \left[{1\over 2}+\delta_{\eta s}^u,1\right] 
\hspace{0.20cm}{\rm and}\hspace{0.20cm}m_s\rightarrow 1-m_{\eta} \, ,
\label{C1cuLlim3}
\end{eqnarray}
respectively. (The terms in Eq. (\ref{C1cuLlim3}) only include the contributions 
up to second order in $(1-m_{\eta}-m_s)\ll 1$.)

\section{Charge stiffness upper bounds within the canonical ensemble}
\label{two}

The function,
\begin{equation}
F^{\rm UB} (m_{\eta},u) \equiv {\vert\langle\hat{J}_{LWS}^{\rm max}(l_{\rm r},S_{\eta},u)\rangle\vert \over 2S_{\eta}}
= t\,(1-m_{\eta})\,C_{m_{\eta},m_s} \, ,
\label{JImaxOverSeta}
\end{equation}
where $C_{m_{\eta},m_s}$ is the coefficient in Eqs. (\ref{JImaxSeta})-(\ref{C1cuLlim3}), is a continuous and decreasing function of $m_{\eta}$
for $u>0$, $m_{\eta}\in [0,1]$, and $m_{s}\in [0,(1-m_{\eta})]$. It has limiting behaviors,
\begin{eqnarray}
F^{\rm UB} (m_{\eta},u)
& = & 4t\,(1-m_{\eta})\hspace{0.20cm}{\rm for}\hspace{0.20cm}m_{\eta}\rightarrow 0 \, , \hspace{0.20cm}m_{s}\rightarrow 0 \, ,
\hspace{0.20cm}{\rm and}\hspace{0.20cm}u\rightarrow 0
\nonumber \\
& = & 2t\,(1-m_{\eta})\hspace{0.20cm}{\rm for}\hspace{0.20cm}m_{\eta}\in [0,1] \, , \hspace{0.20cm}m_{s}\rightarrow 1 - m_{\eta} \, ,
\hspace{0.20cm}{\rm and}\hspace{0.20cm}u\rightarrow 0
\nonumber \\
& = & 2t\,(1-m_{\eta})\hspace{0.20cm}{\rm for}\hspace{0.20cm}m_{\eta}\rightarrow 1 \, , \hspace{0.20cm}m_{s}\rightarrow 0 \, ,
\hspace{0.20cm}{\rm and}\hspace{0.20cm}u\rightarrow 0
\nonumber \\
& = & {2t\sin (\pi m_{\eta})\over\pi m_{\eta}}\hspace{0.20cm}{\rm for}\hspace{0.20cm}m_{\eta} \in [0,1]
\, , \hspace{0.20cm}m_s \in [0,(1 - m_{\eta})] \, ,
\hspace{0.20cm}{\rm and}\hspace{0.20cm}u\rightarrow\infty \, .
\label{JImaxOverSetaLim}
\end{eqnarray}
Its derivative with respect to $m_{\eta}$ is such that,
\begin{eqnarray}
{\partial F^{\rm UB} (m_{\eta},u)\over\partial m_{\eta}} & = & 0
\hspace{0.20cm}{\rm for}\hspace{0.20cm}m_{\eta}\rightarrow 0
\, , \hspace{0.20cm}m_s \in [0,1] \, ,
\hspace{0.20cm}{\rm and}\hspace{0.20cm}u > 0 \, ,
\nonumber \\
{\partial F^{\rm UB} (m_{\eta},u)\over\partial m_{\eta}} & < & 0
\hspace{0.20cm}{\rm for}\hspace{0.20cm}m_{\eta}\in ]0,1]
\, , \hspace{0.20cm}m_s \in [0,(1 - m_{\eta})] \, ,
\hspace{0.20cm}{\rm and}\hspace{0.20cm}u > 0 \, .
\label{derivFUB}
\end{eqnarray}
It has the limiting behaviors,
\begin{eqnarray}
{\partial F^{\rm UB} (m_{\eta},u)\over\partial m_{\eta}} & = & 0
\hspace{0.20cm}{\rm for}\hspace{0.20cm}m_{\eta}\rightarrow 0
\, , \hspace{0.20cm}m_s \in [0,1] \, ,
\hspace{0.20cm}{\rm and}\hspace{0.20cm}u \rightarrow 0 \, ,
\nonumber \\
& = & - 4t
\hspace{0.20cm}{\rm for}\hspace{0.20cm}m_{\eta}\rightarrow 1 \, , \hspace{0.20cm}m_s \rightarrow 0 \, ,
\hspace{0.20cm}{\rm and}\hspace{0.20cm}u \rightarrow 0 \, ,
\nonumber \\
& = & - {2t\over m_{\eta}}\left({\sin (\pi m_{\eta})\over \pi m_{\eta}} - \cos (\pi m_{\eta})\right)
\hspace{0.20cm}{\rm for}\hspace{0.20cm}m_{\eta} \in [0,1] \, , \hspace{0.20cm}m_s \in [0,(1 - m_{\eta})] \, ,
\hspace{0.20cm}{\rm and}\hspace{0.20cm}u \rightarrow\infty \, .
\label{derivFUBexp}
\end{eqnarray}

A first stiffness upper bound, 
\begin{eqnarray}
D^* (T) & = & {(2S_{\eta}^z)^2\over 2 L T}\,\,\sum_{S_{\eta}=\vert S_{\eta}^z\vert}^{L/2}\sum_{l_{\rm r}} 
p_{l_{\rm r},S_{\eta},S_{\eta}^z}(F^{\rm UB} (m_{\eta},u))^2  
\nonumber \\
& = & {t^2 (m_{\eta}^z)^2 L\over 2T}\,\,\sum_{S_{\eta}=\vert S_{\eta}^z\vert}^{L/2}\sum_{l_{\rm r}} 
p_{l_{\rm r},S_{\eta},S_{\eta}^z}\,C_{m_{\eta},m_s}^2\,(1-m_{\eta})^2 \, ,
\label{D-all-T-simp-12}
\end{eqnarray}
is obtained within the canonical ensemble by replacing the moduli of the expectation values $\langle\hat{J}_{LWS} (l_{\rm r},L_{\eta},S_{\eta},u)\rangle$ 
of $\eta$-Bethe states with the same $S_{\eta}$ value in the stiffness expression, Eq. (\ref{D-all-T-simp}),
by the upper bound of the largest absolute value of the charge current, Eq. (\ref{JImaxSeta}).

For each fixed-$S_{\eta}^z$ and $S_s^z=0$ canonical ensemble,
the largest value of $F^{\rm UB} (m_{\eta},u)$ in the $S_{\eta}$ summation of Eq. (\ref{D-all-T-simp-12})  
is that referring to the minimum $S_{\eta}$ and $S_s$ values, $S_{\eta}=\vert S_{\eta}^z\vert = m_{\eta}^z\,L/2$
and $S_s=\vert S_{s}^z\vert = 0$, respectively, such that $m_{\eta} = m_{\eta}^z$ and $m_s= m_{s}^z=0$.
This follows from the function $F^{\rm UB} (m_{\eta},u)$ smoothly decreasing upon increasing $m_{\eta}$.
The same applies upon increasing $m_{s}$ at finite $u$.
A second stiffness upper bound is then reached by replacing in Eq. (\ref{D-all-T-simp-12}) the function $F^{\rm UB} (m_{\eta},u)$ 
by its largest value,
\begin{equation}
F^{\rm UB} (m_{\eta}^z,u) = t\,C_{m_{\eta}^z,0}\,(1-m_{\eta}^z)\hspace{0.20cm}{\rm for}\hspace{0.20cm}m_{\eta}^z \in [0,1] \, .
\label{JmaxUBm2SLV}
\end{equation}
Here $C_{m_{\eta}^z,0}$ is obtained by replacing $m_{\eta}$ and $m_s$ by $m_{\eta}^z$ 
and $m_{s}^z=0$, respectively, in the expression of the coefficient $C_{m_{\eta},m_s}$.
The state summations in Eq. (\ref{D-all-T-simp-12}) can then be performed exactly for all 
finite temperatures $T>0$. Indeed, the probability 
distribution $p_{l_{\rm r},S_{\eta},S_{\eta}^z}$ in each fixed-$S_{\eta}^z$ canonical ensemble is normalized as, 
\begin{equation}
\sum_{S_{\eta}=\vert S_{\eta}^z\vert}^{L/2}\sum_{l_{\rm r}} p_{l_{\rm r},S_{\eta},S_{\eta}^z}=1 \, .
\label{psummation}
\end{equation}
Such state summations account for the subspace dimensions and thus as well for the full $S_s^z=0$ subspace dimension.
For $T>0$ the resulting (larger) upper bound $D^{**} (T) \geq D^* (T) \geq D (T)$, then becomes, 
\begin{equation}
D^{**} (T) = {t^2\,C_{m_{\eta}^z,0}^2\,L\over 2T}\,(1-m_{\eta}^z)^2\,(m_{\eta}^z)^2
\hspace{0.20cm}{\rm for}\hspace{0.20cm}m_{\eta}^z \in [0,1] 
\hspace{0.20cm}{\rm and}\hspace{0.20cm}m_{s}^z = 0 \, .
\label{D**}
\end{equation}

For $m_{\eta}^z\ll 1$ and $m_{s}^z = 0$ its values continuously vary from,
\begin{equation}
D^{**} (T) = {16\,t^2\,L\over 2T}\,(m_{\eta}^z)^2 \, , 
\label{value-D-m01u0}
\end{equation}
for $u\rightarrow 0$ to,
\begin{equation}
D^{**} (T) = {4\,t^2\,L\over 2T}\,(m_{\eta}^z)^2 \, , 
\label{value-D-m01uL}
\end{equation}
for $u\gg 1$ whereas for $(1-m_{\eta}^z)\ll 1$ and $m_{s}^z = 0$ it is given by,
\begin{equation}
D^{**} (T) = {4\,t^2\,L\over 2T}\,(1-m_{\eta}^z)^2 \, , 
\label{value-D-m01allu}
\end{equation}
for all $u>0$ values. 

In the $u\rightarrow\infty$ limit, it has the following simple expression
for $m_{s}^z = 0$ and the whole $m_{\eta}^z \in [0,1]$ interval,
\begin{equation}
D^{**} (T) = {(2t/\pi)^2\,\sin^2 (\pi m_{\eta}^z)\,L\over 2T} 
= {(2t/\pi)^2\,\sin^2 (\pi (1-m_{\eta}^z))\,L\over 2T}
\hspace{0.20cm}{\rm for}\hspace{0.20cm}\hspace{0.20cm}m_{\eta}^z \in [0,1] 
\hspace{0.20cm}{\rm and}\hspace{0.20cm}m_{s}^z = 0 \, .
\label{D**uinf}
\end{equation}
 
The charge stiffness of the 1D Hubbard model was studied in Ref. \cite{PDSC-00} for large $u$, where
it was shown to exactly vanish in the $m_{\eta}^z \rightarrow 0$ limit. As found in that
reference, for $m_{\eta}^z $ finite the charge stiffness of the 1D Hubbard and that
for spinless fermions alone are different even in the $u\rightarrow\infty$ limit, as illustrated
in Fig. 1 of that reference for a finite system. Such a different behavior persists in
the TL and is due to the spinless-fermion phase shifts imposed by the spins $1/2$ \cite{Karlo-97}.\\

The coefficient $c_{\rm c}$ in the upper bound, Eq. (\ref{Dcm0}), then smoothly varies from $c_{\rm c}=16$ for
$u\rightarrow 0$ to $c_{\rm c}=4$ for $u\gg 1$ whereas the coefficient $c_{\rm c}'$ in the upper bound,
Eq. (\ref{Dclinem1}), reads $c_{\rm c}'=4$ for the whole $u>0$ range.
This completes our finding of a vanishing charge stiffness in the TL, $L\to\infty$, within the canonical
ensemble for any fixed range or even distribution of $S_{\eta}^z$, or any distribution of $m_{\eta}^z$ shrinking sufficiently 
fast that $\langle (m_{\eta}^z)^2 \rangle L \to 0$. 

\section{Stiffness upper bounds within the grand-canonical ensemble for $T\rightarrow\infty$}
\label{UPTinf}

The average value of the square of the charge current $\vert\langle\hat{J} (l_{\rm r},L_{\eta},S_{\eta},S_{\eta}^z,u)\rangle\vert^2$, 
Eq. (\ref{currents-gen}), in a fixed-$S_{\eta}^z$ and $S_s^z=0$ subspace that contains the set of S$^z$S subspaces 
with $\eta$-spin values $S_{\eta}=\vert S_{\eta}^z\vert, \vert S_{\eta}^z\vert+1, \vert S_{\eta}^z\vert+2,...$, reads,
\begin{eqnarray}
& & \left\langle\vert\langle\hat{J} (l_{\rm r},L_{\eta},S_{\eta},S_{\eta}^z,u)\rangle\vert^2\right\rangle_{S_{\eta}^z}  
= {(2S_{\eta}^z)^2\sum_{L_{\eta}=2\vert S_{\eta}^z\vert}^{L}\sum_{S_{\eta}=\vert S_{\eta}^z\vert}^{L_{\eta}/2}\sum_{l_{\rm r}} 
{\vert\langle\hat{J}_{LWS} (l_{\rm r},L_{\eta},S_{\eta},u)\rangle\vert^2\over (2S_{\eta})^2}
\over \sum_{L_{\eta}=2\vert S_{\eta}^z\vert}^{L}\sum_{S_{\eta}=\vert S_{\eta}^z\vert}^{L_{\eta}/2}
d_{\rm subspace}^{LWS} (L_{\eta},S_{\eta})}
\nonumber \\
& = & {(2S_{\eta}^z)^2\sum_{L_{\eta}=2\vert S_{\eta}^z\vert}^{L}\sum_{S_{\eta}=\vert S_{\eta}^z\vert}^{L_{\eta}/2}\sum_{l_{\rm r}} 
{\vert\langle\hat{J}_{LWS} (l_{\rm r},L_{\eta},S_{\eta},u)\rangle\vert^2\over (2S_{\eta})^2}
\over \sum_{L_{\eta}=2\vert S_{\eta}^z\vert}^{L}\sum_{S_{\eta}=\vert S_{\eta}^z\vert}^{L_{\eta}/2}
\sum_{S_{s}=0}^{(L-L_{\eta})/2}
{L\choose L_{\eta}}\times 
\left({L_{\eta} \choose L_{\eta}/2-S_{\eta}}-
{L_{\eta}\choose L_{\eta}/2-S_{\eta}-1} \right)\times 
\left({L-L_{\eta} \choose (L-L_{\eta})/2-S_{s}}-
{L-L_{\eta}\choose (L-L_{\eta})/2-S_{s}-1} \right)} 
\nonumber \\
& = & {(2S_{\eta}^z)^2\sum_{L_{\eta}=2\vert S_{\eta}^z\vert}^{L}\sum_{S_{\eta}=\vert S_{\eta}^z\vert}^{L_{\eta}/2}\sum_{l_{\rm r}} 
{\vert\langle\hat{J}_{LWS} (l_{\rm r},L_{\eta},S_{\eta},u)\rangle\vert^2\over (2S_{\eta})^2}
\over \sum_{L_{\eta}=2\vert S_{\eta}^z\vert}^{L}\sum_{S_{\eta}=\vert S_{\eta}^z\vert}^{L_{\eta}/2}
\sum_{S_{s}=0}^{(L-L_{\eta})/2}
{L\choose L_{\eta}}\times
\left(\sum_{\{N_{\eta n}\}}\,\prod_{n =1}^{\infty}\,{L_{\eta n}\choose N_{\eta n}}\right)\times 
\left(\sum_{\{N_{s n'}\}}\,\prod_{n' =1}^{\infty}\,{L_{s n'}\choose N_{s n'}}\right)} \, .
\label{averageINF}
\end{eqnarray}
The subspaces dimensions appearing here are defined in Appendix \ref{Appendix5} and useful information on the 
corresponding states summations was given in Section \ref{SzSS}.

We denote by $\vert\langle\hat{J}_{LWS}\rangle\vert_{A(l_{\eta},m_{\eta},n_{\rho})}$ 
a S$^z$SLN subspace current absolute value average. It is given by,
\begin{equation}
\vert\langle\hat{J}_{LWS}\rangle\vert_{A(l_{\eta},m_{\eta},n_{\rho})} =
{\sum_{l_{n_{\rho}}^{\star}}\vert\langle\hat{J}_{LWS} (l_{n_{\rho}}^{\star},L_{\eta},S_{\eta},u)\rangle\vert\over 
\sum_{S_{s}=0}^{(L-L_{\eta})/2} {L\choose L_{\eta}}\times 
\left(\sum_{\{(N_{\eta n})_{l_{\eta},m_{\eta},n_{\rho}}\}}\, \prod_{n =1}^{\infty}\,{L_{\eta n}\choose N_{\eta n}}\right)
\times \left({L-L_{\eta} \choose (L-L_{\eta})/2-S_{s}}-
{L-L_{\eta}\choose (L-L_{\eta})/2-S_{s}-1} \right)} \, .
\label{JzAveDef}
\end{equation}
Here $l_{n_{\rho}}^{\star}$ denotes $l_{\rm r}^{\star}$ for the set of $\eta$-Bethe states with fixed values for
the densities $l_{\eta}$, $m_{\eta}$, and $n_{\rho}$ that span a S$^z$SLN subspace
and the sum $\sum_{l_{n_{\rho}}^{\star}}$ runs over all $c$-band, $\eta n$-bands of $n=1,...,\infty$ branches, 
and $sn$-band occupancy configurations of spin $S_s = 0,1,...,(L-L_{\eta})/2$ that generate such $\eta$-Bethe states. 
They have the same numbers $M_{\eta}=2S_{\eta}$ of unpaired $\eta$-spins $1/2$ and $N_{\rho} = N_c + N_{\eta}$, 
Eq. (\ref{NchargeP}), of charge pseudoparticles where $N_c = L-L_{\eta}$. The S$^z$SLN subspace dimension
$\sum_{l_{n_{\rho}}^{\star}}1$ is given in the denominator on the right-hand side of Eq. (\ref{JzAveDef}).
Hence the summation $\sum_{\{(N_{\eta n})_{l_{\eta},m_{\eta},n_{\rho}}\}}$ runs over all sets of
$\eta n$ pseudoparticle numbers $\{N_{\eta n}\}$ that obey both
the sum rules $\sum_{n=1}^{\infty}n\,N_{\eta n} = (L_{\eta}-2S_{\eta})=\Pi_{\eta}$, Eq. (\ref{sum-Nseta}) for $\alpha = \eta$,
and $N_{\eta} = \sum_{n=1}^{\infty}N_{\eta n} = (L_{\eta} - N_{\eta 1}^h)/2$,
Eq. (\ref{NpsNapsSR}) for $\alpha = \eta$.

For finite $u$ a reference S$^z$SLN subspace largest charge current absolute value
is proportional to $M_{\eta}\,N_{\rho}$. That in Eq. (\ref{JImax}) it is written as proportional 
to $m_{\eta}\,(1-m_{\eta})$ and thus to $2S_{\eta}\,(L-2S_{\eta})$ follows from the expression given in that equation applying
both to $u\rightarrow 0$ and to finite $u$. Indeed and as justified in Appendix \ref{Appendix3},
the carriers of charge are different for $u\rightarrow 0$ and finite $u$, respectively.
As a result, such a largest charge current absolute value can be written as proportional to
$2S_{\eta}\,(L-2S_{\eta})$ and $M_{\eta}\,N_{\rho}$ for $u\rightarrow 0$ and finite $u$,
respectively. (If one requires it to apply both to the $u\rightarrow 0$ limit and
to finite $u$, then it should be written as given in Eq. (\ref{JImax}).)

That each S$^z$SLN subspace of a S$^z$S subspace is spanned by $\eta$-Bethe states with exactly 
the same number $N_{\rho} = N_c + N_{\eta}$ of charge pseudoparticles simplifies the
form of the current absolute values average, Eq. (\ref{JzAveDef}). Its expression can for $u>0$ be
written in the general form,
\begin{equation}
\vert\langle\hat{J}_{LWS}\rangle\vert_{A(l_{\eta},m_{\eta},n_{\rho})} =
J_{\rho}\,4t\,m_{\eta}\,\sqrt{2N_{\rho}}
= {1\over L}\,J_{\rho}\,4t\,M_{\eta}\,\sqrt{2N_{\rho}} \, .
\label{JzAve}
\end{equation}
The coefficient $J_{\rho}$ obeys the inequality $J_{\rho}\leq 1$, being of the order of unity. 
While for $u>0$ the {\it largest} charge current absolute value of a reference S$^z$SLN subspace 
is proportional to $M_{\eta}\,N_{\rho}$, such a subspace {\it average} current absolute value, Eq. (\ref{JzAveDef}), is 
proportional to $M_{\eta}\,\sqrt{N_{\rho}}$. That $\vert\langle\hat{J}_{LWS}\rangle\vert_{A(l_{\eta},m_{\eta},n_{\rho})}
\propto M_{\eta}\,\sqrt{N_{\rho}}$ stems from the energy and momentum eigenstates that span the S$^z$SLN subspace
being generated by all possible occupancy configurations of the $N_{\rho} = N_c + N_{\eta}$ charge pseudoparticles.

For all S$^z$SLN subspaces contained in a S$^z$S subspace the density $m_{\eta}=m_{\eta}^z$ in Eq. (\ref{JzAve}) has a fixed value.
This combined with $J_{\rho}\leq 1$ being of the order of the unity reveals that the S$^z$SLN subspace in a S$^z$S subspace
whose current absolute value average, Eq. (\ref{JzAveDef}), is largest is that for which $N_{\rho}$ reaches its maximum value. 
One finds from $N_{\rho} = L - (L_{\eta} + N_{\eta 1}^h)/2$ where $N^h_{\eta 1} = 2S_{\eta}+\sum_{n=2}^{\infty}2(n-1)N_{\eta n}$, 
Eq. (\ref{N-h-an}) for $\alpha n=\eta 1$, that the latter maximum value refers to the S$^z$S subspace minimum 
$N^h_{\eta 1}$ value, which for the corresponding fixed $\eta$-spin $S_{\eta}$ reads $N^h_{\eta 1} = 2S_{\eta}$. This 
gives $N_{\rho} = L - (L_{\eta} + 2S_{\eta})/2$. For general S$^z$SLN subspaces of a S$^z$S subspace for which 
$N^h_{\eta 1} = 2S_{\eta}$ and thus $N_{\rho} = L - \Pi_{\eta}$ one has that $N_{\eta n}=0$ for $n>1$, so that 
$N_{\rho} = L - (L_{\eta} + 2S_{\eta})/2 = N_c + N_{\eta 1}$ where $N_c = L - 2S_{\eta} - 2N_{\eta 1}$ and thus 
$N_{\rho} = L - 2S_{\eta} - N_{\eta 1}$. Further maximizing $N_{\rho}$ at the S$^z$S subspace fixed $S_{\eta}$ 
value corresponds to minimizing $L_{\eta}$, which gives $L_{\eta}=2S_{\eta}$ and thus $N_{\eta 1}=0$. This 
corresponds to reference S$^z$SLN subspace 1 of the S$^z$S subspace under consideration for which
$l_{\eta}\rightarrow m_{\eta}$ and thus $n_{\eta}\rightarrow 0$, so that it is indeed spanned by a subset of $\eta$-Bethe states 
for which $N_{\rho} = N_c$. For such states $N_{\rho}$ reaches its maximum value, $N_{\rho} = L - 2S_{\eta}$.
A reference S$^z$SLN subspace 1 current absolute value average, Eq. (\ref{JzAveDef}), can be written as,
\begin{equation}
\vert\langle\hat{J}_{LWS}\rangle\vert_{A(m_{\eta})} =
J_{\rho 1}\,4t\,m_{\eta}\,\sqrt{2N_{\rho}} = {1\over L}\,J_{\rho 1}\,4t\,M_{\eta}\,\sqrt{2N_{\rho}} \, .
\label{JzAve1}
\end{equation}
Since $l_{\eta}=m_{\eta}$ and $n_{\rho}=1-l_{\eta} = 1-m_{\eta}$, the index $A(l_{\eta},m_{\eta},n_{\rho})$
in the general S$^z$SLN subspace current absolute value average, Eq. (\ref{JzAve}), was for the 
particular case of the S$^z$SLN subspace 1 denoted by $A(m_{\eta})$ in Eq. (\ref{JzAve1}).

The spin degrees of freedom do not couple directly to charge probes and the
charge currents do not depend on the spin-singlet $sn$ pseudoparticle 
occupancy configurations associated with the $sn$-bands momentum distribution functions $N_{sn} (q_j)$.
However, for finite $u$ the charge current spectra of the $\eta$-Bethe states that span 
a reference S$^z$SLN subspace 1 depend on the spin density $m_s$ and overall spin $sn$ pseudoparticle density $n_s$.
As a consequence, the corresponding coefficient $J_{\rho}$ in Eq. (\ref{JzAve}) also depends on the densities
$m_s$ and  $n_s$.

One finds that finite-$u$ $\eta$-Bethe states contained in a reference S$^z$SLN subspace 1 
with exactly the same $c$ pseudoparticle occupancy configurations have for any fixed density $m_s$ in the
interval $m_s \in [0,(1-m_{\eta})]$ the largest charge current absolute values for the
S$^z$SLN$_{SN}$ subspaces 1A for which the density $n_s \in [0,(1-m_{\eta}-m_s)/2]$ in Eq. (\ref{densitiesSzSLNN}) 
has its largest value, $n_s=n_s^{\rm max}=(1-m_{\eta}-m_s)/2$. Only in the $u\rightarrow\infty$
limit in which all spin configurations are degenerate have these states the same
charge currents absolute values. Limiting examples are (i) the S$^z$SLN$_{SN}$ subspace 1A 
and (ii) the S$^z$SLN$_{SN}$ subspace 1B. Both such S$^z$SLN$_{SN}$ subspaces of a
reference S$^z$SLN subspace 1 have fixed densities $l_{\eta}\rightarrow m_{\eta}$, $n_{\eta}\rightarrow 0$, 
$m_{\eta} \in [0,1]$, and $m_s \in [0,(1-m_{\eta})]$. Their density $n_s$ is given by (i) its maximum value $n_s \rightarrow (1- m_{\eta}-m_s)/2$ 
and (ii) minimum value $n_s \rightarrow 0$, respectively.

We thus consider here a subspace contained in a reference S$^z$SLN subspace 1 
that we call S$^z$SLN$_{N_1}$ subspace. It is spanned by $\eta$-Bethe states with spin values $S_s = 0, 1,...,L - L_{\eta}$ 
whose overall number of $s n$ pseudoparticles reads $N_{s}=N_{s1}=(L - L_{\eta} - 2S_s)/2$ 
for each such a spin value. Hence the S$^z$SLN$_{N_1}$ subspace corresponds to the
set of reference S$^z$SLN$_{SN}$ subspaces 1A, each with a fixed density $m_s \in [0,(1-m_{\eta})]$. 
Its current absolute value average thus reads, 
\begin{equation}
\vert\langle\hat{J}_{LWS}\rangle\vert_{A_{1N} (m_{\eta})} =
{\sum_{l_{n_{\rho}}^{*}}\vert\langle\hat{J}_{LWS} (l_{n_{\rho}}^{*},L_{\eta},S_{\eta},u)\rangle\vert\over 
\sum_{S_{s}=0}^{(L-L_{\eta})/2} {L\choose L_{\eta}}\times 
\left(\sum_{\{(N_{\eta n})_{l_{\eta},m_{\eta},n_{\rho}}\}}\, \prod_{n =1}^{\infty}\,{L_{\eta n}\choose N_{\eta n}}\right)
\times {(L - 2S_{\eta} + 2S_s)/2\choose 2S_{s}}} \, ,
\label{JzAveDefnsMax}
\end{equation}
where ${(L - 2S_{\eta} + 2S_s)/2\choose 2S_{s}}$ is the number of independent $s1$-band occupancy configurations
for each of the spin values $S_s = 0, 1,...,L - L_{\eta}$. The S$^z$SLN$_{N_1}$ dimension 
$\sum_{l_{n_{\rho}}^{*}}1$ is given by the denominator on the right-hand side of Eq. (\ref{JzAveDefnsMax}).
The current absolute value average, Eq. (\ref{JzAveDefnsMax}), can be written as,
\begin{equation}
\vert\langle\hat{J}_{LWS}\rangle\vert_{A(m_{\eta},n_s^{\rm max})} =
J_{\rho 1N}\,4t\,m_{\eta}\,\sqrt{2N_{\rho}} = {1\over L}\,J_{\rho 1N}\,4t\,M_{\eta}\,\sqrt{2N_{\rho}}\, .
\label{JzAvensMax}
\end{equation}
The difference relative to the current absolute value average, Eq. (\ref{JzAve}), of the reference S$^z$SLN subspace 1
where the S$^z$SLN$_{N_1}$ subspace is contained is that
$J_{\rho 1N}\geq J_{\rho 1}$.

Each S$^z$S subspace only contains one reference S$^z$SLN subspace 1. Since a reference S$^z$SLN subspace 1
only contains one S$^z$SLN$_{N_1}$ subspace, a S$^z$S subspace also only contains one S$^z$SLN$_{N_1}$ subspace.
Let $\langle\hat{J} (l_{\rm r}^{\diamond},S_{\eta},S_{\eta}^z,u)\rangle$ denote the currents of the energy and momentum
eigenstates that span a reduced subspace of the fixed-$S_{\eta}^z$ and $S_s^z=0$ subspace obtained by
replacing each of its S$^z$S subspaces by the corresponding S$^z$SLN$_{N_1}$ subspace.
Here $l_{\rm r}^{\diamond}$ stands for all quantum numbers other than $S_{\eta}$, $S_{\eta}^z$, and $u>0$ 
needed to uniquely define each such an energy and momentum eigenstate. 
An important quantity for our upper-bound procedures is the average value of the current square 
$\vert\langle\hat{J} (l_{\rm r}^{\diamond},S_{\eta},S_{\eta}^z,u)\rangle\vert^2$ in the 
reduced subspace under consideration, which reads, 
\begin{equation}
\left\langle\vert\langle\hat{J} (l_{\rm r}^{\diamond},S_{\eta},S_{\eta}^z,u)\rangle\vert^2\right\rangle_{S_{\eta}^z\,1N}
= {(2S_{\eta}^z)^2\sum_{S_{\eta}=\vert S_{\eta}^z\vert}^{L/2}\sum_{l_{\rm r}^{\diamond}} 
{\vert\langle\hat{J}_{LWS} (l_{\rm r}^{\diamond},S_{\eta},u)\rangle\vert^2\over (2S_{\eta})^2}
\over \sum_{S_{\eta}=\vert S_{\eta}^z\vert}^{L/2}\sum_{S_{s}=0}^{(L-2S_{\eta})/2}
{L\choose 2S_{\eta}}\times {(L - 2S_{\eta} + 2S_s)/2\choose 2S_{s}}} \, .
\label{aveSzSLNINF}
\end{equation}
Here $\sum_{l_{\rm r}^{\diamond}} 1=\sum_{S_{s}=0}^{(L-2S_{\eta})/2}{L\choose 2S_{\eta}}\times {(L - 2S_{\eta} + 2S_s)/2\choose 2S_{s}}$ is 
the dimension of that reduced subspace.

$J_{\rho 1}=$max$\{J_{\rho}\}$ is in Eq. (\ref{JzAve1}) for the reference S$^z$SLN subspace 1 the largest
coefficient $J_{\rho}$ in Eq. (\ref{JzAve}) of all S$^z$SLN subspaces contained in a S$^z$S subspace with density 
$m_{\eta}=m_{\eta}^z$. Moreover, the inequality $J_{\rho 1N}\geq J_{\rho 1}$ involving the coefficients
of the current absolute value averages in Eqs. (\ref{JzAve1}) and (\ref{JzAvensMax}) is valid for
all fixed densities $m_{\eta} \in [0,1]$ of the corresponding reference S$^z$SLN subspace 1
and S$^z$SLN$_{N_1}$ subspace belonging to the same S$^z$S subspace. A consequence of such properties is that the following inequality involving
the average values of the square of the charge current in Eqs. (\ref{averageINF}) and (\ref{aveSzSLNINF}) holds,
\begin{equation}
\left\langle\vert\langle\hat{J} (l_{\rm r}^{\diamond},S_{\eta},S_{\eta}^z,u)\rangle\vert^2\right\rangle_{S_{\eta}^z\,1N} \geq 
\left\langle\vert\langle\hat{J} (l_{\rm r}^{\star},L_{\eta},S_{\eta},S_{\eta}^z,u)\rangle\vert^2\right\rangle_{S_{\eta}^z} \, .
\label{INEaveSzSLNINF}
\end{equation} 

For high temperature $T\rightarrow\infty$, the $T>0$ expression of the charge stiffness, Eq. (\ref{D-all-T-simp}),
simplifies to,
\begin{eqnarray}
D (T) & = & {(2S_{\eta}^z)^2\over 2 L T}\,\,{\sum_{L_{\eta}=2\vert S_{\eta}^z\vert}^{L}\sum_{S_{\eta}=\vert S_{\eta}^z\vert}^{L_{\eta}/2}\sum_{l_{\rm r}} 
{\vert\langle\hat{J}_{LWS} (l_{\rm r},L_{\eta},S_{\eta},u)\rangle\vert^2\over (2S_{\eta})^2}
\over \sum_{L_{\eta}=2\vert S_{\eta}^z\vert}^{L}\sum_{S_{\eta}=\vert S_{\eta}^z\vert}^{L_{\eta}/2}
d_{\rm subspace}^{LWS} (L_{\eta},S_{\eta})}   
\nonumber \\
& = & {\left\langle\vert\langle\hat{J} (l_{\rm r},L_{\eta},S_{\eta},S_{\eta}^z,u)\rangle\vert^2\right\rangle_{S_{\eta}^z}\over 2 L T} \, .
\label{D-all-T-simp-INF}
\end{eqnarray}
A high temperature $T\rightarrow\infty$ charge stiffness upper bound,
\begin{eqnarray}
D^{\diamond} (T) & = & {(2S_{\eta}^z)^2\over 2 L T}\,\,{\sum_{S_{\eta}=\vert S_{\eta}^z\vert}^{L/2}\sum_{l_{\rm r}^{\diamond}} 
{\vert\langle\hat{J}_{LWS} (l_{\rm r}^{\diamond},S_{\eta},u)\rangle\vert^2\over (2S_{\eta})^2}
\over \sum_{S_{\eta}=\vert S_{\eta}^z\vert}^{L/2}\sum_{S_{s}=0}^{(L-2S_{\eta})/2}
{L\choose 2S_{\eta}}\times {(L - 2S_{\eta} + 2S_s)/2\choose 2S_{s}}}
\nonumber \\
& = & {\left\langle\vert\langle\hat{J} (l_{\rm r}^{\diamond},L_{\eta},S_{\eta},S_{\eta}^z,u)\rangle\vert^2\right\rangle_{S_{\eta}^z\,1N}\over 2 L T}\ \, ,
\label{UB-all-T-simp-INF}
\end{eqnarray}
such that $D (T) \leq  D^{\diamond} (T)$ then follows from the inequality, Eq. (\ref{INEaveSzSLNINF}).

As mentioned above, a S$^z$SLN$_{N_1}$ subspace can be divided into 
a set of reference S$^z$SLN$_{SN}$ subspaces 1A, each with a fixed density $m_s \in [0,(1-m_{\eta})]$.
In Appendix \ref{Appendix8} it is shown that a corresponding charge stiffness upper bound only
involving the $S_s=0$ contributions from the $m_s=0$ reference S$^z$SLN$_{SN}$ subspace 1A
is larger than that given in Eq. (\ref{UB-all-T-simp-INF}). This gives our ultimate charge stiffness upper bound within the grand 
canonical ensemble for the TL and high temperature $T\rightarrow\infty$,
\begin{eqnarray}
D^{\diamond\diamond} (T) & = & {(2S_{\eta}^z)^2\over 2 L T}\,\,{\sum_{S_{\eta}=\vert S_{\eta}^z\vert}^{L/2}\sum_{l_{\rm r}^{\diamond}} 
{\vert\langle\hat{J}_{LWS} (l_{\rm r}^{\diamond},S_{\eta},u)\rangle\vert^2\over (2S_{\eta})^2}
\over \sum_{S_{\eta}=\vert S_{\eta}^z\vert}^{L/2}{L\choose 2S_{\eta}}} \, ,
\label{UB-all-T-simp-INF2}
\end{eqnarray}
where $J_c^h (q_j)$ is the general current spectrum in Eq. (\ref{J-partS1}) for $S_s=0$.
Up to $u^{-2}$ order it is given in Eq. (\ref{Jchalletas}) for $S_s=0$. It thus reads,
\begin{equation}
J_c^h (q_j) = 2t \sin q_j - 2t\,{(1-m_{\eta})\ln 2\over u}\sin 2q_j +
6t\,\left({(1-m_{\eta})\ln 2\over u}\right)^2\left(1-{3\over 2}\sin^2 q_j\right)\sin q_j \, .
\label{Jqugg1}
\end{equation}
The general current spectrum in Eq. (\ref{J-partS1}) has up to $u^{-2}$ order
the same universal form for all $\eta$-Bethe states that span the reference S$^z$SLN$_{SN}$ 
subspaces 1A, Eq. (\ref{Jchalletas}) for $m_s \in [0,(1-m_{\eta})]$ and Eq. (\ref{Jqugg1}) at $m_s=0$.

In Appendix \ref{Appendix8} the current spectrum, Eq. (\ref{Jqugg1}), is used in the
charge stiffness upper bound, Eq. (\ref{UB-all-T-simp-INF2}), to derive the following
exact expansion up to $u^{-2}$ order of that upper bound valid in the TL for $m^z_\eta \in [0,1/2]$,
\begin{equation}
D (T) \leq D^{\diamond\diamond}(T) = \frac{c_{\rm gc}\,t^2}{2T} (m^z_\eta)^2 
\hspace{0.20cm}{\rm where}\hspace{0.20cm}
c_{\rm gc} = 2\pi^2\left(1 + \left(\frac{\ln 2}{2u}\right)^2\right) \, .
\label{DdiamTmG}
\end{equation}

On the one hand, this expression applies to the $m^z_\eta\ll 1$ limit. The charge stiffness Mazur's lower bound 
has been derived for $T\rightarrow\infty$ in Ref. \cite{ZNP-97}. 
In the $S_s^z=0$ case considered in the upper-bound studies of 
this paper, one finds that the charge stiffness Mazur's lower bound $D^{\rm Mz}(T)$
can be written as given in Eq. (\ref{DMazurTm}) of Appendix \ref{Appendix8}.
From the combined use of that equation and Eq. (\ref{DdiamTmG}) one finds that in the
$m^z_{\eta}\ll 1$ limit of more interest for our study
and up to ${\cal{O}}(u^{-2})$ order the charge stiffness is of the form 
$D (t) = \frac{c_u\,t^2}{2T} (m^z_{\eta})^2$ where the coefficient $c_u$ obeys the double 
inequality,
\begin{eqnarray}
2\left(1-\left({1/\sqrt{2}\over 2u}\right)^2\right)
& \leq & c_u \leq 2\pi^2\left(1 + \left(\frac{\ln2}{2u}\right)^2\right) 
\hspace{0.2cm}{\rm for}\hspace{0.2cm}m^z_{\eta}\ll 1 \, .
\label{DdoubleIN012}
\end{eqnarray}
Here $1/\sqrt{2}\approx 0.707$ and $\ln 2\approx 0.693$ have near values.

On the other hand, the use of the current spectrum, Eq. (\ref{Jqugg1}), in the upper bound, Eq. (\ref{UB-all-T-simp-INF2}),
trivially leads in the $n_e = (1-m^z_\eta) \ll 1$ limit to,
\begin{equation}
D (T) = D^{\diamond\diamond}(T) = \frac{c_{\rm gc}'\,t^2}{2T} (1-m^z_\eta) 
\hspace{0.20cm}{\rm where}\hspace{0.20cm} c_{\rm gc}' = 2 \, .
\label{DdiamTm1}
\end{equation}
In the $n_e = (1-m^z_\eta) \ll 1$ limit the upper bound, Eq. (\ref{UB-all-T-simp-INF2}),
equals up to ${\cal{O}}(u^{-2})$ order the charge stiffness, so that the expression, Eq. (\ref{DdiamTm1}), gives the exact 
asymptotic behavior in that limit of the charge stiffness for $T\rightarrow\infty$ in the TL.

On the one hand, the coefficient $c_{\rm gc}$ in the upper bound, Eq. (\ref{Dcm0GC}), smoothly slightly increases 
from $c_{\rm gc}=2\pi^2\approx 19.74$ for
$u\rightarrow\infty$ upon decreasing $u$ at least down to $u\approx 3/2$ within the $u>3/2$ range for which its
${\cal{O}}(u^{-2})$ order expansion remains a good approximation. At $u\approx 3/2$ it reads
$c_{\rm gc} = 2\pi^2(1+(\ln2/3)^2) \approx 20.79$. On the other hand, the coefficient $c_{\rm gc}'$ in the upper bound,
Eq. (\ref{Dclinem1GC}), reads $c_{\rm gc}'=2$ up to ${\cal{O}} (u^{-2})$.

This completes our finding of a vanishing charge stiffness in the TL, $L\to\infty$, within the grand-canonical
ensemble for $T\rightarrow\infty$ and any fixed range or even distribution of $S_{\eta}^z$, or any distribution of $m_{\eta}^z$ shrinking sufficiently 
fast that $\langle (m_{\eta}^z)^2 \rangle \to 0$. 

\section{Concluding remarks}
\label{concluding}

At $U=0$ the charge stiffness $D (T)$ of the 1D Hubbard model is a simple problem in terms of the non-interacting 
electron representation. It is found to be finite at $m_{\eta}^z = 0$, both at zero and finite temperature. 
$D (T)>0$ reaches a maximum value at $T=0$, max\,$D(T)=D(0)=2t/\pi$, behaving for low and 
high temperature $T$ as $[D (0)-D(T)]\propto T^2>0$ and $D (T)\propto 1/T$, respectively. 
(The qualitative difference of the $U=0$ and $u> 0$ physics and the related $T>0$ transition that occurs at
$U = U_c = 0$ is an issue discussed in Appendix \ref{Appendix2} for $m_{\eta}^z\rightarrow 0$ and $m_{\eta}^z = 0$
and in Appendix \ref{Appendix3} for $m_{\eta}^z\in [0,1]$.)

In this paper strong evidence is provided that the charge stiffness of the 1D Hubbard model vanishes at $m_{\eta}^z = 0$ for $T>0$ 
and the whole $u>0$ range in the TL within the canonical ensemble. For finite temperatures this leaves out, marginally, the grand 
canonical ensemble in which $\langle (m_{\eta}^z)^2\rangle = {\cal{O}}(1/L)$. However, the following properties lead us 
to expect that our prediction remains valid at finite temperatures in the grand-canonical ensemble case, in accord with the usual expectation 
of the equivalence of ensembles in the TL.

First, we have specifically confirmed the validity of this expectation in the limit of very high temperature $T\rightarrow\infty$.
The corresponding high-temperature charge stiffness upper bound, Eq. (\ref{DdiamTmG}), 
confirms that for $T\rightarrow\infty$ the charge stiffness of the 1D Hubbard model vanishes in the TL 
in the chemical potential $\mu\rightarrow \mu_{u}$ limit where $(\mu -\mu_{u})\geq 0$
and $2\mu_{u}$ is the Mott-Hubbard gap, Eq. (\ref{2mu0}) of Appendix \ref{Appendix1}. That upper bound
was computed up to $u^{-2}$ order, which applies for approximately $u>3/2$, yet it is expected that
similar results apply for $u>0$.

Second, at zero temperature and $m_{\eta}^z = 0$ the charge Drude weight is given in the TL by $D(0) = 2t/\pi$ at $U=0$ and 
vanishes for $u>0$ \cite{Shastry-90,Carmelo-92-C}. That it vanishes at $T=0$ for $u>0$ reveals that
a finite charge stiffness $D (T)$ for $T>0$ at $m_{\eta}^z = 0$ would result from thermal fluctuations alone. That
such fluctuations are largest at high temperature thus provides strong evidence that our $T\rightarrow\infty$
results within the grand-canonical ensemble apply as well to all temperatures $T\geq 0$.

Third, the large overestimate of the charge elementary currents we used in deriving the charge stiffness upper bound, Eq. (\ref{D**}),
is consistent with such an expectation. Our canonical-ensemble charge stiffness upper bounds in Eqs. (\ref{D**})-(\ref{value-D-m01allu})
are also valid for $T\rightarrow\infty$. Their comparison with those provided in Eqs. (\ref{DdiamTmG}) and (\ref{DdiamTm1})
within the grand-canonical ensemble confirms an average charge stiffness upper bound overestimation factor $c_{\rm oe}^2={\cal{O}}(L)$. 
For instance, $c_{\rm oe}^2$ changes for $u\gg 1$ from $c_{\rm oe}^2\approx (2/\pi^2) L$
for $m_{\eta}^z\ll 1$ to $c_{\rm oe}^2= n_{\rho}\,L=N_{\rho}$ for $m_{\eta}^z\rightarrow 1$. 
(For the S$^z$SLN subspace 1 that dominates the contributions to the charge stiffness, one has that $1 -m_{\eta}^z = n_c = n_{\rho}$.)
In terms of the charge current absolute values upper bounds derived for the canonical ensemble relative those constructed for
the grand canonical ensemble, this means for general S$^z$SLN subspaces an average overestimation factor 
$c_{\rm oe}\approx \sqrt{L}$ that for $u\gg 1$ varies from $c_{\rm oe} \approx \sqrt{(2/\pi^2) L}$ for $m_{\eta}^z\ll 1$ 
to $c_{\rm oe} \approx \sqrt{n_{\rho}\,L}=\sqrt{N_{\rho}}$ for $m_{\eta}^z\rightarrow 1$. This huge overestimate of the charge
elementary BA currents used in the computation of the charge Drude weight upper bound, Eq. (\ref{D**}), 
provides additional strong evidence that, as for high temperature $T\rightarrow\infty$, the charge stiffness vanishes for finite temperatures
within the grand-canonical ensemble in the TL for chemical potential $\mu\rightarrow \mu_{u}$ where $(\mu -\mu_{u})\geq 0$.

The use of the general formalism of hydrodynamics introduced in Refs. \cite{Ilievski-17A,Ilievski-17}
provides further strong evidence that the charge or spin stiffnesses vanish at finite temperatures
within the grand canonical ensemble when the corresponding chemical potentials
vanish. (Within our notation, in the case of the charge degrees of freedom the chemical potential 
of such references refers to $(\mu -\mu _{u})$.)
The analysis of Refs. \cite{Ilievski-17A,Ilievski-17} accounts for in the 1D Hubbard model the entire space
of macro-states being in a one-to-one correspondence with particle-hole
invariant commuting (fused) transfer matrices, pertaining
to a discrete family of unitary irreducible representations
of the underlying quantum symmetry. According to the authors of these 
references, this readily implies vanishing finite-temperature charge or spin Drude
weights when the corresponding chemical potentials
vanish, irrespective of the interaction strength. 

The problem studied in this paper refers though to a controversial issue, as different approaches yield contradictory results
\cite{ZP-96,Kawa-98,PDSC-00,PSZL-04,ANI-05,HPZ-11,Ilievski-17A,Ilievski-17,Carmelo-13,Karrasch-16,Karrasch-17}. 
This includes different methods based on the same TBA. Indeed we believe that the problem 
is not the TBA but rather how to use the TBA to access the stiffness of each specific solvable model. 
As mentioned in Section \ref{Introduction}, our $u>0$ and $m_{\eta}^z = 0$ predictions for $D(T)$ 
agree with the conjectures of Ref. \cite{ZP-96} and the exact
large-$u$ results of Ref. \cite{PDSC-00}. The latter disagree with the prediction of Ref. \cite{Kawa-98} that
$D(T)$ should be finite in the TL for $u>0$, $T>0$, and $m_{\eta}^z = 0$. The exact large-$u$ 
results of Ref. \cite{PDSC-00}, which find that $D(T)=0$ in the TL at $m_{\eta}^z = 0$, reveal 
that the results of Ref. \cite{Kawa-98} {\it cannot} be exact. Indeed, it is shown that if $D(T)$ was finite in 
the TL for $T>0$ at $m_{\eta}^z = 0$, the pre-factor of the exponential of the Mott-Hubbard gap, Eq. (\ref{2mu0}) of Appendix \ref{Appendix1}, 
would be, at least, of the order of $t^2/U$, and not of the order one, as found in Ref. \cite{Kawa-98}. 

The method introduced in that reference for the 1D Hubbard model and used in Ref. \cite{Zotos-99}
for the spin-$1/2$ $XXZ$ chain relies on the TBA. In the case of the spin-$1/2$ $XXZ$ chain, 
it leads to completely different results from the phenomenological
method of Ref. \cite{ANI-05}, which however relies on a spinon and anti-spinon particle basis for the same TBA. 
The studies of Refs. \cite{SPA-11,CPC,CP} exclude the large spin stiffness found in Ref. \cite{ANI-05} 
for the spin-$1/2$ $XXX$ chain spin stiffness for zero spin density and $T>0$ in the TL.

On the one hand, the results of Refs. \cite{CPC,CP} provide strong evidence
that those of Ref. \cite{Zotos-99} for the spin-$1/2$ $XXZ$ and $XXX$ chains spin stiffness are correct. On the other hand,
our present results reveal that the results Ref. \cite{Kawa-98} for the charge stiffness of the half-filled 1D Hubbard,
which predict it to be finite at $m_{\eta}^z = 0$ in the TL for $u>0$ and $T>0$, are incorrect. This is despite
such a prediction apparently relying on the TBA-based method that has been used in Ref. \cite{Zotos-99} 
to derive the spin stiffness of the spin-$1/2$ $XXZ$ and $XXX$ chains.

The possible error source of the predictions of Ref. \cite{Kawa-98} is revealed by inspection of separate 
integrals of the individual summands occurring in the integrands of Eq. (25) of that reference. One finds 
that such separate integrals diverge at the hole concentration $m_{\eta}^z = 0$ at which the general Mazur's 
inequality is inconclusive. This turns out to be a fatal problem in that equation. 
Also in the case of the spin-$1/2$ $XXX$ chain the separate integrals of the individual summands 
occurring in the integrands of Eqs. (24) and (25) of Ref. \cite{ANI-05} diverge at zero spin density
or which the general Mazur's inequality is again inconclusive.
However, there is evidence that such divergences  can be removed in the case of
models whose stiffness is finite at zero temperature. This is the case of the spin-$1/2$ $XXX$ chain
in the zero spin density limit \cite{Zotos-99}. They are though a fatal problem 
for the 1D Hubbard model for $m_{\eta}^z\rightarrow 0$, $u>0$, and $T>0$, whose charge stiffness 
vanishes at $T=0$ for $u>0$. This problem deserves though further investigations.
  
Finally, the lack of charge ballistic transport in the 1D Hubbard model for $u>0$ also found
in this paper indicates that charge transport at high temperatures is dominated by a 
diffusive contribution. 
 
\acknowledgements
We thank David K. Campbell, Pedro. D. Sacramento, and Xenophon Zotos for discussions. J. M. P. C. and S. N. thank the support from 
C. S. R. C. (Beijing) and J. M. P. C. and T. P. acknowledge the support of the ERC Advanced Grant 694544 - OMNES. 
J. M. P. C. thanks the support by the Portuguese FCT through the Grant UID/FIS/04650/2013. T. P. acknowledges
the support from the Grants of the Slovenian Research Agency (ARRS) P1-004 and N1-0025.

\appendix

\section{Functional representation for the TBA equations, $\eta$-Bethe states energy eigenvalues, and charge current operator expectation values}
\label{Appendix1}

Here some TBA results needed for the studies of this paper are provided. This includes
the 1D Hubbard model TBA equations within the functional representation used in this paper. Furthermore,
the model's exact energy eigenvalues and other energy scales related to them are also provided and the validity of the
$\beta= c,\eta$ band hole representation of the $\eta$-Bethe states charge currents in Eq. (\ref{J-part}) is confirmed.

Within the pseudoparticle momentum distribution functional notation used in this 
paper, the TBA equations introduced in Ref. \cite{Takahashi} read,
\begin{eqnarray}
q_j & = & k^c (q_j) + {2\over L}\sum_{n =1}^{\infty}
\sum_{j'=1}^{L_{s n}}\,N_{sn}(q_{j'})\arctan\left({\sin
k^c (q_j)-\Lambda^{sn}(q_{j'}) \over n u}\right)
\nonumber \\
& + & {2\over L}\sum_{n =1}^{\infty}
\sum_{j'=1}^{L_{\eta n}}\, N_{\eta n}(q_{j'}) \arctan\left({\sin
k^c (q_j)-\Lambda^{\eta n}(q_{j'}) \over n u}\right) 
\hspace{0.20cm}{\rm for}\hspace{0.20cm}j = 1,...,L \, , 
\label{Tapco1}
\end{eqnarray}
and
\begin{eqnarray}
q_j & = & \delta_{\alpha,\eta}
\sum_{\iota =\pm1}\arcsin (\Lambda^{\alpha n} (q_{j}) - i\,\iota\,u)
+ {2\,(-1)^{\delta_{\alpha,\eta}}\over L} \sum_{j'=1}^{L}\,
N_{c}(q_{j'})\arctan\left({\Lambda^{\alpha n}(q_j)-\sin k^c (q_{j'})\over n u}\right)
\nonumber \\
& - & {1\over L}\sum_{n' =1}^{\infty}\sum_{j'=1}^{L_{\alpha n'}}\, N_{\alpha n'}(q_{j'})\Theta_{n\,n'}
\left({\Lambda^{\alpha n}(q_j)-\Lambda^{\alpha n'}(q_{j'})\over u}\right)
\nonumber \\
& & {\rm for}\hspace{0.20cm}
j = 1,...,L_{\alpha n}\hspace{0.20cm}{\rm where}\hspace{0.20cm}
\alpha = \eta, s\hspace{0.20cm}{\rm and}\hspace{0.20cm}n =1,...,\infty \, .
\label{Tapco2}
\end{eqnarray}
The sets of $j = 1,...,L$ and $j = 1,...,L_{\alpha n}$ quantum numbers $q_j$ in Eqs. (\ref{Tapco1}) and (\ref{Tapco2}), 
respectively, which are defined in Eqs. (\ref{q-j}) and (\ref{Ic-an}), play the role of microscopic momentum values of 
different TBA excitation branches. The corresponding $\beta$-band momentum distribution functions $N_{\beta} (q_j)$ 
where $\beta =c,\eta n,sn$ read $N_{\beta} (q_j)=1$ and $N_{\beta} (q_j)=0$ for occupied and unoccupied discrete momentum values, respectively,
the rapidity functional $\Lambda^{\alpha n}(q_{j})$ is the real part of the complex rapidity, Eq. (\ref{complex-rap}),
and $\Theta_{n\,n'} (x)$ is the function,
\begin{eqnarray}
\Theta_{n\,n'}(x) & = & \delta_{n,n'}\Bigl\{2\arctan\Bigl({x\over 2n}\Bigl) 
+ \sum_{l'=1}^{n -1}4\arctan\Bigl({x\over 2l'}\Bigl)\Bigr\} 
\nonumber \\
& + & (1-\delta_{n,n'})\Bigl\{ 2\arctan\Bigl({x\over \vert\,n-n'\vert}\Bigl)
+ 2\arctan\Bigl({x\over n+n'}\Bigl) 
+ \sum_{l'=1}^{{n+n'-\vert\,n-n'\vert\over 2} -1}4\arctan\Bigl({x\over \vert\, n-n'\vert +2l'}\Bigl)\Bigr\} \, ,
\label{Theta}
\end{eqnarray}
where $n, n' = 1,...,\infty$. The indices $\alpha =\eta,s$ and numbers $n =1,...,\infty$ refer to different 
TBA excitation branches that as discussed in Sections \ref{Introduction3} and \ref{BindingUn}
are associated with the $\alpha n$-pair configurations within which
$n= 1,...,\infty$ $\eta$-spin singlet pairs are bound. They refer to the
internal degrees of freedom of the neutral composite $\alpha n$ pseudoparticles
considered in Section \ref{BindingUn}.

Useful quantities directly related to the rapidity momentum functional 
$k^c (q)$ and rapidity functionals $\Lambda^{\alpha n} (q)$ defined for
each $\eta$-Bethe state by Eqs. (\ref{Tapco1}) and (\ref{Tapco2})
are the general distributions $2\pi\rho_c (k_j)$ and  $2\pi\sigma_{\alpha n} (\Lambda_j)$
where $\alpha = \eta,s$ and $n=1,...,\infty$. (They appear in the current
spectra, Eq. (\ref{jn-fn}).) Such distributions are defined by the following derivatives,
\begin{eqnarray}
2\pi\rho_c (k_j) & = & {\partial q^c (k)\over \partial k}\vert_{k=k_j}
\hspace{0.20cm}{\rm and}\hspace{0.20cm}{\rm thus}\hspace{0.20cm}
{1\over2\pi\rho_c (k^c (q_j))} = {\partial k^c (q)\over \partial q}\vert_{q=q_j} \, ,
\nonumber \\
2\pi\sigma_{\alpha n} (\Lambda_j) & = & {\partial q^{\alpha n} (\Lambda)\over \partial \Lambda}\vert_{\Lambda=\Lambda_j} 
\hspace{0.20cm}{\rm and}\hspace{0.20cm}{\rm thus}\hspace{0.20cm}
{1\over 2\pi\sigma_{\alpha n} (\Lambda^{\alpha n} (q_j))} = {\partial \Lambda^{\alpha n} (q) \over \partial q}\vert_{q=q_j} \, .
\label{sigmsrhoderiv}
\end{eqnarray}
The functions $q^c (k)$ and $q^{\alpha n} (\Lambda)$ in these expressions
stand for the inverse functions of the rapidity momentum functional 
$k^c (q)$ and rapidity functionals $\Lambda^{\alpha n} (q)$, respectively. 

The energy eigenvalues have for the hole concentration range $m_{\eta}^z\in [0,1]$ and 
the spin density interval $m_s^z\in [0,(1-m_{\eta}^z)]$ the following form,
\begin{equation}
E = \sum_{j=1}^{L}\left(N_{c} (q_j)\,E_c (q_j) + U/4 - \mu_{\eta}\right)
+ \sum_{\alpha=\eta,s}\sum_{n=1}^{\infty}\sum_{j=1}^{L_{\alpha n}}\,N_{\alpha n} (q_j)\,E_{\alpha n} (q_j) 
+ \sum_{\alpha=\eta,s}2\mu_{\alpha}\,(S_{\alpha}+S_{\alpha}^z) \, .
\label{E}
\end{equation}
The $\alpha =\eta,s$ energy scales $2\mu_{\alpha}$ are here given by,
\begin{equation}
2\mu_{\eta} = 2\mu\hspace{0.20cm}{\rm and}\hspace{0.20cm}2\mu_{s} = 2\mu_B\,h \, ,
\label{2mu-eta-s}
\end{equation}
where $\mu_B$ is the Bohr magneton and $h$ denotes a magnetic field.
The spectra $E_c (q_j)$ and $E_{\alpha n} (q_j)$ read,
\begin{eqnarray}
E_c (q_j) & = & - 2t\cos k^c (q_j) - U/2 + \mu_{\eta} - \mu_s \, ,
\nonumber \\
E_{\alpha n} (q_j) & = & n\,2\mu_{\alpha} + 
\delta_{\alpha,\eta}\left(2t\,\sum_{\iota=\pm 1}\Bigl\{\sqrt{1-(\Lambda^{\eta n} (q_j) -i\,\iota n u)^2}\Bigr\} - n\,U\right)
\nonumber \\
& &{\rm for}\hspace{0.20cm}\alpha = \eta,s\hspace{0.20cm}{\rm where}\hspace{0.20cm}n =1,...,\infty \, ,
\label{spectra-E-an-c-0}
\end{eqnarray}
respectively. (The corresponding momentum eigenvalues of general $u>0$ energy and momentum 
eigenstates are provided in Eq. (\ref{P}).)	

On the one hand, for the metallic phase densities ranges $m_{\eta}^z\in ]0,1]$ and $m_s^z\in [0,(1-m_{\eta}^z)]$ 
the $\alpha = \eta,s$ energy scales $2\mu_{\alpha}$ in Eq. (\ref{E}) are related to the unpaired physical $\eta$-spin
$(\alpha =\eta)$ and unpaired physical spin $(\alpha =s)$ energies relative to the ground state zero-energy level,
\begin{equation}
\varepsilon_{\alpha,-1/2} = 2\mu_{\alpha}\hspace{0.20cm}{\rm and}\hspace{0.20cm}\varepsilon_{\alpha,+1/2} = 0
\hspace{0.20cm}{\rm for}\hspace{0.20cm}\alpha =\eta,s \, .
\label{energy-eta}
\end{equation}
On the other hand, for the $m_{\eta}^z=0$ Mott-Hubbard insulator phase, $\varepsilon_{s,\pm 1/2}$ is given by Eq. (\ref{energy-eta})
whereas it reads $\varepsilon_{\eta,\pm 1/2} =  (\mu_{u}\mp\mu)$ for the chemical potential interval 
$\mu \in [-\mu_{u},\mu_{u}]$. The energy scale $\mu_{u}$ refers to the Mott-Hubbard gap $2\mu_{u}$, which
for the $S_s^z=0$ subspace considered in some sections of this paper reads \cite{Lieb,Lieb-03,Ovchi},
\begin{equation}
2\mu_{u} = U - 4t + 8t\int_0^{\infty}d\omega {J_1 (\omega)\over\omega\,(1+e^{2\omega u})} 
= {16\,t^2\over U}\int_1^{\infty}d\omega {\sqrt{\omega^2-1}\over\sinh\left({2\pi t\omega\over U}\right)} \, .
\label{2mu0}
\end{equation}
Here $J_1 (\omega)$ is a Bessel function. Its limiting behaviors 
are $2\mu_{u} \approx (8/\pi)\,\sqrt{t\,U}\,e^{-2\pi \left({t\over U}\right)}$ for $u\ll 1$ \cite{Ovchi}
and $2\mu_{u} \approx (U - 4t)$ for $u\gg 1$.

The charge currents absolute values upper bounds and charge stiffness upper bounds
derived in this paper refer to the 1D Hubbard model in the $S_s^z=0$ subspace for which
$h=0$ and thus the energy scale $2\mu_{s} = 2\mu_B\,h$ in
Eq. (\ref{2mu-eta-s}) vanishes, $2\mu_B\,h=0$. The other energy scale $2\mu_{\eta} = 2\mu$
in that equation involves the chemical potential $\mu=\mu (m_{\eta}^z)$. At $m_{\eta}^z=0$ it varies in the range
$\mu \in [-\mu_{u},\mu_{u}]$ in spite of the electronic density remaining constant, which is
a property of the corresponding $m_{\eta}^z=0$ and $u>0$ Mott-Hubbard insulator quantum phase. 
It is an odd function of the hole concentration $m_{\eta}^z$. 
The interval $m_{\eta}^z\in ]0,1[$ refers to the metallic quantum phase for which $\mu = \mu (m_{\eta}^z)$ is a continuous function 
of $m_{\eta}^z$. It smoothly increases from $\mu = \mu_{u}$ for $m_{\eta}^z\rightarrow 0$
to $\mu = (U+4t)/2$ for $m_{\eta}^z\rightarrow 1$ where the Mott-Hubbard gap, Eq. (\ref{2mu0}), obeys
the inequality $2\mu_{u} <(U+4t)$. It is finite for the whole $u>0$ range and vanishes at $U=0$.
 
Concerning the validity of the charge currents hole representation given in Eq. (\ref{J-part}), 
as mentioned in Section \ref{curr-val}, the 1D Hubbard model in a uniform vector potential $\Phi/L$ whose Hamiltonian 
is given in Eq. (4) of Ref. \cite{Gu-07A} remains solvable by the BA. The coupling of the charge degrees of 
freedom to the vector potential is described under the replacement on the left hand side of Eq. (\ref{Tapco1})
of $q_j$ by $q_j+\Phi/L$ and on the left hand side of Eq. (\ref{Tapco2}) for $\alpha n=\eta n$ of $q_j$ by 
$q_j - 2n\,\Phi/L$. This implies that $k^c (q_j)$ and $\Lambda^{\eta n}(q_j)$ remain having the same
expressions provided that their momentum variables $q_j$ are replaced by $q_j+\Phi/L$ and $q_j - 2n\,\Phi/L$,
respectively. Hence the $\eta$-Bethe states charge currents can be derived 
and expressed in the TL in terms of $c$ and $\eta n$-band pseudoparticle occupancies as follows,
\begin{eqnarray}
\langle\hat{J}_{LWS} (l_{\rm r},L_{\eta},S_{\eta},u)\rangle & = & - {d E\over d (\Phi/L)}\vert_{\Phi =0} = 
\sum_{j=1}^{L} N_{c} (q_j)\,\left(- {d\over d (\Phi/L)}E_c (q_j+\Phi/L)\vert_{\Phi =0}\right)
\nonumber \\
& - & \sum_{n=1}^{\infty}\sum_{j=1}^{L_{\eta n}}\,N_{\eta n} (q_j)\,
\left({d\over d (\Phi/L)} E_{\eta n} (q_j-2n\,\Phi/L)\vert_{\Phi =0}\right)
\nonumber \\
& = & \sum_{j=1}^{L}\,N_c (q_j)\,\,J_c (q_j)
+ \sum_{n=1}^{\infty}\sum_{j=1}^{L_{\eta n}}\,N_{\eta n} (q_j)\,\,J_{\eta n} (q_j) \, .
\label{J-partDEF}
\end{eqnarray}
Here $E$ is in $d E/d (\Phi/L)$ the energy functional, Eq. (\ref{E}) for $S_{\alpha}+S_{\alpha}^z=0$ $\eta$-Bethe states,
$E_c (q_j)$ and $E_{\eta n} (q_j)$ are given in Eq. (\ref{spectra-E-an-c-0}), and the current spectra $J_c (q_j)$ 
and $J_{\eta n} (q_j)$ read,
\begin{eqnarray}
J_c (q_j) & = & - {2t\sin k^c (q_j)\over 2\pi\rho_c (k^c (q_j))} 
\hspace{0.20cm}{\rm for}\hspace{0.20cm}q_j \in [-\pi,\pi] \, ,
\nonumber \\
J_{\eta n} (q_j) & = & - 4nt\sum_{\iota=\pm 1}
{\Lambda^{\eta n} (q_j) -i\,\iota n u \over 2\pi\sigma_{\eta n} (\Lambda^{\eta n} (q_j))
\sqrt{1-(\Lambda^{\eta n} (q_j) -i\,\iota n u)^2}}
\hspace{0.20cm}{\rm for}\hspace{0.20cm}q_j \in [-q_{\eta n},q_{\eta n}] \, ,
\label{jn-fnPSEU}
\end{eqnarray} 
respectively. 
 
That Eq. (\ref{J-partDEF}) can be rewritten as in Eq. (\ref{J-part}) requires that,
\begin{equation}
- \sum_{j=1}^{L}\,J_c (q_j) - \sum_{n=1}^{\infty}\sum_{j=1}^{L_{\eta n}}\,J_{\eta n} (q_j) = 0 \, .
\label{zeroQu}
\end{equation}
In the present TL one replaces the discrete momentum values $q_j$ such that $q_{j+1}-q_j = 2\pi/L$ by a 
continuous momentum variable $q$, so that after replacing sums by integrals and
overall multiplication by $-1$ this equation reads,
\begin{equation}
{L\over 2\pi}\int_{-\pi}^{\pi}dq\,J_c (q) + \sum_{n=1}^{\infty}{L\over 2\pi}\int_{-q_{\eta n}}^{q_{\eta n}}dq\,J_{\eta n} (q) = 0 \, .
\label{zeroQucont}
\end{equation} 
By combining Eqs. (\ref{sigmsrhoderiv}) and (\ref{jn-fnPSEU}) one finds that the current
spectra in Eq. (\ref{jn-fnPSEU}) can be written as,
\begin{eqnarray}
J_c (q) & = & 2t{d\over dq}\cos k^c (q)
\hspace{0.20cm}{\rm for}\hspace{0.20cm}q \in [-\pi,\pi] \, ,
\nonumber \\
J_{\eta n} (q) & = & 4nt\sum_{\iota=\pm 1}{d\over dq}\sqrt{1-(\Lambda^{\eta n} (q) -i\,\iota n u)^2}
\hspace{0.20cm}{\rm for}\hspace{0.20cm}q \in [-q_{\eta n},q_{\eta n}] \, .
\label{jn-fnPSEUcont}
\end{eqnarray} 
Another property used in the following is that for all $\eta$-Bethe states the following relations,
\begin{equation}
k^c (\pm\pi) = \pm\pi \hspace{0.20cm}{\rm and}\hspace{0.20cm}
\Lambda^{\eta n} (\pm q_{\eta n}) = \pm \Lambda^{\eta n}_{\rm max} \, ,
\label{useLimits}
\end{equation}
apply. From the use of Eqs. (\ref{jn-fnPSEUcont}) and (\ref{useLimits}) in 
Eq. (\ref{zeroQucont}) one confirms indeed that,
\begin{eqnarray}
& & {L\over 2\pi}2t (\cos (\pi) - \cos (-\pi)) + {L\over 2\pi}4nt 
\nonumber \\
& \times & \left(\sqrt{1-(\Lambda^{\eta n}_{\rm max} - i\,n u)^2}
+ \sqrt{1-(\Lambda^{\eta n}_{\rm max} + i\,n u)^2} - \sqrt{1-(\Lambda^{\eta n}_{\rm max} + i\,n u)^2}
- \sqrt{1-(\Lambda^{\eta n}_{\rm max} - i\,n u)^2}\right) = 0 \, ,
\label{RzeroQucont}
\end{eqnarray} 
where the trivial equality $(-\Lambda^{\eta n}_{\rm max} \mp i\,n u)^2 = (\Lambda^{\eta n}_{\rm max} \pm i\,n u)^2$
was used. (The fact that $\Lambda^{\eta n}_{\rm max}$ is given by $\infty$ is not needed for the vanishing of the 
quantity in Eq. (\ref{RzeroQucont}).)

\section{Qualitative difference of the $U=0$ and $u\rightarrow 0$ physics
and the finite-$T$ transition at $U=U_c=0$ and $m_{\eta}^z = 0$}
\label{Appendix2}

Here the mechanisms behind the behavior of the $m_{\eta}^z = 0$ charge stiffness $D(T)$ in the TL 
under the $T>0$ transition that occurs at $U=U_c=0$ are discussed. Such a transition
is a generalization of the $m_{\eta}^z = 0$ zero-temperature quantum phase transition from a metal to a 
Mott-Hubbard insulator that also occurs at $U=U_c=0$ and is driven by correlation effects.
In addition, it is confirmed that there is no contradiction whatsoever between 
the charge stiffness of the 1D Hubbard model vanishing within
the canonical ensemble for $T>0$ in the TL both at $m_{\eta}^z = 0$ and as $m_{\eta}^z\rightarrow 0$ 
for $u>0$ and being finite at $U=0$ in that limit and at the same hole concentration. 

The general notation $\vert \nu,u\rangle$ for the energy eigenstates used in Eqs. (\ref{DT-gen}) and (\ref{sigma-reg}) 
refers to the whole $u\geq 0$ range. Here we use it for the $m_{\eta}^z=0$ and $m_{\eta}^z\rightarrow 0$
energy and momentum eigenstates. On the one hand, for $u>0$ the 
states $\vert \nu,u\rangle$ correspond in this Appendix to the energy and momentum eigenstates 
$\vert l_{\rm r},S_{\eta},0,u\rangle$, Eq. (\ref{Gstate-BAstate}) for $m_{\eta}^z=0$ and
$m_{\eta}^z\rightarrow 0$. On the other hand, at $U=0$ the states $\vert \nu,u\rangle$ are instead 
chosen to be the common eigenstates of the momentum operator, current operator $\hat{J}$, and $U=0$ tight-binding Hamiltonian,
\begin{equation}
\hat{H}_0 = -t\sum_{\sigma}\sum_{j=1}^{L}\left[c_{j,\sigma}^{\dag}\,c_{j+1,\sigma} + {\rm h.c.}\right]
+ 2\mu\,{\hat{S}}_{\eta}^{z} \, .
\label{H0}
\end{equation}
This Hamiltonian is that of the 1D Hubbard, Eq. (\ref{H}), at $U=0$, which as given in Eq. (\ref{comm-JH})
commutes with the charge current operator, Eq. (\ref{c-s-currents}). 

The unbinding in the $u\rightarrow 0$ limit of the $l = 2,...,n$ $\eta$-spin singlet pairs 
within each $u>0$ $\eta n$-pair configuration that follows from the vanishing of
the imaginary part $i\,(n + 1 - 2l)\,u$ of the set of corresponding $l = 2,...,n$ complex rapidities
with the same real part, Eq. (\ref{complex-rap}) for $\alpha =\eta$, has most severe consequences 
at hole concentration $m_{\eta}^z=0$. As discussed in Section \ref{BindingUn} and Appendix \ref{Appendix3},
that unbinding along with the vanishing also as $u\rightarrow 0$ of the 
commutator $[\hat{J},\hat{H}]$, Eq. (\ref{comm-JH}), is associated with the rearrangement of $\eta$-spin 
and spin degrees of freedom in terms of the noninteracting electrons occupancy configurations that
generate the $U=0$ common eigenstates of the Hamiltonian, momentum operator, and charge current operator.
Those are the $U=0$ states $\vert \nu,u\rangle = \vert\nu ,0\rangle$ considered here. 

On the one hand and as discussed in Section \ref{BindingUn}, the form $i\,(n + 1 - 2l)\,u$ of the 
complex $\eta n$ rapidities imaginary part and of the commutator $[\hat{J},\hat{H}]$, Eq. (\ref{comm-JH}), 
confirms that the $u>0$ physics survives for any arbitrarily small value of $u$.
On the other hand, in the $u\rightarrow 0$ limit the set of energy and momentum eigenstates,
\begin{equation}
\vert l_{\rm r},S_{\eta},0,0^+\rangle \equiv \lim_{u\rightarrow 0}\vert l_{\rm r},S_{\eta},0,u\rangle \, ,
\label{Psi-U0}
\end{equation}
remain being a complete basis in the $m_{\eta}^z=0$ and $m_{\eta}^z\rightarrow 0$ subspaces
of the noninteracting $U=0$ quantum problem, yet are not eigenstates of the $U=0$ Hamiltonian $\hat{H}_0$, Eq. (\ref{H0}). 
This is related to the states, Eq. (\ref{Psi-U0}), being eigenstates of the $c$ lattice $U(1)$ symmetry generator 
${\tilde{N}}^R_{\eta} = \sum_{j=1}^L (1-{\tilde{n}}_{j,\uparrow}{\tilde{n}}_{j,\downarrow})$, 
which counts the number of rotated-electron unoccupied plus doubly occupied sites. 
(Here the $\sigma=\uparrow,\downarrow$ operators ${\tilde{n}}_{j,\sigma}$ are those in Eq. (\ref{rotated-operators}).)
Indeed, that generator only commutes with the 1D Hubbard model Hamiltonian for the $u>0$ range for which
it is well defined. {\it At} $U=0$ the quantum problem described by the Hamiltonian, Eq. (\ref{H0}), lacks the corresponding 
global $c$ lattice $U(1)$ symmetry. 

That the $U=0$ Hamiltonian, momentum operator, and charge current operator eigenstates $\vert \nu,0\rangle$ have quantum 
numbers distinct from those of the states $\vert l_{\rm r},S_{\eta},0,0^+\rangle$, Eq. (\ref{Psi-U0}),
follows from the non-perturbative character of the 1D Hubbard model and
its corresponding different global $SO(4)\otimes Z_2$ and $[SO(4)\otimes U(1)]/Z_2$ symmetries for $U=0$ and $u>0$, respectively.
Indeed, the $U=0$ and $u>0$ energy and momentum eigenstates are in one-to-one correspondence to the
representations of the $SO(4)\otimes Z_2$ and $[SO(4)\otimes U(1)]/Z_2$ symmetry groups, which
have distinct structures. 

The $m_{\eta}^z=0$ and $m_{\eta}^z\rightarrow 0$
common eigenstates $\vert \nu,0\rangle$ of $\hat{H}_0$ and of the current operator 
can be written as the following superposition of the states, Eq. (\ref{Psi-U0}),
\begin{equation}
\vert \nu,0\rangle = \sum_{S_{\eta}=0}^{L/2}\sum_{l_{\rm r}}
C_{l_{\rm r},S_{\eta}}^{\nu,0} \vert l_{\rm r},S_{\eta},0,0^+\rangle\hspace{0.20cm}{\rm where}\hspace{0.20cm}
C_{l_{\rm r},S_{\eta}}^{\nu,0} = \langle l_{\rm r},S_{\eta},0,0^+\vert \nu,0\rangle \, .
\label{m0-ll-C}
\end{equation}
The charge current operator $\hat{J}$ does not commute with both $(\hat{\vec{S}}_{\eta})^2$ and the 
$c$ lattice $U(1)$ symmetry generator ${\tilde{N}}^R_{\eta}$, 
yet commutes with $\hat{S}_{\eta}^{z}$. Hence the set of states $\vert l_{\rm r},S_{\eta},0,0^+\rangle $, Eq. (\ref{Psi-U0}), 
contributing to each state $\vert \nu,0\rangle$ in the expansion of Eq. (\ref{m0-ll-C}) have the same eigenvalue $S_{\eta}^z$ but may
have different $\eta$-spin values $S_{\eta}=0,1,2,...$ and different ${\tilde{N}}^R_{\eta}$ eigenvalues 
$L_{\eta} =2S_{\eta} + 2\Pi_{\eta}=0,2,4,...$. 

The use of the non-interacting basis associated with the representation of the $U=0$ energy
eigenstates $\{\vert \nu,0\rangle\}$ in terms of simple electron occupancy configurations relative to the electronic
vacuum, renders trivial the problem of the calculation of the real part of the 
conductivity $\sigma (\omega,T)$, Eq. (\ref{sigma}), at $m_{\eta}^{z}=0$ and $U=0$. 
That the states $\vert \nu,0\rangle$ are eigenstates of the current operator 
$\hat{J}$, implies that $\langle \nu,0\vert\hat{J}\vert \nu',0\rangle=0$ in Eq. (\ref{sigma-reg}) where $\nu\neq\nu'$, so that 
$\sigma_{reg} (\omega,T)=0$. As is well known, Eq. (\ref{sigma}) becomes then $\sigma (\omega,T) = 2\pi\,D (T)\,\delta (\omega)$.
Trivial calculations relying onto the simple form in terms of
electron creation operators of the generators onto the electronic vacuum of the states
$\vert \nu,0\rangle$ lead to the known result that
$D (T)>0$ reaches a maximum value at $T=0$, max\,$D(T)=D(0)$, behaving for low and 
high temperature $T$ as $[D (0)-D(T)]\propto T^2>0$ and $D (T)\propto 1/T$, respectively. 

Turning on an arbitrarily small infinitesimal $u\ll 1$ value leads to the 
emergence of the complex $\eta n$ rapidities imaginary part $i\,(n + 1 - 2l)\,u$ 
associated with the rearrangement of the $\eta$-spin-singlet configurations and brings about
a nonzero commutator of the charge current operator with the Hamiltonian, Eq. (\ref{comm-JH}).
This prevents the existence of common eigenstates of these two operators. For any finite onsite repulsion value
$u>0$, the energy and momentum eigenstates are the states $\vert l_{\rm r},S_{\eta},0,u\rangle$ on the right-hand side of Eq. (\ref{Psi-U0}),
which are as well eigenstates of the generator ${\tilde{N}}^R_{\eta}$. As found in this
paper, for such eigenstates of both the Hamiltonian, Eq. (\ref{H}), and that generator one has within the canonical
ensemble that $D(T)=0$ for $u>0$ in the TL both at $m_{\eta}^z=0$ and as $m_{\eta}^z\rightarrow 0$.

An interesting property is that the real-part of the conductivity sum rule 
remains invariant under the finite-temperature $U=U_c=0$ transition,
{\it i.e.} it has no discontinuity at $U=0$. Indeed, it has the same value at $U=0$ and for $u\rightarrow 0$,
the following relations holding,
\begin{eqnarray}
\lim_{u\rightarrow 0}2\pi\,D (T) & = & \left[\int_{-\infty}^{\infty} d\omega\,\sigma_{reg} (\omega,T)\right]\vert_{U=0}=0
\hspace{0.20cm}{\rm at}\hspace{0.20cm}m_{\eta}^z = 0 \, ,
\nonumber \\
2\pi\,D (T)\vert_{U=0} & = & \lim_{u\rightarrow 0}\int_{-\infty}^{\infty}d\omega\,\sigma_{reg} (\omega,T) >0
\hspace{0.20cm}{\rm at}\hspace{0.20cm}m_{\eta}^z = 0 \, .
\label{M-MHI-T}
\end{eqnarray}
The singular behavior associated with the $U=U_c=0$ transition is rather that $\{D(T)=0$ and $\sigma_{reg} (\omega,T)>0\}$ and 
$\{D(T)>0$ and $\sigma_{reg} (\omega,T)=0\}$ for any arbitrarily small infinitesimal $u\ll 1$ and at $U=0$, respectively. 

That the real part of the conductivity sum rule is finite and $D(T)=0$ for $u\rightarrow 0$
implies that $\sigma_{reg} (\omega,T)$ does not vanish in the
$u\rightarrow 0$ limit and thus that some of the off-diagonal matrix elements 
in the $\sigma_{reg} (\omega,T)$ expression
between states $\vert l_{\rm r},S_{\eta},0,0^+\rangle$ and
$\vert l_{\rm r},S_{\eta}\pm 1,0,0^+\rangle$ allowed by the selection rule, Eq. (\ref{matrix-0}),
are non vanishing. This indeed occurs provided that the $\eta$-spin value differs by 
$\delta S_{\eta}=\pm 1$ if $S_{\eta}>0$ and by $\delta S_{\eta}=1$ if $S_{\eta}=0$. 
This is consistent with the states $\vert l_{\rm r},S_{\eta},0,0^+\rangle$ not
being eigenstates of the current $\hat{J}$, in contrast to the $U=0$ energy  
eigenstates $\vert \nu,0\rangle$. For the latter states all the off-diagonal current matrix elements 
$\langle \nu,0\vert\hat{J}\vert \nu',0\rangle=0$ in Eq. (\ref{sigma-reg}) vanish, 
so that $\sigma_{reg} (\omega,T)=0$ and $D(T)>0$ at $U=0$, as given in Eq. (\ref{M-MHI-T}). 

Let $\vert \nu,0\rangle$ be a $U=0$ energy and momentum eigenstate that contributes to the charge stiffness $D(T)$,
Eq. (\ref{sigma-reg}), so that $\langle \nu,0\vert\hat{J}\vert \nu,0\rangle\neq 0$.
>From the use of Eq. (\ref{m0-ll-C}) one may express the current operator expectation value
$\langle \nu,0\vert\hat{J}\vert \nu,0\rangle$ of such a state as follows,
\begin{equation}
\langle \nu,0\vert\hat{J}\vert \nu,0\rangle = 
\sum_{S_{\eta}=\vert S_{\eta}^z\vert}^{L/2}\sum_{l_{\rm r}}
\sum_{\iota =\pm 1}\left(C_{l_{\rm r},S_{\eta}}^{\nu,0}\right)^*C_{l_{\rm r},S_{\eta}+\iota}^{\nu,0}
\Theta (S_{\eta}+\iota)\,\langle l_{\rm r},S_{\eta},0,0^+\vert\hat{J}\vert l_{\rm r},S_{\eta}+\iota,0,0^+\rangle \, .
\label{lJr}
\end{equation}
The equality given here confirms that the off-diagonal quantum overlap 
through the current operator of the states $\vert l_{\rm r},S_{\eta},0,0^+\rangle$
and $\vert l_{\rm r},S_{\eta}',0,0^+\rangle$
whose $S_{\eta}$ and $S_{\eta}'$ values differ by $\delta S_{\eta}=\pm 1$
is fully consistent with the $U=0$ Hamiltonian, momentum operator, and charge current operator 
eigenstates $\vert \nu,0\rangle$ having finite current expectation value. 

\section{Effect of varying $u=U/4t$ on the microscopic processes behind the largest charge current absolute value
of a S$^z$S subspace}
\label{Appendix3}

As reported in Section \ref{SzSLNJmax}, for each S$^z$S subspace as
defined in Section \ref{SzSS} the largest charge current absolute value 
is that of the corresponding reference S$^z$SLN subspace 1 introduced in Section \ref{SzSLNJmax}.
This is a result that can be physically understood in terms 
of the microscopic mechanisms that contribute to the charge currents of the $\eta$-Bethe states. Here we present
an analysis of the problem that relies on the exact properties considered in Section \ref{curr-val}. In addition,
we discuss the similarities and differences relative to the case of the spin stiffness and currents
of the spin-$1/2$ $XXX$ chain studied in Refs. \cite{CPC,CP} by the general upper-bound method used in this paper.

Following the analysis and results of Section \ref{curr-val}, the charge currents of the $\eta$-Bethe states result from 
microscopic processes that are easiest to be described in terms of original lattice occupancy configurations.
The corresponding charge carriers naturally describe the charge degrees of freedom of the
rotated electrons whose operators are given in Eq. (\ref{rotated-operators}). Those
are related to the electrons by a unitary transformation uniquely defined 
in Ref. \cite{Carmelo-17}. Specifically, the $\eta$-Bethe-states charge currents result from processes within which a number $M_{\eta}=2S_{\eta}$ 
of unpaired physical $\eta$-spins $1/2$ that couple to charge probes interchange position with 
$N_{\rho}$ charge pseudoparticles belonging to the $c$ and $\eta n$ branches. This occurs
upon the charge pseudoparticles moving along that lattice. This relative motion is associated with
the $c$ and $\eta n$ pseudoparticle momentum $q_j$ and $\pi-q_j$, respectively, as given
in Eq. (\ref{P}). 

The $M_{\eta}=2S_{\eta}$ unpaired physical $\eta$-spins $1/2$ couple to charge probes.
However, the charge current only flows upon them interchanging positions with the charge
$c$ and $\eta n$ pseudoparticles. Consistently, for finite $u$ a S$^z$SLN subspace largest charge current absolute value
is proportional to $M_{\eta}\,N_{\rho}$. (S$^z$SLN subspaces are defined in Section \ref{SzSS}.) 
This refers to $\eta$-Bethe states for which $M_{\eta} = M_{\eta,+1/2} = 2S_{\eta}$.
Upon replacing the numbers of unpaired physical $\eta$-spins $1/2$ and of charge pseudoparticles by those
of unpaired physical spins $1/2$ and of spin pseudoparticles, respectively, this 
factor $M_{\eta}\,N_{\rho}$ is similar to that of the spin-$1/2$ $XXX$ chain
largest spin current absolute values \cite{CP}.

The largest charge current absolute value of each S$^z$SLN subspace is rather
written in Eq. (\ref{JImax}) as proportional to $m_{\eta}\,(1-m_{\eta})$ and thus to
$2S_{\eta}\,(L-2S_{\eta})$. This follows from the expression given in that equation applying
both to $u=U/4t\rightarrow 0$ and to finite $u$. Indeed, as justified in the following,
such a largest charge current absolute value can be written as proportional to
$2S_{\eta}\,(L-2S_{\eta})$ and $M_{\eta}\,N_{\rho}$ for $u\rightarrow 0$ and finite $u$,
respectively. If one requires it to be valid both for $u\rightarrow 0$ limit and
to finite $u$, then it should indeed be written as given in Eq. (\ref{JImax}). 

The charge current absolute value in that equation can be expressed 
as $(C_{l_{\eta},n_{\eta},m_{s},n_{s}}/L)\,t\,M_{\eta}\,(N_c + 2\Pi_{\eta})$ where
$M_{\eta}= 2S_{\eta}$ and $N_c + 2\Pi_{\eta}=L-2S_{\eta}$. The latter number involves
that of a $\eta$-Bethe-state paired rotated $\eta$-spins $1/2$, which reads $2\Pi_{\eta}=\sum_{n=1}^{\infty}2n\,N_{\eta n}$. 
As discussed below for $m_{\eta}\in [0,1]$ and in Appendix \ref{Appendix2} for $m_{\eta}=0$, the physics 
is very different (i) both at $u=0$ and in the $u\rightarrow 0$ limit and 
(ii) for finite $u$. The corresponding numbers of charge carriers will be found to be different in these two cases 
and given by (i) $N_c + 2\Pi_{\eta}=L-2S_{\eta}$ and (ii) $N_{\rho}=N_c + N_{\eta}$, respectively. 
This is why for finite $u$ the largest charge current absolute value of each S$^z$SLN subspace
can be written as proportional to $M_{\eta}\,N_{\rho}=M_{\eta}\,(N_c + N_{\eta})$ and thus expressed in 
terms of the number of charge pseudoparticles $N_{\rho} = N_c + N_{\eta}$, Eq. (\ref{NchargeP}), which 
refers to the $N_c$ $c$ pseudoparticles and $N_{\eta}=\sum_{n=1}^{\infty} N_{\eta n}\leq \Pi_{\eta}$ $\eta n$ pseudoparticles. 
The different charge carriers that within the present representation emerge as $u\rightarrow 0$ is an issue that will 
be clarified below. 

On the one hand, each $c$ and $\eta n$ pseudoparticle occupies one and a number 
$2n=2,4,6,...$, respectively, of original lattice sites. 
Hence the set of charge pseudoparticles of a $\eta$-Bethe state occupy a number $N_c + 2\Pi_{\eta}=N_c + \sum_{n=1}^{\infty}2n\,N_{\eta n}$
of sites of that lattice. On the other hand, each of the $M_{\eta}=2S_{\eta}$ unpaired physical $\eta$-spins $1/2$ occupies a single
site of the same lattice. The corresponding exact sum rule, $M_{\eta} + N_c + 2\Pi_{\eta} = L$, then ensures and confirms that the charge 
degrees of freedom of {\it all} the $L$ original-lattice sites are accounted for within the exact rotated-electron related
representation used in the studies of this paper.

The set of $2n$ original-lattice sites occupied by each $\eta n$ pseudoparticle refers to its internal
degrees of freedom. It is its center of mass that moves with momentum $\pi-q_j$. All $2n$ 
paired rotated $\eta$-spins $1/2$ move coherently with it upon processes within which they
interchange position with the $M_{\eta}$ unpaired physical $\eta$-spins $1/2$ that
singly occupy original-lattice sites. 

For the $\eta$-Bethe states that span a S$^z$SL subspace as defined in Section \ref{SzSS}, both
the number $M_{\eta}=2S_{\eta}$ of unpaired physical $\eta$-spins and $L_{\eta}$ of $\eta$-spins
are fixed. Hence the number $L_{\eta} = N_c + 2\Pi_{\eta}=L-2S_{\eta}$
of original-lattice sites occupied by the charge pseudoparticles is fixed as well. 
In contrast, that of charge pseudoparticles, $N_{\rho} = N_c + N_{\eta}$, varies from 
$N_{\rho}=1$ for the present case of $S_s^z=0$ and thus $M_{\eta}=2S_{\eta}$ even, up to a maximum value $N_{\rho} = N_c = L - 2S_{\eta}$.
The minimum value, $N_{\rho}=1$, refers to a single composite $\eta n$ pseudoparticle with all
the $\eta$-Bethe state $n=\Pi_{\eta}=(L - 2S_{\eta})/2$ $\eta$-spin singlet pairs bound within it.
Each of the possible different numbers $N_{\rho } = N_c + N_{\eta }\in [1,(L-2S_{\eta })]$ of charge pseudoparticles refers to
a S$^z$SLN subspace contained in the S$^z$SL subspace under consideration.

In the case of the 1D Hubbard model in a S$^z$SL subspace,
a second aspect to justify the largest charge current absolute value of that subspace must be accounted for. This in 
addition to the charge current absolute values increasing upon considering its S$^z$SLN subspaces with an 
increasing number $N_{\rho}$ of charge pseudoparticles that interchange position with the fixed number 
$M_{\eta}=2S_{\eta}$ of unpaired physical $\eta$-spins $1/2$.
It refers to the current spectra and elementary currents absolute values $\vert J_{c} (q_j)\vert=\vert J_{c}^h (q_j)\vert$ 
and $\vert j_{c} (q_j)\vert=\vert j_{c}^h (q_j)\vert$ of the charge $c$ pseudoparticles relative to those of 
the $\eta n$ pseudoparticles, $\vert J_{\eta n} (q_j)\vert=\vert J_{\eta n}^h (q_j)\vert$ and 
$\vert j_{\eta n} (q_j)\vert=\vert j_{\eta n}^h (q_j)\vert$, 
respectively. (The (i) $\beta = c, \eta n$ bands current spectra and (ii) $\beta = c, \eta n$ bands elementary currents 
associated with the absolute values (i) $\vert J_{\beta} (q_j)\vert=\vert J_{\beta}^h (q_j)\vert$ and (ii)
$\vert j_{\beta} (q_j)\vert=\vert j_{\beta}^h (q_j)\vert$ are given in (i) Eq. (\ref{jn-fn}) and Eq. (\ref{jn-fnPSEU}) 
of Appendix \ref{Appendix1} and (ii) in Eq. (\ref{jnvf}), respectively.) 

At the fixed $m_{\eta}$ and $l_{\eta}$ densities of a S$^z$SL subspace, such spectra absolute values are in general 
functions of $u$ and of the densities $n_{\rho}$, $m_s$, and $n_s$. It is a simple exercise to show that such absolute values are
larger for $\eta$-Bethe states for which $n_s = n_s^{\rm max} = (1-l_{\eta}-m_s)/2$. Hence in the following
we consider the charge currents of the latter states whose current spectra at fixed $m_{\eta}$ and $l_{\eta}$ densities 
are functions of $u$ and of the densities $n_{\rho}$ and $m_s$. 

We recall that the $\eta$-Bethe states exact charge currents can be expressed either in terms of the spectra
$J_{\beta} (q_j)$ associated with charge $\beta =c,\eta n$ pseudoparticles, Eq. (\ref{J-partDEF})
of Appendix \ref{Appendix1}, or with the related spectra $J_{\beta}^h (q_j)=-J_{\beta} (q_j)$
associated with the $\beta =c,\eta n$ bands holes, Eq. (\ref{J-part}). Here and in the following
analysis it is more convenient to express such charge currents in terms of the 
$\beta =c,\eta n$ pseudoparticles spectra $J_{\beta} (q_j)$. The same applies to
the related alternative particle-like and hole-like elementary currents $j_{\beta} (q_j)$ and $j_{\beta}^h (q_j)$, 
respectively.

It is useful to define the average values $\vert {\bar{J}}_{\beta}\vert$ and 
$\vert {\bar{j}}_{\beta}\vert$ of the $\beta = c,\eta n$ quantities 
$\vert J_{\beta} (q_j)\vert$ and $\vert j_{\beta} (q_j)\vert$, respectively, which read,
\begin{equation}
\vert {\bar{J}}_{\beta}\vert = {1\over 2q_{\beta}}\int_{-q_{\beta}}^{q_{\beta}}dq\,\vert J_{\beta} (q)\vert 
\hspace{0.20cm}{\rm and}\hspace{0.20cm}
\vert {\bar{j}}_{\beta}\vert = {1\over 2q_{\beta}}\int_{-q_{\beta}}^{q_{\beta}}dq\,\vert j_{\beta} (q)\vert 
\hspace{0.20cm} {\rm where}\hspace{0.20cm}
q_c = \pi\hspace{0.20cm}{\rm and}\hspace{0.20cm}q_{\eta n}=\pi n_{\eta n} \, ,
\label{averages}
\end{equation}
where within the TL the discrete momentum values $q_j$ such that $q_{j+1}-q_j =2\pi/L$ were
replaced by a corresponding continuous momentum variable $q$. From manipulations of the TBA equations and 
related $\eta$-Bethe states current spectra and elementary 
currents expressions, one finds that the following exact inequalities hold for the whole $u\geq 0$ range,
\begin{eqnarray}
{\rm max}\{\vert J_{c} (q_j)\vert\} & \geq & C_{u,n}\,{\rm max}\{\vert J_{\eta n} (q_j)\vert\}
\hspace{0.20cm}{\rm and}\hspace{0.20cm}
{\rm max}\{\vert j_{c} (q_j)\vert\} \geq C_{u,n}\,{\rm max}\{\vert j_{\eta n} (q_j)\vert\} \, ,
\nonumber \\
\vert {\bar{J}}_{c}\vert & \geq & C_{u,n}\,\vert {\bar{J}}_{\eta n}\vert
\hspace{0.20cm}{\rm and}\hspace{0.20cm}
\vert {\bar{j}}_{c}\vert \geq C_{u,n}\,\vert {\bar{j}}_{\eta n}\vert \, ,
\label{inequ0}
\end{eqnarray}
where $C_{u,n}\in [1/2n,1]$ is an increasing function of $u$ with limiting behaviors,
\begin{eqnarray}
C_{u,n} & = & {1\over 2n}\hspace{0.20cm}{\rm for}\hspace{0.20cm}
u\rightarrow 0 \, ,
\nonumber \\
& = & 1\hspace{0.20cm}{\rm for}\hspace{0.20cm}
u\rightarrow\infty \, .
\label{Cun}
\end{eqnarray}
According to such inequalities, for large $u$ values and all 
S$^z$SLN subspaces of a S$^z$S subspace,
the maximum and average values of $\vert J_{c} (q_j)\vert$ and $\vert j_{c} (q_j)\vert$ are larger than those of 
$\vert J_{\eta n} (q_j)\vert$ and $\vert j_{\eta n} (q_j)\vert$, respectively. 
Upon decreasing $u$, one finds that for intermediate $u$ 
the maximum and average values of $\vert J_{c} (q_j)\vert$ and $\vert j_{c} (q_j)\vert$ and those of 
$\vert J_{\eta n} (q_j)\vert$ and $\vert j_{\eta n} (q_j)\vert$, respectively, remain of the same order. 

Consistently, for large and intermediate $u$ values maximizing the number of charge pseudoparticles 
$N_{\rho}=N_c + N_{\eta}$ that, upon moving in the original lattice, interchange position with
a fixed number $M_{\eta}= 2S_{\eta}$ of unpaired physical $\eta$-spins $1/2$ that couple to
charge probes, gives the largest charge current absolute value of a S$^z$S subspace.
Indeed, for finite $u$ such an absolute value is proportional to $M_{\eta}\,N_{\rho}=M_{\eta}\,(N_c + N_{\eta})$.
This maximization procedure then leads to the largest charge current absolute value 
being that of the reference S$^z$SLN subspace 1, for which
$N_{\rho}=N_c + N_{\eta}$ reaches its maximum value, $N_{\rho} = N_c = L - 2S_{\eta}$.

Such a procedure is similar to that used for the spin-$1/2$ $XXX$ chain 
concerning the largest spin current absolute value of a fixed-spin subspace \cite{CP}.
There is though a different aspect relative to that chain. For it, all spin pseudoparticles
have internal degrees of freedom involving a number $2n$ of paired spins $1/2$. 
The largest spin current absolute value is then reached for a Bethe state populated by a number $(L - 2S_{s})/2$ 
of $n=1$ spin pseudoparticles whose internal degrees of freedom correspond to one spin-singlet pair.
Within a superficial comparison with the spin-$1/2$ $XXX$ chain, one would then expect that the largest charge 
current absolute value of a 1D Hubbard model S$^z$S subspace
would be that of a $\eta$-Bethe state populated by a number $N_{\rho} = N_{\eta 1} = (L - 2S_{\eta})/2$  
of $\eta 1$ pseudoparticles whose internal degrees of freedom correspond to one $\eta$-spin singlet pair.
This is the largest charge current absolute value of the reference S$^z$SLN subspace 2 as defined in 
Section \ref{SzSLNJmax}. 

However, for the 1D Hubbard model the charge pseudoparticles include both $c$ pseudoparticles without 
internal degrees of freedom and $\eta n$ pseudoparticles whose internal degrees of freedom involve a number $2n$ 
of paired $\eta$-spins $1/2$. This is related to for $u>0$ the charge degrees of freedom
of the 1D Hubbard model being associated with a $U(2)=SU(2)\otimes U(1)$ symmetry whereas the spin degrees
of freedom of the spin-$1/2$ $XXX$ chain are associated only with a $SU(2)$ symmetry. 
As a result, the $\eta$-Bethe state with largest charge current absolute value
is rather found to be populated by $N_{\rho} = N_c = L - 2S_{\eta}$ $c$ pseudoparticles, which thus is 
the largest charge current absolute value of the reference S$^z$SLN subspace 1. 

As one decreases $u$ to reach small $u\ll 1$ values, the coefficient $C_{u,n}$ in Eqs. (\ref{inequ0}) and (\ref{Cun}) tends
to approach the value $1/2n$. The microscopic mechanisms that justify why the largest charge current absolute value
of the reference S$^z$SLN subspace 1 remains being the largest charge current absolute value of
the corresponding S$^z$S subspace become now completely different from those of the spin-$1/2$ $XXX$ chain.
Before addressing that problem, we provide useful information that confirms the validity of the inequalities,
Eq. (\ref{inequ0}), for small and intermediate values of $u$ for which the coefficient $C_{u,n}$ remains being
of the order of the unity. Specifically, we consider densities given in Eqs. (\ref{densitieSzS})-(\ref{spindensities})
in the vicinity of those of the reference S$^z$SLN subspaces 1 to 3 defined in Section \ref{SzSLNJmax}.

The expressions of the $\beta =c,\eta n$ pseudoparticle spectra $J_{\beta} (q_j)$ and elementary currents 
$j_{\beta} (q_j)$ given in the following refer to $\eta$-Bethe states whose $c$ and $\eta n$ band distributions are compact
and asymmetrical, as given in Eq. (\ref{NN}). They refer as well to $\eta$-Bethe states that in the TL are generated from those
by a finite number of $\beta =c,\eta n$ pseudoparticle processes. Furthermore, they are also valid
for $\eta$-Bethe states whose $c$ and $\eta n$ band distributions are compact and symmetrical, Eq. (\ref{CompSymm}), and 
$\eta$-Bethe states generated from those by a finite number of $\beta =c,\eta n$ pseudoparticle processes.

On the one hand, for the S$^z$SLN subspaces for which the $J_{\beta} (q_j)$ and $j_{\beta} (q_j)$ expressions are
independent of $u$ for $u>0$, such expressions are valid for all the classes of $\eta$-Bethe states under consideration. 
On the other hand, for S$^z$SLN subspaces for which they are $u$ dependent for $u>0$, 
their universality only applies to large $u$. Indeed, upon decreasing $u$, the $J_{\beta} (q_j)$ and $j_{\beta} (q_j)$ expressions become different
for different $\eta$-Bethe states, as they become dependent on the specific $q_{F\beta,\iota}^{\tau}$ values in Eq. (\ref{qFcetanpms})
that define such states. In addition to providing $J_{\beta} (q_j)$ and $j_{\beta} (q_j)$ expressions
that are independent of $u$ for $u>0$ and are valid for the $u>0$ range, for simplicity in the case of $u$ dependent
expressions for $u>0$ only their universal expansions up to $u^{-1}$ order are given. 

For $\eta$-Bethe states whose densities, Eqs. (\ref{densitieSzS})-(\ref{spindensities}), are in the vicinity of those of 
a reference S$^z$SLN$_{SN}$ subspace 1B, 
both the $J_{\beta} (q_j)$ and $j_{\beta} (q_j)$ expressions and the corresponding inequalities, Eq. (\ref{inequ0}), 
are independent of $u$ for $u>0$ and remain valid for the whole $u> 0$ range. For 
these $\eta$-Bethe states the following expressions are found for $u> 0$, $m_{\eta} \in [0,1]$,
and $m_s \in [0,(1-m_{\eta})]$,
\begin{eqnarray}
J_c (q_j) & = & j_c (q_j)  = - 2t \sin (q_j)
\hspace{0.20cm}{\rm for}\hspace{0.20cm}q_j\in [-\pi,\pi] \, ,
\nonumber \\
J_{\eta n} (q_j) & = & j_{\eta n} (q_j) = - 2t\,q_j\,{\sin (\pi m_{\eta})\over \pi m_{\eta}}
\hspace{0.20cm}{\rm for}\hspace{0.20cm} q_j\in [-\pi m_{\eta},\pi m_{\eta}] \, ,
\nonumber \\
{\rm max}\{\vert J_{c} (q_j)\vert\} & = & {\rm max}\{\vert j_{c} (q_j)\vert\} = 2t \hspace{0.20cm}{\rm and}\hspace{0.20cm}
\vert {\bar{J}}_{c}\vert = \vert {\bar{j}}_{c}\vert = {4 t\over\pi} \, ,
\nonumber \\
{\rm max}\{\vert J_{\eta n} (q_j)\vert\} & = & {\rm max}\{\vert j_{\eta n} (q_j)\vert\} = 2t\,\sin (\pi m_{\eta}) \hspace{0.20cm}{\rm and}\hspace{0.20cm}
\vert{\bar{J}}_{\eta n}\vert = \vert {\bar{j}}_{\eta n}\vert = t\sin (\pi m_{\eta}) \, .
\label{Jj2}
\end{eqnarray}
These exact expressions and values obey the inequalities, Eqs. (\ref{inequ0}) and (\ref{Cun}). (As discussed below,
for the reference S$^z$SLN$_{SN}$ subspace 1B one has that $C_{u,n}= 1$ for $u>0$ and
$C_{u,n}= 1/2n$ for $u\rightarrow 0$.)

For most S$^z$SLN subspaces of a S$^z$S subspace, the $\beta =c,\eta n$ spectra 
$J_{\beta} (q_j)$ and elementary currents $j_{\beta} (q_j)$ depend on $u$.
This applies for densities, Eqs. (\ref{densitieSzS})-(\ref{spindensities}), in the vicinity of those of a reference 
S$^z$SLN$_{SN}$ subspace 1A, for which the expansions of $J_c (q_j)$ and $J_{\eta n} (q_j)$
are up to $u^{-1}$ order and for $m_{\eta} \in [0,1]$ and $m_s \in [0,(1-m_{\eta})]$ given by,
\begin{eqnarray}
J_c (q_j) & = & - 2t \sin (q_j) + 2t\,{n_{\eta s}\over u}\sin (2q_j )
\hspace{0.20cm}{\rm for}\hspace{0.20cm}q_j\in [-\pi,\pi]
\nonumber \\
J_{\eta n} (q_j) & = & - 2t\,q_j\left(1 + {2\,(1-m_s)\,g_s\over \pi\,u}\sin (\pi m_{\eta})\right){\sin (\pi m_{\eta})\over\pi m_{\eta}}
+ {2t\over u}\sin \left({q_j\over m_{\eta}}\right)
\hspace{0.20cm}{\rm for}\hspace{0.20cm} q_j\in [-\pi m_{\eta},\pi m_{\eta}] \, .
\label{JLcetan}
\end{eqnarray}
As given in Eq. (\ref{Lms}), $n_{\eta s}=(1-m_{\eta}-m_s) \, g_s$ and the function $g_s (m_s)$ 
appearing here continuously increases from $g_s (0)=\ln 2$ to $g_s (1-m_{\eta})=1$ as $m_s$ increases from
$m_s = 0$ to $m_s=(1-m_{\eta})$, respectively.

Up to $u^{-1}$ order and for $m_{\eta} \in [0,1]$ and $m_s \in [0,(1-m_{\eta})]$, 
the corresponding expansions for $j_c (q_j)$ and $j_{\eta n} (q_j)$ are found to read,
\begin{eqnarray}
j_c (q_j) & = & J_c (q_j) - 4t\,{n_{\eta s}\over u}\,
{\sin (\pi (1-m_{\eta}))\over\pi (1-m_{\eta})}\sin (q_j)
\nonumber \\
& = & - 2t \sin (q_j) + 2t\,{n_{\eta s}\over u}\sin (2q_j) - 4t\,{n_{\eta s}\over u}\,
{\sin (\pi (1-m_{\eta}))\over\pi (1-m_{\eta})}\sin (q_j)\hspace{0.20cm}{\rm for}\hspace{0.20cm}q_j\in [-\pi,\pi]
\nonumber \\
j_{\eta n} (q_j) & = & J_{\eta n} (q_j) - {2t\,(1-m_{\eta}-2m_s)\over u\,(1-m_{\eta})}{\sin (2\pi m_{\eta})\over 2\pi m_{\eta}}\sin \left({q_j\over m_{\eta}}\right)
\nonumber \\
& = & - 2t\,q_j\left(1 + {2\,(1-m_s)\,g_s\over \pi\,u}\sin (\pi m_{\eta})\right){\sin (\pi m_{\eta})\over\pi m_{\eta}}
\nonumber \\
& + & {2t\over u}\left(1 - {(1-m_{\eta}-2m_s)\over (1-m_{\eta})}\,{\sin (2\pi m_{\eta})\over 2\pi m_{\eta}}\right)\sin \left({q_j\over m_{\eta}}\right)
\hspace{0.20cm}{\rm for}\hspace{0.20cm}q_j\in [-\pi m_{\eta},\pi m_{\eta}] 
\nonumber \\
& & {\rm where}\hspace{0.20cm}m_{\eta} \in [0,1]\hspace{0.20cm}{\rm and}\hspace{0.20cm}m_s \in [0,(1-m_{\eta})] \, .
\label{jcetan}
\end{eqnarray}
To leading $(1/u)^0$ order, the $\beta =c,\eta n$ values of
${\rm max}\{\vert J_{\beta} (q_j)\vert\}$, ${\rm max}\{\vert j_{\beta} (q_j)\vert\}$,
$\vert{\bar{J}}_{\beta}\vert$, and $\vert{\bar{j}}_{\beta}\vert$ remain being those
given in Eq. (\ref{Jj2}). For simplicity, their $u^{-1}$ corrections are not
given here. The exact expressions and values in Eqs. (\ref{JLcetan}) and (\ref{jcetan})
again obey the inequalities, Eqs. (\ref{inequ0}) and (\ref{Cun}).

For densities in Eqs. (\ref{densitieSzS})-(\ref{spindensities}) in the vicinity of those of the reference S$^z$SLN subspaces 2 and 3,
one finds $J_c (q_j) = j_c (q_j)  = 2t \sin (q_j)$ for $q_j\in [-\pi,\pi]$ and the whole
$u\geq 0$ range. Moreover, $J_{\eta n} (q_j) = J_{\eta n}\sin q_j$ and $j_{\eta n} (q_j) = j_{\eta n}\sin q_j$ 
where $J_{\eta n} \propto 1/u$ and $j_{\eta n} \propto 1/u$ for large $u$, so that
the inequalities, Eq. (\ref{inequ0}), are also obeyed in that limit.

As one decreases $u$ to reach $u\ll 1$ values, the coefficient $C_{u,n}$ in Eqs. (\ref{inequ0}) and (\ref{Cun}) 
approaches its limiting smallest value, $1/2n$. Consistently, for some of the S$^z$SLN subspaces
contained in a S$^z$S subspace the maximum and average absolute values of the
spectra $J_{\eta n} (q_j)$ and elementary currents $j_{\eta n} (q_j)$ become larger than those of 
$J_{c} (q_j)$ and $j_{c} (q_j)$, respectively. This holds particularly for large $n$ values. The question is thus why does 
the largest charge current absolute value of a S$^z$S subspace remains that of
its reference S$^z$SLN subspace 1 for which $N_{\rho}=N_c$ and $N_{\eta}=\sum_{n=1}^{\infty} N_{\eta n}=0$?

To clarify this interesting issue, we account for again that for finite yet arbitrarily small $u$ values the largest charge 
current absolute value of each S$^z$SLN subspace can be written as proportional to 
$M_{\eta}\,N_{\rho}=M_{\eta}\,(N_c + N_{\eta})$. 
This number can be rewritten as $M_{\eta}\,N_{\rho} = R_{\rho}\,M_{\eta}\,(N_c + 2\Pi_{\eta})$ and thus 
$M_{\eta}\,N_{\rho} = R_{\rho}\,2S_{\eta}\,(L-2S_{\eta})$. Here the exact relation
$N_c + 2\Pi_{\eta} = L-2S_{\eta}$ was used and the coefficient $R_{\rho}=N_{\rho}/(L-2S_{\eta})=(N_c + N_{\eta})/(N_c + 2\Pi_{\eta})$
that can be written as $R_{\rho}=n_{\rho}/(1-m_{\eta})=(1-l_{\eta} + n_{\eta})/(1-l_{\eta} + 2\pi_{\eta})$
varies in the range $R_{\rho} \in [1/(L-2S_{\eta}),1]$. Within the TL, that interval reads
$R_{\rho} \in [0,1]$ for $m_{\eta}<1$. 

For the $\eta$-Bethe states that span a S$^z$S subspace, the factor $2S_{\eta}\,(L-2S_{\eta})$
in $M_{\eta}\,N_{\rho} = R_{\rho}\,2S_{\eta}\,(L-2S_{\eta})$ has a fixed value. Thus the changes in $M_{\eta}\,N_{\rho}$
all stem from $R_{\rho}$. On the one hand, the maximum $R_{\rho}$ value, $R_{\rho}=1$, refers to
$N_{\rho}=N_c$ and $N_{\eta}=\sum_{n=1}^{\infty} N_{\eta n}=0$. On the other hand, for 
reference S$^z$SLN$_{SN}$ subspaces for which $N_{\eta}>0$, increasing the population of $\eta n$
pseudoparticles with a large number $n$ of $\eta$-spin singlet pairs leads to a strong decreasing of the
coefficient $R_{\rho} \in [0,1]$ from its maximum value. It turns out that for small yet finite
$u$ values such a decreasing in the charge current factor $M_{\eta}\,N_{\rho} = R_{\rho}\,2S_{\eta}\,(L-2S_{\eta})$
always compensates the increasing of the spectra $J_{\eta n} (q_j)$ and elementary currents $j_{\eta n} (q_j)$
absolute values upon increasing $n$. Indeed, for any arbitrarily small yet finite value of $u$ the first of these
two opposite processes dominates, so that the largest charge current absolute 
value of a S$^z$S subspace remains that of its reference S$^z$SLN$_{SN}$ subspace 1 for which $N_{\rho}=N_c$.

The small-$u$ increasing of the absolute values of $J_{\eta n} (q_j)$ and $j_{\eta n} (q_j)$ upon increasing $n$
turns out to be part of a mechanism that prepares and precedes a qualitative change of the physics that occurs upon reaching the $u\rightarrow 0$
limit. In such a limit the competition between the two above effects becomes critical. However, the largest charge current 
absolute value of a S$^z$S subspace remains that of its reference S$^z$SLN$_{SN}$ subspace 1,
as confirmed in the following.

The $u=0$ problem is easiest to be described directly in terms of electron configurations.
However, the understanding of the corresponding finite-$u$ problem requires the description of the
$u\rightarrow 0$ limit physics in terms of the fractionalized particles emerging from the rotated electrons.
In the case of the 1D Hubbard model, the imaginary
part $i\,(n + 1 - 2l)\,u$ of each set of $\eta $-spin $l = 1,...,n$ rapidities
$\Lambda ^{\eta n,l}(q_{j}) = \Lambda ^{\eta n} (q_{j}) + i\,(n + 1 - 2l)\,u$, Eq. (\ref{complex-rap}) for $\alpha = \eta $,
with the same real part $\Lambda ^{\eta n} (q_{j})$ depends on the interaction $u$ and thus vanishes as $u\rightarrow 0$. This is
in contrast to the $n>1$ spin $l = 1,...,n$ rapidities of the spin $1/2$ $XXX$ chain with the same real part, which are associated with a $n$-band pseudoparticle \cite{CP}. For the  Hubbard model, the imaginary part of such a set of $\eta $-spin $l = 1,...,n$ rapidities describes the bounding of a number $n$
of $\eta $-spin singlet pairs within one $\eta n$ pseudoparticle.

In the case of $n>1$, the vanishing of the $l = 1,...,n$ rapidities imaginary part $i\,(n + 1 - 2l)\,u$
that occurs in the $u\rightarrow 0$ limit has physical consequences. It implies that the $\Pi _{\eta }$ $\eta $-spin singlet pairs
containing $2\Pi _{\eta } = \sum _{n=1}^{\infty } 2n\,N_{\eta n}$ paired rotated $\eta $-spins $1/2$
of each of the corresponding $\eta n$ pseudoparticles of a $\eta $-Bethe state
unbound. Furthermore, although the even number of $2\Pi_{\eta}$ rotated $\eta$-spins $1/2$
remain contributing to $\eta$-spin-singlet configurations, concerning their translational degrees
of freedom they behave as $2\Pi_{\eta}$ independent charge carriers. This means that the $\eta$-spin singlet
configurations associated with both the $n=1$ pair $\eta 1$ pseudoparticles and the $n>1$ pairs composite 
$\eta n$ pseudoparticles are rearranged in the $u\rightarrow 0$ limit in such a away that such 
pseudoparticles cease to exist in that limit. While the $\eta n$ pseudoparticles do not exist both
for $u\rightarrow 0$ and at $u=0$, the $c$ pseudoparticles remain existing in the $u\rightarrow 0$ limit
yet are ill defined at $u=0$.  

A physical consequence of such $c$ and $\eta 1$ pseudoparticle behaviors is that for $u\rightarrow 0$ the charge currents of the 
$\eta$-Bethe states rather result from microscopic processes within which a number $M_{\eta}=2S_{\eta}$ of unpaired physical
$\eta$-spins in the $\eta$-spin multiplet configurations interchange position with 
$N_c$ $c$ pseudoparticles and the $2\Pi_{\eta} =\sum_{n=1}^{\infty} 2n\,N_{\eta n}$
rotated $\eta$-spins $1/2$. Those behave as independent charge carriers in spite of remaining 
participating in $\eta$-spin-singlet configurations. 
Hence in the $u\rightarrow 0$ limit the largest charge current absolute value of each S$^z$SLN subspace can be
written as $(C_{l_{\eta},n_{\eta},m_{s}}/L)\,t\,M_{\eta}\,(N_c + 2\Pi_{\eta})$. 
Now both each of the $M_{\eta}=2S_{\eta}$ unpaired physical $\eta$-spins in the $\eta$-spin multiplet configurations,
$N_c$ $c$ pseudoparticles, and $2\Pi_{\eta}$ rotated $\eta$-spins $1/2$ in $\eta$-spin-singlet configurations
occupy a single site of the original lattice. Indeed, $M_{\eta}+N_c+2\Pi_{\eta}=L$.

The new physics emerging in the $u\rightarrow 0$ limit is accounted for the exact BA solution. Indeed,
the $\eta n$ band spectra $J_{\eta n} (q_j)$ and elementary currents $j_{\eta n} (q_j)$ become 
in the $u\rightarrow 0$ limit of the form,
\begin{equation}
J_{\eta n} (q_j) = 2n\,J_{\eta} (q_j)\hspace{0.20cm}{\rm and}\hspace{0.20cm}
j_{\eta n} (q_j) = 2n\,j_{\eta} (q_j) \, .
\label{Jjdecouple}
\end{equation}
Here $J_{\eta} (q_j)$ and $j_{\eta} (q_j)$ are the corresponding spectra and 
elementary currents carried by each of the $2n$ rotated $\eta$-spins $1/2$ in $\eta$-spin-singlet configurations
resulting in the $u\rightarrow 0$ limit from one $u>0$ $\eta n$ pseudoparticle. Since $N_c + 2\Pi_{\eta}=L-2S_{\eta}$,
the number $N_c + 2\Pi_{\eta}$ of $c$ pseudoparticles plus rotated $\eta$-spins $1/2$ in $\eta$-spin-singlet configurations 
appearing in the expression $(C_{l_{\eta},n_{\eta},m_{s}}/L)\,t\,M_{\eta}\,(N_c + 2\Pi_{\eta})$
of the largest charge current absolute values of each S$^z$SLN subspace
is {\it the same} for all $\eta$-Bethe states that span a S$^z$S subspace. 

The charge current operator expectation values general exact expression, Eq. (\ref{J-partDEF}) of 
Appendix \ref{Appendix1}, remains
valid in the $u\rightarrow 0$ limit yet the corresponding $\eta$-Bethe states are not energy eigenstates
at $u=0$. This issue and its consequences for the charge stiffness is addressed in Appendix
\ref{Appendix2} in the case of the hole concentration $m_{\eta}^z\rightarrow 0$. 
The expressions in Eq. (\ref{Jjdecouple}) of the current spectra and
elementary currents are consistent with at $u=0$ the $\eta n$ pseudoparticles 
not existing. The $\eta n$ pseudoparticle exists for any arbitrarily small yet finite $u$ value.
Hence within the $u\rightarrow 0$ limit the corresponding $2n=2,4,6,...$ rotated $\eta$-spins $1/2$ that emerge from it
continue all moving with momentum $\pi - q_j$. This is why such $2n$ rotated $\eta$-spins $1/2$
have charge current spectra $J_{\eta} (q_j)$ and elementary currents $j_{\eta} (q_j)$ in Eq. (\ref{Jjdecouple})
that depend on the same momentum value $q_j$. 

The increase of the $\eta n$ pseudoparticle current spectrum (and elementary current) absolute value upon decreasing
$u$ is needed for its maximum value $2n\,\vert J_{\eta} (q_j)\vert$ (and $2n\,\vert j_{\eta} (q_j)\vert$),
which is reached in the $u\rightarrow 0$ limit, be compatible with the current spectrum (and elementary current)
of the emerging $2n=2,4,6,...$ independent rotated $\eta$-spins $1/2$ charge carriers reading $J_{\eta} (q_j)$ (and $j_{\eta} (q_j)$.)
That absolute value increase upon decreasing $u$ that occurs in the {\it womb} of the $\eta n$ pseudoparticle 
is a needed preparation for its dead upon it delivering in the $u\rightarrow 0$ limit such $2n=2,4,6,...$ 
independent charge carriers. Due to their emergence from the $\eta n$ pseudoparticle in the $u\rightarrow 0$ limit, 
the corresponding $u\rightarrow 0$ physics is qualitatively different from that of finite $u$. 

From combination of Eqs. (\ref{inequ0}) and (\ref{Jjdecouple}), one finds 
the following exact inequality specific to the $u\rightarrow 0$ limit, which
involves the emerging rotated $\eta$-spins $1/2$ current spectra $J_{\eta} (q_j)$
and elementary currents $j_{\eta} (q_j)$ in Eq. (\ref{Jjdecouple}),
\begin{eqnarray}
{\rm max}\{\vert J_{c} (q_j)\vert\} & \geq & {\rm max}\{\vert J_{\eta} (q_j)\vert\}
\hspace{0.20cm}{\rm and}\hspace{0.20cm}
{\rm max}\{\vert j_{c} (q_j)\vert\} \geq {\rm max}\{\vert j_{\eta} (q_j)\vert\} \, ,
\nonumber \\
\vert {\bar{J}}_{c}\vert & \geq & \vert {\bar{J}}_{\eta}\vert
\hspace{0.20cm}{\rm and}\hspace{0.20cm}
\vert {\bar{j}}_{c}\vert \geq \vert {\bar{j}}_{\eta}\vert \, .
\label{inequU0}
\end{eqnarray}

The number $N_c + 2\Pi_{\eta}=L-2S_{\eta}$ of charge carriers that upon moving in the original lattice
interchange position with the $M_{\eta}=2S_{\eta}$ of unpaired physical $\eta$-spins
in the $\eta$-spin multiplet configurations is {\it the same} for {\it all} the $\eta$-Bethe states that
span a S$^z$S subspace. This is a necessary condition for the largest current absolute value remaining
that of the corresponding reference S$^z$SLN subspace 1. As given in Eq. (\ref{inequU0}), 
for all references S$^z$SLN$_{SN}$ subspaces contained in such a subspace,
the maximum and average values of $\vert J_{c} (q_j)\vert$ and $\vert j_{c} (q_j)\vert$ are
in the $u\rightarrow 0$ limit larger than or equal to those of $\vert J_{\eta} (q_j)\vert$ and $\vert j_{\eta} (q_j)\vert$, respectively.
This is why the largest charge current absolute value of a S$^z$S subspace remains that of its 
reference S$^z$SLN$_{SN}$ subspace 1 for which $N_c + 2\Pi_{\eta}=N_c =L-2S_{\eta}$.

Hence, in contrast to finite $u$, in the $u\rightarrow 0$ limit the number of charge carriers
that upon moving in the original lattice interchange position with $M_{\eta}=2S_{\eta}$ 
unpaired physical $\eta$-spins does not play any role in the maximization of the
charge current absolute value. The only property that
determines its maximization is the relative absolute values
of the charge carriers current spectra and elementary currents. According to the exact
inequalities, Eq. (\ref{inequU0}), the $c$ pseudoparticles indeed win such a competition.

In the case of some S$^z$SLN subspaces, for $u\rightarrow 0$ the expressions of the spectra 
$J_{c} (q_j)$ and $J_{\eta} (q_j)$ and elementary currents $j_{c} (q_j)$ and $j_{\eta} (q_j)$ in Eq. (\ref{inequU0}) depend 
on the specific $q_{F\beta,\iota}^{\tau}$ values in Eq. (\ref{qFcetanpms}).
Those define the $\eta$-Bethe states that have the largest charge current absolute value in that subspace.
The exception for many S$^z$SLN subspaces is for those contained in
S$^z$S subspaces for which $m_{\eta}\ll 1$ and $(1-m_{\eta})\ll 1$, respectively. For them,
such expressions are the same for all $\eta$-Bethe states belonging to the same  
S$^z$SLN subspace. And this applies both to $\eta$-Bethe states whose distributions are asymmetrical, 
Eq. (\ref{NN}), and symmetrical, Eq. (\ref{CompSymm}).

For S$^z$S subspaces corresponding to the $m_{\eta}\ll 1$ and $(1-m_{\eta})\ll 1$ limits, 
the charge currents can be expressed in terms of the elementary currents, as given in Eq. (\ref{deltaJ-partlim}), 
with now $j_{\eta n} (q_j)= 2n\,j_{\eta} (q_j)$. For simplicity, in the present case of the $u\rightarrow 0$ limit 
we only report in the following expressions for the elementary currents $j_{c} (q_j)$ and $j_{\eta} (q_j)$ 
in $j_{\eta n} (q_j)= 2n\,j_{\eta} (q_j)$ that obey the inequalities, Eq. (\ref{inequU0}). Besides controlling the 
exact expressions of the charge currents for $m_{\eta}\ll 1$ and $(1-m_{\eta})\ll 1$, Eq. (\ref{deltaJ-partlim}), such elementary currents
are useful for some of the upper-bound procedures used in this paper. Related expressions apply to the spectra
$J_{c} (q_j)$ and $J_{\eta} (q_j)$ in $J_{\eta n} (q_j)= 2n\,J_{\eta} (q_j)$ also appearing in Eq. (\ref{inequU0}).
Densities in Eqs. (\ref{densitieSzS})-(\ref{spindensities}) in the vicinity of those of the reference S$^z$SLN subspaces 1 to 3 
defined in Section \ref{SzSLNJmax} are again considered. Although the exact inequalities, Eq. (\ref{inequU0}),
apply to S$^z$S subspaces corresponding to the $m_{\eta}\in [0,1]$ and $m_{s}\in [0,(1-m_{\eta})]$ intervals, the 
limit $m_{\eta}\rightarrow 0$ for which $m_{s}\in [0,1]$ plays a central role concerning the charge stiffness issue clarified in this paper.

We start by considering the reference S$^z$SLN$_{SN}$ subspace 1B for which $n_s\rightarrow 0$, in contrast to 
that density maximum value $n_s = n_s^{\rm max} = (1-l_{\eta}-m_s)/2$ generally considered here. Moreover,
for this reference S$^z$SLN$_{SN}$ subspace the coefficient $C_{u,n}$ in Eq. (\ref{Cun}) has as singular behavior, reading
$C_{u,n}=1$ for finite $u$ and $C_{u,n}=1/2n$ for $u\rightarrow 0$. Another property is that for it the expression of the 
elementary currents is independent of the $\eta$-Bethe states that span it, as it is the same for all such states. 
The elementary current $j_{\eta n} (q_j)$ is for
$q_j\in [-\pi m_{\eta},\pi m_{\eta}]$ given by $j_{\eta n} (q_j) = 2t\,q_j\,\sin (\pi m_{\eta})/\pi m_{\eta}$
for $u>0$, as provided in Eq. (\ref{Jj2}). However, following the singular behavior of the coefficient $C_{u,n}$,
its value is different in the $u\rightarrow 0$ limit. For densities in Eqs. (\ref{densitieSzS})-(\ref{spindensities})
in the vicinity of those of a reference S$^z$SLN$_{SN}$ subspace 1B one finds in the $u\rightarrow 0$ limit that,
\begin{eqnarray}
j_c (q_j) & = & - 2t \sin (q_j)
\hspace{0.20cm}{\rm for}\hspace{0.20cm} q_j \in [-\pi,\pi] \, ,
\nonumber \\
j_{\eta n} (q_j) & = & 2n\,j_{\eta} (q_j)\hspace{0.20cm}{\rm where} 
\hspace{0.20cm}j_{\eta} (q_j) = 2t\sin \left({q_j\over 2}\right)
\hspace{0.20cm}{\rm for}\hspace{0.20cm} q_j\in [-\pi m_{\eta},\pi m_{\eta}] \, ,
\nonumber \\
{\rm max}\{\vert j_{c} (q_j)\vert\} & = & 2t \hspace{0.20cm}{\rm and}\hspace{0.20cm}
\vert {\bar{j}}_{c}\vert = {4 t\over\pi} \, ,
\nonumber \\
{\rm max}\{\vert j_{\eta} (q_j)\vert\} & = & 2t\,\sin \left({\pi\over 2} m_{\eta}\right) \hspace{0.20cm}{\rm and}\hspace{0.20cm}
\vert {\bar{j}}_{\eta}\vert = {4 t\over\pi m_{\eta}}\left(1-\cos  \left({\pi\over 2} m_{\eta}\right)\right) \, .
\label{Jj20}
\end{eqnarray}
These exact elementary current expressions and values obey the corresponding inequality in Eq. (\ref{inequU0}).

Moreover, for densities in Eqs. (\ref{densitieSzS})-(\ref{spindensities}) in the vicinity of those 
of a reference S$^z$SLN subspace 2 , the $c$ band elementary current
is for the whole $u\geq 0$ range independent of the $\eta$-Bethe states and reads $j_c (q_j) = 2t\sin (q_j)$, including
in the present $u\rightarrow 0$ limit. The elementary currents $j_{\eta} (q_j)$ in $j_{\eta n} (q_j) = 2n\,j_{\eta} (q_j)$ 
have slightly different expressions for different $\eta$-Bethe states belonging to the subspace under consideration.
For all such states, the corresponding elementary current inequality in Eqs. (\ref{inequU0}) is obeyed. The exception is
as mentioned above for reference S$^z$SLN subspaces 3 contained in S$^z$S subspaces
for which $m_{\eta}\ll 1$ and $(1-m_{\eta})\ll 1$. In that case their expressions are 
for $u\rightarrow 0$ the same for all such states. Specifically, they read  $j_{\eta}(q_j)= - t\sin (q_j)$ for $m_{\eta}\rightarrow 0$ and 
$j_{\eta}(q_j)= - 2t\sin (q_j)$ for $m_{\eta}\rightarrow 1$, so that the corresponding elementary current inequality in Eq. (\ref{inequU0})
is again obeyed. (For a reference S$^z$SLN subspace 2 one has that $l_{\eta}\rightarrow 1$, so that $m_s\rightarrow 0$.)

For densities in Eqs. (\ref{densitieSzS})-(\ref{spindensities})
in the vicinity of those of a reference S$^z$SLN subspace 3, the coefficient $C_{u,n}$ in Eq. (\ref{Cun}) is
a continuous function of $u$. Furthermore, the expressions of the elementary 
currents are for the whole $u\geq 0$ range independent of the $\eta$-Bethe states in that subspace. In the
$u\rightarrow 0$ limit they equal actually those given in Eq. (\ref{Jj20}), which thus obey the corresponding 
inequality in Eqs. (\ref{inequU0}).

For densities in Eqs. (\ref{densitieSzS})-(\ref{spindensities}) near those of a reference S$^z$SLN$_{SN}$ subspace 1A, which
plays the major role in our studies, the elementary currents have in the $u\rightarrow 0$ limit
a universal form common to all $\eta$-Bethe states again only for $m_{\eta}\ll 1$ and $(1-m_{\eta})\ll 1$.
On the one hand, the elementary currents $j_{c} (q_j)$ and $j_{\eta n} (q_j)$ associated with $\eta$-Bethe
states that span reference S$^z$SLN$_{SN}$ subspaces 1A contained in S$^z$S subspaces with densities $m_{\eta}\rightarrow 0$ and 
$m_s \in [0,1]$ are in the $u\rightarrow 0$ limit found to read,
\begin{eqnarray}
j_c (q_j) & = & - 4t\sin \left({q_j\over 2}\right)
\hspace{0.20cm}{\rm for}\hspace{0.20cm} q_j \in [-\pi (1-m_s) ,\pi (1-m_s)] 
\nonumber \\
& = & {\rm sgn}\{q\}\,2t\left(\cos \left(\vert q_j\vert + {\pi\over 2}m_s\right) - \cos \left({\pi\over 2}m_s\right)\right)
\hspace{0.20cm}{\rm for}\hspace{0.20cm} \vert q_j\vert \in [\pi (1-m_s) ,\pi] \, ,
\nonumber \\
j_{\eta n} (q_j) & = & 2n\,j_{\eta} (q_j)\hspace{0.20cm}{\rm where} 
\hspace{0.20cm} j_{\eta} (q_j)= {\rm sgn} (q_j)\,2t\left(\cos \left({\vert q_j\vert - \pi\,m_s\over 2}\right) - {1\over 2n}\cos \left({\pi\over 2}m_s\right)\right) 
\nonumber \\
& & {\rm for}\hspace{0.20cm} q_j\in [-\pi m_{\eta},\pi m_{\eta}] 
\nonumber \\
{\rm max}\{\vert j_{c} (q_j)\vert\} & = & \vert j_{c} \left({\pi\over 2} (2-m_s)\right)\vert = 2t\left(1+\cos \left({\pi\over 2}m_s\right)\right) 
\nonumber \\
{\rm and} && \vert {\bar{j}}_{c}\vert = {8 t\over\pi}\left(1 - {1\over 2}\sin \left({\pi\over 2}m_s\right) + {\pi\over 4}m_s\cos \left({\pi\over 2}m_s\right)\right) \, ,
\nonumber \\
{\rm max}\{\vert j_{\eta} (q_j)\vert\} & = & \vert j_{\eta} (\pi m_{\eta})\vert = \vert {\bar{j}}_{\eta}\vert =
2t\left(2 - {1\over 2n}\right)\cos \left({\pi\over 2}m_s\right)  \, .
\label{jcjetau1metaS}
\end{eqnarray}
Both ${\rm max}\{\vert j_{\beta} (q_j)\vert\}$ and $ \vert {\bar{j}}_{\beta}\vert$
are for $\beta =c,\eta n$ continuous decreasing functions of the spin density $m_s$. For instance, for $m_{\eta}\rightarrow 0$ and the two 
$m_{s}\rightarrow 0$ and $m_{s}\rightarrow 1$ reference S$^z$SLN$_{SN}$ subspaces 1A
such quantities read,
\begin{eqnarray}
\lim_{m_s\rightarrow 0}{\rm max}\{\vert j_{c} (q_j)\vert\} & = & \vert j_{c} (\pi)\vert = 4t\hspace{0.20cm}
{\rm and}\hspace{0.20cm}\lim_{m_s\rightarrow 0}\vert {\bar{j}}_{c}\vert = {8 t\over\pi} \, ,
\nonumber \\
\lim_{m_s\rightarrow 0}{\rm max}\{\vert j_{\eta} (q_j)\vert\} & = & \vert j_{\eta} (\pi m_{\eta})\vert = \vert {\bar{j}}_{\eta}\vert =
2t\left(2 - {1\over 2n}\right)  \, ,
\label{jcjetau1metaS0}
\end{eqnarray}
and
\begin{eqnarray}
\lim_{m_s\rightarrow 1}{\rm max}\{\vert j_{c} (q_j)\vert\} & = & \vert j_{c} \left({\pi\over 2}\right)\vert = 2t\hspace{0.20cm}
{\rm and}\hspace{0.20cm}\lim_{m_s\rightarrow 0}\vert {\bar{j}}_{c}\vert = {4 t\over\pi} \, ,
\nonumber \\
\lim_{m_s\rightarrow 1}{\rm max}\{\vert j_{\eta} (q_j)\vert\} & = & \vert j_{\eta} (\pi m_{\eta})\vert = \vert {\bar{j}}_{\eta}\vert = 0 \, ,
\label{jcjetau1metaS1}
\end{eqnarray}
respectively.

For reference $m_s=0$ S$^z$SLN$_{SN}$ subspace 1A contained in a S$^z$S subspace for which 
$m_{\eta}\rightarrow 1$ the same elementary 
currents are in the $u\rightarrow 0$ limit found to be given by,
\begin{eqnarray}
j_c (q_j) & = & - 2t\sin (q_j)\hspace{0.20cm}{\rm for}\hspace{0.20cm}q_j\in [-\pi,\pi] 
\nonumber \\
j_{\eta n} (q_j) & = & 2n\,j_{\eta} (q_j)\hspace{0.20cm}{\rm where} 
\hspace{0.20cm} j_{\eta} (q_j)= - 2t\sin \left({q_j\over 2}\right)
\nonumber \\
& & {\rm for}\hspace{0.20cm} q_j\in [-\pi,\pi] 
\nonumber \\
{\rm max}\{\vert j_{c} (q_j)\vert\} & = & \vert j_{c} \left({\pi\over 2} m_{\eta}\right)\vert = 4t 
\hspace{0.20cm}{\rm and}\hspace{0.20cm}\vert {\bar{j}}_{c}\vert = {4 t\over\pi} \, ,
\nonumber \\
{\rm max}\{\vert j_{\eta} (q_j)\vert\} & = & \vert j_{\eta} (\pi m_{\eta})\vert = 2t
\hspace{0.20cm}{\rm and}\hspace{0.20cm}\vert {\bar{j}}_{\eta}\vert = {4 t\over\pi} \, .
\label{jcjetau1metaL}
\end{eqnarray}
The exact elementary current expressions and values in Eqs. (\ref{jcjetau1metaS})-(\ref{jcjetau1metaL}) 
obey the corresponding inequality in Eq. (\ref{inequU0}) for all $\eta$-Bethe states under consideration.
(Due to the range $m_s \in [0,(1-m_{\eta})]$, only the $m_s=0$ S$^z$SLN$_{SN}$ subspace 1A
exists in the $m_{\eta}\rightarrow 1$ limit.) 

The number of electronic charges carried by each charge carrier provides 
a complementary physically appealing reason of why for $u>0$ and any
S$^z$S subspace all charge pseudoparticles of the $\eta$-Bethe
state whose charge current has largest absolute value are $c$ pseudoparticles. 
Indeed, the charges carried by the electrons remain invariant under the electron - rotated-electron
unitary transformation. Hence each rotated electrons carries one electronic charge. 
Within the charge - spin degrees of freedom separation of the rotated-electron lattice 
occupancy configurations that generate the $\eta$-Bethe states, each $c$ pseudoparticle carries a single electronic
charge. Moreover, each $\eta n$ pseudoparticle carries a number $2n$ of electronic charges.
The number, $L-2S_{\eta}$, of electronic charges of the $\eta$-Bethe states that span a
S$^z$S subspace is the same for all of them. 
For general $\eta$-Bethe states, a number $N_c$ of such electronic charges are carried  by
$N_c$ $c$ pseudoparticles and the remaining $\sum_{n=1}^{\infty} 2n\,N_{\eta n}$
electronic charges are carried by $N_{\eta} =\sum_{n=1}^{\infty} N_{\eta n}$ $\eta n$ 
pseudoparticles. 

On the one hand, the exact sum rule, $N_c + \sum_{n=1}^{\infty} 2n\,N_{\eta n} = L-2S_{\eta}$,
thus follows from charge conservation. On the other hand, the values of the number of
carriers $N_{\rho}=N_c+N_{\eta}$ of a $\eta$-Bethe state vary in the interval $N_{\rho}\in [1,(L-2S_{\eta})]$
for the corresponding set of S$^z$SLN subspaces contained in a S$^z$S subspace 
for which $2S_{\eta}$ is even. Each $\eta n$ pseudoparticle of such states carries $2n$ electronic charges and
accordingly is a {\it heavier} object than a $c$ pseudoparticle. It is thus a physically expected and appealing result that 
a $\eta$-Bethe state whose $L-2S_{\eta}$ electronic charges are carried by $L-2S_{\eta}$ independent charge carriers,
each carrying a single electronic charge, is that whose charge current absolute value
is the largest of the corresponding S$^z$S subspace. This argument is consistent with 
$N_{\rho}=N_c=L-2S_{\eta}$ for the reference S$^z$SLN subspace 1 of a S$^z$S subspace.

\section{Effects on the charge currents of the deviations from the TBA ideal strings}
\label{Appendix4}

For a large finite system, the $n>1$ complex rapidities with the same real part 
deviate from their TL ideal form, Eq. (\ref{complex-rap}). This affects both
the $\alpha =s$ spin and $\alpha =\eta$ $\eta$-spin complex rapidities.
In some solvable models small effects of such deviations may survive even in the TL.
In the present case of the charge currents of the 1D Hubbard model, only the deviations of the
$n>1$ charge $\eta $ complex rapidities, Eq. (\ref{complex-rap}) for $\alpha =\eta$,
may have effects on the absolute values of the charge currents of some
classes of $\eta$-Bethe states. (For an interesting study on the small effects of the $\alpha =\eta$ deviations
on the charge degrees of freedom of the 1D Hubbard model, see Ref. \cite{Deguchi-00}.)

The set of $l=1,...,n$ distorted charge $\alpha =\eta$ complex rapidities with the same real part under consideration have the general form 
$\Lambda^{\eta n,l}(q_{j}) = \Lambda^{\eta n} (q_{j}) + i\,(n + 1 - 2l)\,u + D_j^{\eta n,l}$.
Here $D_j^{\eta n,l} = R_j^{\eta n,l}+ i \delta_j^{\eta n,l}$, where $R_j^{\eta n,l}$ and $\delta_j^{\eta n,l}$ 
are real numbers, is the fine-structure deviation from the TBA ideal charge $\eta n$ strings.
Importantly, $D_j^{\eta 1,1}=0$ for the $N_{\eta 1}$ rapidities $\Lambda^{\eta 1,1}(q_{j})$ 
of all energy and momentum eigenstates. Indeed, both the $c$ momentum rapidities and such $\eta 1$ rapidities are real
and thus lack such deviations.

The $n>1$ distorted complex rapidities 
$\Lambda^{\eta n,l}(q_{j}) = \Lambda^{\eta n} (q_{j}) + i\,(n + 1 - 2l)\,u + D_j^{\eta n,l}$
remain being labelled by the quantum numbers $n=2,...,\infty$ and $l = 1,...,n$ that refer to the number of $\eta$-spin 
singlet pairs and each of these pairs, respectively. Physically, this means that, as in the case of 
an ideal charge $\eta n$ string, for $n>1$ the distorted charge $\eta n$ string associated with that set of complex rapidities 
also describes an independent configuration within which
$n=2,...,\infty$ $\eta$-spin singlet pairs are bound.

The set of $l=1,...,n$ TBA ideal charge $\eta n$ complex rapidities with the same real part, Eq. (\ref{complex-rap}) for $\alpha =\eta$, 
obey the symmetry relation $\Lambda^{\eta n,l}(q_{j}) = (\Lambda^{\eta n,n+1-l}(q_{j}))^*$.
The two complex rapidities $\Lambda^{\eta n,l}(q_{j})$ and $\Lambda^{\eta n,l'}(q_{j})$ associated with
two $\eta$-spin singlet pairs labelled by the quantum numbers $l$ and $l'=n+1-l$, respectively, are related as
$\Lambda^{\eta n,l}(q_{j})=(\Lambda^{\eta n,l'} (q_{j}))^*$ for $l = 1,...,n$. This is actually a necessary condition for the binding of the
$l = 1,...,n$ $\eta$-spin singlet pairs within the $\eta n$-pair configuration.

Importantly and due to self-conjugacy, the deviations $D_j^{\eta n,l} = R_j^{\eta n,l}+ i \delta_j^{\eta n,l}$
for the set of $l=1,...,n$ complex rapidities with the same real part associated with a $n>1$ distorted 
charge $\eta n$ string are also such that $D_j^{\eta n,l}  = (D_j^{\eta n,n+1-l})^*$. That
the symmetry $\Lambda^{\eta n,l}(q_{j}) = (\Lambda^{\eta n,n+1-l} (q_{j}))^*$ prevails under string deformations
ensures that, as for the ideal $\eta n$ strings, the imaginary part of the $n>1$ real rapidities with the
same real part associated with deformed charge $\eta n$ strings also describe the
binding within the corresponding $n$-pair configurations of $l = 1,...,n$ $\eta$-singlet pairs.

As for the $n>1$ complex spin rapidities of the related spin-$1/2$ $XXX$ chain \cite{CPC,CP},
the {\it collapse of narrow pairs}, which within our representation refers to
$\eta$-singlet pair unbinding processes, is in the TL the only aberration from the $n>1$ ideal 
charge $\eta n$ strings that may have effects on the charge currents. The occurrence for the
1D Hubbard model of two types $c$ and $\eta 1$ of real charge rapidities that are insensitive to such effects 
renders them even less important than for the spin-$1/2$ $XXX$
chain \cite{CPC,CP}. Indeed, for that chain there is only one branch of real rapidities.
As in the case of its spin currents, in the TL the small effects under
consideration have no impact whatsoever in the $T>0$ stiffness upper 
bounds considered in Sections \ref{two} and \ref{UPTinf}.

The very small effects of charge $\eta n$ string deviations occur through the $n>1$ current spectra 
$J_{\eta n} (q_j) = - J_{\eta n}^h (q_j) $, Eq. (\ref{jn-fnPSEU}) of Appendix \ref{Appendix1} and
Eq. (\ref{jn-fn}), in the charge currents general expression, Eq. (\ref{J-partDEF}) of that Appendix and Eq. (\ref{J-part}).
This applies only to $\eta$-Bethe states described by groups of real charge $c$ and $\eta 1$ rapidities and $n>1$ complex charge $\eta n$ rapidities. 
This follows from the dependence of such current 
spectra on the complex charge $\eta n$ rapidities $\Lambda^{\eta n,l}(q_{j}) = 
\Lambda^{\eta n} (q_{j}) + i\,(n + 1 - 2l)\,u + D_j^{\eta n,l}$ associated with 
$n>1$ bound $\eta$-spin singlet pairs. 

The charge $\eta n$ string deviations from the TBA $n>1$ ideal charge $\eta n$ strings do not change the number 
of $\eta$-spin singlet pairs. Their density remains being exactly
given by $\pi_{\eta} = (l_{\eta}-m_{\eta})/2$ for the corresponding $\eta$-Bethe states and non-LWSs.
Narrow pairs refer to a string deformation originated by a deviation $D_j^{\eta n,l} $ 
that renders the separation between two rapidities
$\Lambda^{\eta n,l}(q_{j})$ and $\Lambda^{\eta n,l+1}(q_{j})$ in the imaginary direction less than $i\,u$.
Such a separation may become narrower and eventually merge and split back onto 
the horizontal axis. (This is why such a process is called the collapse of a narrow pair.) 

Within our representation in terms of paired rotated $\eta$-spins $1/2$, each collapse of a narrow 
pair leads to the unbinding of two $\eta$-spin singlet pairs. On the one hand, for the
set of $n>2$ complex charge $\eta n$ rapidities with the same real part associated with $n$ bound 
$\eta$-spin singlet pairs, it leads to the partition of the
corresponding $\eta n$-pair configuration into a $\eta n'$-pair configuration where $n'=n-2$
and two unbound $\eta$-spin singlet pairs described by real charge $\eta 1$ rapidities.
The $\eta n'$-pair configuration is described by a smaller number $n'=n-2$ of complex charge $\eta n$ rapidities with the same
real part in a charge $\eta n'$ string of smaller length $n'=n-2$. On the
other hand, for complex charge $\eta 2$ rapidities with the same real part it leads in turn
to the unbinding of the two $\eta$-spin singlet pairs of the corresponding $\eta 2$-pair configuration.
In this case this gives rise solely to the two unbound $\eta$-spin singlet pairs described by real charge $\eta 1$ rapidities. 

Hence the collapse of a narrow pair is a process that causes an increase in the value of 
the total number of $\eta n$ pseudoparticles $N_{\eta}$, Eq. (\ref{NpsNapsSR}) for $\alpha = \eta$,
and thus in the equal number $N_{\eta}$ of charge $\eta n$ strings of all lengths $n=1,...,\infty$. It does not change though that of 
$\eta$-spin singlet pairs, $\Pi_{\eta} = \sum_{n=1}^{\infty}n\,N_{\eta n} = (L_{\eta} - 2S_{\eta})/2$, Eq. (\ref{sum-Nseta})
for $\alpha = \eta$. The number of rotated $\eta$-spins $L_{\eta} = 2S_{\eta} + 2\Pi_{\eta}$ 
and the corresponding density $l_{\eta}>m_{\eta}$ thus remain unchanged under such charge $\eta n$ string distortions.

The upper bounds used in our procedures within the canonical and grand-canonical ensembles 
rely on the largest current absolute value of $\eta$-Bethe states in each S$^z$S subspace
and on averages of current absolute values of $\eta$-Bethe states described by only real
rapidities, respectively. In both cases, the $\eta$-Bethe states carrying the charge currents 
under consideration have density $l_{\eta}\rightarrow m_{\eta}$. 
Hence they are insensitive to the collapse of narrow pairs. 
Indeed, the corresponding increase in the number $N_{\eta}$ of charge $\eta n$ strings of all lengths does not 
generate states with current absolute values larger than those of such $\eta$-Bethe states. 
This is why the collapse of narrow pairs has no affects
whatsoever in the upper bounds used in the studies of this paper.

\section{The 1D Hubbard model global symmetry group independent state representations and Hilbert space and subspaces dimensions}
\label{Appendix5}

Following the 1D Hubbard model global $[SU(2)\otimes SU(2)\otimes U(1)]/Z_2^2$ symmetry \cite{bipartite}, its 
full Hilbert-space dimension $4^L$ must equal the number of independent state representations of the 
corresponding symmetry group. In each subspace with fixed values for $L_{s} = N_c$, $L_{\eta} = N_c^h$, $S_s$, and $S_{\eta}$ there are   
${\cal{N}}^{\,\eta}(L_{\eta},S_{\eta}) \times {\cal{N}}^{s}(L_s,S_s) \times d_c (N_c)$ such representations. Here,
\begin{equation}
{\cal{N}}^{\,\alpha}(L_{\alpha},S_{\alpha}) = (2S_{\alpha} +1)\,{\cal{N}}_{\rm singlet}^{\,\alpha} (L_{\alpha},S_{\alpha}) 
\hspace{0.20cm}{\rm and}\hspace{0.20cm}d_c ={L\choose L_s}={L\choose L_{\eta}}
\hspace{0.20cm}{\rm for}\hspace{0.20cm}\alpha=\eta, s \, ,
\label{NSalpha}
\end{equation}
where,
\begin{equation}
{\cal{N}}_{\rm singlet}^{\,\alpha} (L_{\alpha},S_{\alpha}) = {L_{\alpha} \choose L_{\alpha}/2-S_{\alpha}}-
{L_{\alpha}\choose L_{\alpha}/2-S_{\alpha}-1}\hspace{0.20cm}{\rm for}\hspace{0.20cm}\alpha=\eta, s \, .
\label{N-singlet}
\end{equation} 
The $\alpha=\eta, s$ $SU(2)$ dimensions are similar to the spin $SU(2)$ dimension of the spin-$1/2$ $XXX$ chain \cite{CPC,CP}.
The dimension $d_c ={L\choose N_c}$ is characteristic of an $U(1)$ symmetry and refers indeed to the $c$ lattice $U(1)$ symmetry.

One finds that,
\begin{eqnarray}
4^{L} & = & \sum_{\substack{N_{c}=0\\({\rm integers})}}^{L} 
\sum_{\substack{2S_{\eta}=0\\({\rm integers})}}^{L_{\eta}}\,
\sum_{\substack{2S_{s}=0\\({\rm integers})}}^{L_{s}} C (L_{\eta},L_{s},S_{\eta},S_{s})\,
{\cal{N}}^{\,\eta}(L_{\eta},S_{\eta}) \times {\cal{N}}^{s}(L_s,S_s) \times d_c (N_c) \, ,
\nonumber \\
& & C (L_{\eta},L_{s},S_{\eta},S_{s}) = \vert\cos\left({\pi\over 2}(2S_{\eta}+L_{\eta})\right)\cos\left({\pi\over 2}(2S_{s}+L_s)\right)\vert \, ,
\label{Ntot-dess}
\end{eqnarray}
where the role of the phase factor, $C (N_c,S_{\eta},S_{s})=0,1$, is to select the allowed independent representations of the global $[SU(2)\otimes SU(2)\otimes U(1)]/Z_2^2$ symmetry.

As for the corresponding spin-singlet dimension of the $XXX$ chain \cite{CPC,CP}, the dimension ${\cal{N}}_{\rm singlet} (S_{\alpha},M_{\alpha}) $
in Eq. (\ref{NSalpha}) can be written as,
\begin{equation}
{\cal{N}}_{\rm singlet}^{\alpha} (L_{\alpha},S_{\alpha}) =
\sum_{\{N_{\alpha n}\}}\, \prod_{n =1}^{\infty}\,{L_{\alpha n}\choose N_{\alpha n}} 
\hspace{0.20cm}{\rm for}\hspace{0.20cm}\alpha=\eta, s \, ,
\label{Ncs-cpb}
\end{equation} 
where $\sum_{\{N_{\alpha n}\}}$  is a summation over all sets of $\{N_{\alpha n}\}$ corresponding to the same number of
pairs, $\Pi_{\alpha} = \sum_{n=1}^{\infty}n\,N_{\alpha n} = (L_{\alpha} - 2S_{\alpha})/2$, Eq. (\ref{sum-Nseta}).
(That the use of the alternative dimension expression, Eq. (\ref{Ncs-cpb}), in Eq. (\ref{Ntot-dess}) leads to the same overall dimension $4^L$ 
is shown in Ref. \cite{Completeness}.)

For the problem studied in this paper, we consider the $S_s^z =0$ subspace. Its
S$^z$SL subspaces as defined in Section \ref{SzSS} are populated by fixed numbers
$L_{\eta} = N_c^h = L-N_c$ rotated $\eta$-spins $1/2$ of which $2\Pi_{\eta} = L_{\eta} - 2S_{\eta}$ are paired and the
remaining $M_{\eta} = 2S_{\eta}$ physical $\eta$-spins $1/2$ are unpaired. They are as well
populated by fixed numbers $L_s = L - L_{\eta}$ of rotated spins $1/2$ of which $2\Pi_{s} = L - L_{\eta} - 2S_{s}$ are paired and the
remaining $M_{s} = 2S_{s}$ physical spins $1/2$ are unpaired. 
Since for the different spin $S_s$ integer values only states for which $S_s^z =0$ contribute,
out of the $2S_s + 1$ states of each spin multiplet tower only that for which $S_s^z =0$
is counted. The dimension of a $S_s^z =0$ subspace thus reads,
\begin{equation}
d_{\rm subspace} (L_{\eta},S_{\eta}) = \sum_{S_{s}=0}^{(L-L_{\eta})/2}(2S_{\eta} +1)\,{\cal{N}}_{\rm singlet}^{\,\eta} (L_{\eta},S_{\eta}) \times 
{\cal{N}}_{\rm singlet}^{s} (L_{\eta},S_s) \times d_c (L_{\eta}) \, ,
\label{NNcharge}
\end{equation}
where accounting for that $L_s = L - L_{\eta}$,
\begin{eqnarray}
{\cal{N}}_{\rm singlet}^{\,\eta} (L_{\eta},S_{\eta}) & = & {L_{\eta} \choose L_{\eta}/2-S_{\eta}}-
{L_{\eta}\choose L_{\eta}/2-S_{\eta}-1} = \sum_{\{N_{\eta n}\}}\, \prod_{n =1}^{\infty}\,{L_{\eta n}\choose N_{\eta n}} \, ,
\nonumber \\
{\cal{N}}_{\rm singlet}^{s} (L_{\eta},S_s) & = & {L - L_{\eta} \choose L/2 - L_{\eta}/2 -S_{s}}-
{L - L_{\eta}\choose L/2 - L_{\eta}/2 -S_{s}-1}
= \sum_{\{N_{sn}\}}\, \prod_{n =1}^{\infty}\,{L_{sn}\choose N_{sn}} \, ,
\nonumber \\
d_c (L_{\eta}) & = & {L\choose L_{\eta}} \, .
\label{Nall}
\end{eqnarray}
The summations $\sum_{\{N_{\eta n}\}}$ and $\sum_{\{N_{sn}\}}$ run again over all sets of 
$\{N_{\eta n}\}$ and $\{N_{sn}\}$, respectively, corresponding to the same number of
singlet pairs. That number is given by $\Pi_{\eta} = \sum_{n=1}^{\infty}n\,N_{\eta n} = (L_{\eta}-2S_{\eta})/2$
for the $\eta$-spin singlet pairs. For the $S_s^z=0$ subspace the number of spin-singlet pairs
reads $\Pi_{s} = \sum_{n=1}^{\infty}n\,N_{sn} = (L-L_{\eta}-2S_s)/2$.

An important subspace of a S$^z$SL subspace
is that spanned by the corresponding $\eta$-Bethe states whose dimension is given by,
\begin{equation}
d_{\rm subspace}^{LWS} (L_{\eta},S_{\eta}) = \sum_{S_{s}=0}^{(L-L_{\eta})/2}{\cal{N}}_{\rm singlet}^{\,\eta} (L_{\eta},S_{\eta}) \times 
{\cal{N}}_{\rm singlet}^{s} (L_{\eta},S_s) \times d_c (L_{\eta}) \, .
\label{dsinglet}
\end{equation}

\section{Quantities in the expression of the pseudoparticle elementary currents}
\label{Appendix6}

The $f$ functions in the elementary currents expression, Eq. (\ref{jnvf}), read,
\begin{eqnarray}
f_{\beta\,\beta'}(q_j,q_{j'}) & = & v_{\beta}(q_{j})\,2\pi \,\Phi_{\beta,\beta'}(q_{j},q_{j'})+
v_{\beta'}(q_{j'})\,2\pi \,\Phi_{\beta',\beta}(q_{j'},q_{j}) 
\nonumber \\
& + & {1\over 2\pi}\sum_{\iota =\pm 1} \theta (N_c)\,\vert v_{c} (q_{Fc,\tau}^{\iota})\vert\,
2\pi\,\Phi_{c,\beta}(q_{Fc,\tau}^{\iota},q_{j})\,2\pi\,\Phi_{c,\beta'} (q_{Fc,\tau}^{\iota},q_{j'}) 
\nonumber \\
& + & {1\over 2\pi}\sum_{\iota =\pm 1}\sum_{n=1}^{\infty}\theta (N_{\eta n})\,\vert v_{\eta n} (q_{F\eta n,\tau_n}^{\iota})\vert\,
2\pi\,\Phi_{\eta n,\beta}(q_{F\eta n,\tau_n}^{\iota},q_{j})\,2\pi\,\Phi_{\eta n,\beta'} (q_{F\eta n,\tau_n}^{\iota},q_{j'}) 
\nonumber \\
& + & {1\over 2\pi}\sum_{\iota =\pm 1}\sum_{n=1}^{\infty}\theta (N_{sn})\,\vert v_{sn} (\iota q_{Fsn})\vert\,
2\pi\,\Phi_{sn,\beta}(\iota q_{Fsn},q_{j})\,2\pi\,\Phi_{sn,\beta'} (\iota q_{Fsn},q_{j'}) \, ,
\label{ff}
\end{eqnarray}
where the $\beta = c, \eta n, sn$ bands group velocities $v_{\beta} (q_j)$ are within the TL continuum 
$q$ representation given by,
\begin{equation}
v_{\beta} (q) = {\partial\varepsilon_{\beta} (q)\over \partial q}\hspace{0.20cm}{\rm where}\hspace{0.20cm}\beta = c, \eta n, sn \, .
\label{vel-beta}
\end{equation}
The $\beta =c,\eta n, sn$ band energy dispersions $\varepsilon_{\beta} (q_j)$ appearing here read,
\begin{eqnarray}
\varepsilon_{\beta} (q_j) & = & E_{\beta} (q_j) + \varepsilon_{\beta}^c (q_j) 
\, ; \hspace{0.50cm}
\varepsilon_{\beta}^c (q_j) =
{t\over \pi}\int_{\{Q^{\iota}_{\tau}\}}dk\,2\pi\,\bar{\Phi }_{c\,\beta}
\left({\sin k\over u}, {\Lambda_{0}^{\beta} (q_j)\over u}\right)\sin k  
\nonumber \\
& & {\rm for}\hspace{0.20cm}j = 1,...,L_{\beta} \, ,
\label{epsilon-q}
\end{eqnarray} 
where the $\beta = c, \eta n, sn$ bands energy spectrum $E_{\beta} (q_j)$ is provided in
Eq. (\ref{spectra-E-an-c-0}) of Appendix \ref{Appendix1} with the rapidity functionals 
$\Lambda_0^{c}(q_j) = \sin k_0^c (q_j)$ and $\Lambda_0^{\alpha n}(q_j)$
being those of a $\eta$-Bethe state with compact distributions of general form, Eq. (\ref{NN}).
The integration $\int_{\{Q^{\iota}_{\tau}\}}dk$ is defined as,
\begin{eqnarray}
\int_{\{Q^{\iota}_{\tau}\}}dk & = & \int_{-\pi}^{Q^{-}_{-}}dk + \int_{Q^{+}_{-}}^{\pi}dk \hspace{0.20cm}{\rm for}
\hspace{0.20cm}m_{\eta}\leq 1/2\hspace{0.20cm}(\tau =-\hspace{0.15cm}{\rm hole}\hspace{0.20cm}{\rm like})
\nonumber \\
& = & \int_{Q^{-}_{+}}^{Q^{+}_{+}}dk\hspace{0.20cm}{\rm for}
\hspace{0.20cm}m_{\eta}\geq 1/2\hspace{0.20cm}(\tau =+\hspace{0.15cm}{\rm particle}\hspace{0.20cm}{\rm like}) \, .
\label{QQ}
\end{eqnarray}
The momentum rapidity variable $k$ integration limiting parameters $Q^{\iota}_{\tau}$ are given by,
\begin{equation}
Q^{\iota}_{\tau} = k^c (q_{Fc,\tau}^{\iota})\hspace{0.20cm}{\rm where}\hspace{0.20cm}\iota =\pm 
\hspace{0.20cm}{\rm and}\hspace{0.20cm}\tau = \pm \, .
\label{QQtaus}
\end{equation}
In this expression $q_{Fc,\tau}^{\iota}$ stands for the $c$ band compact domains limiting momenta, Eq. (\ref{qFcetanpms})
for $\beta =c$.

In the case of excited states of a ground state whose $c$ band distribution is compact and symmetrical as
given in Eq. (\ref{NGS}), the corresponding limiting occupancy momentum rapidities $Q^{\iota}_{\tau}$ in 
Eq. (\ref{QQtaus}) rather read $Q^{\pm}_{+}=\pm Q$ for such a ground state where $Q=k^c (q_{Fc})$ 
and $q_{Fc} = \pi (1-m_{\eta})$. 

For $\eta$-Bethe states generated by a finite number of $c$ band processes 
from those with $c$ and $s1$ bands compact distributions of general form, Eq. (\ref{NN}), belonging 
to the $m_s=0$ reference S$^z$SLN$_{SN}$ subspace 1A, the related group velocity, Eq. (\ref{vel-beta}) for $\beta =c$, 
is up to ${\cal{O}}(u^{-2})$ order and for $m_\eta\in [0,1]$ found to be given by,
\begin{eqnarray}
v_c (q_j)  & = & 2t \sin q_j - 2t\,{(1-m_{\eta})\ln 2\over u}\sin 2q_j
- \tau\,2t\,{\ln 2\over 2\pi u}\left(\sum_{\iota=\pm} (\iota)\cos (q_{Fc,\tau}^{\iota})\right)\cos q
\nonumber \\
& + & 6t\,(1-m_{\eta})^2\left({\ln 2\over u}\right)^2\left(1-{3\over 2}\sin^2 q_j\right)\sin q_j 
- \tau\,2t\,{(1-m_{\eta})\over 2\pi}\left({\ln 2\over u}\right)^2\left(\sum_{\iota=\pm} (\iota)\cos (q_{Fc,\tau}^{\iota})\right)\sin q 
\nonumber \\
& - & \tau\,2t\,{(1-m_{\eta})\over 2\pi}\left({\ln 2\over u}\right)^2\left(\sum_{\iota=\pm} (\iota)\sin^2 (q_{Fc,\tau}^{\iota})\right)\cos q 
- 2t\,{1\over 8\pi^2}\left({\ln 2\over u}\right)^2\left(\sum_{\iota=\pm}\sin (2q_{Fc,\tau}^{\iota})\right)\cos q 
\nonumber \\
& + & 2t\,{1\over 4\pi^2}\left({\ln 2\over u}\right)^2\sin (q_{Fc,\tau}^{-}+q_{Fc,\tau}^{+})\,\cos q \, .
\label{vcallmetaz}
\end{eqnarray}
The terms that depend on the momenta $q_{Fc,\tau}^{\iota}$ where $\tau =\pm$ and $\iota =\pm$ are
state dependent. The expression corresponding to the subset of terms that are state independent gives exactly
the $c$ band current spectrum in Eq. (\ref{Jqugg}).

For the $m_s\rightarrow 1-m_{\eta}$ reference S$^z$SLN$_{SN}$ subspace 1A
and $m_\eta\in [0,1]$ all terms of the $c$-band group velocity under consideration are up to ${\cal{O}}(u^{-2})$ order state
independent and read,
\begin{equation}
v_c (q_j)  = 2t \sin q_j - 2t\,{(1-m_{\eta}-m_s)\over u}\sin 2q_j
+ 6t\,{(1-m_{\eta}-m_s)^2\over u^2}\left(1-{3\over 2}\sin^2 q_j\right)\sin q_j \, .
\label{vcmsL}
\end{equation}
Note that up to that order the group velocity $v_c (q_j)$ exactly equals the current spectrum
$J_c^h (q_j)=-J_c (q_j)$, Eq. (\ref{Jquggmsmax}).

Finally, for $m_{\eta}=0$ and the set of fixed $m_s$ density reference S$^z$SLN$_{SN}$ subspaces 1A
referring to the interval $m_s \in [0,1]$ all terms of the $c$-band group velocity $v_c (q_j)$ 
are up to ${\cal{O}}(u^{-2})$ order state independent and that velocity 
exactly equals the current spectrum $J_c^h (q_j)=-J_c (q_j)$, Eq. (\ref{Jchalletas}),
and thus reads,
\begin{eqnarray}
v_c (q_j) & = & 2t \sin q_j - 2t\,{n_{\eta s}\over u}\sin 2q_j + 
6t\,\left({n_{\eta s}\over u}\right)^2\left(1-{3\over 2}\sin^2 q_j\right)\sin q_j 
\nonumber \\
& & {\rm where}\hspace{0.20cm} n_{\eta s} = (1-m_s)\,g_s
\hspace{0.20cm}{\rm for}\hspace{0.20cm}m_{\eta}=0
\hspace{0.20cm}{\rm and}\hspace{0.20cm}m_s \in [0,1] \, .
\label{vcmeta0}
\end{eqnarray}
Here $g_s = g_s (m_s)$ is the function in Eq. (\ref{Lms}).

The phase shifts $2\pi \,\Phi_{\beta,\beta'}(q_{j},q_{j'})$ in Eq. (\ref{ff}) where $\beta$ and 
$\beta'$ refers to $c$, $\eta n$, and $sn$ branches are a generalization of those considered in Ref. \cite{Carmelo-17} to
$\eta$-Bethe states with compact distributions of general form, Eq. (\ref{NN}), and $\eta$-Bethe
states generated from them by small $c$ and $\eta n$ bands distributions deviations.
For all such states they have the form,
\begin{equation}
2\pi\,\Phi_{\beta,\beta'}(q_j,q_{j'}) = 2\pi\,\bar{\Phi }_{\beta,\beta'} \left(r,r'\right) 
\hspace{0.20cm}{\rm where}\hspace{0.20cm}r = \Lambda_{0}^{\beta}(q_j)/u
\hspace{0.20cm}{\rm and}\hspace{0.20cm}r' = \Lambda_{0}^{\beta'}(q_{j'})/u \, ,
\label{Phi-barPhi}
\end{equation}
and the rapidity functionals are those of the corresponding $\eta$-Bethe states with compact distributions.
The rapidity dressed phase shifts $2\pi\,\bar{\Phi }_{\beta,\beta'}$ on the right-hand side of Eq. (\ref{Phi-barPhi}) 
are the solution of well-defined integral equations. Those are provided in the following for the
reference S$^z$SLN$_{SN}$ subspace 1, as defined in Section \ref{SzSLNJmax}. 
(That subspace plays a central role in the upper bound 
procedures used in the studies of this paper.)

For the case of $\eta$-Bethe states with compact distributions of general 
form, Eq. (\ref{NN}), belonging to the reference S$^z$SLN$_{SN}$ subspace 1 and $\eta$-Bethe states generated 
from them by small $c$ and $\eta n$ bands distributions deviations, a first set of rapidity dressed 
phase shifts $2\pi\,\bar{\Phi }_{\beta,\beta'}$ obey integral equations by their own,
\begin{equation}
\bar{\Phi }_{s1,c}\left(r,r'\right) = -{1\over\pi}\arctan (r-r') + \int_{-B/u}^{B/u}
dr''\,G_{\tau} (r,r'')\,{\bar{\Phi }}_{s1,c}\left(r'',r'\right) \, ,
\label{Phis1c-m}
\end{equation}
\begin{equation}
\bar{\Phi }_{s1,\eta n}\left(r,r'\right) =  -{1\over{\pi^2}\,u}\int_{\{Q^{\iota}_{\tau}\}}dk\,
\cos k\,{\arctan\Bigl({\sin k/u-r'\over n}\Bigr)\over{1+(r-\sin k/u)^2}} +
\int_{-B/u}^{B/u} dr''\,G_{\tau} (r,r'')\,{\bar{\Phi}}_{s1,\eta n}\left(r'',r'\right) \, , 
\label{Phis1cn-m}
\end{equation}
and
\begin{eqnarray}
\bar{\Phi }_{s1,sn}\left(r,r'\right) & = & \delta_{1
,n}\,{1\over\pi}\arctan\Bigl({r-r'\over 2}\Bigl) + (1-\delta_{1
,n}){1\over\pi}\Bigl\{ \arctan\Bigl({r-r'\over n-1}\Bigl) +
\arctan\Bigl({r-r'\over
n+1}\Bigl)\Bigr\} \nonumber \\
& - &  {1\over{\pi^2}\,u}\int_{\{Q^{\iota}_{\tau}\}}dk\,
\cos k\,{\arctan \Bigl({\sin k/u-r'\over n}\Bigr)\over{1+(r-\sin k/u)^2}} +
\int_{-B/u}^{B/u} dr''\,G_{\tau} (r,r'')\,{\bar{\Phi
}}_{s1,s1}\left(r'',r'\right) \, . 
\label{Phis1sn-m}
\end{eqnarray}
The kernel $G_{\tau} (r,r')$ is given by,
\begin{equation}
G_{\tau} (r,r') = - {1\over{2\pi}}\left[{1\over{1+((r-r')/2)^2}}\right]
\left[1 + {\tau\over 2}
\left(t_{\tau}(r)+t_{\tau}(r')+{{l_{\tau}(r)-l_{\tau}(r')}\over{r-r'}}\right)\right] \, .
\label{G}
\end{equation}
Here
\begin{equation}
t_{\tau}(r) = {1\over{\pi}}\sum_{\iota=\pm} (\iota)\arctan\left(r - {\sin Q_{\tau}^{\iota}\over u}\right) 
\, , \label{t}
\end{equation}
and
\begin{equation}
l_{\tau}(r) = {1\over{\pi}}\sum_{\iota=\pm} (\iota)\ln \left(1+\left(r - {\sin Q_{\tau}^{\iota}\over u}\right)^2\right) \, . 
\label{l}
\end{equation}

A second set of rapidity dressed phase shifts are expressed in terms of those in Eqs.
(\ref{Phis1c-m})-(\ref{Phis1sn-m}) as follows,
\begin{equation}
\bar{\Phi }_{c,c}\left(r,r'\right) =
{1\over{\pi}}\int_{-B/u}^{B/u} dr''{\bar{\Phi}_{s1,c}\left(r'',r'\right) \over {1+(r-r'')^2}} \, ,
\label{Phicc-m}
\end{equation}
\begin{equation}
\bar{\Phi }_{c,\eta n}\left(r,r'\right) = -{1\over\pi}\arctan\Bigl({r-r'\over n}\Bigr) +
{1\over{\pi}}\int_{-B/u}^{B/u} dr''{\bar{\Phi}_{s1,\eta n}\left(r'',r'\right) \over {1+(r-r'')^2}} \, ,
\label{Phiccn-m}
\end{equation}
and
\begin{equation}
\bar{\Phi }_{c,sn}\left(r,r'\right) = -{1\over\pi}\arctan\Bigl({r-r'\over n}\Bigr) + {1\over{\pi}}\int_{-B/u}^{B/u} dr''
{\bar{\Phi}_{s1,sn}\left(r'',r'\right) \over {1+(r-r'')^2}} \, .
\label{Phicsn-m}
\end{equation}

The remaining rapidity dressed  phase shifts can be
expressed either in terms of those in Eqs. (\ref{Phicc-m})-(\ref{Phicsn-m}) only,
\begin{equation}
{\bar{\Phi }}_{\eta n,c}\left(r,r'\right) = {1\over\pi}\arctan\Bigl({r-r'\over {n}}\Bigr) 
- {1\over{\pi}\,u}\int_{\{Q^{\iota}_{\tau}\}}dk\,\cos k\,
{{\bar{\Phi}}_{c,c}\left(\sin k/u,r'\right) \over {n[1+({r-\sin k/u\over {n}})^2]}} \, , 
\label{Phicnc-m}
\end{equation}
\begin{equation}
\bar{\Phi }_{\eta n,\eta n'}\left(r,r'\right) = {\Theta_{n,n'}(r-r')\over{2\pi}} -
{1\over{\pi}\,u}\int_{\{Q^{\iota}_{\tau}\}}dk\,\cos k\,{\bar{\Phi }_{c,\eta n'}\left(\sin k/u,r'\right) \over
{n[1+({r-r\sin k/u\over n})^2]}} \, , 
\label{Phicncn-m}
\end{equation}
\begin{equation}
\bar{\Phi }_{\eta n,sn'}\left(r,r'\right) = - {1\over{\pi}\,u}\int_{\{Q^{\iota}_{\tau}\}}dk\,\cos k\,
{\bar{\Phi }_{c,sn'}\left(\sin k/u,r'\right) \over {n[1+({r-\sin k/u\over n})^2]}} \, , 
\label{Phicnsn-m}
\end{equation}
or in terms of both those in Eqs. (\ref{Phis1c-m})-(\ref{Phis1sn-m}) and in Eqs. (\ref{Phicc-m})-(\ref{Phicsn-m}),
\begin{eqnarray}
{\bar{\Phi }}_{sn,c}\left(r,r'\right) & = & - {1\over\pi}\arctan\Bigl({r-r'\over {n}}\Bigr) +
{1\over{\pi}\,u}\int_{\{Q^{\iota}_{\tau}\}}dk\,\cos k\,
{{\bar{\Phi}}_{c,c}\left(\sin k/u,r'\right) \over {n[1+({r-\sin k/u\over n})^2]}} 
\nonumber \\
& - & \int_{-B/u}^{B/u} dr''{\bar{\Phi}}_{s1,c}\left(r'',r'\right){\Theta^{[1]}_{n,1}(r-r'')\over{2\pi}} 
\, ; \hspace{0.5cm} n > 1 \, , 
\label{Phisnc-m}
\end{eqnarray}
\begin{eqnarray}
{\bar{\Phi }}_{sn ,\eta n'}\left(r,r'\right) & = & 
{1\over{\pi}\,u}\int_{\{Q^{\iota}_{\tau}\}}dk\,\cos k\,
{{\bar{\Phi}}_{c,\eta n'}\left(\sin k/u,r'\right) \over {n[1+({r-\sin k/u\over n})^2]}} 
\nonumber \\
& - & \int_{-B/u}^{B/u} dr''{\bar{\Phi}}_{s1,\eta n'}\left(r'',r'\right) {\Theta^{[1]}_{n,1}(r-r'')\over {2\pi}} 
\, ; \hspace{0.5cm} n > 1 \, , 
\label{Phisncn-m}
\end{eqnarray}
\begin{eqnarray}
{\bar{\Phi }}_{sn ,sn'}\left(r,r'\right) & = & {\Theta_{n,n'}(r-r')\over{2\pi}} +
{1\over{\pi}\,u}\int_{\{Q^{\iota}_{\tau}\}}dk\,\cos k\,
{{\bar{\Phi}}_{c,sn'}\left(\sin k/u,r'\right) \over {n[1+({r-\sin k/u\over n})^2]}} 
\nonumber \\
& - & \int_{-B/u}^{B/u} dr''{\bar{\Phi}}_{s1,sn'}\left(r'',r'\right){\Theta^{[1]}_{n,1}(r-r'')\over{2\pi}} \, . 
\label{Phisnsn-m}
\end{eqnarray}

In the above equations, $\Theta_{n\,n'}(x)$ is the function given in Eq. (\ref{Theta}) 
of Appendix \ref{Appendix1} and $\Theta^{[1]}_{n\,n'}(x)$ is its derivative,
\begin{eqnarray}
\Theta^{[1]}_{n,n'}(x) & = & {\partial\Theta_{n,n'}(x)\over
\partial x} = \delta_{n ,n'}\Bigl\{{1\over n[1+({x\over 2n})^2]}+
\sum_{l'=1}^{n -1}{2\over l'[1+({x\over 2l'})^2]}\Bigr\} +
(1-\delta_{n ,n'})\Bigl\{{2\over |n-n'|[1+({x\over
|n-n'|})^2]} \nonumber \\
& + & {2\over (n+n')[1+({x\over n+n'})^2]} +
\sum_{l'=1}^{{n+n'-|n-n'|\over 2} -1}{4\over
(|n-n'|+2l')[1+({x\over |n-n'|+2l'})^2]}\Bigr\} \, .
\label{The1}
\end{eqnarray}

The phase-shift integral equations given here are an extension of those 
provided in Ref. \cite{Carmelo-17}, which refer to $\eta$-Bethe states generated
from ground states by small $c$ and $\eta n$ bands distributions deviations.

The phase shifts suitable to the set of subspaces considered in the analysis of Section 
\ref{SzSLNJmax} other than the reference S$^z$SLN$_{SN}$ subspace 1 to which Eqs. (\ref{Phis1c-m})-(\ref{Phisnsn-m}) apply 
either are easily expressed in terms of those defined by these equations under suitable densities 
interchange or are straightforwardly computed.

\section{Useful $u\gg 1$ expansions in powers of $u^{-l'}$ for $l'=1,...,\infty$}
\label{Appendix7}

The expansion terms of $u^{-j}$ order $j>2$ of the $c$-band current spectrum $J_c^h (q_j)$, Eq. (\ref{J-partS1}),
are state dependent. To illustrate this property, in this Appendix such a current spectrum is expanded 
up to $u^{-3}$ order for the $m_s = 0$ and $m_s\rightarrow 1-m_{\eta}$ reference S$^z$SLN$_{SN}$ subspaces 1A
and the whole $m_{\eta}\in [0,1]$ range. We consider the $\eta$-Bethe states of a general
reference S$^z$SLN$_{SN}$ subspace 1A whose $\beta =c,\eta n,sn$ compact distributions have limiting
momenta of the form given in Eq. (\ref{qFcpm1iota}). The corresponding $c$ band distribution $2\pi\rho_c (k)$ 
is then defined by the equation,
\begin{equation}
2\pi\rho_c (k) = 1 + 
\frac{\cos k}{\pi\,u} \int_{-B}^{B}d\Lambda\,{2\pi\sigma_{s1} (\Lambda)\over 1 +  \left({\sin k - \Lambda\over u}\right)^2} 
\hspace{0.2cm}{\rm where}\hspace{0.2cm}B = \Lambda_{0}^{\beta}(q_{Fs1}) = \Lambda_{0}^{\beta}(k_{F\downarrow}) \, ,
\label{rho}
\end{equation}
and the distribution $2\pi\sigma_{s1} (\Lambda)$ obeys the related equation,
\begin{equation}
2\pi\sigma_{s1} (\Lambda) = {1\over\pi\,u}\int_{\{Q^{\iota}_{\tau}\}}dk\,{2\pi\rho_c (k)\over 1 +  \left({\Lambda-\sin k\over u}\right) ^2} 
- \frac{1}{2\pi\,u} \int_{-B}^{B}d\Lambda^{\prime}\,{2\pi\sigma_{s1} (\Lambda^{\prime})\over 1 +  \left({\Lambda -
\Lambda^{\prime}\over 2u})\right)^2} \, .
\label{sigmas1}
\end{equation}
The integration $\int_{\{Q^{\iota}_{\tau}\}}$ appearing here and its limiting parameters $Q^{\iota}_{\tau}$ are 
defined in Eqs. (\ref{QQ}) and (\ref{QQtaus}) of Appendix \ref{Appendix6}, respectively.

As given in Eq. (\ref{sigmsrhoderiv}) of Appendix \ref{Appendix1}, the distributions $2\pi\rho_c (k)$ and $2\pi\sigma_{s1} (\Lambda)$
are related to the inverse functions $q^c (k)$ and $q^{s1} (\Lambda)$ of the rapidity momentum functional $k^c (q)$
and $s1$ branch rapidity functional $\Lambda^{s1} (q)$, respectively, as,
\begin{equation}
2\pi\rho_c (k) = {\partial q^c (k)\over \partial k}\hspace{0.20cm}{\rm and}\hspace{0.20cm}
2\pi\sigma_{s1} (\Lambda) = {\partial q^{s1} (\Lambda)\over \partial \Lambda} \, .
\label{sigs1rhoderiv}
\end{equation}
Furthermore, for the present subspace the distributions $2\pi\rho_c (k)$ and $2\pi\sigma_{s1} (\Lambda)$ obey
the sum rules,
\begin{equation}
\int_{\{Q^{\iota}_{\tau}\}}dk\,2\pi\rho_c (k) = 2\pi (1-m_{\eta})\hspace{0.20cm}{\rm and}\hspace{0.20cm}
\int_{-B}^{B}d\Lambda\,2\pi\sigma_{s1} (\Lambda) = \pi (1-m_{\eta} -m_s) \, .
\label{Normrhosigma}
\end{equation}

We start by considering the case of the $m_s = 0$ reference S$^z$SLN$_{SN}$ subspace 1A
for densities $m_{\eta}\in [0,1]$. As for a ground state \cite{Call-74}, in the present more general case of energy and momentum
eigenstates whose distributions are of the form, Eq. (\ref{NN}), one can use the Fourier transform
of $2\pi\sigma_{s1} (\Lambda)$ to reach from the use of Eq. (\ref{sigmas1}) the following alternative
exact relation,
\begin{equation}
2\pi\sigma_{s1} (\Lambda) = 
{1\over 4u}\int_{\{Q^{\iota}_{\tau}\}}dk\,{2\pi\rho_c (k)\over\cosh\left({\pi\over 2u}(\Lambda-\sin k)\right)} \, .
\label{sigmaFFT}
\end{equation}
The use of this $2\pi\sigma_{s1} (\Lambda)$ expression in Eq. (\ref{rho}) reveals that the distribution $2\pi\rho_c (k)$ 
obeys the following integral equation,
\begin{equation}
2\pi\rho_c (k) = 1 + \cos k\int_{\{Q^{\iota}_{\tau}\}}dk'\,\Gamma (k,k')\,2\pi\rho_c (k') \, ,
\label{rhocIE}
\end{equation}
where the kernel reads,
\begin{equation}
\Gamma (k,k') = 
{1\over 4\pi}\int_{-\infty}^{\infty}d\Lambda\,\,{1\over (u^2 + \Lambda^2)}\,{1\over\cosh\left({\pi\over 2u}(\Lambda+\sin k-\sin k')\right)} \, .
\label{kernelGamma}
\end{equation}

In order to derive a large-$u$ expansion of the distribution $2\pi\rho_c (k)$, it is convenient to introduce
the following integral representation of the integrand factor $1/\cosh (\pi x/2u)$ in Eq. (\ref{kernelGamma})
where $x=\Lambda+\sin k-\sin k'$,
\begin{equation}
{1\over\cosh\left({\pi x\over 2u}\right)} = {1\over\pi}
\int_{0}^{\infty}dy {\cos\left({xy\over 2u}\right)\over\cosh\left({y\over 2}\right)} \, .
\label{intRepre}
\end{equation}
The use of this expression in Eq. (\ref{kernelGamma}) leads to,
\begin{equation}
\Gamma (k,k') = {1\over 2\pi\,u}\int_{0}^{\infty}dy {\cos\left({(\sin k-\sin k')y\over 2u}\right)\over 1 + e^y} \, .
\label{kernelGRep}
\end{equation}
This expression is suitable for deriving the following $\Gamma (k,k')$ large-$u$ expansion,
\begin{eqnarray}
\Gamma (k,k') & = & {\ln 2\over 2\pi\,u}
+ \sum_{l'=1}^{\infty} {(-1)^{l'}\over 2\pi\,u}\zeta (2l'+1) \left(1-{1\over 2^{2l'}}\right)
\left({\sin k-\sin k'\over 2u}\right)^{2l'} 
\nonumber \\
& = & {\ln 2\over 2\pi\,u}
+ {1\over\pi}\sum_{l'=1}^{\infty} (-1)^{l'}{\zeta (2l'+1)\over (2u)^{2l'+1}} \left(1-{1\over 2^{2l'}}\right)
\sum_{l''=0}^{2l}(-1)^{l''}(\sin k')^{l''}{2l \choose l''}(\sin k)^{2l'-l''} \, .
\label{LuexpG}
\end{eqnarray}

From the use of this large-$u$ expansion in Eq. (\ref{rhocIE}) accounting for the first sum rule
in Eq. (\ref{Normrhosigma}), the following $u^{-l'}$ expansion obeyed by the
distribution $2\pi\rho_c (k)$, which contains all infinite orders $l'=1,2,...,\infty$, is 
straightforwardly derived,
\begin{eqnarray}
2\pi\rho_c (k) & = & 1 - {(1-m_{\eta})\ln 2\over u}\cos k
\nonumber \\
& + & \cos k\sum_{l'=1}^{\infty}{\zeta (2l'+1)\over (2u)^{2l'+1}} \left(1-{1\over 2^{2l'}}\right)
\sum_{l''=0}^{2l}(-1)^{l'+l''} M_{l''} {2l' \choose l''}(\sin k)^{2l'-l''} \hspace{0.2cm}{\rm for}\hspace{0.2cm}
m_s = 0 \, ,
\label{rhoLuExp}
\end{eqnarray}
where
\begin{equation}
M_{l''} = {1\over \pi}\int_{\{Q^{\iota}_{\tau}\}}dk'\,2\pi\rho_c (k')\,(\sin k')^{l''} \, .
\label{Mlline}
\end{equation}

The function $q^c (k)$ in Eq. (\ref{sigs1rhoderiv}), such that $2\pi\rho_c (k) = \partial q^c (k)/\partial k$
and $q^c (0)=0$, which is the inverse function of $k^c (q)$, obeys a corresponding equation expanded in powers of $u^{-l'}$,
\begin{eqnarray}
q^c (k) & = & k - {(1-m_{\eta})\ln 2\over u}\sin k
\nonumber \\
& + & \sum_{l'=1}^{\infty}{\zeta (2l'+1)\over (2u)^{2l'+1}} \left(1-{1\over 2^{2l'}}\right)
\sum_{l''=0}^{2l'}{(-1)^{l'+l''}M_{l''}\over (2l'-l''+1)} {2l \choose l''}(\sin k)^{2l'-l''+1} \, .
\label{qckLuExp}
\end{eqnarray}

In contrast to ground-state particle-like symmetrical and compact $c$ band distributions, Eq. (\ref{NGS}), 
for which the coefficients $M_{l''}$, Eq. (\ref{Mlline}), vanish for $l''$ odd integers, in the case of the $\eta$-Bethe 
states considered here the contributions from such $l''$ odd integers is behind $2\pi\rho_c (k)$ not being 
a pure even function. This also implies that $q^c (k)$ is not a pure odd function. Note though that
$q^c (0) = 0$ and $q^c (\pm\pi) = \pm\pi$, as straightforwardly follows from analysis of Eq. (\ref{qckLuExp}).

Equations (\ref{rhoLuExp}) and (\ref{Mlline}) can be solved order by order in $u^{-1}$.
Accounting for the first sum rule in Eq. (\ref{Normrhosigma}), this gives 
for instance up to $u^{-3}$ order the following expansions for $2\pi\rho_c (k)$ and $q^c (k)$, 
\begin{eqnarray}
2\pi\rho_c (k) & = & 1 + {(1-m_{\eta})\ln 2\over u}\cos k
\nonumber \\
& - & {3\zeta (3)\over 32\,u^3}\left((1-m_{\eta})(1+2\sin^2 k)
+ {\tau\over \pi}\sum_{\iota=\pm} (\iota)\left(2\cos (q_{Fc,\tau}^{\iota})\sin k
- {1\over 4}\sin (2q_{Fc,\tau}^{\iota})\right)\right)\cos k \, ,
\label{rhocugg}
\end{eqnarray}
and 
\begin{eqnarray}
q^c (k) & = & k + {(1-m_{\eta})\ln 2\over u}\sin k
\nonumber \\
& - & {3\zeta (3)\over 32\,u^3}\left((1-m_{\eta})\left(1+{2\over 3}\sin^2 k\right)
+ {\tau\over\pi}\sum_{\iota=\pm} (\iota)\left(\cos (q_{Fc,\tau}^{\iota})\sin k
- {1\over 4}\sin (2q_{Fc,\tau}^{\iota})\right)\right)\sin k \, ,
\label{qckugg}
\end{eqnarray}
respectively.

Inversion of this $q^c (k)$ expansion gives up to $u^{-3}$ order the following expression for
the rapidity momentum functional $k^c (q_j)$,
\begin{eqnarray}
k^c (q_j)  & = & q_j - {(1-m_{\eta})\ln 2\over u}\sin q_j 
\nonumber \\
& + & {((1-m_{\eta})\ln 2)^2\over u^2}\cos q_j\sin q_j - {((1-m_{\eta})\ln 2)^3\over u^3}\left(1-{3\over 2}\sin^2 q_j\right)\sin q_j
\nonumber \\
& + & {3\zeta (3)\over 32\,u^3}\left((1-m_{\eta})\left(1+{2\over 3}\sin^2 q_j\right)
+ {\tau\over\pi}\sum_{\iota=\pm} (\iota)\left(\cos (q_{Fc,\tau}^{\iota})\sin q_j
- {1\over 4}\sin (2q_{Fc,\tau}^{\iota})\right)\right)\sin q_j \, .
\label{kqugg}
\end{eqnarray}

The use of the expansions, Eqs. (\ref{rhocugg}) and (\ref{kqugg}), in the current spectrum $J_c^h (q_j)$ 
expression, Eq. (\ref{J-partS1}), readily leads to the expansion up to $u^{-3}$ order of that spectrum
given in Eq. (\ref{Jqugg}).

In the case of the $m_s\rightarrow 1-m_{\eta}$ reference S$^z$SLN$_{SN}$ subspace 1A also considered here, 
one straightforwardly finds from the use of Eqs. 
(\ref{rho})-(\ref{Normrhosigma}) that for $m_{\eta}\in [0,1]$ and up to second order in $(1-m_{\eta}-m_s)\ll 1$ 
the $c$-band distribution $2\pi\rho_c (k)$ is for $m_{\eta}\in [0,1]$ given by,
\begin{equation}
2\pi\rho_c (k) = 1 + {(1-m_{\eta}-m_s)\over u}{\cos k\over 1 + \left({\sin k\over u}\right)^2} +
{\cal{O}} ((1-m_{\eta}-m_s)^3)\hspace{0.2cm}{\rm for}\hspace{0.2cm}
(1-m_{\eta}-m_s) \ll 1 \, .
\label{rhoLuExpmsmax}
\end{equation}
This result actually holds true as well for any $s1$-band distribution other than that considered 
in Eqs. (\ref{sigmas1}) and (\ref{Normrhosigma}).

That $2\pi\rho_c (k)$, Eq. (\ref{rhoLuExpmsmax}), is an even function of $k$ implies that 
the function $q^c (k)$ in Eq. (\ref{sigs1rhoderiv}) is an odd function of that variable. Hence it reads,
\begin{equation}
q^c (k) = k + (1-m_{\eta}-m_s)\arctan\left({\sin k\over u}\right) +
{\cal{O}} ((1-m_{\eta}-m_s)^3)\hspace{0.2cm}{\rm for}\hspace{0.2cm}
(1-m_{\eta}-m_s) \ll 1 \, .
\label{qckLuExpmsmax}
\end{equation}

Expanding $q^c (k)$ up to $u^{-3}$ order in $u^{-1}$ and second order in $(1-m_{\eta}-m_s)$
and inverting that expansion gives up to $u^{-3}$ order the following expression for
the rapidity momentum functional $k^c (q_j)$,
\begin{eqnarray}
k^c (q_j) & = & q_j - {(1-m_{\eta}-m_s)\over u}\sin q_j + {(1-m_{\eta}-m_s)^2\over u^2}\cos q_j\sin q_j
\nonumber \\
& + & {(1-m_{\eta}-m_s)\over 3u^3}\sin^3 q_j + {\cal{O}} ((1-m_{\eta}-m_s)^3) \, .
\label{kquggmsmax}
\end{eqnarray}

The use of the expansions, Eqs. (\ref{rhoLuExpmsmax}) and (\ref{kquggmsmax}), in the current spectrum $J_c^h (q_j)$ 
expression, Eq. (\ref{J-partS1}), leads to the expansion up to $u^{-3}$ order of that spectrum provided in Eq.
(\ref{Jquggmsmax}).

Finally, the largest charge current absolute value of reference S$^z$SLN$_{SN}$ subspaces 1A in
Eq. (\ref{JImaxSeta}) is derived in some limits of interest. For such subspaces the current deviation functional, 
Eq. (\ref{deltaJ-part}), simplifies to,
\begin{equation}
\delta\langle\hat{J}_{LWS} (l_{\rm r},S_{\eta},u)\rangle = 
\sum_{j=1}^{L}\,\delta N_c (q_j)\,\,j_c (q_j)  \, .
\label{deltaJ-partSN1}
\end{equation}
By combining the exact property that $\langle\hat{J}_{LWS} (l_{\rm r},S_{\eta},u)\rangle=\delta\langle\hat{J}_{LWS} (l_{\rm r},S_{\eta},u)\rangle$
for both $m_{\eta}\rightarrow 0$ and $m_{\eta}\rightarrow 1$ with the use of the $c$-band distribution that
maximizes the corresponding absolute value $\vert\langle\hat{J}_{LWS} (l_{\rm r},S_{\eta},u)\rangle\vert=
\vert\delta\langle\hat{J}_{LWS} (l_{\rm r},S_{\eta},u)\rangle\vert$ in the $u\rightarrow 0$ limit, one finds,
\begin{eqnarray}
\vert\langle\hat{J}_{LWS}^{\rm max}(l_{\rm r},S_{\eta},u)\rangle\vert 
& = & 4t\,L\,m_{\eta}\,(1-m_{\eta})\hspace{0.20cm}{\rm for}\hspace{0.20cm}m_{\eta}\rightarrow 0 \, ,
\hspace{0.20cm} m_s \in [0,1] \, ,
\hspace{0.20cm}{\rm and}\hspace{0.20cm}u\rightarrow 0 
\nonumber \\
& = & 2t\,L\,m_{\eta}\,(1-m_{\eta})\hspace{0.20cm}{\rm for}\hspace{0.20cm}m_{\eta}\in [0,1] \, ,
\hspace{0.20cm} m_s \rightarrow 1 - m_{\eta} \, ,
\hspace{0.20cm}{\rm and}\hspace{0.20cm}u\rightarrow 0 \, ,
\nonumber \\
& = & 2t\,L\,m_{\eta}\,(1-m_{\eta})\hspace{0.20cm}{\rm for}\hspace{0.20cm}m_{\eta}\rightarrow 1 \, ,
\hspace{0.20cm} m_s \rightarrow 0 \, ,
\hspace{0.20cm}{\rm and}\hspace{0.20cm}u\rightarrow 0 \, .
\label{JImaxSetaLimu0}
\end{eqnarray}

The use of the expansions up to $u^{-3}$ order of $J_c^h (q_j)$, Eqs. (\ref{Jqugg}) and (\ref{Jquggmsmax}),
in the general charge current expression, Eq. (\ref{J-partS1}), for $c$ and $s1$ bands compact distributions 
of general form, Eq. (\ref{NN}), belonging to reference S$^z$SLN$_{SN}$ subspaces 1A leads up to
$u^{-2}$ order to the $\langle\hat{J}_{LWS} (l_{\rm r},S_{\eta},u)\rangle$ expansion, Eq. (\ref{J-partS1-comp3}),
in the cases of $m_s=0$ and $m_s\rightarrow 1-m_{\eta}$ reference S$^z$SLN$_{SN}$ subspaces 1A, respectively. Up to $u^{-3}$ order
that leads to the expansions in Eqs. (\ref{J-only3ms0}) and (\ref{J-only3msmax}), respectively.

Moreover, the use of the specific limiting occupancy momenta $q_{Fc,\tau}^{\iota}$, Eq. (\ref{qFcpm1iota}),
that maximize the absolute values of the $\langle\hat{J}_{LWS} (l_{\rm r},S_{\eta},u)\rangle$ expansions
in Eqs. (\ref{J-partS1-comp3}), (\ref{J-only3ms0}), and (\ref{J-only3msmax}) leads for
general reference S$^z$SLN$_{SN}$ subspaces 1A up to $u^{-2}$ order to,
\begin{eqnarray}
\vert\langle\hat{J}_{LWS}^{\rm max}(l_{\rm r},S_{\eta},u)\rangle\vert 
& = & {2t\,L\,\sin (\pi m_{\eta})\over\pi}\left(1 - {7\over 2}\left({n_{\eta s}\over u}\right)^2
\left(1 - {8\over 7}\cos (\pi m_{\eta}) - {3\over 7}\sin^2 (\pi m_{\eta})\right)\right) + {\cal{O}} (u^{-4})
\nonumber \\
& & \hspace{0.20cm}{\rm for}\hspace{0.20cm}m_{\eta} \in \left[0,{1\over 2}-\delta_{\eta s}^u\right]
\hspace{0.20cm}{\rm and}\hspace{0.20cm}m_s \in [0,(1-m_{\eta})] 
\nonumber \\
& = & {2t\,L\,\sin^2 (\pi m_{\eta})\over\pi} + {t\,L\,n_{\eta s}\,\sin^2 (2\pi m_{\eta})\over\pi\,u}
\left(1+ {3n_{\eta s}\over 2u}\cos (2\pi m_{\eta})\right) + {\cal{O}} (u^{-3})
\nonumber \\
& & \hspace{0.20cm}{\rm for}\hspace{0.20cm}m_{\eta} \in \left[{1\over 2}-\delta_{\eta s}^u,{1\over 2}+\delta_{\eta s}^u\right]
\hspace{0.20cm}{\rm and}\hspace{0.20cm}m_s \in [0,(1-m_{\eta})]  
\nonumber \\
& = & {2t\,L\,\sin (\pi m_{\eta})\over\pi}\left(1 - {7\over 2}\left({n_{\eta s}\over u}\right)^2
\left(1 + {8\over 7}\cos (\pi m_{\eta}) - {3\over 7}\sin^2 (\pi m_{\eta})\right)\right) + {\cal{O}} (u^{-4})
\nonumber \\
& & \hspace{0.20cm}{\rm for}\hspace{0.20cm}m_{\eta} \in \left[{1\over 2}+\delta_{\eta s}^u,1\right]
\hspace{0.20cm}{\rm and}\hspace{0.20cm}m_s \in [0,(1-m_{\eta})]  \, .
\label{JImaxSetaLim}
\end{eqnarray}
Both for $m_{\eta} \in [0,1/2-\delta_{\eta s}^u]$ and $m_{\eta} \in [1/2+\delta_{\eta s}^u,1]$
the terms of orders $u^{-1}$, $u^{-3}$, and remaining odd orders $u^{-j}$ where $j=5,7,...$ 
of this maximum current absolute value expansion exactly vanish. 

In the case of the expansions of $u^{-3}$ order in Eqs. (\ref{J-only3ms0}) and (\ref{J-only3msmax}) 
specific to the $m_s=0$ and $m_s\rightarrow 1-m_{\eta}$ reference S$^z$SLN$_{SN}$ subspaces 1A,
respectively, this leads to,
\begin{eqnarray}
\vert\langle\hat{J}_{LWS}^{\rm max (3)}(l_{\rm r},S_{\eta},u)\rangle\vert
& = & 0 + {\cal{O}} (u^{-4})\hspace{0.20cm}{\rm for}\hspace{0.20cm}m_{\eta} \in \left[0,{1\over 2}-\delta_{\eta s}^u\right]
\hspace{0.20cm}{\rm and}\hspace{0.20cm}m_s = 0
\nonumber \\
& = & {2t\,L\,\sin^2 (2\pi m_{\eta})\over\pi\,u^3}\{((1-m_{\eta})\ln 2)^3
\left(1 - {4\over 3}\sin^2 (2\pi m_{\eta})\right)
\nonumber \\
& + & {3\zeta (3)\over 64}\left((1-m_{\eta})\left(1+{2\over 3}\sin^2 (2\pi m_{\eta})\right)
+ \left(1 - {1\over 2}\cos (2\pi m_{\eta})\right){\sin (2\pi m_{\eta})\over \pi}\right)\} 
\nonumber \\
& + & {\cal{O}} (u^{-4})\hspace{0.20cm}{\rm for}\hspace{0.20cm}m_{\eta} \in \left[{1\over 2}-\delta_{\eta s}^u,{1\over 2}+\delta_{\eta s}^u\right]
\hspace{0.20cm}{\rm and}\hspace{0.20cm}m_s = 0  
\nonumber \\
& = & 0 + {\cal{O}} (u^{-4})\hspace{0.20cm}{\rm for}\hspace{0.20cm}m_{\eta} \in \left[{1\over 2}+\delta_{\eta s}^u,1\right]
\hspace{0.20cm}{\rm and}\hspace{0.20cm}m_s = 0  \, ,
\label{JImaxSetaLim3}
\end{eqnarray}
and
\begin{eqnarray}
\vert\langle\hat{J}_{LWS}^{\rm max (3)}(l_{\rm r},S_{\eta},u)\rangle\vert 
& = & 0 + {\cal{O}} ((1-m_{\eta}-m_s)^3)\hspace{0.20cm}{\rm for}\hspace{0.20cm}m_{\eta} \in \left[0,{1\over 2}-\delta_{\eta s}^u\right]
\hspace{0.20cm}{\rm and}\hspace{0.20cm}m_s \rightarrow 1 - m_{\eta}
\nonumber \\
& = & - {t\,L\sin^4 (2\pi m_{\eta}) \over 3\pi u^3}\,(1-m_{\eta}-m_s) + {\cal{O}} ((1-m_{\eta}-m_s)^3)
\nonumber \\
& & \hspace{0.20cm}{\rm for}\hspace{0.20cm}m_{\eta} \in \left[{1\over 2}-\delta_{\eta s}^u,{1\over 2}+\delta_{\eta s}^u\right]
\hspace{0.20cm}{\rm and}\hspace{0.20cm}m_s \rightarrow 1 - m_{\eta} \, ,
\nonumber \\
& = & 0 + {\cal{O}} ((1-m_{\eta}-m_s)^3)\hspace{0.20cm}{\rm for}\hspace{0.20cm}m_{\eta} \in \left[{1\over 2}+\delta_{\eta s}^u,1\right]
\hspace{0.20cm}{\rm and}\hspace{0.20cm}m_s \rightarrow 1 - m_{\eta} \, ,
\label{JImaxSetaLimmsmax3}
\end{eqnarray}
respectively.

\section{Derivation of the $T\rightarrow\infty$ charge stiffness upper bound
and comparison to the Mazur's lower bound}
\label{Appendix8}

The first goal of this appendix is to confirm that the charge stiffness upper bound
in Eq. (\ref{UB-all-T-simp-INF2}) is larger than that given in Eq. (\ref{UB-all-T-simp-INF}).
At fixed density $m_s$, the charge currents $\langle\hat{J}_{LWS} (l_{\rm r},S_{\eta},u)\rangle = \sum_{j=1}^{L}\,N_c^h (q_j)\,\,J_c^h (q_j)$
in Eq. (\ref{J-partS1}) are the same for all 
${L/2-S_{\eta}+S_s\choose 2S_{s}}$ independent spin configurations with
the same spin $S_s = 0,1,...,(L-2S_{\eta})/2$. Those generate
the spin degrees of freedom of the $\eta$-Bethe states that span a S$^z$SLN$_{N_1}$ subspace as defined in Section \ref{UPTinf} 
and thus contribute to the charge stiffness upper bound, Eq. (\ref{UB-all-T-simp-INF}). Indeed, such currents only
depend on the density $m_s$ common to all such spin configurations through the dependence
on that density of the charge current spectrum $J_c^h (q_j)=-J_c (q_j)$ in Eq. (\ref{J-partS1}).

At fixed density $m_{\eta}$, the charge currents in Eq. (\ref{J-partS1}) directly depend on the occupancy
configurations of the $N_c = (L-2S_{\eta})=L\,(1-m_{\eta})$ charge $c$ pseudoparticles over the
available $j=1,...,L$ $c$-band discrete momentum values $q_j$ of which $N_c^h = 2S_{\eta} = L\,m_{\eta}$
are unoccupied. Such currents also depend on the density $m_{\eta}$ common to all such 
charge configurations through the dependence on that density of the charge current spectrum
$J_c^h (q_j)=-J_c (q_j)$ in Eq. (\ref{J-partS1}). The set of $j=1,...,L$ $c$-band discrete momentum values
$\{q_j\}$ are exactly the same for all $\eta$-Bethe states that span the S$^z$SLN$_{N_1}$ subspace 
under consideration.

Hence concerning the spin occupancy configurations and $c$-band occupancy configurations
that generate such states, only the latter determine the form of the charge currents in Eq. (\ref{J-partS1}).
The only effect of the spin degrees of freedom onto the charge currents is the $m_s$
dependence of the current spectrum $J_c^h (q_j)=-J_c (q_j)$ in Eq. (\ref{J-partS1}).

The S$^z$SLN$_{N_1}$ subspace considered here can be divided into references S$^z$SLN$_{SN}$ subspaces 1A,
each with a fixed spin belonging to the set $S_s = 0,1,...,(L-2S_{\eta})/2$. One can then choose the fixed-$S_s$ 
reference S$^z$SLN$_{SN}$ subspace 1A whose corresponding charge stiffness upper bound is larger
than that in Eq. (\ref{UB-all-T-simp-INF}), which is defined within the whole S$^z$SLN$_{N_1}$ subspace.
It follows from the above properties that such a fixed-$S_s$ subspace
is that for which the average $\vert {\bar{J}}_c\vert = {1\over L}\sum_{j=1}^L\vert J_c (q_j)\vert$, 
Eq. (\ref{averages}) of Appendix \ref{Appendix3} for $\beta =c$,
of the absolute value $\vert J_c (q_j)\vert$ of the charge current spectrum
$J_c^h (q_j)=-J_c (q_j)$ in Eq. (\ref{J-partS1}) is largest.
Such an average is independent of the $q_j$ occupancies and runs over all 
$j=1,...,L$ such momentum values. Within the TL, one
replaces the discrete momentum values $q_j$ such that $q_{j+1}-q_j=2\pi/L$
by a continuum momentum variable $q$. By replacing the sum $\sum_{j=1}^L$ 
by an integral it is found that up to $u^{-2}$ order such an average value reads,
\begin{equation}
\vert {\bar{J}}_c\vert = {4t\over\pi}\left(1 + 3\left({(1-m_{\eta}-m_s)\,g_s\over u}\right)^2\right) \, ,
\label{AveJqugg3}
\end{equation}
where $g_s = g_s (m_s)$ is the function in Eq. (\ref{Lms}).

Suitable analysis of this expression reveals that $\vert {\bar{J}}_c (q_j)\vert $ is largest for the $S_s=0$ 
reference S$^z$SLN$_{SN}$ subspace 1A. This follows from the inequality,
\begin{equation}
\ln 2 > {(1-m_{\eta}-m_s)\over (1-m_{\eta})}\,g_s \hspace{0.35cm}{\rm for}\hspace{0.35cm}
m_s \in [0,1-m_{\eta}] \hspace{0.35cm}{\rm where}\hspace{0.35cm}g_s \in [\ln 2,1] \, .
\label{Ineln2}
\end{equation}
One finds that the derivative $\partial g_s (m_s)/\partial m_s$ is such that
$\partial g_s (m_s)/\partial m_s\vert_{m_s=0} = 0$ at $m_s=0$ and
$\partial g_s (m_s)/\partial m_s\vert_{m_s} > 0$ for $m_s\in ]0,(1-m_{\eta})]$.
The use of such derivative behaviors confirms the validity of the inequality, Eq. (\ref{Ineln2}).

Although for simplicity here it was confirmed that the charge stiffness upper bound
in Eq. (\ref{UB-all-T-simp-INF2}) is larger than that given in Eq. (\ref{UB-all-T-simp-INF})
for the approximate $u>3/2$ range for which the $u^{-2}$ order expansions of these upper bounds
apply, the validity of the result can be shown to apply to the whole $u>0$ range.
That here its validity was confirmed up to $u^{-2}$ order stems from the charge stiffness
upper bound, Eq. (\ref{UB-all-T-simp-INF2}), being computed in the following to that order.

The second goal of this Appendix is to determine an exact expression for that charge stiffness
upper bound up to $u^{-2}$ order and in the TL for the hole concentration range
$m_{\eta}^z\in [0,1/2]$. We start by decomposing the current $J_c (q_j)=-J_c^h (q_j)$, Eq. (\ref{Jqugg1}), 
into a polynomial in $N_c=L-2S_{\eta}$,
\begin{equation}
J_c (q_j) = \sum_{\mu=0}^2 \alpha_\mu(q_j) N_c^\mu \, ,
\end{equation}
where
\begin{equation}
\alpha_0(q_j) = -2t \sin q_j,\quad
\alpha_1(q_j) = 2t \,\frac{\ln 2}{L \,u} \,\sin 2 q_j,\quad
\alpha_2(q_j) = -6t \,\frac{(\ln 2)^2}{L^2 \,u^2}\, \left(1- \frac{3}{2}\sin^2 q_j\right) \sin q_j \, .
\end{equation}

We use a simplified notation within which $N_c (q_j)=:\nu_j \in\{0,1\}$ are the binary occupation numbers and $N_c = \sum_j \nu_j = L - 2S_\eta$.
The main technical step is is the evaluation of the many-body sum (which is just the sum over $\sum_{l^*}$ in 
Eq. (\ref{UB-all-T-simp-INF2}) without the constant $1/(2S_\eta)^2$ prefactor),
\begin{equation}
I_{N_c} = \sum_{\{ \nu_j\}\in \{0,1\}^L} \left( \sum_{j=1}^L \nu_j J_j\right)^2 \delta_{N_c,\sum_j \nu_j} = 
\sum_{\mu,\mu'=0}^2 \phi_{\mu,\mu'}(N_c) \,N^{\mu+\mu'} \, ,
\end{equation}
where $J_j = J_c(q_j)$ and,
\begin{equation}
\phi_{\mu,\mu'}(N_c) =  \sum_{\{ \nu_j\}\in \{0,1\}^L}  \delta_{N_c,\sum_j \nu_j} \sum_{j,k=1}^L \alpha_\mu(q_j) \,\alpha_{\mu'}(q_k)\, \nu_j\, \nu_k \, .
\end{equation}
This object is in turn evaluated via writing its discrete Laplace transform,
\begin{eqnarray}
\tilde{\phi}_{\mu,\mu'}(\lambda) &=& \sum_{N_c=0}^L e^{\lambda N_c} \phi_{\mu,\mu'}(N_c) \\
&=& \sum_{\{\nu_j\}} \left( \prod_{n=1}^L e^{\lambda \nu_n}\right) \sum_{j,k} \alpha_\mu(q_j) \,\alpha_{\mu'}(q_k) \,  \nu_j \,\nu_k  \\
&=& (1+e^\lambda)^{L-2} e^{2\lambda}
\sum_{j,k}^{j \neq k}  \alpha_\mu(q_j) \,\alpha_{\mu'}(q_k)   + 
(1+e^\lambda)^{L-1} e^\lambda
\sum_j \alpha_\mu(q_j)\, \alpha_{\mu'}(q_j)  \\
&=& (1+e^\lambda)^{L-2} e^\lambda A_{\mu,\mu'} \, ,
\end{eqnarray}
where we accounted for that $\sum_j \alpha_\mu(q_j) = 0$ for all $\mu$ and defined the sums, or integrals,
\begin{equation}
A_{\mu,\nu} := \sum_j \alpha_\mu(q_j) \,\alpha_{\mu'}(q_j) \simeq \frac{L}{2\pi} \int_{-\pi}^\pi dq\,\alpha_\mu(q) \,\alpha_{\mu'}(q) \, ,
\end{equation}
which can be straightforwardly computed. From there we read,
\begin{equation}
\phi_{\mu,\mu'}(N_c) = {L-2\choose N_c-1} A_{\mu,\mu'}
\hspace{0.20cm}{\rm and}\hspace{0.20cm}\phi_{\mu,\mu'}(0) = \phi_{\mu,\mu'}(L-1) = 0 \, .
\end{equation}

Plugging all that to the charge stiffness upper bound, Eq. (\ref{UB-all-T-simp-INF2}), we find,
\begin{equation}
D^{\diamond\diamond} (T) =  {(2S_{\eta}^z)^2\over 2 L T}\,\,{\sum_{S_{\eta}=\vert S_{\eta}^z\vert}^{L/2} \sum_{\mu,\mu'=0}^2 \frac{1}{(2S_\eta)^2}{L-2 \choose L-2S_\eta-1} A_{\mu,\mu'}(L-2S_\eta)^{\mu+\mu'} \over \sum_{S_{\eta}=\vert S_{\eta}^z\vert}^{L/2}{L\choose 2S_{\eta}}} \, .
\label{ratio}
\end{equation}

Next we perform an asymptotic analysis of this expression, accounting for that $L\rightarrow\infty$ within the present TL. First we note that,
\begin{equation}
{L-2 \choose L-2S_\eta-1} \simeq \frac{2 S_\eta (L-2 S_\eta)}{L^2} {L \choose 2S_\eta} \, .
\end{equation}
Then we realize that both sums over the binomial symbols in the numerator and the denominator of the expression, Eq. (\ref{ratio}), 
become sharply peaked around $S_\eta = S^* = L/4$, under the condition that,
\begin{equation} 
|S^z_\eta| < \frac{L}{4} \, ,
\end{equation}
and thus $m_{\eta}^z \in [0,1/2]$.

Finally, the charge stiffness upper bound, Eq. (\ref{UB-all-T-simp-INF2}), then reads within the TL and
for $m_{\eta}^z \in [0,1/2]$,
\begin{equation}
D^{\diamond\diamond} (T) =  {(2S_{\eta}^z)^2\over 2 L T} \sum_{\mu,\mu'=0}^2 A_{\mu,\mu'} L^{-2} (L/2)^{\mu+\mu'}=
\left(\frac{2 S^z_\eta}{L}\right)^2 \frac{\pi^2 t^2}{T} \left(1 + \left(\frac{\ln 2}{2u}\right)^2 + {\cal O}(u^{-4})\right) \, ,
\end{equation}
which can indeed be written as given in Eq. (\ref{DdiamTmG}).

The charge stiffness Mazur's lower bound has been derived for $T\rightarrow\infty$ in Ref. \cite{ZNP-97}. 
In the zero-spin case considered in the upper-bound studies of this paper one finds that 
the charge stiffness Mazur's lower bound $D^{\rm Mz}(T)$ is such that,
\begin{equation}
D (t) \geq D^{\rm Mz}(T) = \frac{c_{\rm Mz}\,t^2}{2T} (m^z_{\eta})^2 
\hspace{0.2cm}{\rm where}\hspace{0.2cm}
c_{\rm Mz} = {2\,(1-(m^z_{\eta})^2)\over
(1+(m^z_{\eta})^2)\left(1+\left({1\over u}\right)^2{1\over 8(1+(m^z_{\eta})^2)}\right)} \, .
\label{DMazurTm}
\end{equation}

On the one hand, for $(1-m^z_{\eta}) \ll 1$ and up to ${\cal{O}}(u^{-2})$ order the upper bound, Eq. (\ref{DdiamTm1}), 
equals the charge stiffness. Hence one finds to such an order that in that limit for which 
$c_{\rm Mz} \approx 2(1-m^z_\eta)(1-(1/4u)^2)$ the use of the Mazur's lower bound leads to the inequality,
\begin{equation}
D (T) = \frac{2t^2}{2T} (1-m^z_\eta) \geq \frac{2t^2}{2T} (1-m^z_\eta)\left(1-\left({1\over 4u}\right)^2\right) 
\hspace{0.2cm}{\rm for}\hspace{0.2cm}(1-m^z_{\eta})\ll 1\, .
\label{ineqmeta1}
\end{equation}
For $(1-m^z_{\eta}) \ll 1$ the Mazur's lower bound thus equals the charge stiffness only in the $u\rightarrow\infty$ limit.
 
On the other hand, for $m^z_{\eta}\ll 1$ one finds up to ${\cal{O}}(u^{-2})$ order that 
$c_{\rm Mz} \approx 2(1- ((1/\sqrt{2})/2u)^2)$ and the charge stiffness is of the form $D (t) = \frac{c_u\,t^2}{2T} (m^z_{\eta})^2$ 
where the coefficient $c_u$ obeys the double inequality given in Eq. (\ref{DdoubleIN012}).

\end{document}